%% file: paper.tex
\documentclass[10pt,letterpaper]{article}
\pdfoutput=1

\usepackage[margin=1in]{geometry}
\usepackage[utf8]{inputenc}
\usepackage[T1]{fontenc}
\usepackage[nopatch=footnote]{microtype}
\usepackage{amsfonts}
\usepackage{amssymb}
\usepackage{amsmath}
\usepackage{amsthm}
\usepackage{etoolbox}
\usepackage{booktabs}
\usepackage[inline]{enumitem}
\usepackage{setspace}
\usepackage{titlesec}
\usepackage{needspace}
\usepackage[hidelinks,bookmarks,bookmarksopen,bookmarksnumbered]{hyperref}
\usepackage[capitalize]{cleveref}
\usepackage{mathtools}
\usepackage{tikz}
\usepackage[font=small,labelfont=bf]{caption}
\usepackage{thmtools}
\usepackage{wrapfig}
\usepackage{tablefootnote}
\usepackage{thm-restate}
\usepackage{soul}
\usepackage[draft]{fixme}
\usepackage{framed}
\usepackage{tikz}
\usepackage{subcaption}
\usepackage[norefs,nocites,nomsgs]{refcheck}
\usepackage{xspace}
\usepackage[most]{tcolorbox}
\usepackage{xcolor}

\usetikzlibrary{trees}
\usetikzlibrary{arrows}
\usetikzlibrary{arrows.meta}
\usetikzlibrary{decorations.pathreplacing}
\usetikzlibrary{positioning}
\usetikzlibrary{calc}

\input{settings}

\input{macros}

\begin{document}

\newgeometry{top=0.7in,bottom=0.7in,left=1in,right=1in}

\title{Cell-Probe Lower Bounds and Complexity-Preserving Reductions\\ for Suffix Array Queries}

\author{
  \large Dominik Kempa\thanks{Partially funded by the
  NSF CAREER Award 2337891.}\\[-0.3ex]
  \normalsize Department of Computer Science,\\[-0.3ex]
  \normalsize Stony Brook University,\\[-0.3ex]
  \normalsize Stony Brook, NY, USA\\[-0.3ex]
  \normalsize \texttt{kempa@cs.stonybrook.edu}
  \and
  \large Tomasz Kociumaka\\[-0.3ex]
  \normalsize Max Planck Institute for Informatics,\\[-0.3ex]
  \normalsize Saarland Informatics Campus,\\[-0.3ex]
  \normalsize Saarbrücken, Germany\\[-0.3ex]
  \normalsize \texttt{tomasz.kociumaka@mpi-inf.mpg.de}
}

\hypersetup{
  pdftitle = {Cell-Probe Lower Bounds and Complexity-Preserving Reductions
    for Suffix Array Queries},
  pdfauthor = {Dominik Kempa, Tomasz Kociumaka}
}
\date{\vspace{-1.0cm}}
\hypersetup{pageanchor=false}
\maketitle

\input{abstract}

\thispagestyle{empty}
\restoregeometry
\hypersetup{pageanchor=true}
\setcounter{page}{1}
\input{intro}

\input{prelim}
\input{overview}

\input{tools}

\input{space-lower-bounds}

\input{prefix-select}

\input{prefix-special-rank}

\input{acknowledgements}

\bibliographystyle{alphaurl}
\bibliography{paper}

\input{appendix}

\end{document}

%% file: macros.tex
\newcommand{\bigO}{\mathcal{O}}
\newcommand{\Oh}{\bigO}

\newcommand{\ceil}[1]{\left\lceil #1 \right\rceil}
\newcommand{\floor}[1]{\left\lfloor #1 \right\rfloor}
\newcommand{\dd}{\mathinner{.\,.}}
\newcommand{\probname}[1]{\text{\sc #1}}

\newcommand{\Pat}{P}
\newcommand{\Text}{T}
\newcommand{\Textinf}{\Text^{\infty}}
\newcommand{\Textlen}{n}
\newcommand{\Seqlen}{m}
\newcommand{\AlphabetSize}{\sigma}
\newcommand{\IntegerAlphabet}{[0 \dd \AlphabetSize)}
\newcommand{\BinaryAlphabet}{\{{\tt 0}, {\tt 1}\}}
\newcommand{\emptystring}{\varepsilon}
\newcommand{\deltatext}{\delta_{\rm text}}
\newcommand{\SSS}{\mathsf{S}}

\newcommand{\Z}{\mathbb{Z}}
\newcommand{\Zz}{\Z_{\ge 0}}
\newcommand{\Zn}{\Zz}

\newcommand{\SA}[1]{\mathrm{SA}_{#1}}
\newcommand{\ISA}[1]{\mathrm{SA}^{-1}_{#1}}

\newcommand{\Successor}[2]{\mathrm{succ}_{#1}(#2)}

\newcommand{\LCE}[3]{\mathrm{LCE}_{#1}(#2,#3)}
\newcommand{\lcp}[2]{\mathrm{lcp}(#1,#2)}
\newcommand{\per}[1]{\mathrm{per}(#1)}
\newcommand{\revstr}[1]{\overline{#1}}
\newcommand{\increase}[2]{\mathrm{increase}_{#1}(#2)}

\newcommand{\Val}[2]{\mathrm{val}_{#1}(#2)}
\newcommand{\AlphabetMap}[3]{\mathrm{map}_{#1,#2}(#3)}
\newcommand{\ExtAlphabetMap}[3]{\mathrm{extmap}_{#1,#2}(#3)}

\newcommand{\PackedRepresentation}[3]{\mathrm{packed}_{#1,#2}(#3)}
\newcommand{\PackedSeqRepresentation}[3]{\PackedRepresentation{#1}{#2}{#3}}
\newcommand{\SeqToString}[2]{\mathrm{seq\mbox{-}to\mbox{-}str}_{#1}(#2)}
\newcommand{\Row}[2]{\mathrm{row}_{#1}(#2)}
\newcommand{\Col}[2]{\mathrm{col}_{#1}(#2)}

\newcommand{\OccTwo}[2]{\mathrm{Occ}(#1, #2)}
\newcommand{\RangeBegTwo}[2]{\mathrm{RangeBeg}(#1, #2)}
\newcommand{\RangeEndTwo}[2]{\mathrm{RangeEnd}(#1, #2)}

\newcommand{\DistPrefixPos}[4]{\mathrm{DistPrefix}(#1,#2,#3,#4)}
\newcommand{\DistPrefixes}[3]{\mathcal{D}(#1, #2, #3)}

\newcommand{\RName}{\mathsf{R}}
\newcommand{\RTwo}[2]{\RName(#1, #2)}
\newcommand{\RThree}[3]{\RName_{#1}(#2, #3)}
\newcommand{\RFour}[4]{\RName_{#1,#2}(#3, #4)}
\newcommand{\RFive}[5]{\RName_{#1,#2,#3}(#4, #5)}
\newcommand{\RMinusTwo}[2]{\RName^{-}(#1, #2)}
\newcommand{\RPlusTwo}[2]{\RName^{+}(#1, #2)}
\newcommand{\RSignedFour}[5]{\RName^{#1}_{#2,#3}(#4, #5)}
\newcommand{\RSignedFive}[6]{\RName^{#1}_{#2,#3,#4}(#5, #6)}
\newcommand{\RPrimName}{\RName'}
\newcommand{\RPrimTwo}[2]{\RPrimName(#1, #2)}
\newcommand{\RPrimSignedTwo}[3]{\RPrimName^{#1}(#2, #3)}
\newcommand{\RPrimSignedThree}[4]{\RPrimName^{#1}_{#2}(#3, #4)}

\newcommand{\RootPos}[3]{\mathrm{root}(#1,#2,#3)}
\newcommand{\HeadPos}[3]{\mathrm{head}(#1,#2,#3)}
\newcommand{\ExpPos}[3]{\mathrm{exp}(#1,#2,#3)}
\newcommand{\TailPos}[3]{\mathrm{tail}(#1,#2,#3)}
\newcommand{\TypePos}[3]{\mathrm{type}(#1,#2,#3)}
\newcommand{\RunEndPos}[3]{e(#1,#2,#3)}
\newcommand{\RunEndFullPos}[3]{e^{\rm full}(#1,#2,#3)}

\newcommand{\Rank}[3]{\mathsf{rank}_{#1}(#2,#3)}
\newcommand{\SpecialRank}[2]{\mathsf{special\mbox{-}rank}_{#1}(#2)}
\newcommand{\Select}[3]{\mathsf{select}_{#1}(#2,#3)}

\newcommand{\PrefixRank}[3]{\mathsf{prefix\mbox{-}rank}_{#1}(#2,#3)}
\newcommand{\PrefixSpecialRank}[3]{\mathsf{prefix\mbox{-}special\mbox{-}rank}_{#1}(#2,#3)}
\newcommand{\PrefixSelect}[3]{\mathsf{prefix\mbox{-}select}_{#1}(#2,#3)}

\newcommand{\TwoSidedRangeCount}[3]{\mathsf{range\mbox{-}count}_{#1}(#2, #3)}
\newcommand{\RangeSelect}[3]{\mathsf{range\mbox{-}select}_{#1}(#2, #3)}

\newcommand{\LexSorted}[2]{\mathrm{LexSorted}(#1,#2)}
\newcommand{\RunsLexSorted}[3]{\mathrm{RunsLexSorted}^{#1}(#2,#3)}

\newcommand{\zero}{{\tt 0}}
\newcommand{\one}{{\tt 1}}
\newcommand{\two}{{\tt 2}}
\newcommand{\three}{{\tt 3}}

%% file: abstract.tex
\begin{abstract}
  For a text $\Text$ of length $\Textlen$ over alphabet $[0 \dd
  \AlphabetSize)$, its suffix array $\SA{\Text}[1 \dd \Textlen]$
  stores the starting positions of all suffixes of $\Text$ in
  lexicographic order. The suffix array is one of the central data
  structures in string processing and underlies dozens of applications
  ranging from text search and data compression to finding repeats and
  common/unique substrings. Since the breakthrough works of Ferragina
  and Manzini (FOCS 2000) and of Grossi and Vitter (STOC 2000), who
  developed the \emph{FM-Index} and \emph{Compressed Suffix Array
  (CSA)}, respectively, it has been known that suffix array (and
  inverse suffix array) queries---given $i \in [1 \dd \Textlen]$,
  return $\SA{\Text}[i]$ (resp.\ $\SA{\Text}^{-1}[i]$)---can be
  supported in $\bigO(\log^{\epsilon}_\AlphabetSize \Textlen)$ time
  using $\bigO(\Textlen \log \AlphabetSize)$ bits for any constant
  $\epsilon>0$. Very recently, Thankachan (FOCS 2026) also showed how
  to support inverse suffix array queries in
  $\bigO(\log\log\Textlen/\log\log\AlphabetSize)$ time using
  $\bigO(\Textlen \log \AlphabetSize)$ bits of space. However, despite these advances on the upper-bound side and
  despite two decades of intense work on suffix arrays, the
  lower-bound side has remained far less developed. In particular,
  the question asked by Grossi and Vitter in their original CSA paper
  (STOC 2000)---whether there exists an $\bigO(\Textlen \log
  \AlphabetSize)$-bit data structure that answers suffix-array queries
  in $\bigO(1)$ time---remains unanswered.

  In this work, we address the above question and, more broadly, ask
  what governs the time-space tradeoffs for suffix array and inverse
  suffix array queries. We present two contributions:

  \begin{itemize}
  \item First, we prove that, already for the binary alphabet (i.e.,
    when $\AlphabetSize = 2$), every deterministic data structure using
    $\bigO(\Textlen (\log\log\Textlen)^{\bigO(1)})$ bits and answering
    suffix-array queries has worst-case query time
    \vspace{-0.1ex}
    \[
      \Omega\left(\frac{\log\log \Textlen}{\log\log\log \Textlen}\right).
    \]
    \vspace{-0.6ex}

    \noindent
    In particular, when $\AlphabetSize = 2$, there is
    no deterministic data structure using $\bigO(\Textlen)$ bits that
    answers suffix-array queries in $\bigO(1)$ worst-case time. This is the first nontrivial
    space-time lower bound for suffix array queries in compact space, and it answers
    the 25-year-old question of Grossi and Vitter. We prove this result by
    developing a more general lower bound on the achievable time-space
    tradeoffs, which provides a more detailed answer to Grossi and
    Vitter's question. In particular, achieving $\bigO(1)$ worst-case
    query time requires
    $\Omega(\Textlen \log^{\epsilon} \Textlen)$ bits
    for some constant $\epsilon>0$, which matches the known upper bounds.

  \vspace{1ex}
  \item Our second result addresses the understanding of the suffix
    array time-space tradeoffs on the \emph{upper bound} side. We
    prove an exact characterization of suffix array and inverse suffix
    array queries in terms of simpler \emph{prefix range
    queries}. These types of queries have recently been shown to
    capture suffix array and inverse suffix-array queries up to an
    additive $\bigO(\log\log \Textlen)$ term in the query time when
    $\AlphabetSize = 2$. We improve these reductions, and prove that
    indexing for suffix array queries is in fact \emph{perfectly}
    equivalent to indexing for prefix select queries in four
    fundamental measures: (1) space, (2) query time, (3) preprocessing
    time, and (4) preprocessing space. Likewise, indexing for inverse
    suffix array queries is fully equivalent to indexing for prefix
    special rank queries. We also generalize the equivalences to hold
    uniformly for every alphabet size $\AlphabetSize$ satisfying~$2
    \leq \AlphabetSize \leq \Textlen$. These equivalences simplify
    the problem of designing future data structures answering SA and
    inverse SA queries, and they leave a clean bit-level problem,
    eliminating string-specific intricacies (such as periodicity).

  \end{itemize}
  Taken together, our results identify a clean core problem underlying
  suffix-array access. They give the first nontrivial cell-probe
  space-time lower bounds for suffix-array queries in compact space, and they show that
  future upper and lower bounds for SA and inverse-SA access can be
  studied through the corresponding prefix-range primitives, with no
  loss in the standard efficiency measures.
\end{abstract}

%% file: intro.tex
\section{Introduction}\label{sec:intro}

\begin{wrapfigure}{R}{0.275\textwidth}
  \vspace{-.75cm}
  \small
  \setlength{\FrameSep}{1.2ex}
  \begin{framed}\centering
  \begin{tikzpicture}[yscale=0.35]
    \foreach \x [count=\i] in {a, aabba, abaabba, abba, abbabaabba,
        ba, baabba, babaabba, bba, bbabaabba}
      \draw (2.0, -0.92*\i) node[right]
        {$\texttt{\x}$};
    \draw(2.0,0) node[right] {\scriptsize $\Text[\SA{\Text}[i]\dd \Textlen]$};
    \foreach \x [count=\i] in {a, a, a, a, a, a, a, a, a, a}
      \draw (0.8, -0.92*\i) node {\footnotesize $\i$};
    \draw(0.8,0) node{\scriptsize $i$};
    \foreach \x [count=\i] in {10,6,4,7,1,9,5,3,8,2}
      \draw (1.5, -0.92*\i) node {$\x\vphantom{\textbf{\underline{7}}}$};
    \draw(1.5,0) node{\scriptsize $\SA{\Text}[i]$};
  \end{tikzpicture}
  \vspace{-1.5ex}
  \caption{\small A list of sorted suffixes of $\Text=
    \texttt{abbabaabba}$ along with
    the suffix array.}\label{fig:sa-example}
  \end{framed}
  \vspace{-0.6cm}
\end{wrapfigure}

For a text $\Text$ of length $\Textlen$, the \emph{suffix array}
$\SA{\Text}[1 \dd \Textlen]$ is the permutation of integers in
$\{1,\ldots,\Textlen\}$ that orders the suffixes of $\Text$
lexicographically, that is, $\Text[\SA{\Text}[1] \dd \Textlen] \prec
\Text[\SA{\Text}[2] \dd \Textlen] \prec \cdots \prec
\Text[\SA{\Text}[\Textlen] \dd \Textlen]$~\cite{ManberM90}; see
\cref{fig:sa-example}. The canonical application of a suffix array is
text indexing: for any pattern $\Pat$ of length $m>0$, all suffixes of
$\Text$ having $\Pat$ as a prefix form a contiguous block
$\SA{\Text}[b \dd e)$. All occurrences of $\Pat$ can be thus found by
locating the endpoints of this block using binary search in $\bigO(m
\log \Textlen)$ time. The block length $e-b$ is then the number of
occurrences, and $\SA{\Text}[b \dd e)$ gives the starting positions of
all occurrences~of~$\Pat$~in~$\Text$~\cite{ManberM90}.

The suffix array is widely considered to be one of the most important
data structures in text processing, data compression, and
bioinformatics. For over three and a half decades, its simplicity,
strong theoretical bounds, and space efficiency have made it a
centerpiece~in~numerous classes of problems:

\begin{itemize}
\item \emph{Search:} Suffix arrays and their variants underlie exact pattern
  matching~\cite{ManberM90}, approximate matching~\cite{Navarro01},
  wildcard matching~\cite{ManberB91}, gapped
  matching~\cite{CaceresPZ20}, parameterized
  matching~\cite{FujisatoNIBT19}, regular-expression
  search~\cite{baeza1999modern,ArroyueloBCMN14}, and compressed
  text indexing~\cite{Gagie2020}.

\item \emph{Analysis:} Combined with standard augmentations such as
  LCP array~\cite{AbouelhodaKO04}, suffix arrays can compute the
  longest common substring~\cite{gusfield,Charalampopoulos21}, longest
  repeated substring~\cite{gusfield}, maximal repeats~\cite{gusfield},
  tandem repeats~\cite{gusfield2004linear}, maximal exact
  matches~\cite{gusfield}, maximal unique
  matches~\cite{kurtz2004versatile}, minimal unique
  substrings~\cite{IlieS11}, minimal~absent~words~\cite{BartonHMP14}.

\item \emph{Compression:} Suffix arrays are a central component in
  numerous data compression algorithms, including the Burrows--Wheeler
  transform~\cite{bwt,OkanoharaS09,sss}, Lempel--Ziv (LZ77)
  factorization~\cite{CrochemoreI08,OhlebuschG11,KempaP13,GotoB13,sublinearlz}.
  They also underlie many recent powerful compressed text indexes such
  as~\cite{Gagie2020,collapsing}. Through these connections, they
  shape the complexity of a large body of work on compressed indexing
  and data compression~\cite{NavarroIndexes,NavarroMeasures,bwtbook}.

\item \emph{Generalizations:} Suffix arrays have been generalized in
  many different ways, including enhanced suffix
  arrays~\cite{AbouelhodaKO04}, sparse suffix arrays~\cite{Prezza18},
  and dynamic suffix arrays~\cite{dynsa}. Techniques based on suffix
  arrays and the Burrows--Wheeler transform have also been applied to
  weighted strings~\cite{Charalampopoulos20}, circular
  strings~\cite{HonLST11}, labeled trees~\cite{FerraginaLMM05},
  alignments~\cite{NaPLHLMP13}, and temporal graphs~\cite{BrisaboaCFR18}.
\end{itemize}
Due to this wealth of applications, understanding the complexity of
suffix-array queries is one of the most consequential problems in
string
processing:
any \emph{improvement} or \emph{limitation} at this level propagates
to many fundamental problems.
In this paper, we focus on the most basic operation supported by space-efficient suffix array
representations: given an index $i$, return the $i$th entry $\SA{\Text}[i]$
of the suffix array.

In its basic form, a suffix array $\SA{\Text}[1 \dd \Textlen]$ is
simply a permutation using $\Textlen \lceil \log \Textlen \rceil$ bits
supporting the computation of $\SA{\Text}[i]$ in $\bigO(1)$ time. One
of the most profound results in string processing is due to Ferragina
and Manzini~\cite{FerraginaM00,FerraginaM05} and, independently, Grossi and
Vitter~\cite{GrossiV00,GrossiV05}, who, at the turn of the millennium, proposed
suffix array representations that, for any string $\Text \in
\IntegerAlphabet^{\Textlen}$ over an integer alphabet of size
$\AlphabetSize$, use only $\bigO(\Textlen \log \AlphabetSize)$ bits
(which is asymptotically precisely the space needed to store the text
itself) and support suffix array (SA) and inverse suffix array
(SA$^{-1}$) queries in only $\bigO(\log_\AlphabetSize^{\epsilon}
\Textlen)$ time (where $\epsilon > 0$ is any constant). Known as the
\emph{FM-Index}~\cite{FerraginaM00} and the \emph{Compressed Suffix
Array (CSA)}~\cite{GrossiV00}, these structures transformed the
field of string processing by enabling complex string processing to be
executed in $\bigO(\Textlen \log \AlphabetSize)$ bits of
space~\cite{bwtbook,Kingsford26,MBCT2023}.

Recent progress has further improved the bounds for inverse suffix array
queries. Thankachan~\cite[Theorem~1.2]{ISA26} showed that these queries can be supported in
$\bigO(\log\log\Textlen/\log\log\AlphabetSize)$ time using
$\bigO(\Textlen \log \AlphabetSize)$ bits of space.\footnote{In subsequent
work~\cite{SaTradeoffs}, we further improve the query time to $\bigO(1)$
while retaining $\bigO(\Textlen\log\AlphabetSize)$ bits of space.
We discuss this application of our equivalence near the end of the
introduction.}

For suffix array queries, whether the optimal space of
$\bigO(\Textlen \log \AlphabetSize)$ bits could be combined with
$\bigO(1)$ query time was a longstanding open problem.
Already for $\AlphabetSize = 2$, Grossi and Vitter asked in their
2000 paper~\cite{GrossiV00}:
  \emph{``An interesting open problem is to improve upon our $\bigO(n)$-bit
compressed suffix array so that each call to lookup takes constant time.''}

Before this work, no nontrivial lower bound on suffix array query time in
$\bigO(\Textlen \log \AlphabetSize)$ bits of space was known, and the
minimum space needed to achieve $\bigO(1)$ query time was also unknown.
This motivates our first question:

\begin{center}
\begin{tcolorbox}[
  width=0.65\textwidth,
  colback=black!2,
  colframe=black!100,
  boxrule=0.5pt,
  arc=0mm
]
\centering
  \textbf{Question 1:}
  \emph{Are there any nontrivial lower bounds\\ for the time-space tradeoff
  for suffix array queries?}
\end{tcolorbox}
\end{center}

The lack of space-time lower bounds for suffix-array queries in compact space makes it
unclear what upper-bound tradeoffs should be considered the right
target. At the same time, existing suffix-array indexes are based on
rather different primitives: the FM-index relies on rank queries and
the Burrows--Wheeler transform~\cite{bwt}, whereas compressed suffix
arrays rely on Elias--Fano-type encodings and queries on the $\Psi$ function.

Recently, a new approach has emerged in the study of suffix
arrays. In~\cite{breaking}, a structure with $\bigO(\log^\epsilon\Textlen)$ query time
(for any constant $\epsilon>0$) was developed, but also equipped with a
fast construction time of $\bigO((\Textlen \log \AlphabetSize) /
\sqrt{\log \Textlen})$ (which is, for example, $\bigO(\Textlen / \sqrt{\log
\Textlen}) = o(\Textlen)$ for $\AlphabetSize = 2$) while using
$\bigO(\Textlen \log \AlphabetSize)$ bits of space. The key idea
in~\cite{breaking} was to reduce SA and inverse SA queries to a new
type of query called \emph{prefix rank and select queries}, incurring an additive overhead of $\bigO(\log \log \Textlen)$ in query time.
In~\cite{PrefixEquiv}, it was then proved that, up to the same overhead, these reductions
can be reversed for prefix select queries and for a restricted form of prefix rank queries,
which yields equivalences with SA and ISA queries, respectively. Still, the
possibility of a tighter reduction remained, since the reductions
in~\cite{breaking,PrefixEquiv} introduce an extra $\bigO(\log \log \Textlen)$
term in query time. Furthermore, the partial equivalences were developed
only for the case of a binary alphabet $\AlphabetSize = 2$.
This left open the existence of other---possibly more
precise---equivalences.
The power of such reductions is that the corresponding non-string problem
eliminates string-specific intricacies, such as periodicity analysis and
heavy text-indexing machinery, leaving a cleaner ``bit-level'' problem.
This can make both the design of upper-bound tradeoffs and the transfer
of lower-bound tradeoffs substantially more transparent.
We thus pose our second question as follows:

\begin{center}
\begin{tcolorbox}[
  width=0.95\textwidth,
  colback=black!2,
  colframe=black!100,
  boxrule=0.5pt,
  arc=0mm
]
\centering
\textbf{Question 2:} \emph{What governs the query time, space usage,
  and construction time/space\\ of suffix array and inverse suffix array queries?}
\end{tcolorbox}
\end{center}
\vspace{-2ex}

\myparagraph{Our Results}
In this paper, we make significant progress on both questions.
Our first main result
is the first nontrivial lower bound on the query time of data
structures for suffix array queries in compact space, answering Question 1.
We prove the following general lower bound:

\begin{theorem}[\cref{th:lb-suffix-array} simplified]\label{th:lb-simple}
  In the cell probe model with $\Theta(\log N)$-bit words, every deterministic
  $S$-bit data structure with $S\geq N$ that answers suffix array queries for binary strings of
  length at most $N$ has worst-case query time
  \vspace{1ex}
  \[
    \Omega\left(
      \frac{\log \log N}{\log((S/N)\log \log N)}
    \right).
  \]
\end{theorem}

In particular, indexes for binary texts using
$\Oh(N (\log \log N)^{\bigO(1)})$ bits, including compact indexes, must have worst-case query time
$\Omega(\log \log N / \log \log \log N)$.
Conversely, achieving $\Oh(1)$ worst-case query time requires
$\Omega(N \log^\epsilon N)$ bits for some constant $\epsilon > 0$.
For $S \ge N \log^{1+\delta}\log N$ with any constant $\delta>0$, our bound matches the
trade-off of~\cite[Theorem~5]{Rao02}.
To derive \cref{th:lb-simple}, we interpret
\emph{ball inheritance}~\cite{ChanLP11} as a special case of prefix select and transfer the lower bound of Gr{\o}nlund and Larsen~\cite{GronlundL16} to
SA queries via
our prefix-select-to-suffix-array construction, which uses a constant number
of suffix-array structures, $\bigO(N)$ additional bits, and
$\bigO(1)$ additional probes per query in the cell-probe model
(see \cref{th:lb-suffix-array}).
The same lower bound in terms of total space $S$ holds for every
$2\leq\AlphabetSize\leq N$ by a simple relabeling and padding argument.

Our second main result proves
that suffix array queries are perfectly equivalent to
\emph{prefix select} queries, and inverse suffix array queries to
\emph{prefix special rank} queries.
Our equivalences hold for every alphabet size and preserve,
up to constant factors, all standard measures:
\begin{enumerate*}[label=(\roman*)]
\item space usage,
\item query time,
\item preprocessing time, and
\item preprocessing~space.
\end{enumerate*}
This improves previous reductions~\cite{breaking,PrefixEquiv} in three ways:
\begin{enumerate}
\item We eliminate the extra $\bigO(\log \log \Textlen)$ term in query time. To
  this end, we optimally solve the problem of range counting queries on
  bounded-sum instances (\cref{th:range-count-for-small-sum}),
  which was responsible for the predecessor-search bottleneck in the reductions.
\item We
  generalize the equivalences to hold uniformly for every
  alphabet size $\AlphabetSize$ with $2 \leq \AlphabetSize \leq \Textlen$.
  In contrast to previous work, which provided
  partial equivalences only for binary alphabets ($\AlphabetSize = 2$), we
  directly relate the full families of suffix-array and
  prefix-range-query problems over arbitrary alphabets.
\item We prove
  space lower bounds of $\Omega(\Textlen \log \AlphabetSize)$ bits for
  all four problems (suffix array, inverse suffix array,
  prefix select, and prefix special rank queries).
  As a result, we eliminate $\Theta(\Textlen \log \AlphabetSize)$ additive space overheads that would otherwise appear when extending our reductions to super-constant alphabets.
\end{enumerate}

Let us first focus on the equivalence of suffix array and prefix select queries.
The two equivalent problems expect different types of input: one takes a single
string and the other a sequence of short strings.
To unify both problems,
we introduce a
parameter $N$, defined as an upper bound on the total number of symbols
from $[0 \dd \AlphabetSize)$ in the input. With this parameter, we can uniformly
express the complexities of both problems as functions of
$\AlphabetSize$ and $N$. Thus, let us fix parameters $\AlphabetSize,
N \in \Z_{\geq 2}$ and assume $\AlphabetSize \leq N$. Let us also
consider a word RAM model with word size $w$ satisfying
$1+\log N \leq w = \bigO(\log N)$.\footnote{As usual in the word
  RAM model, we assume $w = \Theta(\log N)$. The explicit lower
  bound $w\geq1+\log N$ ensures that the positions, lengths, and
  counters used by our constructions fit in one word.}
The two problems under consideration can then be defined as follows:

\vspace{1ex}
\begin{bfdescription}
\item[\probname{Indexing for Suffix Array Queries over Alphabet
  $[0 \dd \AlphabetSize)$:}] The input consists of the length $\Textlen$ and the packed representation
  $\PackedRepresentation{w}{\AlphabetSize}{\Text}$\footnote{A \emph{packed
    representation} is a standard word-RAM encoding of strings over
    an integer alphabet: for a string $S \in [0 \dd \AlphabetSize)^{*}$,
    each symbol $S[i]$ is written using $\lceil \log \AlphabetSize \rceil$
    bits, and consecutive symbols are stored contiguously in machine words.
    Consequently, $S$ occupies $\bigO(1 + (|S|\log\AlphabetSize)/w)$ words
    of space; see
    \cref{def:packed-representation}.}
  of a string $\Text \in
  \IntegerAlphabet^{\Textlen}$ such that $\AlphabetSize \leq \Textlen \leq N$.
  Given any $i \in [1 \dd \Textlen]$, the query returns the value
  $\SA{\Text}[i]$, i.e., the starting position of the
  lexicographically $i$th smallest suffix of $\Text$ (see
  \cref{def:suffix-array}).
\end{bfdescription}

\begin{bfdescription}
\item[\probname{Indexing for Prefix Select Queries over Alphabet
  $[0 \dd \AlphabetSize)$:}] The input consists of the integer $m$ and the packed sequence representation
  $\PackedSeqRepresentation{w}{\AlphabetSize}{W}$
  (\cref{def:packed-sequence-representation}) of a
  sequence $W[1 \dd m]$ of $m \geq \AlphabetSize$ nonempty strings of
  common length $\ell = \lfloor \log_{\AlphabetSize} m \rfloor$ such
  that $m \cdot \ell \leq N$. Given any $r \in [1 \dd m]$, the
  length $|X|$, and the packed representation
  $\PackedRepresentation{w}{\AlphabetSize}{X}$ of any string
  $X \in [0 \dd \AlphabetSize)^{\leq \ell}$, the query returns
  $\PrefixSelect{W}{r}{X}$ (\cref{def:prefix-rank-and-select}),
  i.e., the $r$th smallest element of the set $\{j \in [1 \dd m] :
  X\text{ is a prefix of }W[j]\}$ (if the set has at least $r$
  elements) or $\infty$ (otherwise); see \cref{fig:prefix-select-example-small} for an example.
\end{bfdescription}

\begin{wrapfigure}{r}{0.30\textwidth}
  \vspace{-0.05cm}
  \normalsize
  \setlength{\FrameSep}{1.1ex}
  \begin{framed}\centering
  \begin{tikzpicture}[yscale=0.32]
    \fill[blue!10] (0.22,7.2) rectangle (0.75,8.3);
    \fill[blue!10] (0.22,2.8) rectangle (0.75,3.9);
    \fill[blue!10] (0.22,0.6) rectangle (0.75,1.7);
    \draw (0, 8.8) node {\normalsize $W[1] = \texttt{0\,1\,0}$};
    \draw (0, 7.7) node {\normalsize $W[2] = \texttt{1\,0\,1}$};
    \draw (0, 6.6) node {\normalsize $W[3] = \texttt{0\,0\,1}$};
    \draw (0, 5.5) node {\normalsize $W[4] = \texttt{1\,1\,0}$};
    \draw (0, 4.4) node {\normalsize $W[5] = \texttt{1\,1\,1}$};
    \draw (0, 3.3) node {\normalsize $W[6] = \texttt{1\,0\,0}$};
    \draw (0, 2.2) node {\normalsize $W[7] = \texttt{0\,1\,1}$};
    \draw (0, 1.1) node {\normalsize $W[8] = \texttt{1\,0\,1}$};
    \draw (-2.0, -1) node[right] {\normalsize $\PrefixSelect{W}{2}{\texttt{10}} = 6$};
    \draw (-2.0, -2.2) node[right] {\normalsize $\PrefixSelect{W}{3}{\texttt{10}} = 8$};
    \draw (-2.0, -3.4) node[right] {\normalsize $\PrefixSelect{W}{4}{\texttt{10}} = \infty$};
  \end{tikzpicture}
  \captionsetup{skip=4pt}
  \caption{\small Example of prefix select queries
    with $m = 8$, $\AlphabetSize = 2$, and $\ell = 3$. We show three example
    queries with $X = \texttt{10}$ and highlight the relevant occurrences of
    $X$ as prefixes.}\label{fig:prefix-select-example-small}
  \end{framed}
\end{wrapfigure}

Our main equivalence theorem can then be stated as follows:

\begin{theorem}[Perfect equivalence of prefix select and suffix array queries for every alphabet size]\label{th:equivalence-of-prefix-select-and-sa-for-all-alphabet-sizes}
  Let $\AlphabetSize, N \in \Z_{\geq 2}$ be such that $\AlphabetSize
  \leq N$. In the word RAM model with word size $w$ satisfying
  $1+\log N \leq w = \bigO(\log N)$, the problems of
  \probname{Indexing}
  \probname{for}
  \probname{Prefix}
  \probname{Select}
  \probname{Queries}
  \probname{over}
  \probname{Alphabet}
  $[0 \dd \AlphabetSize)$ and
  \probname{Indexing}
  \probname{for}
  \probname{Suffix}
  \probname{Array}
  \probname{Queries}
  \probname{over}
  \probname{Alphabet}
  $[0 \dd \AlphabetSize)$
  are asymptotically equivalent:
  If there exists a data structure for one of these two problems that, given
  $N$, $\AlphabetSize$, and a valid input of at most $N$ symbols over alphabet
  $\IntegerAlphabet$,
  achieves:
  \vspace{-0.4ex}
  \begin{itemize}
  \item space usage of $S(\AlphabetSize,N)$ bits,
  \item preprocessing time $P_t(\AlphabetSize,N)$,
  \item preprocessing space $P_s(\AlphabetSize,N)$ bits,
  \item query time $Q(\AlphabetSize,N)$,
  \end{itemize}
  \vspace{-0.4ex}
  then there exists a data structure for the other problem that, given
  $N$, $\AlphabetSize$, and a valid input of at most $N$ symbols over alphabet $\IntegerAlphabet$, achieves:
  \vspace{-0.4ex}
  \begin{itemize}
  \item space usage of $\bigO(S(\AlphabetSize,N))$ bits,
  \item preprocessing time $\bigO(P_t(\AlphabetSize,N))$,
  \item preprocessing space $\bigO(P_s(\AlphabetSize,N))$ bits,
  \item query time $\bigO(Q(\AlphabetSize,N))$.
  \end{itemize}
\end{theorem}
\WFclear

The significance of these equivalences comes from the central role of suffix
arrays in string algorithms. As explained above, suffix arrays form the
backbone of numerous such algorithms, and understanding the complexity of
their queries is at the heart of understanding string processing and data
compression.

More practically, such equivalences can simplify the design and construction
of data structures answering suffix array queries (recall that our equivalence
captures not only query time and index space, but also construction time and
construction space). Suffix-array structures and related string-indexing
algorithms often rely on nontrivial data structures, string combinatorics, or
periodicity analysis; see, e.g.,~\cite{HonSS03,Bille0GKSV16,GawrychowskiK17,Gagie2020,MatsudaSST20,Pissis26a}.

Indeed, this advantage is already visible in~\cite{breaking}: the tradeoff for prefix rank and prefix select queries matching the state-of-the-art bounds for suffix-array queries, while improving the construction time and working space, takes only around one page to describe. The reason is that prefix select replaces the suffixes of one highly structured text by an array of short strings, while preserving exactly the difficulty relevant to SA access. Thus, the reductions handle the string-specific complications, and the remaining data-structure problem becomes substantially cleaner.

We next consider inverse suffix array queries. Here the right counterpart is
not fully general prefix rank but its restriction to patterns that are
prefixes of the string at the queried position, called
\emph{prefix special rank} in~\cite{PrefixEquiv}
(see \cref{def:prefix-rank-and-select}). Although the inverse suffix array
structure of~\cite{breaking} is built on a general prefix rank data structure,
all prefix rank queries it issues are of this special form. We prove that
these two problems are again perfectly equivalent in the same strong sense
as above.

\begin{theorem}[Perfect equivalence of prefix special rank and inverse suffix array queries for every alphabet size]\label{th:equivalence-of-prefix-special-rank-and-isa-for-all-alphabet-sizes}
  Let $\AlphabetSize, N \in \Z_{\geq 2}$ be such that $\AlphabetSize \leq N$.
  In the word RAM model with word size $w$ satisfying
  $1+\log N \leq w = \bigO(\log N)$, the problems of
  \probname{Indexing}
  \probname{for}
  \probname{Prefix}
  \probname{Special}
  \probname{Rank}
  \probname{Queries}
  \probname{over}
  \probname{Alphabet}
  $[0 \dd \AlphabetSize)$ and
  \probname{Indexing}
  \probname{for}
  \probname{Inverse}
  \probname{Suffix}
  \probname{Array}
  \probname{Queries}
  \probname{over}
  \probname{Alphabet}
  $[0 \dd \AlphabetSize)$
  are asymptotically equivalent:
  If there exists a data structure for one of these two problems that,
  given $N$, $\AlphabetSize$, and a valid input of at most $N$ symbols over alphabet
  $\IntegerAlphabet$,\footnote{See
  \cref{th:equivalence-of-prefix-special-rank-and-isa-summary} and its
  footnote for the precise definition of a valid input to each of the
  problems.} achieves:
  \begin{itemize}
  \item space usage of $S(\AlphabetSize,N)$ bits,
  \item preprocessing time $P_t(\AlphabetSize,N)$,
  \item preprocessing space $P_s(\AlphabetSize,N)$ bits,
  \item query time $Q(\AlphabetSize,N)$,
  \end{itemize}
  then there exists a data structure for the other problem that, given
  $N$, $\AlphabetSize$, and a valid input of at most $N$ symbols over alphabet $\IntegerAlphabet$, achieves:
  \begin{itemize}
  \item space usage of $\bigO(S(\AlphabetSize,N))$ bits,
  \item preprocessing time $\bigO(P_t(\AlphabetSize,N))$,
  \item preprocessing space $\bigO(P_s(\AlphabetSize,N))$ bits,
  \item query time $\bigO(Q(\AlphabetSize,N))$.
  \end{itemize}
\end{theorem}

\myparagraph{Broader View of Prefix Rank and Select Queries}
At a high level, prefix range queries (i.e., prefix rank, prefix
select, and prefix special rank) occupy a natural middle ground in the
landscape of sequence queries. They are richer than basic rank and select
queries, yet more structured than the fully general orthogonal range
queries. For example, focusing on the counting variants of these
queries
(\cref{def:rank-select,def:prefix-rank-and-select,def:range-count}),
and using $\lesssim$ to denote that one type of query is a special
case of another, the~relationship~is:
\begin{center}
  \textsc{Rank Queries}\quad $\lesssim$\quad \textsc{Prefix Rank Queries}\quad $\lesssim$\quad \textsc{Range Counting}.
\end{center}

To illustrate this, let $2 \leq \AlphabetSize \leq \Textlen$, and
assume for simplicity that $\AlphabetSize = 2^{\ell}$ for some $\ell
\in \Z_{\geq 1}$. Then:

\begin{itemize}
\item If $S[1 \dd \Textlen]$ is a string over alphabet
  $[0 \dd \AlphabetSize)$, then for every
  $i \in [0 \dd \Textlen]$ and every
  $a \in [0 \dd \AlphabetSize)$, standard rank is a special case of
  prefix rank:
  \[
    \Rank{S}{i}{a} = \PrefixRank{W}{i}{X},
  \]
  where $W[1 \dd \Textlen]$ is obtained by replacing each symbol of
  $S$ with its length-$\ell$ binary encoding (padding with zeros if necessary),
  and $X$ is the length-$\ell$ binary encoding of $a$.

\item Moreover, if $W[1 \dd \Textlen]$ is a sequence of binary
  strings of length $\ell$, then for every
  $i \in [0 \dd \Textlen]$ and every
  $X \in \BinaryAlphabet^{\leq \ell}$, prefix rank can be expressed
  using range counting:
  \[
    \PrefixRank{W}{i}{X}
      = \TwoSidedRangeCount{A}{i}{v_1}
        - \TwoSidedRangeCount{A}{i}{v_2+1},
  \]
  where $A[1 \dd \Textlen]$ is such that, for
  every $i \in [1 \dd \Textlen]$, the length-$\ell$ binary encoding of $A[i]$
  equals $W[i]$, and
  $v_1$ (resp.\ $v_2$) is an integer whose length-$\ell$
  binary encoding is $X \zero^{\ell-|X|}$ (resp.\ $X \one^{\ell-|X|}$).
\end{itemize}

This hierarchy is not merely formal; both extremes have already played
major roles in suffix-array data structures. On the one hand,
rank and select are among the main primitives behind the FM-index and
compressed suffix arrays~\cite{FerraginaM00,GrossiV00}. On the other
hand, fully general range counting and selection play a key role in
recent compressed indexes that answer SA queries using space that is optimal with respect to
the substring-complexity measure~\cite{collapsing}. Our results show that the
right level of abstraction for understanding SA and ISA queries lies
between these two~worlds.

\myparagraph{Applications}
The equivalences established in this paper have already led to new data
structures for suffix array and inverse suffix array queries. In subsequent
work~\cite{SaTradeoffs}, we construct a data structure supporting prefix
special rank queries in constant time and using space linear in the size of
the packed input. Applying
\cref{th:equivalence-of-prefix-special-rank-and-isa-for-all-alphabet-sizes}
yields an $\bigO(\Textlen\log\AlphabetSize)$-bit data structure answering
inverse suffix array queries in $\bigO(1)$ time, for every alphabet size
$2\leq\AlphabetSize\leq\Textlen$. The absence of an additive query-time
overhead in our equivalence is essential for obtaining constant-time queries
through this reduction. The structure can be constructed from the packed
text in $\bigO(\Textlen\min\{1,\log\AlphabetSize/\sqrt{\log\Textlen}\})$
time using $\bigO(\Textlen\log\AlphabetSize)$ bits of space. Thus, index
size, query time, and construction space are simultaneously optimal.
The construction time is also conditionally optimal for binary texts:
improving it to $o(\Textlen/\sqrt{\log\Textlen})$, while retaining
constant-time queries, would yield an $o(N/\sqrt{\log N})$-time algorithm
for the Dictionary Matching problem studied in~\cite{hierarchy}, on
inputs of $N$ bits. Together with our suffix array lower bound
(\cref{th:lb-simple}), the constant-time ISA result also establishes a
separation between the two types of queries for compact deterministic
data structures.

The same work applies our equivalence for prefix select
(\cref{th:equivalence-of-prefix-select-and-sa-for-all-alphabet-sizes}) to
obtain two families of suffix array data structures over arbitrary
alphabets. These include, for every fixed $\epsilon\in(0,1)$, a structure
using $\bigO(\Textlen\log\AlphabetSize)$ bits and answering queries in
$\bigO(\log_\AlphabetSize^\epsilon\Textlen)$ time, and a structure using
$\bigO(\Textlen\log\AlphabetSize\log_\AlphabetSize^\epsilon\Textlen)$ bits
with constant query time. The former matches the classical compressed
suffix array bounds~\cite{GrossiV05}, while the latter extends Rao's
constant-query-time result~\cite[Corollary~6]{Rao02} from binary strings to arbitrary
alphabets. Both structures admit the same construction-time bound as the
ISA structure above, with construction space satisfying the respective
index-size bound. In particular, both can be constructed in
$\bigO(\Textlen/\sqrt{\log\Textlen})$ time for binary texts. For the
constant-query-time structure, this construction time is again optimal
under the Dictionary Matching hardness assumption.

\myparagraph{Related Work}
Alongside his improvements to inverse suffix array queries,
Thankachan~\cite[Lemma~1.1]{ISA26} investigates the difficulty of suffix array queries
in compact space. Building on the reduction from two-dimensional orthogonal
range reporting to text indexing by Chien et al.~\cite{ChienHSTV15}, he
shows that substantially faster suffix array queries in compact space would
improve the time per reported point in geometric range reporting.

In the revised arXiv version, Thankachan~\cite[Theorem~6.1]{ISA26}
also proves a cell-probe lower bound for deterministic suffix array
indexes that store the text separately in packed form. With
$\Theta(\log\Textlen)$-bit cells,
$\bigO((\Textlen/\tau)\log\AlphabetSize)$ auxiliary bits require
$\Omega(\tau)$ query time for
$1\leq\tau=\bigO(\min\{\log\AlphabetSize,\log_\AlphabetSize\Textlen\})$.

A different line of work bounds the space of an index in terms of the
text's compressibility. In earlier work~\cite{dichotomy}, we proved tight
$\Omega(\log\Textlen/\log\log\Textlen)$ query-time lower bounds for both
suffix array and inverse suffix array queries using
$\bigO(\delta(\Text)\log^{\bigO(1)}\Textlen)$ words of space, where
$\delta(\Text)$ is the substring complexity of the text~\cite{delta}. These bounds hold
even over binary alphabets. They concern indexes whose space is bounded in
terms of compressibility, rather than allowing $\bigO(\Textlen)$ bits for
every binary text.

\myparagraph{Organization of the Paper}
In \cref{sec:prelim}, we introduce the notation and definitions used in the paper.
In \cref{sec:overview}, we give an overview of our results and techniques.
In \cref{sec:tools}, we present the tools used in our reductions.
In \cref{sec:space-lower-bounds}, we then provide our space lower bounds for
  suffix array (\cref{sec:suffix-array-space-lower-bound}),
  inverse suffix array (\cref{sec:inverse-suffix-array-space-lower-bound}),
  select (\cref{sec:select-space-lower-bound}),
  prefix select (\cref{sec:prefix-select-space-lower-bound}),
  special rank (\cref{sec:special-rank-space-lower-bound}), and
  prefix special rank (\cref{sec:prefix-special-rank-space-lower-bound}) queries.
In \cref{sec:equiv-prefix-select-and-sa}, we prove the equivalence of suffix array and prefix select queries (\cref{th:equivalence-of-prefix-select-and-sa-for-all-alphabet-sizes}) and derive our cell-probe lower bound (\cref{th:lb-simple}) for suffix array queries.
Finally, in \cref{sec:equiv-prefix-special-rank-and-isa}, we prove the equivalence of inverse suffix array and prefix special rank queries (\cref{th:equivalence-of-prefix-special-rank-and-isa-for-all-alphabet-sizes}).

%% file: prelim.tex
\section{Preliminaries}\label{sec:prelim}

\subsection{Basic Definitions}\label{sec:prelim-basic}

A \emph{string} is a finite sequence of characters from a given
\emph{alphabet} $\Sigma$.  The length of a string $S$ is denoted $|S|$. For $i
\in [1\dd |S|]$,\footnote{For $i,j\in \mathbb{Z}$, we let
  $[i \dd j] = \{k \in \Z : i \leq k \leq j\}$,
  $[i \dd j) = \{k \in \Z : i \leq k < j\}$, and
  $(i \dd j] = \{k \in \Z: i < k \leq j\}$.}
the $i$th character of $S$ is denoted $S[i]$. A~\emph{substring} or a
\emph{factor} of $S$ is a string of the form
$S[i \dd j) = S[i]S[{i+1}] \cdots S[{j-1}]$ for some
$1 \leq i \leq j \leq |S| + 1$.
Substrings of the form $S[1 \dd j)$ and $S[i \dd |S|{+}1)$ are called
\emph{prefixes} and \emph{suffixes}, respectively. We use
$\revstr{S}$ to denote the \emph{reverse} of $S$, i.e.,
$S[|S|]\cdots S[2]S[1]$.
We denote the \emph{concatenation} of two strings $U$ and
$V$, that is, the string $U[1]\cdots U[|U|]V[1]\cdots V[|V|]$, by $UV$
or $U\cdot V$. Furthermore, $S^k = \bigodot_{i=1}^k S$ is the
concatenation of $k \in \Zz$ copies of $S$; note that $S^0 =
\emptystring$ is the \emph{empty string}. An integer $p \in [1\dd |S|]$ is
a \emph{period} of $S$ if $S[i] = S[i + p]$ holds for every
$i \in [1 \dd |S|-p]$. We denote the smallest period of $S$ as
$\per{S}$.  For every $S \in \Sigma^{+}$, we define the infinite power
$S^{\infty}$ so that $S^{\infty}[i] = S[1 + (i-1) \bmod |S|]$ for
$i \in \Z$.  In particular, $S = S^{\infty}[1 \dd |S|]$.
For $U,V\in\Sigma^*$, we denote by $\lcp{U}{V}$ the length of their
longest common prefix. For
any string $S \in \Sigma^{*}$ and any $j_1, j_2 \in [1 \dd |S|]$, we
denote $\LCE{S}{j_1}{j_2} = \lcp{S[j_1 \dd |S|]}{S[j_2 \dd |S|]}$. We
use $\preceq$ to denote the order on $\Sigma$, extended to the
\emph{lexicographic} order on $\Sigma^*$, so that $U,V\in \Sigma^*$
satisfy $U \preceq V$ if and only if either
\begin{enumerate*}[label=(\alph*)]
  \item $U$ is a prefix of $V$, or
  \item $U[1 \dd i) = V[1 \dd i)$ and
    $U[i]\prec V[i]$ holds for some $i\in [1\dd \min(|U|,|V|)]$.
\end{enumerate*}
We write $U\prec V$ if $U\preceq V$ and $U\neq V$.
All logarithms are base two unless a different base is specified.

\begin{definition}[Pattern occurrences and SA-interval]\label{def:occ}
  For any pattern $\Pat \in \Sigma^{*}$ and text $\Text \in \Sigma^*$,
  we define
  \begin{align*}
    \OccTwo{\Pat}{\Text}
      &= \{j \in [1 \dd |\Text|] : j + |\Pat| \leq |\Text| + 1\text{ and }\Text[j \dd j + |\Pat|) = \Pat\},\\
    \RangeBegTwo{\Pat}{\Text}
      &= |\{j \in [1 \dd |\Text|] : \Text[j \dd |\Text|] \prec \Pat\}|,\\
    \RangeEndTwo{\Pat}{\Text}
      &= \RangeBegTwo{\Pat}{\Text} + |\OccTwo{\Pat}{\Text}|.
  \end{align*}
\end{definition}

\subsection{Model of Computation}\label{sec:prelim-model}

For our upper bounds, we use the standard word RAM model of computation~\cite{Hagerup98},
in which all standard bitwise and arithmetic operations on $w$-bit \emph{machine words}
take $\bigO(1)$ time. We assume $w=\Omega(\log N)$, where $N$ is the size of the input.
We assume that the word size is sufficiently large to address all memory used
by the algorithm, including its input and output.
If the unit of space is not specified (e.g., it is stated as \emph{space usage} or
\emph{space complexity}), then
the space is measured in $w$-bit words. In many situations, however, we explicitly
specify the unit; most commonly in this paper we measure the space \emph{in bits}.

Our query-time lower bounds hold in the cell-probe model with
$w=\Theta(\log N)$-bit cells. In this model, preprocessing is unrestricted,
and query time counts only accesses to stored cells; all other computation
is free. Thus, these lower bounds also apply in the word RAM model with
the same word size.

\begin{remark}\label{rm:space}
  In this paper, we adopt the standard notion that the space usage/complexity of an algorithm is its peak memory usage.
  Since all our algorithms operate entirely in memory, this quantity includes the input and output size of an algorithm.
  Throughout the paper, the algorithms and data structures we construct and use in the word RAM model have explicit representations: auxiliary
  memory and output data structures are materialized word by word during the computation. Thus, an algorithm running in
  $t$ time creates and retains at most $\bigO(t)$ words of auxiliary or constructed space. In particular, our data
  structures constructed in $t$ time occupy $\bigO(t)$ words.
\end{remark}

\subsection{Packed Representation of Strings and String Sequences}\label{sec:prelim-packed}

\begin{definition}[Packed representation of a string]\label{def:packed-representation}
  In the word RAM model with word size $w$, let $\AlphabetSize \in \Z_{\geq 2}$
  be an alphabet size satisfying $\log \AlphabetSize \leq w$.
  The \emph{packed representation} of a string $S \in \IntegerAlphabet^{*}$,
  denoted $\PackedRepresentation{w}{\AlphabetSize}{S}$,
  is the unique sequence of integers $(x_1, x_2, \ldots, x_k)$ encoding $S$,
  satisfying the following properties:
  \begin{enumerate}
  \item Each $x_i$ is a $w$-bit word, i.e., $x_i \in [0 \dd 2^{w})$.
  \item Each symbol of $S$ is encoded using $b := \lceil \log \AlphabetSize \rceil$ bits.
  \item Each word except possibly the last stores $s := \lfloor w / b \rfloor$ symbols.
  \item The sequence consists of $k := \lceil |S| / s \rceil$ words.
  \item Any bits in a word not used to store a symbol from $S$ are set to
    $0$.
  \end{enumerate}
  Formally, for every $i \in [1 \dd k]$,
  \[
    x_i = \sum_{\delta=1}^{\min(s, |S|-(i-1)s)} S[(i-1)s+\delta] \cdot 2^{(\delta-1)b}.
  \]
  In other words, for every $j \in [1 \dd |S|]$, the symbol $S[j]$ is stored
  in word $x_i$ where $i = \lceil j/s \rceil$, at the relative position
  $\delta = j - (i-1)s$. The retrieval formula is thus given by:
  \[
    S[j] = \left\lfloor \frac{x_i}{2^{(\delta-1)b}} \right\rfloor \bmod 2^b.
  \]
  Using the inequalities $\lceil x \rceil < x+1$ and, for $x \geq 1$,
  $\lfloor x \rfloor \geq x/2$, the number of words
  $k$ can be bounded as follows:
  \[
    k =
    \left\lceil \frac{|S|}{\lfloor w / b \rfloor} \right\rceil
    \leq 1 + \frac{|S|}{\lfloor w / b \rfloor}
    \leq 1 + \frac{2 |S| b}{w}
    = \bigO\left(1 + \frac{|S| \log \AlphabetSize}{w} \right).
  \]
\end{definition}
\vspace{1ex}

\begin{definition}[Packed representation of a sequence of strings]\label{def:packed-sequence-representation}
  In the word RAM model with word size $w$, let
  $\AlphabetSize \in \Z_{\geq 2}$ satisfy $\log \AlphabetSize \leq w$, and
  let $m,\ell \in \Z_{\geq 0}$. For any sequence $W[1 \dd m]$ of $m$
  strings, each of length $\ell$ over alphabet $[0\dd\AlphabetSize)$, we
  define the \emph{packed sequence representation} of $W$ as the packed
  representation of the concatenation of its strings:
  \[
    \PackedSeqRepresentation{w}{\AlphabetSize}{W}
      :=\PackedRepresentation{w}{\AlphabetSize}{\textstyle\bigodot_{i=1}^m W[i]}.
  \]
  In particular, word boundaries need not coincide with string boundaries.
  The symbol $W[i][j]$ occupies position $(i-1)\ell+j$ in the represented
  concatenation. By \cref{def:packed-representation}, the representation
  consists of
  $\bigO(1+m\ell\log\AlphabetSize/w)$ words.
\end{definition}
\vspace{1ex}

\begin{example}\label{ex:packed-representation}
  Let $w = 8$ and $\AlphabetSize = 3$. Then
  $b = \lceil \log \AlphabetSize \rceil = 2$ and
  $s = \lfloor w/b \rfloor = 4$. Each word therefore stores $4$ symbols,
  each using $2$ bits. Consider the string
  $S = (2,0,1,0,2)\in [0\dd 3)^5$, corresponding to $\texttt{cabac}$
  assuming mapping $0 = \texttt{a}$, $1 = \texttt{b}$, and $2 = \texttt{c}$.
  Since $|S|=5$, its packed representation has
  $k = \lceil 5/4\rceil = 2$ words:
  \[
    \PackedRepresentation{8}{3}{S}
      = (x_1, x_2)
      = \big((00010010)_2, (00000010)_2\big)
      = \big(18, 2\big).
  \]
  Indeed, the first word stores $S[1], S[2], S[3], S[4]$ in consecutive
  $2$-bit fields starting from the least significant bits. Consequently,
  $(00010010)_2$ encodes $(2,0,1,0)$, while the second word stores $S[5] = 2$
  and pads all remaining bits with $0$. For example, to retrieve $S[3]$, we
  have $i = \lceil 3/4 \rceil = 1$ and $\delta = 3$. Thus
  \[
    S[3]
      = \left\lfloor \frac{x_1}{2^{(\delta-1)b}} \right\rfloor \bmod 2^b
      = \left\lfloor \frac{18}{2^4} \right\rfloor \bmod 4
      = 1.
  \]
\end{example}
\vspace{1ex}

\begin{remark}\label{rm:packed-representation}
  We use the operations in \cref{pr:packed-representation} to keep the
  implementation of packed strings separate from our algorithms.
  Short strings can also be represented by their lengths and integer
  values (\cref{def:val}), using the conversions from
  \cref{th:val-encoding,th:val-decoding}.
  A different packing convention requires adapting only these operations
  and conversions, provided that they retain their guarantees and the
  representation retains the same asymptotic size.
\end{remark}

\begin{proposition}[{\cite{hierarchy}}]\label{pr:packed-representation}
  Let $\AlphabetSize,N\in\Z_{\geq2}$, and assume that $\AlphabetSize$
  is known and that $b = \lceil \log \AlphabetSize \rceil$ has been computed.
  In the word RAM model with word size $w$ satisfying $w\geq b$ and
  $w>\log N$, the following operations on packed representations
  (\cref{def:packed-representation}) are supported:
  \begin{enumerate}[itemsep=1ex]

  \item\label{pr:packed-representation-initialize}
    \ul{Initialize:}
    Given any $m \in [0 \dd N]$ and any symbol
    $a \in \IntegerAlphabet$, the packed representation
    $\PackedRepresentation{w}{\AlphabetSize}{a^m}$ of $a^m$ can be
    constructed in $\bigO(1 + m\log\AlphabetSize/w)$ time.

  \item\label{pr:packed-representation-access}
    \ul{Access:}
    Given the packed representation $\PackedRepresentation{w}{\AlphabetSize}{S}$
    of $S$ and its length $|S|$, where
    $S \in \IntegerAlphabet^{\leq N}$, and any
    $i \in [1 \dd |S|]$, $S[i]$ can be computed in $\bigO(1)$ time.

  \item\label{pr:packed-representation-update}
    \ul{Update:}
    Given the packed representation $\PackedRepresentation{w}{\AlphabetSize}{S}$
    of $S$ and its length $|S|$, where
    $S \in \IntegerAlphabet^{\leq N}$, any $i \in [1 \dd |S|]$,
    and any $a \in \IntegerAlphabet$,
    the packed representation $\PackedRepresentation{w}{\AlphabetSize}{S}$
    can be updated in $\bigO(1)$ time to reflect the assignment $S[i] := a$.

  \item\label{pr:packed-representation-substring}
    \ul{Substring:}
    Given the packed representation $\PackedRepresentation{w}{\AlphabetSize}{S}$
    of $S$ and its length $|S|$, where
    $S \in \IntegerAlphabet^{\leq N}$, any $\ell \in [0 \dd |S|]$,
    and any $p \in [1 \dd |S|-\ell+1]$,
    the packed representation $\PackedRepresentation{w}{\AlphabetSize}{S[p \dd p+\ell)}$
    can be computed in
    $\bigO(1 + \ell\log\AlphabetSize/w)$ time.

  \item\label{pr:packed-representation-concat}
    \ul{Concatenate:}
    Given $m \in [1 \dd N]$ and, for every $i \in [1 \dd m]$,
    the packed representation $\PackedRepresentation{w}{\AlphabetSize}{S_i}$
    of $S_i$ and its length $|S_i|$, where
    $S_i \in \IntegerAlphabet^{\leq N}$ and
    $\sum_{i=1}^{m}|S_i| \leq N$,
    the packed representation $\PackedRepresentation{w}{\AlphabetSize}{S}$
    of $S = S_1 \cdots S_m$ can be computed in
    $\bigO(m + |S|\log\AlphabetSize/w)$ time.

  \item\label{pr:packed-representation-split}
    \ul{Split:}
    Given the packed representation $\PackedRepresentation{w}{\AlphabetSize}{S}$
    of $S$ and its length $|S|$, where
    $S \in \IntegerAlphabet^{\leq N}$, an integer
    $m \in [1 \dd N]$, and nonnegative integers
    $\ell_1, \ldots, \ell_m$ summing to $|S|$, the packed representations
    $\PackedRepresentation{w}{\AlphabetSize}{S_i}$ of the strings
    $S_i \in \IntegerAlphabet^{\ell_i}$, $i \in [1 \dd m]$, satisfying
    $S = S_1 \cdots S_m$ can be computed in
    $\bigO(m + |S|\log\AlphabetSize/w)$ time.

  \item\label{pr:packed-representation-increase}
    \ul{Increase:}
    Given the packed representation $\PackedRepresentation{w}{\AlphabetSize}{S}$
    of $S$, its length $|S|$, and an integer $\delta\in[0\dd\AlphabetSize)$,
    where $S\in\IntegerAlphabet^{\leq N}$ and
    $S[i]+\delta<\AlphabetSize$ for every $i\in[1\dd|S|]$,
    the packed representation $\PackedRepresentation{w}{\AlphabetSize}{S'}$
    of the string $S'$ satisfying $|S'|=|S|$ and
    $S'[i]=S[i]+\delta$ for every $i\in[1\dd|S|]$ can be computed in
    $\bigO(1+|S|\log\AlphabetSize/w)$ time.

  \item\label{pr:packed-representation-compare}
    \ul{Compare:}
    Given the lengths $|X|,|Y|$ and the packed representations
    $\PackedRepresentation{w}{\AlphabetSize}{X}$ and
    $\PackedRepresentation{w}{\AlphabetSize}{Y}$ of strings
    $X,Y\in\IntegerAlphabet^{\leq N}$, we can determine
    whether $X\prec Y$, $X=Y$, or $Y\prec X$ in
    $\bigO(1+\min\{|X|,|Y|\}\log\AlphabetSize/w)$ time.

  \end{enumerate}
\end{proposition}

\subsection{Suffix Arrays}\label{sec:prelim-suffix-array}

\begin{definition}[Suffix array]\label{def:suffix-array}
  Let $\Textlen\in\Z_{\geq1}$ and $\Text\in\Sigma^{\Textlen}$.
  The \emph{suffix array} $\SA{\Text}[1 \dd \Textlen]$
  of $\Text$ is a permutation of $[1 \dd \Textlen]$ such that
  $\Text[\SA{\Text}[1] \dd \Textlen] \prec
  \Text[\SA{\Text}[2] \dd \Textlen] \prec \cdots \prec
  \Text[\SA{\Text}[\Textlen] \dd \Textlen]$, i.e., $\SA{\Text}[i]$ is
  the starting position of the lexicographically $i$th nonempty suffix
  of $\Text$, for every $i\in[1\dd\Textlen]$.
\end{definition}

\begin{definition}[Inverse suffix array]\label{def:inverse-suffix-array}
  Let $\Textlen\in\Z_{\geq1}$ and $\Text\in\Sigma^{\Textlen}$.
  The \emph{inverse suffix array} of $\Text$ is an array $\ISA{\Text}[1 \dd \Textlen]$ containing the inverse
  permutation of $\SA{\Text}$ (\cref{def:suffix-array}), i.e., for every
  $i,j\in[1\dd\Textlen]$, it holds $\ISA{\Text}[j] = i$ if and only if
  $\SA{\Text}[i] = j$. Equivalently, for every $j\in[1\dd\Textlen]$,
  $\ISA{\Text}[j]$ stores
  the lexicographic \emph{rank} of $\Text[j \dd \Textlen]$ among the
  nonempty suffixes of $\Text$, that is,
  $\ISA{\Text}[j] = 1 + \RangeBegTwo{\Text[j \dd \Textlen]}{\Text}$
  (\cref{def:occ}).
\end{definition}

\begin{definition}[Sequence of positions in lex-order]\label{def:lex-sorted}
  Let $\Text \in \Sigma^{\Textlen}$, $Q \subseteq [1 \dd \Textlen+1]$, and $q = |Q|$. By
  $\LexSorted{Q}{\Text}$, we denote a sequence $(a_i)_{i \in [1 \dd q]}$
  containing all positions from $Q$ such that for every
  $i, j \in [1 \dd q]$, $i < j$ implies
  $\Text[a_i \dd \Textlen] \prec \Text[a_j \dd \Textlen]$.
  Position $\Textlen+1$ represents the empty suffix.
\end{definition}

\begin{example}\label{ex:sa-and-isa}
  The suffix array for the example string
  $\Text = \texttt{abbabaabba}$
  is shown in \cref{fig:sa-example}. The inverse suffix array for the same
  example string is
  \[
    \ISA{\Text} = [5, 10, 8, 3, 7, 2, 4, 9, 6, 1].
  \]
\end{example}

\begin{remark}\label{rm:occ}
  Note that the two values $\RangeBegTwo{\Pat}{\Text}$ and
  $\RangeEndTwo{\Pat}{\Text}$ (\cref{def:occ}) are the endpoints of
  the so-called \emph{SA-interval} representing the occurrences of
  $\Pat$ in $\Text$, i.e.,
  \[
    \OccTwo{\Pat}{\Text} =
      \{\SA{\Text}[i] : i \in (\RangeBegTwo{\Pat}{\Text} \dd \RangeEndTwo{\Pat}{\Text}]\}
  \]
  holds for every $\Text \in \Sigma^{+}$ and $\Pat \in \Sigma^{*}$,
  including when $\Pat = \emptystring$ and when
  $\OccTwo{\Pat}{\Text} = \emptyset$.
  In particular, $\RangeBegTwo{\emptystring}{\Text}=0$ and
  $\RangeEndTwo{\emptystring}{\Text}=|\Text|$.
\end{remark}

\begin{example}\label{ex:occ-and-ranges}
  For the example text $\Text$ in \cref{fig:sa-example} and pattern
  $\Pat = \texttt{abb}$,
  it holds
  $\OccTwo{\Pat}{\Text} = \{7,1\} = \{\SA{\Text}[i] : i \in (3 \dd 5]\}$,
  which gives $\RangeBegTwo{\Pat}{\Text} = 3$ and $\RangeEndTwo{\Pat}{\Text} = 5$.
\end{example}

\begin{theorem}[{\cite{Farach97,KoA03,KarkkainenSB06}}]\label{th:sa-and-isa-construction}
  Let $\AlphabetSize, \Textlen \in \Z_{\geq 2}$ be such that $\AlphabetSize \leq \Textlen$.
  In the word RAM model
  with word size $w > \log \Textlen$, the suffix array $\SA{\Text}$ (\cref{def:suffix-array}) and
  the inverse suffix array $\ISA{\Text}$ (\cref{def:inverse-suffix-array}) of
  any text $\Text \in [0 \dd \AlphabetSize)^{\Textlen}$ can
  be computed in $\bigO(\Textlen)$ time using $\bigO(\Textlen)$ words of space,
  given $\Textlen$, $\AlphabetSize$, and the string $\Text$ represented as
  an array of $\Textlen$ integers.
\end{theorem}

\subsection{Rank and Selection Queries}\label{sec:prelim-rank-select}

\begin{definition}[Rank and selection queries]\label{def:rank-select}
  Let $m\in\Z_{\geq0}$ and $S\in\Sigma^m$.
  \begin{description}[style=sameline,itemsep=1ex]
  \item[Rank query:]
    For every $a \in \Sigma$ and $j \in [0 \dd m]$, we define
    $\Rank{S}{j}{a} := |\{i \in [1\dd j]: S[i] = a\}|$.
  \item[Special rank query:]
    For every $j \in [1 \dd m]$, we define
    $\SpecialRank{S}{j} := \Rank{S}{j}{S[j]}$.
  \item[Select query:]
    For every $a \in \Sigma$ and $r \in \Z_{\geq 1}$,
    we define $\Select{S}{r}{a}$ as the $r$th smallest element of
    the set $\{i \in [1 \dd m] : S[i] = a\}$ (if $r \leq \Rank{S}{m}{a}$),
    and $\Select{S}{r}{a} := \infty$ (otherwise).
  \end{description}
\end{definition}

\begin{example}\label{ex:rank-and-select-queries}
  Let $S = \texttt{abbaabaababbabbaabb}$. Then
  \[
    \Rank{S}{11}{\texttt{a}} = 6,\qquad
    \SpecialRank{S}{14} = 7,\qquad
    \Select{S}{5}{\texttt{a}} = 8,\qquad
    \Select{S}{11}{\texttt{b}} = \infty.
  \]
  Indeed, among the first $11$ symbols of $S$, the letter $\texttt{a}$
  occurs $6$ times. Moreover, $S[14] = \texttt{b}$, and $\texttt{b}$
  occurs $7$ times in the prefix of length $14$. Finally, the
  occurrences of $\texttt{a}$ in $S$ are at positions
  $1,4,5,7,8,10,13,16,17$. Hence, the fifth occurrence is at position $8$,
  while $\texttt{b}$ occurs only $10$ times in total.
\end{example}

\begin{theorem}[Rank and selection queries on
    bitvectors~{\cite[Lemma~2.3]{WaveletSuffixTree};
    see also~\cite{Clark96,Jac89,MunroNV16}}]\label{th:bin-rank-select}
  Let $N \in \Z_{\geq 2}$. Consider a word RAM model with word size $w > \log N$.
  For every binary string $S \in \BinaryAlphabet^{\leq N}$, there exists a data structure of
  $\bigO(|S|)$ bits answering rank and selection queries (\cref{def:rank-select}) in
  $\bigO(1)$ time. Moreover, given $N$, $m$, and the lengths and packed representations
  (\cref{def:packed-representation}) of any $m$ binary strings of total length at most $N$, the data structures for all
  these strings can be constructed in $\bigO(m + N / \log N)$ time.
\end{theorem}

\subsection{Prefix Rank and Selection Queries}\label{sec:prelim-prefix-queries}

\begin{definition}[Prefix rank and selection queries~\cite{breaking,PrefixEquiv}]\label{def:prefix-rank-and-select}
  Let $S \in (\Sigma^{*})^m$ be a sequence of strings over alphabet $\Sigma$.
  \begin{description}[style=sameline,itemsep=1ex]
  \item[Prefix rank query:]
    For every $X \in \Sigma^{*}$ and $j \in [0 \dd m]$, we define
    $\PrefixRank{S}{j}{X} := |\{i \in [1 \dd j]: X\text{ is a prefix of }S[i]\}|$.
  \item[Prefix special rank query:]
    For every $j \in [1 \dd m]$ and $p \in [0 \dd |S[j]|]$, we let
    $\PrefixSpecialRank{S}{j}{p} :=\allowbreak \PrefixRank{S}{j}{S[j][1 \dd p]}$.
  \item[Prefix select query:]
    For every $X \in \Sigma^{*}$ and $r \in \Z_{\geq 1}$,
    we define $\PrefixSelect{S}{r}{X}$ as the $r$th smallest element of
    the set $\{i \in [1 \dd m] : X\text{ is a prefix of }S[i]\}$
    (if $r \leq \PrefixRank{S}{m}{X}$),
    and $\PrefixSelect{S}{r}{X} := \infty$ (otherwise).
  \end{description}
\end{definition}

\noindent
\begin{minipage}[t]{\dimexpr\linewidth-0.18\linewidth\relax}
  \begin{example}\label{ex:prefix-queries}
    Let $S \in (\Sigma^*)^9$ be the sequence of strings shown on the right, and let
    $X = \texttt{bb}$. Then
    \begin{itemize}
    \item $\PrefixRank{S}{8}{X} = 2$,
    \item $\PrefixSpecialRank{S}{7}{2} = 2$,
    \item $\PrefixSelect{S}{3}{X} = 9$, and
    \item $\PrefixSelect{S}{4}{X} = \infty$.
    \end{itemize}
    Indeed, exactly the strings $S[4]$, $S[7]$, and $S[9]$ have prefix $X$.
    Moreover, $S[7][1 \dd 2] = \texttt{bb}$. Thus, among the first $7$ strings,
    exactly two have prefix $S[7][1 \dd 2]$, namely $S[4]$ and $S[7]$.
  \end{example}
\end{minipage}\hfill%
\begin{minipage}[t]{0.18\linewidth}
  \vspace{0pt}
  \renewcommand{\arraystretch}{0.95}
  \makebox[\linewidth][r]{%
    \begin{tabular}[t]{@{}r@{\ }c@{\ }l@{}}
      $S[1]$ & = & \texttt{caba}\\
      $S[2]$ & = & \texttt{baba}\\
      $S[3]$ & = & \texttt{abba}\\
      $S[4]$ & = & \texttt{bbab}\\
      $S[5]$ & = & \texttt{baaa}\\
      $S[6]$ & = & \texttt{aabb}\\
      $S[7]$ & = & \texttt{bbaa}\\
      $S[8]$ & = & \texttt{abab}\\
      $S[9]$ & = & \texttt{bbba}
    \end{tabular}%
  }
\end{minipage}

\begin{observation}[{\cite{PrefixEquiv}}]\label{ob:prefix-rank-and-select-padding}
  Let $W[1 \dd m]$ be a sequence of $m \geq 1$ strings of length $\ell \geq 1$ over alphabet $\Sigma$ and let
  $W'[1 \dd m]$ be any sequence of $m$ strings of length $\ell' \geq \ell$ over alphabet $\Sigma$
  such that for every $j \in [1 \dd m]$, $W[j]$ is a prefix of $W'[j]$.
  Consider any string $X \in \Sigma^{\leq \ell}$. Then,
  \begin{itemize}
  \item For every $i \in [0 \dd m]$, it holds \[\PrefixRank{W}{i}{X} = \PrefixRank{W'}{i}{X}.\]
  \item For every $r \geq 1$, it holds
    \[\PrefixSelect{W}{r}{X} = \PrefixSelect{W'}{r}{X}.\]
  \end{itemize}
  Furthermore, for every $i \in [1 \dd m]$ and every $p \in [0 \dd \ell]$, it holds
  \[\PrefixSpecialRank{W}{i}{p} = \PrefixSpecialRank{W'}{i}{p}.\]
\end{observation}

\subsection{Range Counting and Selection Queries}\label{sec:prelim-range-queries}

\begin{definition}[Range counting]\label{def:range-count}
  Let $A[1 \dd m]$ be an array of $m \geq 0$ nonnegative integers.
  For every $j \in [0 \dd m]$ and $v \in \Z_{\geq 0}$, we define
  \[
    \TwoSidedRangeCount{A}{j}{v} := |\{i \in [1 \dd j] : A[i] \geq v\}|.
  \]
\end{definition}

\begin{definition}[Range selection]\label{def:range-select}
  Let $A[1 \dd m]$ be an array of $m \geq 0$ nonnegative integers.
  For every $v \in \Z_{\geq 0}$ and every $r \in \Z_{\geq 1}$,
  we define $\RangeSelect{A}{r}{v}$ to be the $r$th smallest element of
  $\{i \in [1 \dd m] : A[i] \geq v\}$ (if $r \leq \TwoSidedRangeCount{A}{m}{v}$)
  or $\infty$ (otherwise).
\end{definition}

\begin{example}\label{ex:range-queries}
  Let $A = [5, 1, 2, 8, 4, 7, 6, 2, 9]$. Then
  \[
    \TwoSidedRangeCount{A}{6}{4} = 4,\qquad
    \RangeSelect{A}{4}{5} = 7,\qquad
    \RangeSelect{A}{6}{5} = \infty.
  \]
  Indeed, among the first $6$ entries of $A$, namely
  $[5,1,2,8,4,7]$, exactly four are at least $4$, occurring at
  positions $1$, $4$, $5$, and $6$. Moreover,
  $\{i \in [1 \dd 9] : A[i] \geq 5\} = \{1,4,6,7,9\}$.
  The fourth smallest element of this set is $7$, and there is no sixth such element.
\end{example}

\subsection{Periodic Positions}\label{sec:prelim-periodic-positions}

\begin{definition}[$\tau$-periodic positions~\cite{sss}]
  \label{def:periodic-positions}
  Let $\Textlen\in\Z_{\geq0}$, $\Text\in\Sigma^{\Textlen}$, and
  $\tau\in\Z_{\geq1}$. We define the set of \emph{$\tau$-periodic
  positions} of $\Text$ as
  \[
    \RTwo{\tau}{\Text}
      := \{i \in [1 \dd \Textlen - 3\tau + 2] :
        \per{\Text[i \dd i + 3\tau - 1)} \leq \tfrac{1}{3}\tau\}.
  \]
\end{definition}

\begin{definition}[$\tau$-periodic and $\tau$-nonperiodic patterns]\label{def:periodic-pattern}
  Let $m\in\Z_{\geq0}$, $\Pat \in \Sigma^{m}$, and $\tau\in\Z_{\geq1}$.
  We say that $\Pat$ is
  \emph{$\tau$-periodic} if it holds $m \geq 3\tau - 1$ and
  $\per{\Pat[1 \dd 3\tau - 1]} \leq \tfrac{1}{3}\tau$. Otherwise, it
  is called \emph{$\tau$-nonperiodic}.
\end{definition}

\subsection{String Synchronizing Sets}\label{sec:prelim-sss}

\begin{definition}[$\tau$-synchronizing set~\cite{sss}]\label{def:sss}
  Let $\Textlen\in\Z_{\geq2}$, $\Text\in \Sigma^{\Textlen}$, and $\tau \in
  [1 \dd \lfloor\frac{\Textlen}{2}\rfloor]$. A set $\SSS
  \subseteq [1 \dd \Textlen - 2\tau + 1]$ is called a
  \emph{$\tau$-synchronizing set} of $\Text$ if it satisfies the
  following \emph{consistency} and \emph{density} conditions:
  \begin{enumerate}
  \item\label{def:sss-consistency}
    If $\Text[i \dd i + 2\tau) = \Text[j\dd j + 2\tau)$, then $i \in \SSS$
    holds if and only if $j \in \SSS$
    (for $i, j \in [1 \dd \Textlen - 2\tau + 1]$),
  \item\label{def:sss-density}
    $\SSS\cap[i \dd i + \tau) = \emptyset$ if and only if
    $i \in \RTwo{\tau}{\Text}$ (for
    $i \in [1 \dd \Textlen - 3\tau + 2]$; see
    \cref{def:periodic-positions}).
  \end{enumerate}
\end{definition}

\begin{example}\label{ex:sss}
  Let $\tau = 3$ and let $\Text$ be the string shown below, where
  positions in $\SSS$ are underlined and positions in $\RTwo{3}{\Text}$
  are overlined:
  \[
    \begin{tikzpicture}[x=0.48cm,y=0.80cm,baseline=(current bounding box.center)]
      \tikzset{sym/.style={anchor=base,inner sep=0pt,minimum width=0.48cm,text height=1.8ex,text depth=.45ex,font=\ttfamily}}
      \foreach \i in {1,...,25}
        \node[anchor=base,font=\scriptsize] at (\i,0.55) {$\i$};
      \foreach \x [count=\i] in {c,d,e,f,g,h,x,a,a,a,a,a,a,a,a,a,a,a,y,c,d,e,f,g,h}
        \node[sym] at (\i,0) {\x};
      \foreach \i in {8,9,10,11}
        \draw[line width=1.2pt] (\i-0.20,0.34) -- (\i+0.20,0.34);
      \foreach \i in {1,4,7,14,17,20}
        \draw[line width=1.2pt] (\i-0.20,-0.14) -- (\i+0.20,-0.14);
    \end{tikzpicture}
  \]
  Let $\SSS = \{1,4,7,14,17,20\}$. We claim that $\SSS$ is a
  $3$-synchronizing set of $\Text$:
  \begin{itemize}

  \item Note that $2\tau = 6$ and $3\tau - 1 = 8$.
    The only length-$8$ factors of $\Text$ having period at most $1$ are
    the factors equal to $\texttt{aaaaaaaa}$, which start exactly at
    positions $8,9,10,11$. Hence,
    $\RTwo{3}{\Text} = \{8,9,10,11\}$.
    Therefore, the density condition requires
    $\SSS \cap [i \dd i+3) = \emptyset$ exactly for
    $i \in \{8,9,10,11\}$. This is satisfied:
    \begin{enumerate*}[label=(\alph*)]
    \item since $\SSS \cap [8 \dd 14) = \emptyset$, the intervals
      $[8 \dd 11)$, $[9 \dd 12)$, $[10 \dd 13)$, and $[11 \dd 14)$ contain no element of $\SSS$, and
    \item every other interval $[i \dd i+3)$ with
      $i \in [1 \dd 18] \setminus \{8,9,10,11\}$ contains at least one
      element of $\SSS$.
    \end{enumerate*}

  \item For the consistency condition, we consider length-$6$ factors.
    The factor $\texttt{cdefgh}$ occurs exactly twice, namely at positions
    $1$ and $20$, and both positions belong to $\SSS$.
    Moreover, the factor $\texttt{aaaaaa}$ occurs at positions
    $8,9,10,11,12,13$, and none of these positions belongs to $\SSS$.
    All remaining length-$6$ factors are distinct. Hence, whenever two
    length-$6$ factors are equal, their starting positions are either both
    in $\SSS$ or both outside $\SSS$, as required by the consistency condition.
  \end{itemize}
\end{example}

\begin{theorem}[{\cite{sss}}]\label{th:sss-existence-and-construction}
  Let $\Textlen\in\Z_{\geq2}$. For every $\Text\in\Sigma^{\Textlen}$ and $\tau \in
  [1 \dd \lfloor\frac{\Textlen}{2}\rfloor]$, there exists a $\tau$-synchronizing
  set $\SSS$ (\cref{def:sss}) of size at most $30\Textlen/\tau$.
  Moreover, in the word RAM model with word size $w\geq1+\log\Textlen$,
  if $\Text \in \IntegerAlphabet^{\Textlen}$, where
  $\AlphabetSize\in\Z_{\geq1}$ satisfies $\AlphabetSize = \Textlen^{\bigO(1)}$,
  then, given $\Textlen$, $\AlphabetSize$, $\tau$, and $\Text$, such $\SSS$
  can be deterministically constructed in $\bigO(\Textlen)$ time.
\end{theorem}

\begin{theorem}[{\cite{sss}}]\label{th:sss-packed-construction}
  Let $\Textlen,\AlphabetSize\in\Z_{\geq2}$ satisfy
  $\AlphabetSize\leq\Textlen$, let $\mu\in(0,\tfrac15)$ be a
  constant, and let $\tau\in\Z_{\geq1}$ satisfy
  $\tau\leq\mu\log_{\AlphabetSize}\Textlen$.
  Consider the word RAM model with word size $w\geq1+\log\Textlen$.
  Given $\Textlen$, $\AlphabetSize$, $\tau$, and the packed
  representation $\PackedRepresentation{w}{\AlphabetSize}{\Text}$
  (\cref{def:packed-representation}) of a text
  $\Text\in\IntegerAlphabet^{\Textlen}$, we can deterministically
  construct an array containing a $\tau$-synchronizing set of
  $\Text$ (\cref{def:sss}) of size at most $30\Textlen/\tau$ in
  $\bigO(\Textlen/\tau)$ time using
  $\bigO(\Textlen/\tau)$ words of space.
\end{theorem}

\begin{theorem}[{\cite{sss}}]\label{th:sss-lex-sort}
  Let $\AlphabetSize,\Textlen\in\Z_{\geq2}$ satisfy
  $\AlphabetSize\leq\Textlen$, and let
  $\tau\in[1\dd\lfloor\tfrac{\Textlen}{2}\rfloor]$ satisfy
  $\tau=\bigO(\log_{\AlphabetSize}\Textlen)$.
  Consider the word RAM model with word size $w\geq1+\log\Textlen$.
  Let $\SSS$ be a $\tau$-synchronizing set (\cref{def:sss}) of a text
  $\Text\in\IntegerAlphabet^{\Textlen}$, with
  $|\SSS|=\bigO(\Textlen/\tau)$. Given $\Textlen$, $\AlphabetSize$, $\tau$,
  the packed representation
  $\PackedRepresentation{w}{\AlphabetSize}{\Text}$
  (\cref{def:packed-representation}) and an array listing the elements
  of $\SSS$, we can deterministically compute the sequence
  $\LexSorted{\SSS}{\Text}$ (\cref{def:lex-sorted}) in
  $\bigO(\Textlen/\tau)$ time.
\end{theorem}

%% file: overview.tex
\section{Technical Overview}\label{sec:overview}

In this overview, we explain the main ideas behind our reductions and how
they yield both the equivalence theorems and the lower bound for suffix array
queries. We start with the two pairs of problems involved in the equivalences:
suffix array queries and prefix select queries, and inverse suffix array
queries and prefix special rank queries
(\cref{th:equivalence-of-prefix-select-and-sa-for-all-alphabet-sizes,th:equivalence-of-prefix-special-rank-and-isa-for-all-alphabet-sizes}).
For each pair, our goal is to turn a data structure for either problem into
one for the other, while preserving query time, index space, and preprocessing
time and space up to constant factors. The alphabet and the input-size bound
are preserved exactly.

To see how these connections arise, we first explain the direction from
prefix queries to suffix array and inverse suffix array queries
(\cref{sec:overview-prefix-to-suffix}). Both reductions construct a text from a sequence of strings, arranging the relevant
suffixes so that their lexicographic order reflects the order of the strings
in the original sequence. In \cref{sec:overview-cell-probe}, we explain
how this construction also provides the connection needed for our query-time
lower bound: we reduce Ball Inheritance to prefix
select and then to suffix array queries (\cref{th:lb-suffix-array}).

We next explain the reverse reductions in
\cref{sec:overview-suffix-to-prefix}, following the framework
of~\cite{breaking}. The main additional difficulty in this direction is a
range-counting operation used in the treatment of periodic substrings. We
explain why this operation arises and, in \cref{sec:overview-range-count},
present a constant-time solution that removes the extra query-time term from the earlier reductions
(\cref{th:range-count-for-small-sum}).

Finally, in \cref{sec:overview-space}, we account for the auxiliary
structures used by the reductions.
Their space and construction costs appear as additive terms in the initial
bounds. The space lower bounds from \cref{sec:space-lower-bounds} allow us to
absorb these terms into the space and preprocessing bounds of the data
structures to which we reduce, completing the equivalences.

\subsection{From Prefix Queries to SA and ISA Queries}
\label{sec:overview-prefix-to-suffix}

We begin by explaining how prefix select and prefix special rank queries can
be answered using suffix array and inverse suffix array queries, respectively.
Both reductions use the same construction of a text from a sequence of strings.
Let $W[1\dd m]$ be a sequence of strings over $\IntegerAlphabet$, with
$\AlphabetSize\leq m$ and common string length
$\ell=\lfloor\log_{\AlphabetSize}m\rfloor$, as in our prefix-query problems.
Recall that prefix select asks for the $r$th index $i$ such that a given string
$X$ is a prefix of $W[i]$. Prefix special rank instead specifies $i$ and a
prefix length $p$, and counts the indices $j\leq i$ such that $W[j]$ has
prefix $W[i][1\dd p]$ (\cref{def:prefix-rank-and-select}). In both cases,
the order of the strings in the sequence matters. Our aim is to construct a
text in which the relevant suffixes appear in this same order in the suffix
array.

\myparagraph{The Text Construction}
The first step is to turn prefixes into substrings whose occurrences can be
found through a suffix-array interval. Reversing $W[i]$ turns a prefix $X$
into a suffix $\revstr{X}$ of $\revstr{W[i]}$. We place the symbol
$c=\AlphabetSize$ immediately after each reversed string. Since $c$ does not
occur in $W[i]$, the string $\revstr{X}\cdot c$ occurs at the end of
$\revstr{W[i]}\cdot c$ precisely when $X$ is a prefix of $W[i]$. It remains
to ensure that the suffixes starting at these occurrences are ordered by
their original indices $i$.

For this purpose, denote $k=\lceil\log_{\AlphabetSize}m\rceil$. For each
$i\in[0\dd m)$, let $s(i)$ be the length-$k$ base-$\AlphabetSize$
representation of $i$ with every digit increased by $\AlphabetSize$. Thus,
all symbols of $s(i)$ lie in $[\AlphabetSize\dd2\AlphabetSize)$, and
$s(i)$ precedes $s(j)$ lexicographically whenever $i<j$. We construct
\[
  \Text =
    \revstr{W[1]}\cdot c\cdot s(0)\cdot
    \revstr{W[2]}\cdot c\cdot s(1)\cdots
    \revstr{W[m]}\cdot c\cdot s(m-1).
\]
This is the sequence-to-string mapping of \cref{def:seq-to-string}.
Every block has length $\gamma=\ell+k+1$. Consider a nonempty string $X$
of length at most $\ell$. The occurrences of $\revstr{X}\cdot c$ in
$\Text$ correspond exactly to the indices $i$ for which $X$ is a prefix of
$W[i]$: the symbols from $\IntegerAlphabet$ occur only inside the reversed
strings, and $c$ immediately follows each such string. Moreover, the suffix
associated with index $i$ begins with $\revstr{X}\cdot c\cdot s(i-1)$.
The strings $s(i-1)$ are distinct and have equal length. Consequently, the
lexicographic order of these suffixes is determined by the order of the
corresponding indices $i$, regardless of the remaining symbols of $\Text$.
These properties are established in \cref{lm:seq-to-string}.

\myparagraph{Answering Queries}
We can now use the correspondence to answer both types of prefix queries.
The suffixes beginning with $\revstr{X}\cdot c$ occupy a contiguous
suffix-array interval, which we denote by $\SA{\Text}[b_X+1\dd e_X]$.
Thus, $b_X=\RangeBegTwo{\revstr{X}\cdot c}{\Text}$ and
$e_X=\RangeEndTwo{\revstr{X}\cdot c}{\Text}$ in the notation of
\cref{def:occ}. The two reductions use this interval as follows:
\begin{itemize}
\item For prefix select with arguments $r\in[1\dd m]$ and $X$, first
  check whether $r\leq e_X-b_X$. If this holds, the value
  $\SA{\Text}[b_X+r]$ is the position of the occurrence in the $r$th
  block containing $\revstr{X}\cdot c$. Since every block has length
  $\gamma$, we recover its index from this position in constant time as
  $\frac{1}{\gamma}\cdot(\SA{\Text}[b_X+r]+k+|X|)$. If $r>e_X-b_X$, the
  answer is $\infty$ (\cref{lm:reduce-prefix-select-to-sa}).
\item For prefix special rank, we retain the packed sequence $W$. Given
  $i$ and $p\geq1$, we extract $X=W[i][1\dd p]$ in constant time.
  The corresponding occurrence starts at the known
  position $q=\gamma i-(k+p)$ in the $i$th block. Its rank among the
  suffixes beginning with $\revstr{X}\cdot c$ equals the number of
  indices $j\leq i$ for which $X$ is a prefix of $W[j]$. Therefore, the answer is
  $\ISA{\Text}[q]-b_X$
  (\cref{lm:reduce-prefix-special-rank-to-isa}).
\end{itemize}
For the empty prefix, the answers are $r$ and $i$, respectively.

\myparagraph{Auxiliary Tables and Alphabet Conversion}
To obtain constant query overhead, we also need to find $b_X$ and $e_X$ in
constant time. Since $\ell=\lfloor\log_{\AlphabetSize}m\rfloor$, there are only
$\sum_{p=1}^{\ell}\AlphabetSize^p=\bigO(m)$ possible nonempty query
strings. We can therefore store the two interval boundaries for every such
string using $\bigO(m)$ words.
\Cref{pr:seq-to-str-range-beg-and-range-end} shows how to construct these
tables in $\bigO(m)$ time and support each lookup in $\bigO(1)$ time.

Finally, the construction increases the text length only by a constant factor
relative to the total length $m\ell$ of the input strings, but it uses
alphabet $[0\dd2\AlphabetSize)$. To obtain reductions over the original
alphabet $\IntegerAlphabet$, we apply our suffix-array and
inverse-suffix-array alphabet reductions
(\cref{th:general-suffix-array-alphabet-reduction-2,th:general-inverse-suffix-array-alphabet-reduction-2}).
They increase the length by another constant factor and allow the required
SA or ISA value to be recovered from one query with $\bigO(1)$ additional
time. Constructing the packed text and the auxiliary structures takes
$\bigO(m)$ time and $\bigO(m)$ words, in addition to constructing the data
structure for the resulting text; the packed text construction is given in
\cref{pr:seq-to-string-construction}.

We also need to account for these constant-factor increases in length while
preserving the input-size bound $N$. The main idea is to divide $W$ into a
constant number of consecutive blocks and apply the construction separately
to each block. Each resulting text has length at most $N$, including the
alphabet conversion. The stored interval boundaries let us count the strings
having a given prefix in each block. Using these counts, we can answer either
type of prefix query with at most one SA or ISA query, respectively, and
$\bigO(1)$ additional time. The complete reductions are given in
\cref{th:reduce-prefix-select-to-sa-for-all-alphabet-sizes,th:reduce-prefix-special-rank-to-isa-for-all-alphabet-sizes}.

\subsection{The Cell-Probe Lower Bound for SA Queries}
\label{sec:overview-cell-probe}

Next, we explain how the preceding reduction allows us to obtain a lower
bound for suffix-array queries from the Ball Inheritance problem. This
problem was introduced by Chan, Larsen, and P\u{a}tra\textcommabelow{s}cu~\cite{ChanLP11}.
Consider a perfect binary tree with $m$ leaves, where $m$ is a power of two,
and an ordered list $L$ of $m$ balls, with one ball assigned to each leaf.
For every node $v$, let $L_v$ be the subsequence of $L$ containing the balls
assigned to leaves in the subtree rooted at $v$, in their original order.
A query specifies $v$ and an index $i\in[1\dd|L_v|]$, and asks for the leaf
assigned to the $i$th ball in $L_v$.

To obtain a lower bound for suffix-array queries, we show how a suffix-array
data structure can be used to answer Ball Inheritance queries. We need this
reduction to use only $\bigO(m\log m)$ additional bits and $\bigO(1)$
additional probes per query. The preceding subsection already explains how
to answer prefix-select queries using a suffix-array data structure. It
therefore remains to show how to answer Ball Inheritance queries using
prefix select. We describe this step next, and then check the bounds for
the combined reduction before applying the known lower bound for Ball
Inheritance.

For this purpose, label left edges by $0$ and right edges by $1$. Let $I_v$
denote the binary string on the path from the root
to $v$. Number the leaves from $0$ to $m-1$ in left-to-right order. The
identifier of each leaf is then the length-$\log m$ binary representation
of its index. Let $W[j]$ be the identifier of the leaf assigned to
$L[j]$. Every string in $W$ has length $\log m$, exactly the length required
in our binary prefix-select problem.

The connection follows from the relation between tree paths and string
prefixes: a leaf lies below $v$ if and only if its identifier has prefix
$I_v$. Consequently, a Ball Inheritance query $(v,i)$ is answered by
computing $p=\PrefixSelect{W}{i}{I_v}$ and returning the integer represented
by $W[p]$. Here $p$ is the position in the original list $L$ of the $i$th
ball in $L_v$. Storing $W$ and the node identifiers requires
$\bigO(m\log m)$ additional bits, and retrieving the query string and the
requested leaf index takes $\bigO(1)$ additional probes. This reduction is
proved in \cref{lm:prefix-select-to-balls}.

We now apply the preceding reduction from prefix select to suffix-array
queries, including its conversion back to the binary alphabet. It constructs
a constant number of binary texts, each of length $\bigO(m\log m)$, and
builds a suffix-array structure for each text. The composed reduction uses
$\bigO(m\log m)$ bits in addition to these structures, and
each Ball Inheritance query makes at most one suffix-array query and
$\bigO(1)$ additional memory probes. In particular, the reduction applies
in the cell-probe model: the auxiliary values accessed by a query occupy
$\bigO(1)$ cells each, and the remaining arithmetic requires no probes.
The direct accounting of the additional probes is given in the proof of
\cref{th:lb-suffix-array}.

For a text-length bound $N$, choose the largest power of two $m$ satisfying
$m\log m\leq N$. Then $m\log m=\Theta(N)$, and the preceding reduction
ensures that every constructed text has length at most $N$.
Thus, a deterministic suffix-array data structure for
all binary texts of length at most $N$, using $S(N)=\Omega(N)$ bits and
answering queries in $Q(N)$ probes to cells of $\Theta(\log N)$ bits,
yields a Ball Inheritance data structure using $\bigO(S(N))$ bits and
answering queries in $\bigO(Q(N))$ probes. Applying the lower bound of
Gr{\o}nlund and Larsen~\cite{GronlundL16}, stated in \cref{th:lb-bi}, yields
\[
  Q(N)=\Omega\left(
    \frac{\log\log N}{\log((S(N)/N)\log\log N)}
  \right).
\]
This is the suffix-array lower bound of \cref{th:lb-suffix-array}.
In particular, a suffix-array data structure using $\bigO(N)$ bits must take
$\Omega(\log\log N/\log\log\log N)$ time in the worst case.

\subsection{From SA and ISA Queries to Prefix Queries}
\label{sec:overview-suffix-to-prefix}

We now turn to the reverse reductions. Given a text $\Text$ of length
$\Textlen$ over $\IntegerAlphabet$, our goal is to answer suffix array
and inverse suffix array queries using prefix queries on a sequence of
short strings. The construction follows the framework
of~\cite{breaking}, which treats periodic and nonperiodic positions
separately. We first explain this division and the role of prefix queries
in the nonperiodic case. We then describe the range-counting problem
that arises at periodic positions and explain how our new structure
removes its contribution to the query time. The formal reductions are
given in \cref{sec:reduce-sa-to-prefix-select,sec:reduce-isa-to-prefix-special-rank}.

For the main construction, assume that $\Text[\Textlen]$ occurs only
once in $\Text$, that $\AlphabetSize<\Textlen^{1/7}$, and choose
$\tau=\lfloor\tfrac17\log_{\AlphabetSize}\Textlen\rfloor\geq1$.
A position $j\leq\Textlen-3\tau+2$ is periodic when
$\Text[j\dd j+3\tau-1)$ has period at most $\tau/3$
(\cref{def:periodic-positions}). All other positions are nonperiodic.
Suffixes shorter than $3\tau-1$ are handled directly by the
short-string structures in \cref{sec:tools-packed-suffix-array-range-queries,sec:tools-packed-prefix-extraction-by-suffix-array-rank};
we first concentrate on the remaining nonperiodic positions.

\subsubsection{Nonperiodic Positions}
\label{sec:overview-nonperiodic}

The main tool used in our reductions from SA and ISA queries to prefix
queries is the notion of a string synchronizing set, introduced
in~\cite{sss}. A $\tau$-synchronizing set $\SSS$ is a
subset of text positions satisfying density and consistency conditions
(\cref{def:sss}). The density condition states that, for every
$j\in[1\dd\Textlen-3\tau+2]$, the interval $[j\dd j+\tau)$ contains a
position of $\SSS$ if and only if $j$ is nonperiodic. The consistency
condition requires membership in $\SSS$ to be the same for the starting
positions of equal length-$2\tau$ substrings. Consequently,
when two suffixes starting at nonperiodic positions have equal
length-$(3\tau-1)$ prefixes, the first position of $\SSS$ in each suffix
occurs at the same offset. We construct $\SSS$ of size
$\bigO(\Textlen/\tau)$ using \cref{th:sss-packed-construction}.

To define the sequence on which we will use prefix queries, we order
the positions of $\SSS$ by their suffixes. Let $s_1,\ldots,s_q$ be
these positions, ordered so that
$\Text[s_1\dd\Textlen]\prec\cdots\prec\Text[s_q\dd\Textlen]$. This
order can be computed in $\bigO(\Textlen/\tau)$ time using
\cref{th:sss-lex-sort}. We store $s_1,\ldots,s_q$ in an array and
define the sequence $W[1\dd q]$ by
\[
  W[i]
    =\revstr{\Textinf[s_i-\tau\dd s_i+2\tau)},
  \qquad i\in[1\dd q].
\]
Thus, $W[i]$ reverses a length-$3\tau$ substring
containing $\tau$ symbols before $s_i$ and $2\tau$ symbols starting
at $s_i$. The periodic extension $\Textinf$ supplies the symbols
when this interval extends to the left of the text. Its right endpoint
does not exceed $\Textlen+1$, since
$\SSS\subseteq[1\dd\Textlen-2\tau+1]$.

Before explaining the correspondence with suffixes, it is useful to
clarify how an SA query obtains a short prefix when only the suffix
rank is known. The structure from \cref{pr:rank-to-pref} returns, in
$\bigO(1)$ time, the first $3\tau-1$ symbols of the suffix of a
given rank, or the entire suffix if it is shorter. In the latter case,
its length immediately determines its starting position. In the former
case, these symbols will suffice to find the distinguishing prefix
used below. Thus, the query algorithm can obtain the needed prefix
information before it knows the suffix's starting position.

We now explain why prefix matches in $W$ identify the
required text suffixes. Consider a nonperiodic position
$j\leq\Textlen-3\tau+2$. Let $s$ be the first position of $\SSS$
at or after $j$ in text order, denote $\deltatext=s-j<\tau$, and
let $D=\Text[j\dd s+2\tau)$. This string has length
$\deltatext+2\tau$ and is called the \emph{distinguishing prefix} of
$\Text[j\dd\Textlen]$ (\cref{def:dist-prefixes}). The consistency
of $\SSS$ ensures that every occurrence of $D$ has its first
position of $\SSS$ at the same offset $\deltatext$. Consequently,
the indices $i$ for which $\revstr{D}$ is a prefix of
$W[i]$ correspond exactly to the occurrences of $D$
starting at $s_i-\deltatext$
(\cref{lm:reduce-suffix-array-to-prefix-select}). Moreover, each
corresponding text suffix is
\[
  D[1\dd\deltatext]\cdot\Text[s_i\dd\Textlen].
\]
Since the first factor is fixed, these suffixes have the same lexicographic
order as their indices in $W$.

Let $b=\RangeBegTwo{D}{\Text}$ be the number of text suffixes
lexicographically smaller than $D$. The two query algorithms use
this correspondence as follows:
\begin{itemize}
\item For an SA query with rank $r$, use the short prefix obtained
  from \cref{pr:rank-to-pref} to find $D$. The existence and uniqueness
  of this distinguishing prefix follow from
  \cref{lm:dist-prefixes,lm:dist-prefix-existence}, and
  \cref{pr:dist-prefix-lookup} finds it in $\bigO(1)$ time.
  Compute $\deltatext=|D|-2\tau$ and obtain $b$ using
  \cref{pr:range-beg-and-end}. A prefix select query with rank
  $r-b$ and string $\revstr{D}$ returns the index
  $i=\PrefixSelect{W}{r-b}{\revstr{D}}$.
  By \cref{lm:reduce-suffix-array-to-prefix-select}, the selected string
  $W[i]$ corresponds to the suffix of rank $r$, which starts
  $\deltatext$ positions before $s_i$. We therefore retrieve $s_i$ from
  the stored array and return $\SA{\Text}[r]=s_i-\deltatext$.
\item For an inverse SA query at position $j$, find the first
  position $s$ of $\SSS$ at or after $j$ in text order, using a bitvector
  marking $\SSS$. Then map $s$ to its index $i$ in suffix order, so that
  $s=s_i$, using a permutation from text order to suffix order. Both
  steps take $\bigO(1)$ time. Set $\deltatext=s_i-j$, extract
  $D=\Text[j\dd s_i+2\tau)$, and obtain $b$ using
  \cref{pr:range-beg-and-end}. Since $\revstr{D}$ is the
  length-$|D|$ prefix of $W[i]$,
  a prefix special rank query with arguments $i$ and $|D|$ counts
  the matching strings through index $i$. By
  \cref{lm:reduce-inverse-suffix-array-to-prefix-special-rank},
  adding $b$ gives
  $\ISA{\Text}[j]=b+\PrefixSpecialRank{W}{i}{|D|}$.
\end{itemize}
The detailed query algorithms, including the short-suffix cases, are
given in \cref{pr:nonperiodic-sa,pr:nonperiodic-isa}.

It remains to put this sequence into the input format required by the
prefix-query structures. The sequence $W$ contains
$\bigO(\Textlen/\tau)$ strings of length $3\tau$, and therefore has
total length $\bigO(\Textlen)$. The constructions in
\cref{pr:nonperiodic-sa,pr:nonperiodic-isa} extend the strings and
divide the sequence into a constant number of padded blocks satisfying
the exact length and cardinality constraints of the intermediate
prefix-query problems. Auxiliary prefix-count tables combine the
information from these blocks in $\bigO(1)$ time, leaving only one
prefix select or prefix special rank query. The sequence $W$ and auxiliary
structures can be constructed in
$\bigO(\Textlen/\log_{\AlphabetSize}\Textlen)$ time using
$\bigO(\Textlen\log\AlphabetSize)$ bits, in addition to the
construction costs of the prefix-query structures.

\subsubsection{Periodic Positions}
\label{sec:overview-periodic}

The periodic positions require a different argument because they have
no position of $\SSS$ within the next $\tau$ symbols. The construction
in~\cite{breaking} describes these positions using \emph{$\tau$-runs}:
substrings of $\Text$ of length at least $3\tau-1$ whose shortest period
is at most $\tau/3$ and which cannot be extended to either side without
increasing this period. Each maximal block of consecutive periodic
positions corresponds to a $\tau$-run. There are
$\bigO(\Textlen/\tau)$ such runs, and the sum of their lengths is
$\bigO(\Textlen)$~\cite{breaking}.

Using these runs, the construction reduces SA and inverse SA queries
at periodic positions to range counting and selection queries on arrays
of positive integers, with $\bigO(1)$ additional work
(\cref{pr:reduce-small-sum-range-count-to-periodic-sa-queries,pr:reduce-small-sum-range-count-to-periodic-isa-queries}).
Each array contains at most
$m=\Theta(\Textlen/\log_{\AlphabetSize}\Textlen)$ integers, and their
sum is $\bigO(\Textlen)$, and hence also $\bigO(m\log m)$. The counting
queries ask for
\[
  \TwoSidedRangeCount{A}{j}{v}
    =|\{x\in[1\dd j]:A[x]\geq v\}|.
\]
A periodic SA query uses one range counting query and one range
selection query, while a periodic inverse SA query uses a constant
number of range counting queries. The selection query already takes
$\bigO(1)$ time in~\cite{breaking}, but its counting structure takes
$\bigO(\log\log m)$ time. This is the source of the additional term
in the earlier reductions.

Our bounded-sum structure (\cref{th:range-count-for-small-sum})
answers both counting and selection queries in $\bigO(1)$ time
with $\bigO(m)$ preprocessing time and words of space.
This gives constant-time
periodic SA and inverse SA queries within the required preprocessing
and space bounds
(\cref{pr:periodic-pos-optimal-sa,pr:periodic-pos-optimal-isa}).

\subsubsection{Combining the Reductions}
\label{sec:overview-combining}

To combine the two cases, we use auxiliary structures that determine
in $\bigO(1)$ time whether a query belongs to the periodic or
nonperiodic case, including when the query
specifies only a suffix rank (\cref{pr:sa-core,pr:isa-core}). Thus,
an arbitrary SA or inverse SA query requires at most one query of the
corresponding prefix type and $\bigO(1)$ additional time.

Finally, the assumptions used to describe the construction do not
restrict the final results. We supply a unique final symbol by
enlarging the alphabet and use our alphabet reductions to return to
the original alphabet and the canonical prefix-query input format;
see \cref{th:reduce-sa-to-prefix-select-for-nearly-all-alphabet-sizes,th:reduce-isa-to-prefix-special-rank-for-nearly-all-alphabet-sizes}.
For large alphabets, with
$\log\AlphabetSize=\Theta(\log\Textlen)$, constructing and storing
the full SA or inverse SA already fits the required
$\bigO(\Textlen/\log_{\AlphabetSize}\Textlen)$ time and
$\bigO(\Textlen\log\AlphabetSize)$ bits. This completes the
reductions for all alphabet sizes
(\cref{th:reduce-sa-to-prefix-select-for-all-alphabet-sizes,th:reduce-isa-to-prefix-special-rank-for-all-alphabet-sizes}).

\subsection{Range Counting on Bounded-Sum Instances}
\label{sec:overview-range-count}

We now describe the range-counting structure used in the periodic case
(\cref{th:range-count-for-small-sum}).
Specifically, we describe a data structure that, for integers
$2\leq m\leq N$ and a nonempty array $A[1\dd m']$ of positive
integers satisfying $m'\leq m$, $\sum_{t=1}^{m'}A[t]\leq N$, and
$\sum_{t=1}^{m'}A[t]=\bigO(m\log m)$, answers range-counting queries
in $\bigO(1)$ time. The structure occupies $\bigO(m)$ words of space
and can be constructed in $\bigO(m)$ time, assuming $w\geq1+\log N$.
Each query specifies $j\in[0\dd m']$ and a threshold $v\in\Zn$,
and asks how many of $A[1],\ldots,A[j]$ are at least $v$
(\cref{def:range-count}).

\myparagraph{The Data Structure}
The main idea is to group the possible values of the threshold $v$ into
intervals $[2^i\dd2^{i+1})$. For every threshold $v$ in such an interval, positions $t$ with $A[t]<2^i$
can be ignored. We store the answers to comparisons with these thresholds
only for the remaining positions, retaining their original order.
This reduces the space required for large thresholds, because the bound
on the sum limits the number of large values in $A$.

Denote $a_{\max}=\max_{t\in[1\dd m']}A[t]$ and
$i_{\max}=\lfloor\log_2 a_{\max}\rfloor$.
For each $i\in[0\dd i_{\max}]$, let $B_i[1\dd m']$ be the bitvector
satisfying $B_i[t]=\one$ if and only if $A[t]\geq2^i$.
Denote its number of ones by $m_i$, and let $P_i[1\dd m_i]$ contain
the positions of these ones in ascending order.
A rank query on $B_i$ translates a prefix of $A$ into a prefix of the
subsequence indexed by $P_i$: exactly $\Rank{B_i}{j}{\one}$ of the
first $j$ positions remain.

For every $v\in[2^i\dd2^{i+1})$, let $C_v[1\dd m_i]$ satisfy
$C_v[t]=\one$ if and only if $A[P_i[t]]\geq v$.
These vectors all have the same length, and we store their concatenation
\[
  C'_i=C_{2^i}C_{2^i+1}\cdots C_{2^{i+1}-1}.
\]
Thus, $C_v$ begins at offset $(v-2^i)m_i$ in $C'_i$.
A rank query on $C_v$ reduces to two rank queries on $C'_i$.

The assumption on the sum bounds the total length of these concatenations.
For each $i$, the length of $C'_i$ is $m_i2^i$, and each original
position $t$ contributes $2^i$ to this length precisely when $2^i\leq A[t]$.
This occurs for $i=0,\ldots,\lfloor\log_2 A[t]\rfloor$. Therefore,
\[
  \sum_{i=0}^{i_{\max}}|C'_i|
    =\sum_{i=0}^{i_{\max}}m_i2^i
      =\sum_{t=1}^{m'}\sum_{i=0}^{\lfloor\log_2 A[t]\rfloor}2^i
    <2\sum_{t=1}^{m'}A[t]=\bigO(m\log m).
\]
There are $\bigO(\log m)$ vectors $B_i$, each of length at most $m$,
and hence their total length is also $\bigO(m\log m)$.

We need one additional bitvector to determine the group containing a
query threshold $v$. Let $B[1\dd a_{\max}]$ mark the powers of two.
For $1\leq v\leq a_{\max}$, we have
$\lfloor\log_2 v\rfloor=\Rank{B}{v}{\one}-1$.
The length of $B$ is $\bigO(m\log m)$.
We store $B$, the $B_i$, and the $C'_i$, all with constant-time rank
support from \cref{th:bin-rank-select}, together with $a_{\max}$ and
the lengths $m_i$. Together they occupy $\bigO(m)$ words.

\myparagraph{Answering Queries}
For a query $(j,v)$, the answer is $j$ when $v=0$ and zero when
$v>a_{\max}$. Otherwise, we proceed in three steps:
\begin{enumerate}
\item Compute $i=\Rank{B}{v}{\one}-1$, identifying the interval
  $[2^i\dd2^{i+1})$ containing $v$.
\item Compute $j'=\Rank{B_i}{j}{\one}$, the number of positions
  among the first $j$ whose values are at least $2^i$.
\item Set $d=(v-2^i)m_i$ and return
  $\Rank{C'_i}{d+j'}{\one}-\Rank{C'_i}{d}{\one}$.
  This gives the number of ones in the first $j'$ positions of $C_v$,
  i.e., the number of values at least $v$ among $A[1],\ldots,A[j]$.
\end{enumerate}
The query uses four rank operations and $\bigO(1)$ arithmetic operations.
For example, suppose $A$ contains $5,1,9,3,6$, and consider $j=4$
and $v=6$. The relevant group starts at $4$, and its subsequence is
$5,9,6$. Only its first two positions come from $A[1\dd4]$.
Since $C_6=\texttt{011}$, the answer is one, corresponding to $A[3]=9$.

\myparagraph{Construction}
It remains to construct the packed bitvectors in $\bigO(m)$ time.
Their total length can be $\Theta(m\log m)$, and therefore we cannot
construct them one bit at a time. The construction in the proof of
\cref{th:range-count-for-small-sum} instead processes changes in their
bits and constructs repeated parts in batches.
After radix sorting the pairs $(A[t],t)$ in ascending order of their
first coordinates in $\bigO(m)$ time, we construct the two families
as follows:
\begin{itemize}
\item For $B_0,B_1,\ldots$, begin with an all-one bitvector.
  As the threshold $2^i$ increases, clear the positions whose values
  have fallen below it, and save a packed copy for each $i$.
  Each position is cleared at most once, and copying all
  $\bigO(\log m)$ packed bitvectors takes $\bigO(m)$ time.
  Build rank support for this family before proceeding to the construction
  of the $C'_i$. We account for this cost below.
\item For $C'_i$, begin with $C_{2^i}=\one^{m_i}$.
  As the threshold $v$ increases within $[2^i\dd2^{i+1})$,
  a bit changes to zero when $v$ passes the corresponding value of $A$.
  A change caused by $A[t]$ occurs at position
  $\Rank{B_i}{t}{\one}$ in the current bitvector.
  For each maximal interval of $r\geq1$ consecutive thresholds with
  identical bitvectors $C_v$, we use packed string repetition (\cref{pr:packed-string-repetition})
  to construct their concatenation in one call. Each position of
  $A$ causes a bit change for at most one index $i$. Concatenate the resulting packed blocks to obtain $C'_i$.
\end{itemize}

Repeating a length-$m_i$ bitvector $r\geq1$ times takes
$\bigO(1+\log r+rm_i/w)$ time by
\cref{pr:packed-string-repetition}.
The repetition calls produce a total of
$\sum_{i=0}^{i_{\max}}|C'_i|=\bigO(m\log m)$ bits. Since $w\geq\log m$,
their $rm_i/w$ terms sum to $\bigO(m\log m/w)=\bigO(m)$.
To bound the remaining $1+\log r$ terms, distinguish two cases.
\begin{itemize}
\item For $2^i<m$, denote the numbers of copies produced by the
  calls for $C'_i$ by $r_1,\ldots,r_h$. These calls together
  produce all $2^i$ bitvectors $C_v$ in $C'_i$, and hence
  $\sum_{j=1}^h r_j=2^i$. Since $1+\log r_j\leq2r_j$, their
  $1+\log r_j$ terms sum to at most $2^{i+1}$. Summing over
  all indices $i$ with $2^i<m$ gives $\bigO(m)$.
\item For $2^i\geq m$, only positions $t$ with $A[t]\geq m$ can
  cause bit changes. The bound
  $\sum_{t=1}^{m'}A[t]=\bigO(m\log m)$ implies that there are
  $\bigO(\log m)$ such positions in total. Each causes changes
  for at most one $i$. Thus, there are $\bigO(\log m)$ repetition
  calls, including one final call per index $i$. Each call has
  $r\leq2^i\leq a_{\max}=\bigO(m\log m)$ and therefore
  $1+\log r=\bigO(\log m)$. The total is
  $\bigO(\log^2m)=\bigO(m)$.
\end{itemize}
Initializing the bitvectors, applying the changes, and concatenating
the blocks also take $\bigO(m)$ time.

The bitvector $B$ is constructed by initializing its packed representation
to zeros and setting its $\bigO(\log m)$ marked positions.
Rank support is built for each family once its packed bitvectors are
available. Each of the $\bigO(\log m)$ bitvectors $B_i$ has length at
most $m$ and is augmented separately in $\bigO(1+m/\log m)$ time.
For the bitvectors $C'_i$, of total length $\bigO(m\log m)$, we use the
batched construction in \cref{th:bin-rank-select}, which takes
$\bigO(\log m+(m\log m)/\log(m\log m))$ time.
Both bounds are $\bigO(m)$.
This completes the $\bigO(m)$-time preprocessing.

\subsection{Space Lower Bounds}\label{sec:overview-space}

We now return to the question of preserving all four complexity bounds in
our equivalences. The reductions described above require auxiliary structures
in addition to the data structure for the problem to which we reduce. For
inputs containing at most $N$ symbols over $\IntegerAlphabet$, these auxiliary
structures occupy $\bigO(N\log\AlphabetSize)$ bits and take
$\bigO(N/\log_{\AlphabetSize}N)$ time to construct. Consequently, starting
with a data structure using $S(\AlphabetSize,N)$ bits initially gives a bound
of $\bigO(S(\AlphabetSize,N)+N\log\AlphabetSize)$ bits. To obtain the claimed
equivalences, we must show that the additive terms do not increase the
asymptotic complexities.

It is tempting to regard the space term as unavoidable simply because the
input contains $N$ symbols. However, after preprocessing, a data structure
need only retain enough information to answer its queries. It need not retain
the original input, and different inputs may have identical answers to every
query. For example, the texts $\zero^{\Textlen}$ and $\one^{\Textlen}$ have
the same suffix array. In \cref{sec:space-lower-bounds}, we therefore prove
an $\Omega(N\log\AlphabetSize)$-bit worst-case space lower bound for each
of the four query problems.

The common principle is to count distinct sequences of query answers. If
there are $M$ such sequences, a data structure must have at least $M$
different stored representations, requiring at least $\log M-1$ bits in the
worst case. For some query types, the answers determine the input itself.
For others, we need to show that sufficiently many inputs remain
distinguishable even though not all of them are. The time required to
reconstruct an input from its answers does not affect this counting argument.

\subsubsection{Reconstruction Arguments}\label{sec:overview-reconstruction}

We first show how to reconstruct the input from the answers to
prefix-select queries.
Consider queries whose prefix has the full length $\ell$ of a string in $W$.
For every $X\in\IntegerAlphabet^{\ell}$, these queries list the positions
at which $W[i]=X$. Thus, the answers determine the entire sequence $W$, giving
a lower bound of $\Omega(\Seqlen\ell\log\AlphabetSize)$ bits for $\Seqlen$
strings of length $\ell$. The formal argument expresses this as a reduction
from select over alphabet $[0\dd\AlphabetSize^{\ell})$
(\cref{lm:reduce-select-to-prefix-select,th:prefix-select-space-lower-bound}).

For suffix-array and inverse-suffix-array queries, we can reconstruct a text
$\Text\in\IntegerAlphabet^{\Textlen}$ if we additionally know the number
$C(a)$ of occurrences of each symbol $a\in\IntegerAlphabet$. Given a position
$j$, let $r=\ISA{\Text}[j]$. The value $\Text[j]$ is the unique symbol $c$
satisfying $\sum_{a=0}^{c-1}C(a)<r\leq\sum_{a=0}^{c}C(a)$, because suffixes
beginning with the same symbol form a consecutive block in lexicographic
order. An inverse-suffix-array structure returns $r$ directly. A suffix-array
structure also determines $r$, by finding the index at which its answer is $j$.

The symbol counts can be stored in
$\bigO(\min\{\AlphabetSize\log\Textlen,\Textlen\})$ bits
(\cref{pr:symbol-counts}), which is $o(\Textlen\log\AlphabetSize)$ uniformly
for $2\leq\AlphabetSize\leq\Textlen$. In particular, they require only
$\bigO(\log\Textlen)$ bits for a binary alphabet. A representation of SA or
ISA using substantially fewer than $\Textlen\log\AlphabetSize$ bits would
therefore give a representation of every text using fewer than
$\Textlen\log\AlphabetSize$ bits, contradicting the number
$\AlphabetSize^{\Textlen}$ of possible texts. This proves the required lower
bounds (\cref{sec:suffix-array-space-lower-bound,sec:inverse-suffix-array-space-lower-bound}).
Appendix~\ref{app:stronger-lower-bound} gives stronger explicit bounds by counting
distinct suffix arrays directly.

\subsubsection{Special Rank and Young Tableaux}\label{sec:overview-special-rank}

We next turn to special rank, where the reconstruction argument needs more
care. Recall that a query at position $j$ returns
$\SpecialRank{\Text}{j}=|\{i\in[1\dd j]:\Text[i]=\Text[j]\}|$
(\cref{def:rank-select}). Let $R_{\Text}$ denote the sequence
$(\SpecialRank{\Text}{1},\ldots,\SpecialRank{\Text}{\Textlen})$
of all answers. This sequence gives the occurrence
number of the symbol at each position, but does not identify the symbol.
Even after fixing the order in which distinct symbols first appear, many
different texts can have the same answer sequence: for example,
$\texttt{abcabc}$ and $\texttt{abccba}$ both give $(1,1,1,2,2,2)$. We need
a large family of texts for which this ambiguity does not occur.

We obtain such a family from rectangular Young tableaux
(\cref{lm:tableau-special-rank-family}). To explain the construction, suppose
that $\AlphabetSize$ divides $\Textlen$ and $\AlphabetSize\leq\sqrt{\Textlen}$,
and denote $q=\Textlen/\AlphabetSize$. Let $\mathcal M$ be the set of
$\AlphabetSize\times q$ matrices containing the numbers $1,\ldots,\Textlen$
exactly once, in ascending order in every row and every column. For each
$M\in\mathcal M$ and $t\in[1\dd\Textlen]$, we define $\Text_M[t]$ to be the
index of the row containing $t$, minus one. Because the numbers in a row are
increasing, the answer $\SpecialRank{\Text_M}{t}$ is precisely the index of
the column containing $t$. For example,
\[
  M=\begin{pmatrix}
    1&3&4\\
    2&5&8\\
    6&7&9
  \end{pmatrix}
\]
gives $\Text_M=\texttt{010012212}$ and
$R_{\Text_M}=(1,1,2,3,2,1,2,3,3)$.

We can reconstruct $M$ from either the text $\Text_M$ or the answer
sequence $R_{\Text_M}$ as follows:
\begin{itemize}
\item The text $\Text_M$ specifies the numbers belonging to each row.
  Placing those numbers in ascending order reconstructs $M$.
\item The answer sequence $R_{\Text_M}$ specifies the numbers belonging to
  each column. Placing those numbers in ascending order also reconstructs $M$.
\end{itemize}
Thus, different tableaux give different texts and different answer sequences.
It remains to count the tableaux. The hook-length formula
(see \cref{lm:tableau-special-rank-family}), together with $q\geq\AlphabetSize$
and $\Textlen!\geq(\Textlen/e)^{\Textlen}$, gives
\[
  |\mathcal M|
    =\frac{\Textlen!}
      {\textstyle\prod_{i=0}^{\AlphabetSize-1}\prod_{j=0}^{q-1}(i+j+1)}
    \geq\frac{\Textlen!}{(q+\AlphabetSize)^{\Textlen}}
    \geq\left(\frac{\AlphabetSize}{2e}\right)^{\Textlen}.
\]
For sufficiently large $\AlphabetSize$, this is
$\AlphabetSize^{\Omega(\Textlen)}$, yielding the
$\Omega(\Textlen\log\AlphabetSize)$-bit space lower bound. For bounded
$\AlphabetSize$, a separate binary family already gives
$2^{\Omega(\Textlen)}$ distinct answer sequences
(\cref{lm:binary-special-rank-family}). For $\AlphabetSize>\sqrt{\Textlen}$,
we use tableaux with $\lfloor\sqrt{\Textlen}\rfloor$ rows and enough columns
to contain $\Theta(\Textlen)$ numbers, and append zeros to make the resulting
texts have length $\Textlen$. Since
$\log\sqrt{\Textlen}=\Theta(\log\AlphabetSize)$ throughout this range, the
same asymptotic lower bound follows. \Cref{th:special-rank-space-lower-bound}
handles these cases and removes the divisibility assumption.

To transfer this bound to prefix special rank, we represent every symbol
$A[i]$ of a string $A\in[0\dd\AlphabetSize^{\ell})^{\Seqlen}$ by its
length-$\ell$ base-$\AlphabetSize$ representation $W[i]$. A prefix-special-rank
query with prefix length $\ell$ then returns
$|\{j\in[1\dd i]:W[j]=W[i]\}|=\SpecialRank{A}{i}$. Thus, any
prefix-special-rank structure is also a special-rank structure for $A$,
without any additional stored information
(\cref{lm:reduce-special-rank-to-prefix-special-rank}). For
$\ell=\lfloor\log_{\AlphabetSize}\Seqlen\rfloor$, we have
$\AlphabetSize^{\ell}\leq\Seqlen$, and the special-rank lower bound gives
$\Omega(\Seqlen\ell\log\AlphabetSize)$ bits, matching the total size of the
strings (\cref{th:prefix-special-rank-asymptotic-space-lower-bound}).

\myparagraph{Completing the Equivalences}
We can now use these lower bounds to remove the additive terms from our
reductions. The corresponding bounded-input-size statements establish
$S(\AlphabetSize,N)=\Omega(N\log\AlphabetSize)$ for all four problems
(\cref{th:suffix-array-bounded-length-space-lower-bound,th:inverse-suffix-array-bounded-length-space-lower-bound,th:prefix-select-bounded-total-length-space-lower-bound,th:prefix-special-rank-bounded-total-length-space-lower-bound}).
Hence, the auxiliary space in our reductions is $\bigO(S(\AlphabetSize,N))$.
These lower bounds also account for the preprocessing overhead: with
$w=\Theta(\log N)$, producing an explicit structure of
$\Omega(N\log\AlphabetSize)$ bits requires
$\Omega(N/\log_{\AlphabetSize}N)$ time, and its size also lower-bounds
preprocessing space. Together with the constant query overhead, this yields
reductions preserving index space, preprocessing time and space, and query
time up to constant factors, while preserving the alphabet and the input-size
bound exactly (\cref{sec:prefix-select-and-sa-summary,sec:prefix-special-rank-and-isa-summary}).

%% file: tools.tex
\section{Tools}\label{sec:tools}

\input{tools/packed-integer-conversion}

\input{tools/packed-alphabet-lifting}

\input{tools/packed-string-reversal}
\input{tools/packed-string-repetition}
\input{tools/packed-pointwise-concatenation-and-partition}

\input{tools/packed-alphabet-mapping}

\input{tools/packed-extended-alphabet-mapping}
\input{tools/packed-morphisms}
\input{tools/packed-prefix-frequencies}
\input{tools/packed-substring-frequencies}
\input{tools/packed-suffix-array-range-queries}

\input{tools/packed-prefix-extraction-by-suffix-array-rank}
\input{tools/periodicity-testing}
\input{tools/recognizing-periodic-positions}

\input{tools/packed-sequence-to-string-mapping}
\input{tools/sparse-suffix-sorting}

\input{tools/optimal-range-count-for-bounded-sum-instances}

%% file: tools/packed-integer-conversion.tex
\subsection{Integer Values of Packed Strings}\label{sec:tools-integer-coding-of-variable-length-strings}

\begin{definition}[Integer value of a string]\label{def:val}
  Let $\AlphabetSize\in\Z_{\geq2}$. For every
  $X\in\IntegerAlphabet^+$, we define
  $\Val{\AlphabetSize}{X}:=\sum_{i=1}^{|X|}X[i]\cdot
    \AlphabetSize^{|X|-i}\in[0\dd\AlphabetSize^{|X|})$.
  Thus, $X$ is the base-$\AlphabetSize$ representation of
  $\Val{\AlphabetSize}{X}$, padded on the left with zeros to
  length $|X|$. We also define $\Val{\AlphabetSize}{\emptystring}=0$.
\end{definition}

\begin{example}\label{ex:val}
  Let $\AlphabetSize=4$ and $X=(0,2,1)$. Then
  $\Val{4}{X}=0\cdot4^2+2\cdot4+1=9$; the base-$4$ representation of
  $9$, padded to length $|X|=3$, is $021$.
  We also have $\Val{4}{\emptystring}=0$.
\end{example}

\begin{theorem}[{\cite{hierarchy}}]\label{th:val-encoding}
  Let $N,\AlphabetSize\in\Z_{\geq2}$ satisfy $\AlphabetSize\leq N$,
  and denote $\ell=\lfloor\log_{\AlphabetSize}N\rfloor$.
  In the word RAM model with word size $w\geq1+\log N$, given $N$ and
  $\AlphabetSize$, we can in $\bigO(\sqrt N)$ time construct a data structure of size
  $\bigO(\sqrt N)$ words that, given $|X|$ and the packed representation
  $\PackedRepresentation{w}{\AlphabetSize}{X}$ (\cref{def:packed-representation})
  of any string $X\in\IntegerAlphabet^{\leq\ell}$, returns
  $\Val{\AlphabetSize}{X}$ (\cref{def:val}) in $\bigO(1)$ time.
\end{theorem}

\begin{theorem}[{\cite{hierarchy}}]\label{th:val-decoding}
  Let $N,\AlphabetSize\in\Z_{\geq2}$ satisfy $\AlphabetSize\leq N$,
  and denote $\ell=\lfloor\log_{\AlphabetSize}N\rfloor$.
  In the word RAM model with word size $w\geq1+\log N$, given $N$ and
  $\AlphabetSize$, we can in $\bigO(\sqrt N)$ time construct a data structure of size
  $\bigO(\sqrt N)$ words that, given $|X|$ and the integer
  $\Val{\AlphabetSize}{X}$ (\cref{def:val}) for any string
  $X\in\IntegerAlphabet^{\leq\ell}$, returns the packed representation
  $\PackedRepresentation{w}{\AlphabetSize}{X}$
  (\cref{def:packed-representation}) in $\bigO(1)$ time.
\end{theorem}

%% file: tools/packed-alphabet-lifting.tex
\subsection{Packed Alphabet Lifting}\label{sec:tools-alphabet-lifting}

\begin{definition}[Increase operation]\label{def:increase}
  Let $\AlphabetSize\in\Z_{\geq2}$, $S\in\IntegerAlphabet^*$, and
  $\delta\in\Zn$. The string
  $\increase{\delta}{S}\in
    [\delta\dd\delta+\AlphabetSize)^{|S|}$ is obtained by adding
  $\delta$ to every symbol of $S$, i.e.,
  \[
    \increase{\delta}{S}:=
      \textstyle\bigodot_{i=1}^{|S|}(S[i]+\delta).
  \]
\end{definition}

\begin{example}\label{ex:increase}
  Let $\AlphabetSize=2$, $S=(0,1,1,0)$, and $\delta=2$. Then
  $\increase{2}{S}=(2,3,3,2)$.
\end{example}

\begin{theorem}[{\cite{hierarchy}}]\label{th:alphabet-lifting}
  Let $N,\AlphabetSize_1,\AlphabetSize_2\in\Z_{\geq2}$ satisfy
  $\AlphabetSize_1\leq\AlphabetSize_2\leq N$.
  In the word RAM model with word size $w\geq1+\log N$, given $N$,
  $\AlphabetSize_1$, and $\AlphabetSize_2$, we can in
  $\bigO(\sqrt N)$ time construct a data structure of size
  $\bigO(\sqrt N)$ words that, given $|S|$ and the packed representation
  $\PackedRepresentation{w}{\AlphabetSize_1}{S}$
  (\cref{def:packed-representation}) of any string
  $S\in[0\dd\AlphabetSize_1)^{\leq N}$, returns the packed representation
  $\PackedRepresentation{w}{\AlphabetSize_2}{S}$ of $S$ interpreted
  over alphabet $[0\dd\AlphabetSize_2)$ in
  $\bigO(1+|S|/\log_{\AlphabetSize_2}N)$ time.
  More generally, given also an integer
  $\delta\in[0\dd\AlphabetSize_2-\AlphabetSize_1]$, the data structure
  returns $\PackedRepresentation{w}{\AlphabetSize_2}{R}$ within the same
  time bound, where $R\in[0\dd\AlphabetSize_2)^{|S|}$ satisfies
  $R[j]=S[j]+\delta$ for every $j\in[1\dd|S|]$.
\end{theorem}

%% file: tools/packed-string-reversal.tex
\subsection{Packed String Reversal}\label{sec:tools-string-reversal}

\begin{theorem}[{\cite{hierarchy}}]\label{th:string-reversal}
  Let $N,\AlphabetSize\in\Z_{\geq2}$ satisfy $\AlphabetSize\leq N$.
  In the word RAM model with word size $w\geq1+\log N$, given $N$ and
  $\AlphabetSize$, we can in $\bigO(\sqrt N)$ time construct a data structure of size
  $\bigO(\sqrt N)$ words that, given $|S|$ and the packed representation
  $\PackedRepresentation{w}{\AlphabetSize}{S}$
  (\cref{def:packed-representation}) of any string
  $S\in\IntegerAlphabet^{\leq N}$, returns the packed representation
  $\PackedRepresentation{w}{\AlphabetSize}{\revstr{S}}$ of $\revstr{S}$
  (\cref{sec:prelim-basic}) in $\bigO(1+|S|/\log_{\AlphabetSize}N)$ time.
\end{theorem}

%% file: tools/packed-string-repetition.tex
\subsection{Packed String Repetition}
\label{sec:tools-packed-string-repetition}

\begin{proposition}[{\cite{hierarchy}}]\label{pr:packed-string-repetition}
  Let $N,\AlphabetSize\in\Z_{\geq2}$ satisfy $\AlphabetSize\leq N$,
  and assume that $\AlphabetSize$ is known and that
  $b=\lceil\log\AlphabetSize\rceil$ has been computed.
  Consider the word RAM model with word size $w>\log N$.
  Given the length $|S|$ and packed representation
  $\PackedRepresentation{w}{\AlphabetSize}{S}$ of a nonempty string
  $S\in\IntegerAlphabet^{\leq N}$ (\cref{def:packed-representation}),
  and an integer $m\in[0\dd N]$, we can compute the packed
  representation of the length-$m$ prefix
  \[
    R=S^\infty[1\dd m]\in\IntegerAlphabet^m,
  \]
  where $S^\infty$ denotes the infinite repetition of $S$
  (\cref{sec:prelim-basic}).
  The algorithm takes
  \[
    \bigO\left(1+\log\left(1+\frac{m}{|S|}\right)
      +\frac{m\log\AlphabetSize}{w}\right)
  \]
  time and uses
  $\bigO(1+(|S|+m)\log\AlphabetSize/w)$ words of space,
  including the input and output.
\end{proposition}

%% file: tools/packed-pointwise-concatenation-and-partition.tex
\subsection{Packed Pointwise Concatenation and Partition}
  \label{sec:tools-packed-pointwise-concatenation-partition}

\begin{theorem}[{\cite{hierarchy}}]\label{th:packed-pointwise-concatenation-partition}
  Let $N,\AlphabetSize\in\Z_{\geq2}$ satisfy $\AlphabetSize\leq N$.
  Consider the word RAM model with word size $w\geq1+\log N$.
  Given $N$ and $\AlphabetSize$, we can construct in $\bigO(\sqrt N)$ time
  a data structure of size
  $\bigO(\sqrt N)$ words supporting pointwise concatenation and partition
  as follows. Each query specifies $m\in[0\dd N]$ and
  $\ell_1,\ell_2\in\Z_{\geq0}$ satisfying
  $1\leq\ell_1+\ell_2\leq N$ and $m(\ell_1+\ell_2)\leq N$. For sequences
  $U\in(\IntegerAlphabet^{\ell_1})^m$ and
  $V\in(\IntegerAlphabet^{\ell_2})^m$, let
  $W\in(\IntegerAlphabet^{\ell_1+\ell_2})^m$ be the sequence satisfying
  \[
    W[i]=U[i]V[i]\qquad\text{for every }i\in[1\dd m].
  \]
  The structure supports the following two operations on packed representations of string sequences
  (\cref{def:packed-sequence-representation}):
  \begin{enumerate}
  \item Given $\PackedSeqRepresentation{w}{\AlphabetSize}{U}$ and
    $\PackedSeqRepresentation{w}{\AlphabetSize}{V}$, construct
    $\PackedSeqRepresentation{w}{\AlphabetSize}{W}$.
  \item Given $\PackedSeqRepresentation{w}{\AlphabetSize}{W}$, construct
    $\PackedSeqRepresentation{w}{\AlphabetSize}{U}$ and
    $\PackedSeqRepresentation{w}{\AlphabetSize}{V}$.
  \end{enumerate}
  Each operation takes
  $\bigO(1+m(\ell_1+\ell_2)/\log_{\AlphabetSize}N)$ time.
\end{theorem}

\begin{theorem}[{\cite{hierarchy}}]\label{th:packed-pointwise-concatenation-partition-offline}
  Let $N,\AlphabetSize\in\Z_{\geq2}$ satisfy $\AlphabetSize\leq N$.
  Consider the word RAM model with word size $w\geq1+\log N$.
  Let $m\in[0\dd N]$ and $\ell_1,\ell_2\in\Z_{\geq0}$ satisfy
  $1\leq\ell_1+\ell_2\leq N$ and $m(\ell_1+\ell_2)\leq N$. For sequences
  $U\in(\IntegerAlphabet^{\ell_1})^m$ and
  $V\in(\IntegerAlphabet^{\ell_2})^m$, let
  $W\in(\IntegerAlphabet^{\ell_1+\ell_2})^m$ be the sequence satisfying
  \[
    W[i]=U[i]V[i]\qquad\text{for every }i\in[1\dd m].
  \]
  Given $N$, $\AlphabetSize$, and $m,\ell_1,\ell_2$, we can perform the following
  two tasks on packed representations of string sequences
  (\cref{def:packed-sequence-representation}):
  \begin{enumerate}
  \item Given $\PackedSeqRepresentation{w}{\AlphabetSize}{U}$ and
    $\PackedSeqRepresentation{w}{\AlphabetSize}{V}$, construct
    $\PackedSeqRepresentation{w}{\AlphabetSize}{W}$.
  \item Given $\PackedSeqRepresentation{w}{\AlphabetSize}{W}$, construct
    $\PackedSeqRepresentation{w}{\AlphabetSize}{U}$ and
    $\PackedSeqRepresentation{w}{\AlphabetSize}{V}$.
  \end{enumerate}
  Each task takes
  $\bigO(\sqrt N+m(\ell_1+\ell_2)/\log_{\AlphabetSize}N)$ time
  and uses
  $\bigO(\sqrt N+m(\ell_1+\ell_2)/\log_{\AlphabetSize}N)$ words of space,
  including the input and output representations.
\end{theorem}

%% file: tools/packed-alphabet-mapping.tex
\Needspace{7\baselineskip}
\subsection{Packed Alphabet Mapping}\label{sec:tools-alphabet-mapping}

\paragraph{Definitions}\label{sec:tools-alphabet-mapping-definitions}

\begin{definition}[Alphabet mapping]\label{def:alphabet-map}
  Let $\AlphabetSize_1, \AlphabetSize_2 \in \Z_{\geq 2}$ be such that $\AlphabetSize_1 \geq \AlphabetSize_2$.
  Denote $k = \lceil \log_{\AlphabetSize_2} \AlphabetSize_1 \rceil$. For any $x \in [0 \dd \AlphabetSize_1)$,
  by $\AlphabetMap{\AlphabetSize_1}{\AlphabetSize_2}{x} \in [0 \dd \AlphabetSize_2)^k$ we denote the
  string of length $k$ containing the base-$\AlphabetSize_2$ representation of $x$
  (padded on the left side with leading zeros, if needed).
  For any string $S \in [0 \dd \AlphabetSize_1)^{*}$, we then let
  \[
    \AlphabetMap{\AlphabetSize_1}{\AlphabetSize_2}{S} :=
      \textstyle\bigodot_{i=1,\ldots,|S|} \AlphabetMap{\AlphabetSize_1}{\AlphabetSize_2}{S[i]}
      \in [0 \dd \AlphabetSize_2)^{|S|\cdot k}.
  \]
\end{definition}

\begin{example}\label{ex:alphabet-map}
  Let $\AlphabetSize_1=5$ and $\AlphabetSize_2=2$. Then $k=3$.
  For example, $\AlphabetMap{5}{2}{4}=\one\zero\zero$ and
  $\AlphabetMap{5}{2}{1}=\zero\zero\one$.
  For $S=(4,1,3)$, we obtain
  \[
    \AlphabetMap{5}{2}{S}
      =\one\zero\zero\cdot\zero\zero\one\cdot\zero\one\one.
  \]
\end{example}

\begin{remark}\label{rm:alphabet-map}
  Note that the choice
  $k = \lceil \log_{\AlphabetSize_2} \AlphabetSize_1 \rceil$ in \cref{def:alphabet-map} guarantees that every
  $x \in [0 \dd \AlphabetSize_1)$ admits a base-$\AlphabetSize_2$ representation of length $k$.
  Indeed, any $k$-digit base-$\AlphabetSize_2$ string represents a value in $[0 \dd \AlphabetSize_2^k)$, and since
  \[
    \AlphabetSize_2^{k}
    = \AlphabetSize_2^{\lceil \log_{\AlphabetSize_2} \AlphabetSize_1 \rceil}
    \geq \AlphabetSize_2^{\log_{\AlphabetSize_2} \AlphabetSize_1}
    = \AlphabetSize_1,
  \]
  it follows that $[0 \dd \AlphabetSize_1) \subseteq [0 \dd \AlphabetSize_2^k)$.
\end{remark}

\begin{observation}\label{ob:alphabet-map-properties}
  Let $\AlphabetSize_1,\AlphabetSize_2\in\Z_{\geq2}$ satisfy
  $\AlphabetSize_1\geq\AlphabetSize_2$.
  For every $A,B\in[0\dd\AlphabetSize_1)^*$, the following equivalences
  hold (see \cref{def:alphabet-map}):
  \begin{enumerate}
  \item\label{ob:alphabet-map-prefix-property}
    $A\text{ is a prefix of }B
      \quad\Longleftrightarrow\quad
      \AlphabetMap{\AlphabetSize_1}{\AlphabetSize_2}{A}
      \text{ is a prefix of }
      \AlphabetMap{\AlphabetSize_1}{\AlphabetSize_2}{B}$,
  \item\label{ob:alphabet-map-lex-property}
    $A\prec B
      \quad\Longleftrightarrow\quad
      \AlphabetMap{\AlphabetSize_1}{\AlphabetSize_2}{A}\prec
      \AlphabetMap{\AlphabetSize_1}{\AlphabetSize_2}{B}$.
  \end{enumerate}
\end{observation}

\paragraph{Small Alphabets}\label{sec:tools-alphabet-mapping-small-alphabet}

\begin{proposition}\label{pr:alphabet-map-small-sigma1}
  Let $N, \AlphabetSize_1, \AlphabetSize_2 \in \Z_{\geq 2}$ be such that
  $\AlphabetSize_2 \leq \AlphabetSize_1 < N^{1/4}$, and denote
  $k=\lceil\log_{\AlphabetSize_2}\AlphabetSize_1\rceil$.
  In the word RAM model with word size $w \geq 1+\log N$, given $N$,
  $\AlphabetSize_1$, and $\AlphabetSize_2$, we can in $\bigO(\sqrt{N})$ time construct a data structure of size
  $\bigO(\sqrt{N})$ words that, given $|S|$ and the packed
  representation $\PackedRepresentation{w}{\AlphabetSize_1}{S}$ (\cref{def:packed-representation})
  of any string $S \in [0 \dd \AlphabetSize_1)^*$ satisfying $k|S|\leq N$, returns the packed representation $\PackedRepresentation{w}{\AlphabetSize_2}{R}$
  of the string $R = \AlphabetMap{\AlphabetSize_1}{\AlphabetSize_2}{S} \in [0 \dd \AlphabetSize_2)^{*}$ (\cref{def:alphabet-map})
  in $\bigO(1 + |S| / \log_{\AlphabetSize_1} N)$ time.
\end{proposition}
\begin{proof}

  We use the following definitions.
  Let $b_1 = \lceil \log \AlphabetSize_1 \rceil$ and $b_2 = \lceil \log \AlphabetSize_2 \rceil$ denote
  the number of bits needed to store a symbol over alphabet $[0 \dd \AlphabetSize_1)$ and $[0 \dd \AlphabetSize_2)$,
  respectively.
  Let $m = \lfloor \tfrac{1}{4}\log_{\AlphabetSize_1} N \rfloor$.
  Since we assumed $\AlphabetSize_1 < N^{1/4}$, it follows that
  $N > \AlphabetSize_1^{4} \geq 2^{4} = 16$.
  Note also that $m \geq 1$ holds since we assumed
  $\AlphabetSize_1 < N^{1/4}$
  (which is equivalent to $\log_{\AlphabetSize_1} N > 4$).
  We use the following bounds on the representations and integers stored
  by the data structure:
  \begin{itemize}
  \item For every $X \in [0 \dd \AlphabetSize_1)^{\leq m}$, the packed
    representations of $X$ and
    $X'=\AlphabetMap{\AlphabetSize_1}{\AlphabetSize_2}{X}$ each occupy
    $\bigO(1)$ words. Indeed, $|X'|=|X|k$, and
    \[
      k\log\AlphabetSize_2
        \leq \log\AlphabetSize_1+\log\AlphabetSize_2
        \leq 2\log\AlphabetSize_1.
    \]
    Hence $|X|\log\AlphabetSize_1\leq\tfrac14\log N$ and
    $|X'|\log\AlphabetSize_2\leq\tfrac12\log N$.
    The space bound follows from \cref{def:packed-representation} and
    $w\geq1+\log N$.
  \item For every $X \in [0 \dd \AlphabetSize_1)^{\leq m}$, the integer
    $\Val{\AlphabetSize_1}{X}$ fits in one word, since
    $\Val{\AlphabetSize_1}{X}<\AlphabetSize_1^m\leq N^{1/4}$
    (\cref{def:val}).
  \item For every input string $S$, the output length
    $|R|=k|S|\leq N<2^w$ fits in one word.
  \end{itemize}
  For every $r\in[0\dd m]$, let $L_r[0\dd\AlphabetSize_1^r)$ be an array
  such that, for every $X\in[0\dd\AlphabetSize_1)^r$, the entry
  $L_r[\Val{\AlphabetSize_1}{X}]$ stores the packed representation of
  $\AlphabetMap{\AlphabetSize_1}{\AlphabetSize_2}{X}$. In particular,
  $L_0[0]$ stores the packed representation of the empty string.
  Each representation occupies $\bigO(1)$ words by the bounds above.
  The length $r$ and the value $\Val{\AlphabetSize_1}{X}$ uniquely determine
  $X$, including its leading zeros.

  \DSComponents
  The data structure consists of the following components:
  \begin{enumerate}

  \item We store the data structure from \cref{th:val-encoding} with
    parameters $N$ and $\AlphabetSize_1$. It uses
    $\bigO(\sqrt{N})$ words of space.

  \item We also store the arrays $L_0,\ldots,L_m$ and the integers
    $b_1$, $b_2$, $k$, and $m$.
    Their total space is
    $\bigO(\sum_{r=0}^m\AlphabetSize_1^r)
      =\bigO(\AlphabetSize_1^m)=\bigO(N^{1/4})$ words.
  \end{enumerate}

  In total, the structure needs $\bigO(\sqrt{N})$ words of space.

  \DSQueries
  If $|S|=0$, return the empty packed representation in $\bigO(1)$ time.
  Otherwise, we construct the packed representation of
  $R=\AlphabetMap{\AlphabetSize_1}{\AlphabetSize_2}{S}$ in three steps:
  \begin{enumerate}

  \item First, we obtain the packed representations of consecutive
    blocks $S_1,\ldots,S_q$ whose concatenation is $S$.
    To this end, compute $q:=\lceil |S|/m\rceil$ and the block lengths
    $\ell_i:=m$ for $i\in[1\dd q)$ and
    $\ell_q:=|S|-(q-1)m$.
    Apply \cref{pr:packed-representation}\eqref{pr:packed-representation-split}
    to the supplied packed representation with these lengths, obtaining
    the packed representation of $S_i$ with $|S_i|=\ell_i$ for every
    $i\in[1\dd q]$.
    This takes $\bigO(q+|S|\log\AlphabetSize_1/w)$ time.

  \item Next, we retrieve the packed representation of
    $R_i=\AlphabetMap{\AlphabetSize_1}{\AlphabetSize_2}{S_i}$ for every
    $i\in[1\dd q]$. To this end, for each $i$, compute
    $y_i=\Val{\AlphabetSize_1}{S_i}$ using the structure from
    \cref{th:val-encoding}, and retrieve the representation stored in
    $L_{\ell_i}[y_i]$.
    By the definition of $L_{\ell_i}$, this is the packed representation
    of $R_i$, which has length $k\ell_i$.
    Each conversion takes $\bigO(1)$ time because
    $\ell_i\leq m\leq\lfloor\log_{\AlphabetSize_1}N\rfloor$,
    and each retrieved representation occupies $\bigO(1)$ words by the
    bounds above. Thus, this step takes $\bigO(q)$ time.

  \item Finally, we concatenate the packed representations of
    $R_1,\ldots,R_q$ in this order, as follows.
    Compute their lengths $\ell'_i:=k\ell_i$ for $i\in[1\dd q]$, and
    apply \cref{pr:packed-representation}\eqref{pr:packed-representation-concat}
    with these lengths.
    Since
    \[
      R_1\cdots R_q
        =\AlphabetMap{\AlphabetSize_1}{\AlphabetSize_2}{S}
        =R,
    \]
    the result is the required packed representation of $R$.
    This step takes $\bigO(q+|R|\log\AlphabetSize_2/w)$ time.
  \end{enumerate}

  The split and the concatenation use length bound $N$, which applies
  because $|S|\leq|R|=k|S|\leq N$ and $w>\log N$.
  The inequality $k\log\AlphabetSize_2\leq2\log\AlphabetSize_1$
  established above gives
  $|R|\log\AlphabetSize_2\leq2|S|\log\AlphabetSize_1$.
  Consequently, the total query time is
  $\bigO(q+|S|\log\AlphabetSize_1/w)$.
  Since $m=\Theta(\log_{\AlphabetSize_1}N)$ and
  $w\geq1+\log N$, this is
  $\bigO(1+|S|/\log_{\AlphabetSize_1}N)$.

  \DSConstruction
  The components of the data structure are constructed as follows:
  \begin{enumerate}

  \item We construct the data structure from \cref{th:val-encoding} in
    $\bigO(\sqrt{N})$ time.

  \item We construct the arrays $L_0,\ldots,L_m$ and their parameters as follows:
    \begin{enumerate}

    \item Compute and store $b_1$, $b_2$, $k$, and $m$ in
      $\bigO(\log N)$ time as follows. To compute $k$ while keeping all
      intermediate integers within one word, repeatedly divide
      $\AlphabetSize_1-1$ by $\AlphabetSize_2$, taking floors, until the
      quotient is zero, and count the divisions. Similarly, compute
      $\lfloor\log_{\AlphabetSize_1}N\rfloor$ by successive divisions of
      $N$ by $\AlphabetSize_1$, and take the floor of one quarter of this integer to obtain $m$.

    \item Next, we construct an auxiliary array $P[0\dd\AlphabetSize_1)$
      whose entry $P[c]$ stores the packed representation of
      $\AlphabetMap{\AlphabetSize_1}{\AlphabetSize_2}{c}$ for every
      $c\in[0\dd\AlphabetSize_1)$.
      To this end, construct the data structure from \cref{th:val-decoding}
      in $\bigO(\sqrt{N})$ time.
      Since $\AlphabetSize_2^k<\AlphabetSize_1\AlphabetSize_2<\sqrt{N}$,
      it holds $k\leq\lfloor\log_{\AlphabetSize_2}N\rfloor$.
      For every $c\in[0\dd\AlphabetSize_1)$, query this structure with
      length $k$ and integer $c$, and store the returned packed
      representation in $P[c]$.
      By \cref{def:alphabet-map,def:val}, the string
      $\AlphabetMap{\AlphabetSize_1}{\AlphabetSize_2}{c}$ has length $k$
      and integer value $c$ over alphabet $[0\dd\AlphabetSize_2)$.
      Thus, each query returns the required representation in
      $\bigO(1)$ time, and this step takes $\bigO(\sqrt{N})$ time.

    \item We are now ready to populate the arrays $L_0,\ldots,L_m$.
      Allocate the arrays and initialize $L_0[0]$ with the packed
      representation of the empty string using
      \cref{pr:packed-representation}\eqref{pr:packed-representation-initialize}.
      For $\ell=0,\ldots,m-1$, in increasing order, iterate over all
      $x\in[0\dd\AlphabetSize_1^\ell)$ and $c\in[0\dd\AlphabetSize_1)$.
      Let $X\in[0\dd\AlphabetSize_1)^\ell$ satisfy
      $\Val{\AlphabetSize_1}{X}=x$.
      Set $x':=\AlphabetSize_1 x+c=\Val{\AlphabetSize_1}{X\cdot c}$
      (\cref{def:val}).
      By \cref{def:alphabet-map},
      \[
        \AlphabetMap{\AlphabetSize_1}{\AlphabetSize_2}{X\cdot c}
          = \AlphabetMap{\AlphabetSize_1}{\AlphabetSize_2}{X}
            \cdot\AlphabetMap{\AlphabetSize_1}{\AlphabetSize_2}{c}.
      \]
      Apply \cref{pr:packed-representation}\eqref{pr:packed-representation-concat}
      to the representations stored in $L_\ell[x]$ and $P[c]$, with
      lengths $\ell k$ and $k$, respectively, and store the resulting
      packed representation in $L_{\ell+1}[x']$.
      Each entry takes $\bigO(1)$ time. The total number of entries filled is
      $\sum_{\ell=0}^{m-1} \AlphabetSize_1^{\ell} \cdot \AlphabetSize_1
        = \sum_{j=1}^{m} \AlphabetSize_1^{j}
        < 2\AlphabetSize_1^{m}
        = \bigO(N^{1/4})$.
      Thus, this step takes $\bigO(N^{1/4})$ time.
    \end{enumerate}

    In total, constructing the arrays $L_0,\ldots,L_m$ and their parameters takes
    $\bigO(\sqrt{N})$ time.
  \end{enumerate}

  The operations from \cref{pr:packed-representation} used in this
  construction have length bound $N$. The decoding structure uses size
  parameter $N$ and alphabet size $\AlphabetSize_2$. All constructed strings
  have length at most $mk\leq\tfrac12\log N<N$, and their
  packed representations occupy $\bigO(1)$ words. Hence each initialization
  and concatenation above takes $\bigO(1)$ time.
  In total, the construction takes $\bigO(\sqrt{N})$ time.
\end{proof}

\Needspace{5\baselineskip}
\paragraph{Large Alphabets}\label{sec:tools-alphabet-mapping-large-alphabet}

\begin{proposition}\label{pr:alphabet-map-large-sigma2}
  Let $N, \AlphabetSize_1, \AlphabetSize_2 \in \Z_{\geq 2}$ be such that
  $N^{1/4} \leq \AlphabetSize_2 \leq \AlphabetSize_1 \leq N$, and denote
  $k=\lceil\log_{\AlphabetSize_2}\AlphabetSize_1\rceil$.
  In the word RAM model with word size $w \geq 1+\log N$, given $N$,
  $\AlphabetSize_1$, and $\AlphabetSize_2$, we can in $\bigO(\log N)$ time construct a data structure of size
  $\bigO(1)$ words that, given $|S|$ and the packed
  representation $\PackedRepresentation{w}{\AlphabetSize_1}{S}$ (\cref{def:packed-representation})
  of any string $S \in [0 \dd \AlphabetSize_1)^*$ satisfying $k|S|\leq N$, returns the packed representation
  $\PackedRepresentation{w}{\AlphabetSize_2}{R}$ of the string
  $R = \AlphabetMap{\AlphabetSize_1}{\AlphabetSize_2}{S} \in [0 \dd \AlphabetSize_2)^{*}$
  (\cref{def:alphabet-map}) in $\bigO(1 + |S| / \log_{\AlphabetSize_1} N)$ time.
\end{proposition}
\begin{proof}

  We use the following definitions. Denote
  $b_1 = \lceil \log \AlphabetSize_1 \rceil$ and
  $b_2 = \lceil \log \AlphabetSize_2 \rceil$.
  Observe that since $\AlphabetSize_2 \geq N^{1/4}$ and $\AlphabetSize_1 \leq N$,
  it holds that
  $k
    =     \lceil \log_{\AlphabetSize_2} \AlphabetSize_1 \rceil
    \leq  \lceil \log_{N^{1/4}} N \rceil
    =     \lceil 4 \rceil
    =     4$.
  For every input string $S$, the output length
  $|\AlphabetMap{\AlphabetSize_1}{\AlphabetSize_2}{S}|=k|S|\leq N<2^w$
  fits in one word.

  \DSComponents
  We store $N$, $\AlphabetSize_1$, $\AlphabetSize_2$, $b_1$, $b_2$, and $k$.
  These integers occupy $\bigO(1)$ words.

  \DSQueries
  If $|S|=0$, return the empty packed representation in $\bigO(1)$ time.
  Otherwise, we construct the packed representation of
  $R=\AlphabetMap{\AlphabetSize_1}{\AlphabetSize_2}{S}$ in three steps:
  \begin{enumerate}

  \item First, we extract the symbols $c_i=S[i]$ for every
    $i\in[1\dd|S|]$, using
    \cref{pr:packed-representation}\eqref{pr:packed-representation-access}.
    This takes $\bigO(|S|)$ time.

  \item Next, for each $i\in[1\dd|S|]$, we construct the packed
    representation of
    $\AlphabetMap{\AlphabetSize_1}{\AlphabetSize_2}{c_i}$ as follows.
    Initialize the packed representation of $R_i:=\zero^k$
    over alphabet $[0\dd\AlphabetSize_2)$ using
    \cref{pr:packed-representation}\eqref{pr:packed-representation-initialize}.
    For $j=1,\ldots,k$, compute
    $d:=\lfloor c_i/\AlphabetSize_2^{k-j}\rfloor\bmod\AlphabetSize_2$
    and update the representation to reflect $R_i[j]:=d$ using
    \cref{pr:packed-representation}\eqref{pr:packed-representation-update}.
    The formula gives the $j$th base-$\AlphabetSize_2$ digit of $c_i$,
    including leading zeros, and hence the resulting string is
    $R_i=\AlphabetMap{\AlphabetSize_1}{\AlphabetSize_2}{c_i}$.
    All powers used in the digit formula are at most
    $\AlphabetSize_2^{k-1}<\AlphabetSize_1\leq N$.
    Since $k\leq4$ and $\log\AlphabetSize_2\leq\log N<w$,
    the initialization, digit computations, and updates together take
    $\bigO(1)$ time per symbol. Thus, this step takes $\bigO(|S|)$ time.

  \item Finally, we concatenate the packed representations of
    $R_1,\ldots,R_{|S|}$ in this order.
    To this end, construct an array containing $|S|$ lengths equal to $k$
    and apply
    \cref{pr:packed-representation}\eqref{pr:packed-representation-concat}
    with these lengths. Since
    \[
      R_1\cdots R_{|S|}
        =\AlphabetMap{\AlphabetSize_1}{\AlphabetSize_2}{S}
        =R,
    \]
    the result is the required packed representation of $R$.
    This step takes
    $\bigO(|S|+|R|\log\AlphabetSize_2/w)=\bigO(|S|)$ time,
    because $|R|=k|S|$, $k\leq4$, and $\log\AlphabetSize_2<w$.
  \end{enumerate}

  All operations from \cref{pr:packed-representation} use length bound
  $N$, since $|S|\leq|R|=k|S|\leq N$ and $w>\log N$.
  The total query time is $\bigO(1+|S|)$.
  Since $N^{1/4}\leq\AlphabetSize_1\leq N$, we have
  $1\leq\log_{\AlphabetSize_1}N\leq4$, giving the claimed
  $\bigO(1+|S|/\log_{\AlphabetSize_1}N)$ bound.

  \DSConstruction
  Compute and store $b_1$, $b_2$, and $k$ in $\bigO(\log N)$ time
  by successive integer divisions to keep all intermediate integers
  within one word. In particular, $k$ is the number of
  divisions of $\AlphabetSize_1-1$ by $\AlphabetSize_2$ needed to reach zero.
\end{proof}

\paragraph{Mixed Alphabets}\label{sec:tools-alphabet-mapping-mixed-alphabet}

\begin{proposition}\label{pr:alphabet-map-mixed-sigma}
  Let $N, \AlphabetSize_1, \AlphabetSize_2 \in \Z_{\geq 2}$
  be such that $\AlphabetSize_2 < N^{1/4} \leq \AlphabetSize_1 \leq N$, and denote
  $k=\lceil\log_{\AlphabetSize_2}\AlphabetSize_1\rceil$.
  In the word RAM model with word size $w \geq 1+\log N$, given $N$,
  $\AlphabetSize_1$, and $\AlphabetSize_2$,
  we can in $\bigO(\sqrt{N})$ time construct a data structure of size
  $\bigO(\sqrt{N})$ words that, given $|S|$ and the packed representation $\PackedRepresentation{w}{\AlphabetSize_1}{S}$ (\cref{def:packed-representation})
  of any string $S \in [0 \dd \AlphabetSize_1)^*$ satisfying $k|S|\leq N$,
  returns the packed representation $\PackedRepresentation{w}{\AlphabetSize_2}{R}$ of the string
  $R = \AlphabetMap{\AlphabetSize_1}{\AlphabetSize_2}{S} \in [0 \dd \AlphabetSize_2)^{*}$ (\cref{def:alphabet-map})
  in $\bigO(1 + |S| / \log_{\AlphabetSize_1} N)$ time.
\end{proposition}
\begin{proof}

  First, note that the assumptions imply that $N>16$.

  The central idea of the data structure is to bridge the
  gap between $\AlphabetSize_1$ and $\AlphabetSize_2$ using an
  intermediate alphabet size $\AlphabetSize_m$. We choose
  $\AlphabetSize_m$ as a power of $\AlphabetSize_2$, small enough that
  \cref{pr:alphabet-map-small-sigma1} applies and large enough that every
  symbol over $[0\dd\AlphabetSize_1)$ has a constant-length
  base-$\AlphabetSize_m$ representation. At query time,
  for each such symbol $a$, we
  first construct $\AlphabetMap{\AlphabetSize_1}{\AlphabetSize_m}{a}$
  and then apply \cref{pr:alphabet-map-small-sigma1} to this string.
  Since $\AlphabetSize_m$ is a power of $\AlphabetSize_2$, the result is
  the base-$\AlphabetSize_2$ representation of $a$, possibly padded with
  more leading zeros than
  $\AlphabetMap{\AlphabetSize_1}{\AlphabetSize_2}{a}$. Thus, the desired
  representation is obtained by extracting an appropriate suffix.

  We now specify the parameters. Denote
  \[
    p:=\lceil\log_{\AlphabetSize_2}(N^{1/4})\rceil-1,\qquad
    \AlphabetSize_m:=\AlphabetSize_2^p,\qquad
    d:=\lceil\log_{\AlphabetSize_m}\AlphabetSize_1\rceil.
  \]
  Since $\log_{\AlphabetSize_2}(N^{1/4})>1$, it holds $p\geq1$.
  The inequalities $x/2\leq\lceil x\rceil-1<x$ for $x>1$ also give
  $N^{1/8}\leq\AlphabetSize_m<N^{1/4}$.
  Hence, $\AlphabetSize_2\leq\AlphabetSize_m<N^{1/4}$, and
  \cref{pr:alphabet-map-small-sigma1} applies with source alphabet size
  $\AlphabetSize_m$ and target alphabet size $\AlphabetSize_2$.
  Its expansion factor is exactly $p$, since
  $\AlphabetSize_m=\AlphabetSize_2^p$. Moreover,
  $\AlphabetSize_m\geq N^{1/8}$ and $\AlphabetSize_1\leq N$
  imply that $d\leq\lceil\log_{N^{1/8}}N\rceil=8$.

  Applying these two mappings to a symbol from $[0\dd\AlphabetSize_1)$
  produces a string of length $dp$. Since
  $\AlphabetSize_2^{dp}=\AlphabetSize_m^d\geq\AlphabetSize_1$, it holds
  $k\leq dp$. Furthermore, $d\leq8$ and
  $p<\frac{1}{4}\log_{\AlphabetSize_2}N
    \leq\frac{1}{4}\log N$ give $dp<2\log N<N$.
  Thus, the length-$k$ suffix used below is well-defined, and the packed
  substring operation is applicable with length bound $N$.

  To identify this suffix, let $a\in[0\dd\AlphabetSize_1)$ and denote
  $C=\AlphabetMap{\AlphabetSize_1}{\AlphabetSize_m}{a}$.
  Since $\AlphabetSize_m=\AlphabetSize_2^p$, both
  $\AlphabetMap{\AlphabetSize_m}{\AlphabetSize_2}{C}$ and
  $\AlphabetMap{\AlphabetSize_1}{\AlphabetSize_2}{a}$ are
  base-$\AlphabetSize_2$ representations of $a$, padded with leading zeros
  to lengths $dp$ and $k$, respectively. Consequently,
  \[
    \AlphabetMap{\AlphabetSize_m}{\AlphabetSize_2}{C}
      =\zero^{dp-k}\cdot
        \AlphabetMap{\AlphabetSize_1}{\AlphabetSize_2}{a}.
  \]

  Finally, denote $b_1:=\lceil\log\AlphabetSize_1\rceil$,
  $b_2:=\lceil\log\AlphabetSize_2\rceil$, and
  $b_m:=\lceil\log\AlphabetSize_m\rceil$. Let
  $A_{\rm pow}[0\dd d)$ denote an array satisfying
  $A_{\rm pow}[j]=\AlphabetSize_m^j$ for every $j\in[0\dd d)$.
  The array stores the powers used below to construct the
  base-$\AlphabetSize_m$ representations. The inequalities above imply
  that $b_1$, $b_2$, $b_m$, $k$, $p$, $\AlphabetSize_m$, and $d$ fit in
  one $w$-bit word. Moreover,
  $\AlphabetSize_m^{d-1}<\AlphabetSize_1\leq N$. Thus, every entry of
  $A_{\rm pow}$ fits in one $w$-bit word.

  \DSComponents
  The data structure consists of the following components:
  \begin{enumerate}

  \item The integers $b_1$, $b_2$, $b_m$, $k$, $p$,
    $\AlphabetSize_m$, and $d$, and the array $A_{\rm pow}$.
    This component uses $\bigO(1)$ words of space.

  \item The data structure from \cref{pr:alphabet-map-small-sigma1}
    instantiated with size parameter $N$, source alphabet size
    $\AlphabetSize_m$, and target alphabet size $\AlphabetSize_2$.
    The application is valid because
    $\AlphabetSize_2\leq\AlphabetSize_m<N^{1/4}$.
    This component uses $\bigO(\sqrt{N})$ words of space.
  \end{enumerate}

  In total, the data structure uses $\bigO(\sqrt{N})$ words of
  space.

  \DSQueries
  Denote $n:=|S|$.
  If $n=0$, return in $\bigO(1)$ time the packed representation of
  $\AlphabetMap{\AlphabetSize_1}{\AlphabetSize_2}{S}=\emptystring$.
  Assume henceforth that $n\geq1$ and that we are given $n$ and the
  packed representation $\PackedRepresentation{w}{\AlphabetSize_1}{S}$ of $S$.
  For every $i\in[1\dd n]$, denote
  $C_i:=\AlphabetMap{\AlphabetSize_1}{\AlphabetSize_m}{S[i]}$,
  $A_i:=\AlphabetMap{\AlphabetSize_m}{\AlphabetSize_2}{C_i}$, and
  $R_i:=A_i[dp-k+1\dd dp]$. The identity above implies that
  $R_i=\AlphabetMap{\AlphabetSize_1}{\AlphabetSize_2}{S[i]}$.
  We proceed as follows:
  \begin{enumerate}

  \item First, we compute the packed representation
    $\PackedRepresentation{w}{\AlphabetSize_2}{R_i}$ of $R_i$ for every
    $i\in[1\dd n]$. For each $i$, we proceed as follows:
    \begin{enumerate}

    \item We first compute the packed representation of $C_i$.
      To this end, retrieve $S[i]$ using
      \cref{pr:packed-representation}\eqref{pr:packed-representation-access},
      and initialize the packed representation of $P_i:=\zero^d$ over
      $[0\dd\AlphabetSize_m)$ using
      \cref{pr:packed-representation}\eqref{pr:packed-representation-initialize}.
      For $j=1,\ldots,d$, set
      $P_i[j]:=\lfloor S[i]/A_{\rm pow}[d-j]\rfloor\bmod\AlphabetSize_m$
      using \cref{pr:packed-representation}\eqref{pr:packed-representation-update}.
      These are the base-$\AlphabetSize_m$ digits of $S[i]$, including
      leading zeros. Thus, after the updates, $P_i=C_i$.
      Since $d\leq8$, this takes $\bigO(d)=\bigO(1)$ time.

    \item Next, we obtain the packed representation
      $\PackedRepresentation{w}{\AlphabetSize_2}{A_i}$ of $A_i$ from that
      of $C_i$ using the stored data structure from
      \cref{pr:alphabet-map-small-sigma1}. This takes
      $\bigO(1+d/\log_{\AlphabetSize_m}N)=\bigO(1)$ time.

    \item We then extract the packed representation
      $\PackedRepresentation{w}{\AlphabetSize_2}{R_i}$ of the length-$k$
      suffix $R_i=A_i[dp-k+1\dd dp]$ using
      \cref{pr:packed-representation}\eqref{pr:packed-representation-substring}.
      Since $k\log\AlphabetSize_2
        \leq\log\AlphabetSize_1+\log\AlphabetSize_2
        \leq2\log N=\bigO(w)$,
      this takes $\bigO(1+k\log\AlphabetSize_2/w)=\bigO(1)$ time.
    \end{enumerate}

    All packed operations in this step use length bound $N$ and
    the precomputed values $b_1$, $b_m$, and $b_2$ for the respective alphabets.
    The computation of $A_i$ and the extraction of $R_i$ are valid
    because $k\leq dp<N$, as shown above.
    In total, this step takes $\bigO(n)$ time.

  \item Finally, we concatenate the packed representations of
    $R_1,\ldots,R_n$, in this order, to obtain the packed representation
    $\PackedRepresentation{w}{\AlphabetSize_2}{R}$ of
    \[
      R=R_1\cdots R_n
       =\AlphabetMap{\AlphabetSize_1}{\AlphabetSize_2}{S}.
    \]
    To this end, apply
    \cref{pr:packed-representation}\eqref{pr:packed-representation-concat}
    with length $k$ for each string; it uses length bound $N$ because
    $nk=k|S|\leq N$.
    By $k\log\AlphabetSize_2\leq2\log N=\bigO(w)$, this step takes
    $\bigO(n+nk\log\AlphabetSize_2/w)=\bigO(n)$ time.
  \end{enumerate}

  In total, the query takes $\bigO(1+n)$ time. Since
  $N^{1/4}\leq\AlphabetSize_1\leq N$, it holds
  $1\leq\log_{\AlphabetSize_1}N\leq4$. Consequently, this running
  time is
  $\bigO(1+n/\log_{\AlphabetSize_1}N)$, as required.

  \DSConstruction
  The components of the data structure are constructed as follows:
  \begin{enumerate}

  \item Compute $b_1$, $b_2$, $b_m$, $k$, $p$, $\AlphabetSize_m$, $d$,
    and $A_{\rm pow}$ in $\bigO(\log N)$ time as follows, keeping all
    intermediate integers within one word.
    To compute $p$, first compute $L:=\lfloor\log_{\AlphabetSize_2}N\rfloor$
    by successive integer divisions of $N$ by $\AlphabetSize_2$.
    Set $p:=\lfloor L/4\rfloor$, and decrease $p$ by one if
    $\AlphabetSize_2^{4p}=N$; the tested power is at most $N$.
    Compute $\AlphabetSize_m=\AlphabetSize_2^p$ by repeated multiplication.
    To compute $k$ and $d$, count the divisions of $\AlphabetSize_1-1$
    by $\AlphabetSize_2$ and $\AlphabetSize_m$, respectively, needed to
    reach zero. The remaining parameters are obtained by integer
    divisions and multiplications whose results are at most $N$.

  \item Construct the data structure from
    \cref{pr:alphabet-map-small-sigma1}.
    This takes $\bigO(\sqrt{N})$ time.
  \end{enumerate}

  In total, the construction takes $\bigO(\sqrt{N})$ time.
\end{proof}

\paragraph{Summary}\label{sec:tools-alphabet-mapping-summary}

\begin{theorem}\label{th:alphabet-map}
  Let $N, \AlphabetSize_1, \AlphabetSize_2 \in \Z_{\geq 2}$ be such that
  $\AlphabetSize_2 \leq \AlphabetSize_1 \leq N$, and denote
  $k=\lceil\log_{\AlphabetSize_2}\AlphabetSize_1\rceil$.
  In the word RAM model with word size $w \geq 1+\log N$, given $N$,
  $\AlphabetSize_1$, and $\AlphabetSize_2$,
  we can in $\bigO(\sqrt{N})$ time construct a data structure of size
  $\bigO(\sqrt{N})$ words that, given $|S|$ and the packed representation $\PackedRepresentation{w}{\AlphabetSize_1}{S}$ (\cref{def:packed-representation})
  of any string $S \in [0 \dd \AlphabetSize_1)^*$ satisfying $k|S|\leq N$,
  returns the packed representation $\PackedRepresentation{w}{\AlphabetSize_2}{R}$ of the string
  $R = \AlphabetMap{\AlphabetSize_1}{\AlphabetSize_2}{S} \in [0 \dd \AlphabetSize_2)^{*}$ (\cref{def:alphabet-map})
  in $\bigO(1 + |S| / \log_{\AlphabetSize_1} N)$ time.
\end{theorem}
\begin{proof}
  We first decide which case applies, in $\bigO(1)$ time. For
  $\AlphabetSize\in\{\AlphabetSize_1,\AlphabetSize_2\}$, the condition
  $\AlphabetSize<N^{1/4}$ is equivalent to $\AlphabetSize^4<N$. To test it,
  we compute the powers $\AlphabetSize^2,\AlphabetSize^3,\AlphabetSize^4$
  in turn. We multiply a power $a$ by $\AlphabetSize$ only if
  $a\leq\lfloor N/\AlphabetSize\rfloor$, and otherwise conclude that
  $\AlphabetSize^4>N$. All computed values are at most $N$.
  We consider three cases:
  \begin{enumerate}
  \item If $\AlphabetSize_1 < N^{1/4}$,
    we use the data structure from \cref{pr:alphabet-map-small-sigma1}.
  \item If $\AlphabetSize_2 \geq N^{1/4}$,
    we use the data structure from \cref{pr:alphabet-map-large-sigma2}.
  \item If $\AlphabetSize_2 < N^{1/4} \leq \AlphabetSize_1$,
    we use the data structure from \cref{pr:alphabet-map-mixed-sigma}.
  \end{enumerate}
  In all cases, the structure occupies $\bigO(\sqrt{N})$ words,
  the construction takes $\bigO(\sqrt{N})$ time, and queries take
  $\bigO(1 + |S| / \log_{\AlphabetSize_1} N)$ time.
\end{proof}

%% file: tools/packed-extended-alphabet-mapping.tex
\subsection{Packed Extended Alphabet Mapping}\label{sec:tools-extended-alphabet-mapping}

\begin{definition}[Extended alphabet mapping]\label{def:ext-alphabet-map}
  Let $\AlphabetSize_1, \AlphabetSize_2 \in \Z_{\geq 2}$ be such that $\AlphabetSize_1 \geq \AlphabetSize_2$.
  Denote $k = \lceil \log_{\AlphabetSize_2} \AlphabetSize_1 \rceil$. For any $x \in [0 \dd \AlphabetSize_1)$,
  we define
  \[
    \ExtAlphabetMap{\AlphabetSize_1}{\AlphabetSize_2}{x} :=
      c^{k+1} \cdot \zero \cdot \AlphabetMap{\AlphabetSize_1}{\AlphabetSize_2}{x} \cdot \zero
      \in [0 \dd \AlphabetSize_2)^{2k+3},
  \]
  where $c = \AlphabetSize_2 - 1$ and $\AlphabetMap{\AlphabetSize_1}{\AlphabetSize_2}{x}$ is as in \cref{def:alphabet-map}.
  For any string $S \in [0 \dd \AlphabetSize_1)^{*}$, we then let
  \[
    \ExtAlphabetMap{\AlphabetSize_1}{\AlphabetSize_2}{S} :=
      \textstyle\bigodot_{i=1,\ldots,|S|} \ExtAlphabetMap{\AlphabetSize_1}{\AlphabetSize_2}{S[i]}
      \in [0 \dd \AlphabetSize_2)^{|S|\cdot (2k+3)}.
  \]
\end{definition}

\begin{example}\label{ex:ext-alphabet-map}
  Let $\AlphabetSize_1=5$ and $\AlphabetSize_2=2$. Then $k=3$ and
  $c=\one$. For example,
  $\ExtAlphabetMap{5}{2}{4}
    =\one\one\one\one\cdot\zero\cdot\one\zero\zero\cdot\zero$
  and
  $\ExtAlphabetMap{5}{2}{1}
    =\one\one\one\one\cdot\zero\cdot\zero\zero\one\cdot\zero$.
  For $S=(4,1,3)$, we obtain
  \[
    \ExtAlphabetMap{5}{2}{S}
      =\one\one\one\one\zero\one\zero\zero\zero\cdot
       \one\one\one\one\zero\zero\zero\one\zero\cdot
       \one\one\one\one\zero\zero\one\one\zero.
  \]
\end{example}

\begin{observation}\label{ob:ext-alphabet-map-properties}
  Let $\AlphabetSize_1,\AlphabetSize_2\in\Z_{\geq2}$ satisfy
  $\AlphabetSize_1\geq\AlphabetSize_2$.
  For every $A,B\in[0\dd\AlphabetSize_1)^*$, the following equivalences
  hold (see \cref{def:ext-alphabet-map}):
  \begin{enumerate}
  \item\label{ob:ext-alphabet-map-prefix-property}
    $A\text{ is a prefix of }B
      \quad\Longleftrightarrow\quad
      \ExtAlphabetMap{\AlphabetSize_1}{\AlphabetSize_2}{A}
      \text{ is a prefix of }
      \ExtAlphabetMap{\AlphabetSize_1}{\AlphabetSize_2}{B}$,
  \item\label{ob:ext-alphabet-map-lex-property}
    $A\prec B
      \quad\Longleftrightarrow\quad
      \ExtAlphabetMap{\AlphabetSize_1}{\AlphabetSize_2}{A}\prec
      \ExtAlphabetMap{\AlphabetSize_1}{\AlphabetSize_2}{B}$.
  \end{enumerate}
\end{observation}

\begin{lemma}\label{lm:general-suffix-array-alphabet-reduction}
  Let $\AlphabetSize_1, \AlphabetSize_2 \in \Z_{\geq 2}$ be such that $\AlphabetSize_1 \geq \AlphabetSize_2$.
  Let $\Text_1 \in [0 \dd \AlphabetSize_1)^{+}$. For every $i \in [1 \dd |\Text_1|]$, it holds
  \[
    \SA{\Text_1}[i] = \frac{\SA{\Text_2}[\Delta + i] - 1}{\delta} + 1,
  \]
  where
  $\Text_2 = \ExtAlphabetMap{\AlphabetSize_1}{\AlphabetSize_2}{\Text_1} \in [0 \dd \AlphabetSize_2)^{+}$
  (\cref{def:ext-alphabet-map}),
  $\Delta = |\Text_2| - |\Text_1|$,
  $k = \lceil \log_{\AlphabetSize_2} \AlphabetSize_1 \rceil$, and
  $\delta = 2k + 3$.
\end{lemma}
\begin{proof}
  Let $n=|\Text_1|$ and, for every $i\in[1\dd n]$, denote
  $q_i=(\SA{\Text_1}[i]-1)\delta+1$.
  Since $\SA{\Text_1}$ is a permutation of $[1\dd n]$, the positions
  $q_1,\ldots,q_n$ are precisely the starting positions of the encoded
  symbols in $\Text_2$. We establish two properties of these suffixes.
  \begin{enumerate}
  \item First, we prove that
    \[
      \Text_2[q_1\dd|\Text_2|]\prec\cdots\prec
      \Text_2[q_n\dd|\Text_2|].
    \]
    For every $i\in[1\dd n]$, \cref{def:ext-alphabet-map} gives
    $\Text_2[q_i\dd|\Text_2|]
      =\ExtAlphabetMap{\AlphabetSize_1}{\AlphabetSize_2}
        {\Text_1[\SA{\Text_1}[i]\dd n]}$.
    The suffixes $\Text_1[\SA{\Text_1}[i]\dd n]$ occur in ascending
    lexicographic order as $i$ increases. The required order therefore
    follows from \cref{ob:ext-alphabet-map-properties}\eqref{ob:ext-alphabet-map-lex-property}.
  \item Next, we show that every suffix starting outside
    $\{q_1,\ldots,q_n\}$ is smaller than every suffix starting in this set.
    Denote $c=\AlphabetSize_2-1$. Each encoded symbol has the form
    $c^{k+1}\zero\AlphabetMap{\AlphabetSize_1}{\AlphabetSize_2}{x}\zero$,
    where $\AlphabetMap{\AlphabetSize_1}{\AlphabetSize_2}{x}$ has length
    $k$. The zeros on both sides of this length-$k$ string ensure that
    the occurrences of $c^{k+1}$ in $\Text_2$ start precisely at
    $q_1,\ldots,q_n$. Since $\Text_2$ ends in zero, every suffix starting
    elsewhere begins with at most $k$ copies of $c$ followed by a smaller
    symbol, whereas each suffix starting at a position $q_i$ begins with
    $k+1$ copies of $c$. This proves the required comparison.
  \end{enumerate}
  There are exactly $|\Text_2|-n=\Delta$ other suffixes. Hence, for every
  $i\in[1\dd n]$, the two properties give
  \[
    \SA{\Text_2}[\Delta+i]=q_i=(\SA{\Text_1}[i]-1)\delta+1.
  \]
  Rearranging this equality yields the claim.
\end{proof}

\begin{theorem}\label{th:ext-alphabet-map}
  Let $N, \AlphabetSize_1, \AlphabetSize_2 \in \Z_{\geq 2}$ be such that
  $\AlphabetSize_2 \leq \AlphabetSize_1 \leq N$, and denote
  $k=\lceil\log_{\AlphabetSize_2}\AlphabetSize_1\rceil$.
  In the word RAM model with word size $w \geq 1+\log N$, given $N$,
  $\AlphabetSize_1$, and $\AlphabetSize_2$,
  we can in $\bigO(\sqrt{N})$ time construct a data structure of size
  $\bigO(\sqrt{N})$ words that, given $|S|$ and the packed representation $\PackedRepresentation{w}{\AlphabetSize_1}{S}$ (\cref{def:packed-representation})
  of any string $S \in [0 \dd \AlphabetSize_1)^*$ satisfying $(2k+3)|S|\leq N$,
  returns the packed representation $\PackedRepresentation{w}{\AlphabetSize_2}{R}$ of the string
  $R = \ExtAlphabetMap{\AlphabetSize_1}{\AlphabetSize_2}{S} \in [0 \dd \AlphabetSize_2)^{*}$ (\cref{def:ext-alphabet-map})
  in $\bigO(1 + |S| / \log_{\AlphabetSize_1} N)$ time.
\end{theorem}
\begin{proof}
  If $N<32$, store the alphabet sizes and answer queries as follows.
  For $|S|=0$, return the empty packed representation. Otherwise,
  $(2k+3)|S|\leq N$ implies $2k+3\leq N$, and we compute the output
  symbol by symbol in $\bigO(1)$ time.
  Assume henceforth that $N\geq32$. Denote
  $b_2:=\lceil\log\AlphabetSize_2\rceil$ and
  $c:=\AlphabetSize_2-1$. It holds
  $k\leq\lceil\log N\rceil$, and hence
  $2k+3\leq2\lceil\log N\rceil+3<N$.

  \DSComponents
  The data structure consists of the following components:
  \begin{enumerate}

  \item The data structure $\mathcal D_{\rm map}$ and the integers
    $b_2$, $k$, $c$, and $N$.
    The integers use $\bigO(1)$ words. The structure is obtained by applying
    \cref{th:alphabet-map} with size parameter $N$, input alphabet
    size $\AlphabetSize_1$, and output alphabet size $\AlphabetSize_2$.
    It uses $\bigO(\sqrt{N})$ words of space.

  \item The data structure $\mathcal D_{\rm con}$ obtained by applying
    \cref{th:packed-pointwise-concatenation-partition} with size parameter
    $N$ and alphabet size $\AlphabetSize_2$.
    It uses $\bigO(\sqrt{N})$ words of space.
  \end{enumerate}

  In total, the data structure uses $\bigO(\sqrt{N})$ words of space.

  Every application of \cref{pr:packed-representation} below uses alphabet
  size $\AlphabetSize_2$ and length bound $N$. These parameters
  satisfy $\AlphabetSize_2\leq N$ and $w>\log N$.

  \DSQueries
  Let $m:=|S|$. If $m=0$, return the empty packed representation.
  Otherwise, let
  $C\in([0\dd\AlphabetSize_2)^{k+1})^m$,
  $M\in([0\dd\AlphabetSize_2)^k)^m$, and
  $Z\in([0\dd\AlphabetSize_2)^1)^m$ denote the sequences satisfying
  $C[i]=c^{k+1}$,
  $M[i]=\AlphabetMap{\AlphabetSize_1}{\AlphabetSize_2}{S[i]}$, and
  $Z[i]=\zero$ for every $i\in[1\dd m]$. We construct the packed representation of
  $R=\ExtAlphabetMap{\AlphabetSize_1}{\AlphabetSize_2}{S}$ as follows:
  \begin{enumerate}

  \item First, we compute the packed sequence representation of $M$.
    To this end, query $\mathcal D_{\rm map}$ with $m$ and the supplied
    packed representation of $S$. Its output represents
    $\AlphabetMap{\AlphabetSize_1}{\AlphabetSize_2}{S}=M[1]\cdots M[m]$,
    and hence is also the packed sequence representation of $M$
    (\cref{def:packed-sequence-representation}). The query is valid
    because $mk\leq m(2k+3)\leq N$. This takes
    $\bigO(1+m/\log_{\AlphabetSize_1}N)$ time.

  \item We now construct the packed sequence representations of $C$ and $Z$.
    Use \cref{pr:packed-representation}\eqref{pr:packed-representation-initialize}
    to construct the packed representations of $c^{m(k+1)}$ and $\zero^m$.
    These are precisely the packed sequence representations of $C$ and $Z$,
    respectively. The initialized strings have lengths at most
    $m(2k+3)\leq N$, and the construction takes
    $\bigO(1+m(k+1)\log\AlphabetSize_2/w)$ time.

  \item Finally, we assemble the packed representation of $R$.
    Let $A$, $B$, and $E$ denote the sequences satisfying
    \[
      A[i]=C[i]Z[i],\qquad
      B[i]=A[i]M[i],\qquad
      E[i]=B[i]Z[i]
    \]
    for every $i\in[1\dd m]$. The strings in these sequences have
    lengths $k+2$, $2k+2$, and $2k+3$, respectively, and use alphabet
    $[0\dd\AlphabetSize_2)$.
    Using $\mathcal D_{\rm con}$, pointwise concatenate $C$ and $Z$
    to obtain the packed sequence representation of $A$.
    Next, pointwise concatenate $A$ and $M$ to obtain that of $B$.
    A final query concatenates $B$ and $Z$ pointwise to obtain the
    packed sequence representation of $E$.
    All three queries are valid because their output sequences contain
    at most $m(2k+3)\leq N$ symbols in total and each string has
    positive length.
    Together, they take
    $\bigO(1+m(2k+3)/\log_{\AlphabetSize_2}N)$ time.
    By \cref{def:ext-alphabet-map}, we have
    $E[i]=\ExtAlphabetMap{\AlphabetSize_1}{\AlphabetSize_2}{S[i]}$
    for every $i\in[1\dd m]$. Thus, $E[1]\cdots E[m]=R$, and the
    packed sequence representation of $E$ is also the packed
    representation of $R$, which we return.
  \end{enumerate}

  Since $k\leq\log_{\AlphabetSize_2}\AlphabetSize_1+1$ and
  $\AlphabetSize_2\leq\AlphabetSize_1$, it holds
  $
    \frac{2k+3}{\log_{\AlphabetSize_2}N}
      =\frac{(2k+3)\log\AlphabetSize_2}{\log N}
      \leq\frac{7\log\AlphabetSize_1}{\log N}
      =\frac{7}{\log_{\AlphabetSize_1}N}
  $.
  By $w\geq1+\log N$, the bounds above yield total query time
  $\bigO(1+m\log\AlphabetSize_1/\log N)
    =\bigO(1+m/\log_{\AlphabetSize_1}N)$.

  \DSConstruction
  The components of the data structure are constructed as follows:
  \begin{enumerate}

  \item Compute and store $b_2$, $k$, $c$, and $N$ in
    $\bigO(\log N)$ time as follows. To keep all intermediate integers
    within one word, compute $k$ by counting the successive integer divisions of $\AlphabetSize_1-1$ by $\AlphabetSize_2$ needed to reach zero.
    Construct $\mathcal D_{\rm map}$ using \cref{th:alphabet-map} in
    $\bigO(\sqrt{N})$ time.

  \item Construct $\mathcal D_{\rm con}$ using
    \cref{th:packed-pointwise-concatenation-partition} in
    $\bigO(\sqrt{N})$ time.
  \end{enumerate}

  The total construction time is $\bigO(\sqrt{N})$.
\end{proof}

%% file: tools/packed-morphisms.tex
\subsection{Packed Morphisms}
\label{sec:tools-packed-morphisms}

\begin{theorem}[{\cite{hierarchy}}]\label{th:packed-morphism}
  Let $d\in\Z_{\geq1}$ be a fixed constant, and let
  $N,\AlphabetSize\in\Z_{\geq2}$ satisfy $\AlphabetSize\leq N$.
  Consider the word RAM model with word size $w\geq1+\log N$, and let
  $\mu:\IntegerAlphabet\to\IntegerAlphabet^d$.
  Given $N$, $\AlphabetSize$, and an array containing the packed representations of
  $\mu(0),\ldots,\mu(\AlphabetSize-1)$
  (\cref{def:packed-representation}), we can in
  $\bigO(\sqrt N+\AlphabetSize)$ time construct a data structure of
  size $\bigO(\sqrt N+\AlphabetSize)$ words supporting the following query.
  Given $|S|$ and the packed representation
  $\PackedRepresentation{w}{\AlphabetSize}{S}$ of
  $S\in\IntegerAlphabet^*$ satisfying $d|S|\leq N$, it computes in
  $\bigO(1+|S|/\log_{\AlphabetSize}N)$ time the packed representation of
  \[
    \mu(S)=\mu(S[1])\cdots\mu(S[|S|])
      \in\IntegerAlphabet^{d|S|}.
  \]
\end{theorem}

%% file: tools/packed-prefix-frequencies.tex
\subsection{Packed Prefix Frequencies}
  \label{sec:tools-packed-prefix-frequencies}

\begin{theorem}[Packed prefix frequencies]
  \label{th:packed-prefix-frequencies}
  Let $N,\AlphabetSize,m\in\Z_{\geq2}$ and $\ell\in\Z_{\geq1}$ satisfy
  $\AlphabetSize^\ell\leq m$ and $m\ell\leq N$. Consider the word RAM model with word size
  $w\geq1+\log N$.
  Given $N$, $\AlphabetSize$, $m$, $\ell$, and the packed sequence representation
  $\PackedSeqRepresentation{w}{\AlphabetSize}{W}$
  (\cref{def:packed-sequence-representation}) of a sequence
  $W\in([0\dd\AlphabetSize)^\ell)^m$, we can construct, for every
  $p\in[0\dd\ell]$, an array $F_p[0\dd\AlphabetSize^p)$ satisfying
  (see \cref{def:val,def:prefix-rank-and-select}):
  \[
    F_p[\Val{\AlphabetSize}{X}]
      =\PrefixRank{W}{m}{X}
    \qquad
    \text{for every }
      X\in[0\dd\AlphabetSize)^p.
  \]
  The construction takes $\bigO(m\ell\log\AlphabetSize/\log m)$ time and has a peak space
  usage of $\bigO(m\ell\log\AlphabetSize/\log m)$ words.
\end{theorem}
\begin{proof}
  If $m<16$, scan every input string character by character, count the
  occurrences of each length-$\ell$ string, and sum these counts bottom-up
  over their prefixes. All arrays then have constant size, which gives the claimed
  bounds. Thus, assume that $m\geq16$.

  Compute $b:=\ceil{\log\AlphabetSize}$ by repeated doubling and
  $\AlphabetSize^\ell$ by repeated multiplication. This takes
  $\bigO(\log m)$ time because
  $\AlphabetSize\leq\AlphabetSize^\ell\leq m$ and $\ell\leq\log m$.
  Moreover, $b\leq2\log\AlphabetSize$. Using
  $\log(m/\AlphabetSize^\ell)\leq m/\AlphabetSize^\ell$, we obtain
  \begin{equation}\label{eq:packed-prefix-frequencies-universe}
    \AlphabetSize^\ell\log m
      =\AlphabetSize^\ell\left(
        \ell\log\AlphabetSize
        +\log\frac{m}{\AlphabetSize^\ell}
      \right)
      \leq m\ell\log\AlphabetSize+m
      =\bigO(m\ell\log\AlphabetSize),
    \qquad
    \AlphabetSize^\ell
      =\bigO\left(\frac{m\ell\log\AlphabetSize}{\log m}\right).
  \end{equation}
  Computing $b$ and $\AlphabetSize^\ell$ is also bounded by
  $\bigO(m\ell\log\AlphabetSize/\log m)$ because
  $m/\log m\geq\log m$ and $\ell\log\AlphabetSize\geq1$.

  Every application of
  \cref{pr:packed-representation}\eqref{pr:packed-representation-substring}
  below uses length bound $N$. Its assumptions hold because
  $\AlphabetSize\leq m\leq N$ and $w\geq1+\log N$.
  The supplied packed sequence representation is the packed representation
  of $W[1]\cdots W[m]$, whose length is $m\ell\leq N$.
  We construct the arrays $F_0,\ldots,F_\ell$ in two steps.
  \begin{enumerate}

  \item We first compute the frequencies of the length-$\ell$ strings.
    To this end, we allocate the arrays $F_0,\ldots,F_\ell$ and initialize
    every entry to zero. Each counter fits in one word. Since
    $\sum_{p=0}^{\ell}\AlphabetSize^p=\bigO(\AlphabetSize^\ell)$,
    \cref{eq:packed-prefix-frequencies-universe} bounds the initialization
    time and the space occupied by these arrays by
    $\bigO(m\ell\log\AlphabetSize/\log m)$. Next, construct the
    integer-value encoding structure from \cref{th:val-encoding} with
    parameters $m$ and $\AlphabetSize$. It supports strings of length
    at most $\floor{\log_{\AlphabetSize}m}$, and its assumptions hold
    because $\AlphabetSize\leq m\leq N$ and
    $w\geq1+\log N\geq1+\log m$.
    Its $\bigO(\sqrt m)$ construction time and space fit the same bound,
    since $\sqrt m\leq m/\log m$ for $m\geq16$ and
    $\ell\log\AlphabetSize\geq1$.
    We compute the frequencies using one of the following two methods.
    Since $8\ell b$ and $4\ell b$ are integers, the test between them and
    the value $t$ in the second method depend on $\log m$ only through
    $\lceil\log m\rceil$ and $\lfloor\log m\rfloor$, respectively,
    which we obtain by repeatedly halving $m$ in $\bigO(\log m)$ time.
    \begin{enumerate}

      \item Suppose that $\ell b\geq(\log m)/8$.
        For every $j\in[1\dd m]$, apply
        \cref{pr:packed-representation}\eqref{pr:packed-representation-substring}
        to the supplied representation, with starting position
        $(j-1)\ell+1$ and substring length $\ell$. Query the structure from
        \cref{th:val-encoding} with $\ell$ and the returned representation to compute
        $y:=\Val{\AlphabetSize}{W[j]}$, and increment $F_\ell[y]$.
        The substring and encoding queries take constant time. Since
        $b\leq2\log\AlphabetSize$, the assumption in this case gives
        $\ell\log\AlphabetSize\geq(\log m)/16$.
        Thus, the scan takes $\bigO(m\ell\log\AlphabetSize/\log m)$ time.

      \item Suppose that $\ell b<(\log m)/8$. Compute
        $t:=\floor{\tfrac{1}{4\ell b}\log m}
        =\Theta(\log m/(\ell b))$. In particular, $t\geq2$ and
        $t\ell b\leq(\log m)/4$. Hence
        $t\ell\leq\floor{\log_{\AlphabetSize}m}$, and the encoding
        structure supports concatenations of $t$ input strings.
        Compute the frequencies in three substeps.
        \begin{enumerate}

          \item We first compute a frequency array for the strings
            $G_i=W[it+1]\cdots W[(i+1)t]$, where
            $i\in[0\dd\floor{m/t})$.
            Compute $\AlphabetSize^{t\ell}=(\AlphabetSize^\ell)^t$
            in $\bigO(t)$ time and allocate an array
            $H[0\dd\AlphabetSize^{t\ell})$ of counters, initially zero.
            For every $i\in[0\dd\floor{m/t})$, apply
            \cref{pr:packed-representation}\eqref{pr:packed-representation-substring}
            to the supplied representation, with starting position
            $it\ell+1$ and substring length $t\ell$, to obtain the packed
            representation of $G_i$. Query the structure from
            \cref{th:val-encoding} with $t\ell$ and this representation
            to compute $z:=\Val{\AlphabetSize}{G_i}$, and increment $H[z]$.
            Each substring and encoding query takes $\bigO(1)$ time.

          \item Next, we update $F_\ell$ using the counts in $H$.
            For every $z\in[0\dd\AlphabetSize^{t\ell})$ with $H[z]>0$,
            set $v:=z$ and repeat the following operations exactly $t$
            times: compute $y:=v\bmod\AlphabetSize^\ell$, add $H[z]$ to
            $F_\ell[y]$, and set $v:=\floor{v/\AlphabetSize^\ell}$.
            By \cref{def:val}, the successive values of $y$ are the
            integer values of the $t$ strings in the group, in reverse
            order. If a string occurs $r$ times in a group with integer
            value $z$, its frequency increases by $rH[z]$, as required.
            Performing exactly $t$ iterations
            includes initial strings equal to $\zero^\ell$, even when
            $v$ has already become zero.

          \item Finally, count the fewer than $t$ remaining strings.
            For every $j\in[t\floor{m/t}+1\dd m]$, extract the packed
            representation of $W[j]$, compute
            $y:=\Val{\AlphabetSize}{W[j]}$ using the structure from
            \cref{th:val-encoding}, and increment $F_\ell[y]$, as in
            the first case above. This takes $\bigO(t)$ time.
        \end{enumerate}

        The three substeps take
        \[
          \bigO\left(m/t+\AlphabetSize^{t\ell}(1+t)+t\right)
            =\bigO\left(
              \frac{m\ell b}{\log m}+m^{1/4}\log m
            \right)
            =\bigO\left(
              \frac{m\ell\log\AlphabetSize}{\log m}
            \right)
        \]
        time. Here, we use $1/t=\bigO(\ell b/\log m)$,
        $\AlphabetSize^{t\ell}\leq m^{1/4}$, $t=\bigO(\log m)$,
        $b\leq2\log\AlphabetSize$, and
        $m^{1/4}\log m=\bigO(m/\log m)$. The array $H$ uses
        $\bigO(m^{1/4})=\bigO(m\ell\log\AlphabetSize/\log m)$ words.
    \end{enumerate}

    In either case,
    $F_\ell[\Val{\AlphabetSize}{X}]$ now equals the number of
    occurrences of $X$ in $W$ for every
    $X\in[0\dd\AlphabetSize)^\ell$.

  \item Finally, we compute the frequencies of the shorter prefixes by
    processing $p=\ell-1,\ell-2,\ldots,0$ in decreasing order. For every
    $y\in[0\dd\AlphabetSize^p)$, set
    \[
      F_p[y]
        :=\sum_{c=0}^{\AlphabetSize-1}F_{p+1}[\AlphabetSize y+c].
    \]
    There is a unique $X\in[0\dd\AlphabetSize)^p$ satisfying
    $y=\Val{\AlphabetSize}{X}$, and
    $\AlphabetSize y+c=\Val{\AlphabetSize}{Xc}$ for every
    $c\in[0\dd\AlphabetSize)$. When length $p$ is processed, the entries
    corresponding to the extensions $Xc$ already contain their correct prefix
    frequencies. Their sum is the number of strings in $W$ having prefix $X$.
    The total number of additions is
    $\sum_{p=0}^{\ell-1}\AlphabetSize^{p+1}
      =\bigO(\AlphabetSize^\ell)$. By
    \cref{eq:packed-prefix-frequencies-universe}, this step takes
    $\bigO(m\ell\log\AlphabetSize/\log m)$ time. In particular,
    $F_0[0]=m$, as required for the empty prefix.
  \end{enumerate}

  The input representation occupies
  $\bigO(1+m\ell\log\AlphabetSize/w)
    =\bigO(m\ell\log\AlphabetSize/\log m)$ words. The arrays
  $F_0,\ldots,F_\ell$, the integer-value encoding structure from
  \cref{th:val-encoding}, and, in the second case, $H$ are stored
  along with this representation. Together, they occupy
  $\bigO(m\ell\log\AlphabetSize/\log m)$ words. The total running time is
  $\bigO(m\ell\log\AlphabetSize/\log m)$.
\end{proof}

%% file: tools/packed-substring-frequencies.tex
\subsection{Packed Substring Frequencies}
\label{sec:tools-packed-substring-frequencies}

\begin{theorem}[{\cite{sss}; see also~\cite{hierarchy}}]
\label{th:packed-substring-frequencies}
  Let $N,\AlphabetSize\in\Z_{\geq2}$ satisfy $\AlphabetSize\leq N$,
  and let $\ell\in[1\dd\lfloor\frac12\log_{\AlphabetSize}N\rfloor]$.
  Consider the word RAM model with word size $w\geq1+\log N$.
  Given $N$, $\AlphabetSize$, $\ell$, $|\Text|$, and the packed representation
  $\PackedRepresentation{w}{\AlphabetSize}{\Text}$
  (\cref{def:packed-representation}) of any nonempty string
  $\Text\in\IntegerAlphabet^{\leq N}$, we can
  deterministically compute arrays $F_p[0\dd\AlphabetSize^p)$,
  for $p\in[1\dd\ell]$, satisfying
  \[
    F_p[\Val{\AlphabetSize}{X}]=|\OccTwo{X}{\Text}|
    \qquad\text{for every }X\in\IntegerAlphabet^p
  \]
  (\cref{def:val,def:occ}). The algorithm takes
  $\bigO(N/\log_{\AlphabetSize}N)$ time and uses
  $\bigO(N/\log_{\AlphabetSize}N)$ words of space.
\end{theorem}

%% file: tools/packed-suffix-array-range-queries.tex
\subsection{Packed Suffix-Array Range Queries}
\label{sec:tools-packed-suffix-array-range-queries}

\begin{proposition}[Detailed version adapted from~\cite{breaking}]\label{pr:range-beg-and-end}
  Let $N,\AlphabetSize\in\Z_{\geq2}$ satisfy $\AlphabetSize\leq N$.
  Consider the word RAM model with word size $w\geq1+\log N$.
  Given $N$, $\AlphabetSize$, and the packed representation
  $\PackedRepresentation{w}{\AlphabetSize}{\Text}$
  (\cref{def:packed-representation})
  of any string $\Text \in \IntegerAlphabet^{N}$,
  we can in $\bigO(N / \log_{\AlphabetSize} N)$ time and using
  $\bigO(N / \log_{\AlphabetSize} N)$ words of space construct a data structure
  of size $\bigO(\sqrt N)$ words that,
  given $|X|$ and the packed representation
  $\PackedRepresentation{w}{\AlphabetSize}{X}$ of any
  string $X\in\IntegerAlphabet^*$ satisfying
  $|X|\leq\tfrac12\log_{\AlphabetSize}N$,
  returns the following values in $\bigO(1)$ time (see \cref{def:occ}):
  \begin{itemize}
  \item $\RangeBegTwo{X}{\Text}$,
  \item $\RangeEndTwo{X}{\Text}$.
  \end{itemize}
\end{proposition}
\begin{proof}

  Compute $h=\lfloor\tfrac12\log_{\AlphabetSize}N\rfloor$ in
  $\bigO(\log N)$ time by enumerating the powers of $\AlphabetSize$
  that do not exceed $N$.
  We multiply the current power $a$ by $\AlphabetSize$ only if
  $a\leq\lfloor N/\AlphabetSize\rfloor$.
  If $h=0$, store $N$ and answer every query by returning $(0,N)$,
  since the only permitted query string is $\emptystring$.
  Henceforth, assume that $h\geq1$. For every $p\in[0\dd h]$, let
  $L_p[0\dd\AlphabetSize^p)$ be the array satisfying
  \[
    L_p[\Val{\AlphabetSize}{X}]
      =(\RangeBegTwo{X}{\Text},\RangeEndTwo{X}{\Text})
    \qquad\text{for every }X\in\IntegerAlphabet^p
  \]
  (\cref{def:val,def:occ}). In particular, $L_0[0]=(0,N)$.

  \DSComponents
  The data structure consists of the following components:
  \begin{enumerate}

  \item The integer-value encoding structure from \cref{th:val-encoding}
    for $N$ and $\AlphabetSize$. It occupies
    $\bigO(\sqrt N)$ words.

  \item We store the arrays $L_0,\ldots,L_h$. Their total size is
    $\bigO(\sum_{p=0}^{h}\AlphabetSize^p)
      =\bigO(\AlphabetSize^h)=\bigO(\sqrt N)$ words.
  \end{enumerate}

  Together, these components occupy $\bigO(\sqrt N)$ words.

  \DSQueries
  Given $|X|$ and $\PackedRepresentation{w}{\AlphabetSize}{X}$, query
  the structure from \cref{th:val-encoding} to obtain
  $x=\Val{\AlphabetSize}{X}$, and return $L_{|X|}[x]$.
  Since $|X|\leq h\leq\lfloor\log_{\AlphabetSize}N\rfloor$,
  this takes $\bigO(1)$ time, including when $X=\emptystring$.

  \DSConstruction
  We construct the data structure in three steps.
  \begin{enumerate}

  \item We first construct the integer-value encoding structure from
    \cref{th:val-encoding}.
    We also compute an auxiliary array $A_{\rm pow}[0\dd h]$
    satisfying $A_{\rm pow}[p]=\AlphabetSize^p$ by repeated multiplication.
    This takes $\bigO(\sqrt N+h)$ time and words of space.

  \item Next, we compute the occurrence counts of all nonempty strings
    of length at most $h$. Apply \cref{th:packed-substring-frequencies}
    to $\Text$ with $\ell=h$.
    This gives arrays $F_p[0\dd\AlphabetSize^p)$, for $p\in[1\dd h]$,
    satisfying $F_p[\Val{\AlphabetSize}{X}]=|\OccTwo{X}{\Text}|$
    for every $X\in\IntegerAlphabet^p$.
    The theorem applies with size parameter $N$ and alphabet size
    $\AlphabetSize$ because $1\leq h\leq\frac12\log_{\AlphabetSize}N$
    and $w\geq1+\log N$.
    This step takes $\bigO(N/\log_{\AlphabetSize}N)$ time and words of space.

  \item Finally, we construct the range tables $L_0,\ldots,L_h$.
    Set $L_0[0]:=(0,N)$. For $k=0,1,\ldots,h-1$ and every
    $x\in[0\dd A_{\rm pow}[k])$, retrieve $(b,e):=L_k[x]$ and
    set $\delta:=0$. Then process
    $c=\AlphabetSize-1,\AlphabetSize-2,\ldots,0$, setting
    \[
      L_{k+1}[\AlphabetSize x+c]
        :=(e-\delta-F_{k+1}[\AlphabetSize x+c],\ e-\delta)
    \]
    and increasing $\delta$ by $F_{k+1}[\AlphabetSize x+c]$.
    To see why this is correct, let $X\in\IntegerAlphabet^k$ have
    value $x$. Within the suffix-array interval of $X$, a suffix
    equal to $X$, if one exists, precedes all its proper extensions.
    The remaining suffixes are ordered by the next character.
    Scanning these characters in decreasing order therefore recovers
    each child's interval from the right endpoint $e$, using its
    occurrence count as the interval length. This also gives the
    correct empty intervals for strings that do not occur.
    The total time is $\bigO(\AlphabetSize^h)$.
  \end{enumerate}

  Overall, construction takes $\bigO(N/\log_{\AlphabetSize}N)$ time
  and uses $\bigO(N/\log_{\AlphabetSize}N)$ words of space.
  All values fit in a machine word: occurrence counts and range endpoints are at most $N$, and
  all powers and encoded values are also at most $N$.
  The encoder applies because $\AlphabetSize\leq N$ and
  $w\geq1+\log N$.
\end{proof}

%% file: tools/packed-prefix-extraction-by-suffix-array-rank.tex
\subsection{Packed Prefix Extraction by Suffix-Array Rank}
\label{sec:tools-packed-prefix-extraction-by-suffix-array-rank}

\begin{proposition}[Detailed version adapted from~\cite{breaking}]\label{pr:rank-to-pref}
  Let $N,\AlphabetSize\in\Z_{\geq2}$ satisfy $\AlphabetSize\leq N$.
  Consider the word RAM model with word size $w\geq1+\log N$.
  Given $N$, $\AlphabetSize$, and the packed representation
  $\PackedRepresentation{w}{\AlphabetSize}{\Text}$ (\cref{def:packed-representation})
  of any string $\Text \in \IntegerAlphabet^{N}$,
  we can in $\bigO(N / \log_{\AlphabetSize} N)$ time
  and using $\bigO(N / \log_{\AlphabetSize} N)$ words of space construct
  a data structure of size $\bigO(N / \log N)$ words
  that supports the following query in $\bigO(1)$ time.
  Given $i\in[1\dd N]$ and an integer $\ell\geq1$ satisfying
  $\ell\leq\tfrac12\log_{\AlphabetSize}N$, it returns $|X|$ and the
  packed representation $\PackedRepresentation{w}{\AlphabetSize}{X}$, where
  (see \cref{def:suffix-array})
  \[
    X=\Text[\SA{\Text}[i]\dd\min(\SA{\Text}[i]+\ell,N+1)).
  \]
\end{proposition}
\begin{proof}

  Compute $h=\lfloor\tfrac12\log_{\AlphabetSize}N\rfloor$ in
  $\bigO(\log N)$ time by enumerating the powers of $\AlphabetSize$
  that do not exceed $N$. We multiply the current power $a$ by
  $\AlphabetSize$ only if $a\leq\lfloor N/\AlphabetSize\rfloor$.
  If $h=0$, store only $N$; there are no
  permitted query lengths. Henceforth, assume that $h\geq1$.
  For every $j \in [1 \dd N]$, let $X_j := \Text[j \dd \min(j + h, N + 1))$.
  Let $B_h[1 \dd N]$ be a bitvector defined such that, for every
  $i \in [1 \dd N]$, $B_h[i] = \one$ holds if and only if $i = N$, or $i < N$
  and $X_{\SA{\Text}[i]} \neq X_{\SA{\Text}[i+1]}$.
  Let $k := \Rank{B_h}{N}{\one}$ (\cref{def:rank-select}).
  Let $A_{\rm short}[1\dd k]$ and $A_{\rm len}[1\dd k]$ be arrays
  defined as follows. For every $i\in[1\dd k]$, let
  $i'=\Select{B_h}{i}{\one}$ (\cref{def:rank-select}).
  Entry $A_{\rm short}[i]$ stores the packed representation of
  $X_{\SA{\Text}[i']}$, and $A_{\rm len}[i]=|X_{\SA{\Text}[i']}|$.
  Denote $b_{\AlphabetSize}=\lceil\log\AlphabetSize\rceil$.

  \DSComponents
  The data structure consists of the following components:
  \begin{enumerate}

  \item The bitvector $B_h[1 \dd N]$, augmented with support for $\bigO(1)$-time rank
    queries using \cref{th:bin-rank-select}. The bitvector needs
    $\bigO(N / \log N)$ words of space, and the augmentation with
    \cref{th:bin-rank-select} does
    not increase the space.

  \item We store the arrays $A_{\rm short}$ and $A_{\rm len}$, and the
    integer $b_{\AlphabetSize}$. Each string has length at most $h$,
    and hence its packed representation occupies $\bigO(1)$ words.
    Since $k\leq\sum_{t\in[1\dd h]}\AlphabetSize^t<2\AlphabetSize^h$,
    these objects use $\bigO(\AlphabetSize^h)=\bigO(\sqrt N)$
    words of space.
  \end{enumerate}

  In total, the structure uses
  $\bigO(N/\log N+\sqrt N)=\bigO(N/\log N)$ words of space.

  \DSQueries
  Given $i\in[1\dd N]$ and $\ell\in[1\dd h]$, denote
  $X=\Text[\SA{\Text}[i]\dd\min(\SA{\Text}[i]+\ell,N+1))$.
  We compute $|X|$ and $\PackedRepresentation{w}{\AlphabetSize}{X}$
  in two steps:
  \begin{enumerate}

  \item In $\bigO(1)$ time, we compute
    $j := \Rank{B_h}{i-1}{\one}+1$.

  \item\label{step:rank-to-pref-query-return}
    Compute $d:=\min(\ell,A_{\rm len}[j])$, and extract the length-$d$
    prefix of the string represented in $A_{\rm short}[j]$ using
    \cref{pr:packed-representation}\eqref{pr:packed-representation-substring}.
    Return $d$ and the extracted representation.
  \end{enumerate}

  Let $p=\SA{\Text}[i]$. By the definitions of $B_h$,
  $A_{\rm short}$, and $A_{\rm len}$, entry $A_{\rm short}[j]$ stores
  the packed representation of $X_p$, and $A_{\rm len}[j]=|X_p|$.
  Since $\ell\leq h$, the string $X$ is the length-$d$ prefix of
  $X_p$. Thus, Step~\ref{step:rank-to-pref-query-return} returns
  $|X|=d$ and the packed representation of $X$, including when the
  suffix has length smaller than $\ell$.
  The substring extraction uses length bound $N$ and takes
  $\bigO(1+d\log\AlphabetSize/w)=\bigO(1)$ time. Consequently, the entire query
  takes $\bigO(1)$ time.

  \DSConstruction
  Given $N$, $\AlphabetSize$, and the packed representation
  $\PackedRepresentation{w}{\AlphabetSize}{\Text}$ of $\Text$, we
  construct the above data structure as follows:
  \begin{enumerate}

  \item We first prepare the supporting structures and integer parameters.
    Construct the structure from \cref{pr:range-beg-and-end} with
    the packed representation of $\Text$.
    Construct also the integer-value decoding structure from
    \cref{th:val-decoding}.
    Compute $\AlphabetSize^h$ and
    $b_{\AlphabetSize}=\lceil\log\AlphabetSize\rceil$ in $\bigO(\log N)$ time.

  \item Next, we construct $B_h$ and its rank support.
    Initialize the packed representation of $B_h[1\dd N]$ to zeros
    using \cref{pr:packed-representation}\eqref{pr:packed-representation-initialize},
    and proceed as follows:
    \begin{enumerate}

    \item\label{step:rank-to-pref-mark-full-strings}
      For every $x\in[0\dd\AlphabetSize^h)$, query the structure
      from \cref{th:val-decoding} with $(h,x)$ to obtain the packed
      representation of the string $X\in\IntegerAlphabet^{h}$
      satisfying $\Val{\AlphabetSize}{X}=x$.
      Using $h$ and this representation, query the structure from
      \cref{pr:range-beg-and-end} to compute
      $b=\RangeBegTwo{X}{\Text}$ and $e=\RangeEndTwo{X}{\Text}$.
      If $b<e$, set $B_h[e]:=\one$ using
      \cref{pr:packed-representation}\eqref{pr:packed-representation-update}.

    \item\label{step:rank-to-pref-mark-short-strings}
      We now mark the positions corresponding to suffixes shorter than
      $h$. For every $d\in[1\dd h)$, extract the packed representation
      of $X=\Text[N-d+1\dd N]$. Query the structure from
      \cref{pr:range-beg-and-end} with $d$ and this representation to
      compute $b=\RangeBegTwo{X}{\Text}$. Set $r:=b+1$ and
      $B_h[r]:=\one$.

    \item Augment $B_h$ with $\bigO(1)$-time rank queries using
      \cref{th:bin-rank-select}. This takes $\bigO(N/\log N)$ time
      and words of space.
    \end{enumerate}
    For every occurring $X\in\IntegerAlphabet^{h}$ with suffix-array
    interval $(b,e]$, it holds
    $(b,e]=\{i\in[1\dd N]:X_{\SA{\Text}[i]}=X\}$.
    Thus, Step~\ref{step:rank-to-pref-mark-full-strings} marks the last
    position of every maximal interval whose common string
    $X_{\SA{\Text}[i]}$ has length $h$.
    For every $d\in[1\dd h)$, the suffix $X=\Text[N-d+1\dd N]$
    has suffix-array rank $r=\RangeBegTwo{X}{\Text}+1$
    (\cref{def:occ}), because it is the lexicographically smallest
    suffix of $\Text$ with prefix $X$. It satisfies
    $X_{\SA{\Text}[r]}=X$, and no other string $X_j$ has length $d$.
    Thus, $\{i\in[1\dd N]:X_{\SA{\Text}[i]}=X\}=\{r\}$, and
    Step~\ref{step:rank-to-pref-mark-short-strings} marks this singleton
    interval.
    Together, the two cases account for every $X_j$ and establish that
    the constructed bitvector is $B_h$.

  \item Finally, we construct $A_{\rm short}$ and $A_{\rm len}$.
    Compute $k=\Rank{B_h}{N}{\one}$, allocate both arrays with
    index range $[1\dd k]$, and fill them as follows:
    \begin{enumerate}

    \item\label{step:rank-to-pref-store-full-strings}
      For every $x\in[0\dd\AlphabetSize^h)$, use the structure
      from \cref{th:val-decoding} with $(h,x)$ to obtain the packed
      representation of $X\in\IntegerAlphabet^{h}$ satisfying
      $\Val{\AlphabetSize}{X}=x$.
      Compute $b=\RangeBegTwo{X}{\Text}$ and $e=\RangeEndTwo{X}{\Text}$
      using the structure from \cref{pr:range-beg-and-end}.
      If $b<e$, compute $i=\Rank{B_h}{e}{\one}$, store the
      packed representation of $X$ in $A_{\rm short}[i]$, and set
      $A_{\rm len}[i]:=h$.

    \item\label{step:rank-to-pref-store-short-strings}
      For every $d\in[1\dd h)$, extract the packed representation
      of $X=\Text[N-d+1\dd N]$. Query the structure from
      \cref{pr:range-beg-and-end} with $d$ and this representation to
      compute $b=\RangeBegTwo{X}{\Text}$, and set $r:=b+1$.
      Compute $i=\Rank{B_h}{r}{\one}$, store the packed
      representation of $X$ in $A_{\rm short}[i]$, and set
      $A_{\rm len}[i]:=d$.
    \end{enumerate}
    The preceding step marks exactly the last position of each maximal
    interval of equal strings $X_{\SA{\Text}[i]}$. Applying a rank query
    at this position gives the index of the interval in suffix-array order.
    For length-$h$ strings, this position is $e$ in
    Step~\ref{step:rank-to-pref-store-full-strings}; for shorter suffixes,
    it is $r$ in Step~\ref{step:rank-to-pref-store-short-strings}.
    The two substeps therefore store the packed representation of each
    interval's common string and its length at the required index.
  \end{enumerate}

  The first step takes $\bigO(N/\log_{\AlphabetSize}N)$ time and words of
  space. Every decoded string has length
  $h\leq\lfloor\log_{\AlphabetSize}N\rfloor$, each decoding query takes
  $\bigO(1)$ time, and extracting a substring of length at most $h$ using
  \cref{pr:packed-representation}\eqref{pr:packed-representation-substring}
  takes $\bigO(1)$ time. Consequently, each iteration
  in Steps~\ref{step:rank-to-pref-mark-full-strings}--\ref{step:rank-to-pref-mark-short-strings}
  and~\ref{step:rank-to-pref-store-full-strings}--\ref{step:rank-to-pref-store-short-strings}
  takes $\bigO(1)$ time. Their total time is
  $\bigO(\AlphabetSize^h+h)$, while initializing $B_h$
  and constructing its rank support take $\bigO(N/\log N)$ time.
  All strings stored in $A_{\rm short}$ have length at most $h$ and occupy
  $\bigO(1)$ words in packed form. Since $\AlphabetSize^h\leq\sqrt N$ and
  $h=\bigO(\log N)$, the entire construction takes $\bigO(N/\log_{\AlphabetSize}N)$ time
  and uses $\bigO(N/\log_{\AlphabetSize}N)$ words of space.
  All packed operations use length bound $N$. Their assumptions and
  those of the decoding structure follow from $\AlphabetSize\leq N$
  and $w\geq1+\log N$.
  The structures from \cref{pr:range-beg-and-end,th:val-decoding} use
  size parameter $N$ and alphabet size $\AlphabetSize$, and the rank
  support for $B_h$ uses size parameter $N$.
  All calls to \cref{pr:range-beg-and-end} use strings of length at most
  $h\leq\tfrac12\log_{\AlphabetSize}N$.
\end{proof}

%% file: tools/periodicity-testing.tex
\subsection{Periods of Short Packed Strings}
\label{sec:tools-periodicity-testing}

\begin{proposition}[{\cite{hierarchy}}]\label{pr:periods-of-short-strings}
  Let $N,\AlphabetSize\in\Z_{\geq2}$ satisfy $\AlphabetSize\leq N$, and
  let $d\in[1\dd\lfloor\log_{\AlphabetSize}N\rfloor]$. Consider the word
  RAM model with word size $w\geq1+\log N$. Given $N$, $\AlphabetSize$,
  and $d$, in $\bigO(\AlphabetSize^d d)$ time and using
  $\bigO(\AlphabetSize^d)$ words of space, we can construct a data
  structure of size $\bigO(\AlphabetSize^d)$ words that, given $|X|$
  and the packed representation
  $\PackedRepresentation{w}{\AlphabetSize}{X}$
  (\cref{def:packed-representation}) of any nonempty string
  $X\in\IntegerAlphabet^+$ satisfying $|X|\leq d$, returns
  $\per{X}$ in $\bigO(1)$ time.
\end{proposition}

%% file: tools/recognizing-periodic-positions.tex
\subsection{Recognizing Periodic Positions}
\label{sec:tools-packed-suffix-array}

\begin{proposition}[Adapted from~\cite{breaking}]\label{pr:sa-core}
  Let $N,\AlphabetSize\in\Z_{\geq2}$ satisfy $\AlphabetSize\leq N$.
  Consider the word RAM model with word size $w\geq1+\log N$.
  Given $N$, $\AlphabetSize$, and the packed representation
  $\PackedRepresentation{w}{\AlphabetSize}{\Text}$
  (\cref{def:packed-representation}) of any string
  $\Text\in\IntegerAlphabet^N$, we can in
  $\bigO(N/\log_{\AlphabetSize}N)$ time and using
  $\bigO(N/\log_{\AlphabetSize}N)$ words of space construct a data
  structure of size $\bigO(N/\log N)$ words that supports
  the following query in $\bigO(1)$ time. Given $i\in[1\dd N]$
  and an integer $\tau\geq1$ satisfying
  $\tau\leq\tfrac16\log_{\AlphabetSize}N$, it determines whether
  $\SA{\Text}[i]\in\RTwo{\tau}{\Text}$
  (\cref{def:suffix-array,def:periodic-positions}).
\end{proposition}

\begin{proof}
  Compute $h=\lfloor\tfrac12\log_{\AlphabetSize}N\rfloor$ in
  $\bigO(\log N)$ time by enumerating the powers of $\AlphabetSize$
  up to $N$. We multiply the current power $a$ by $\AlphabetSize$
  only if $a\leq\lfloor N/\AlphabetSize\rfloor$.
  If $h=0$, store only $N$; there are no
  permitted query values of $\tau$. Henceforth, assume that $h\geq1$.

  \DSComponents
  We store the following two data structures:
  \begin{enumerate}
  \item The short-prefix structure from \cref{pr:rank-to-pref} for
    $\Text$. It occupies $\bigO(N/\log N)$ words.
  \item The shortest-period structure from \cref{pr:periods-of-short-strings}
    for strings of length at most $h$. It occupies $\bigO(\AlphabetSize^h)=\bigO(\sqrt N)$ words.
  \end{enumerate}
  Together, they occupy
  $\bigO(N/\log N+\sqrt N)=\bigO(N/\log N)$ words.

  \DSQueries
  Given $(i,\tau)$, query the structure from \cref{pr:rank-to-pref}
  with $i$ and $3\tau-1$ to obtain $|X|$ and the packed representation
  of $X=\Text[\SA{\Text}[i]\dd\min(\SA{\Text}[i]+3\tau-1,N+1))$.
  We then distinguish two cases:
  \begin{itemize}
  \item If $|X|<3\tau-1$, return that
    $\SA{\Text}[i]\notin\RTwo{\tau}{\Text}$.
  \item Otherwise, compute $p=\per{X}$ using the structure from
    \cref{pr:periods-of-short-strings}. Return that
    $\SA{\Text}[i]\in\RTwo{\tau}{\Text}$ if and only if $p\leq\tau/3$.
  \end{itemize}
  In the first case, the suffix is shorter than $3\tau-1$, and hence
  $\SA{\Text}[i]>N-3\tau+2$. In the second case,
  $X=\Text[\SA{\Text}[i]\dd\SA{\Text}[i]+3\tau-1)$ has length
  $3\tau-1$. Thus, both cases follow directly from
  \cref{def:periodic-positions}. Each query takes $\bigO(1)$ time.

  \DSConstruction
  Given $N$, $\AlphabetSize$, and $\PackedRepresentation{w}{\AlphabetSize}{\Text}$,
  construct the structures from
  \cref{pr:rank-to-pref,pr:periods-of-short-strings}.
  The first construction takes $\bigO(N/\log_{\AlphabetSize}N)$ time
  and words of space. The second takes
  $\bigO(\AlphabetSize^h h)=\bigO(\sqrt N\log N)
    =\bigO(N/\log_{\AlphabetSize}N)$ time and
  $\bigO(\AlphabetSize^h)=\bigO(\sqrt N)$ words of space.
  These bounds also cover computing $h$.
  Both structures use size parameter $N$ and alphabet size
  $\AlphabetSize$. They apply because $\AlphabetSize\leq N$,
  $1\leq h\leq\lfloor\log_{\AlphabetSize}N\rfloor$, and
  $w\geq1+\log N$. For every permitted query,
  $1\leq3\tau-1\leq\tfrac12\log_{\AlphabetSize}N$;
  since $3\tau-1$ is an integer, it is at most $h$.
  Thus, both structures support every query used above.
\end{proof}

\begin{proposition}[Adapted from~\cite{breaking}]\label{pr:isa-core}
  Let $N,\AlphabetSize\in\Z_{\geq2}$ satisfy $\AlphabetSize\leq N$.
  Consider the word RAM model with word size $w\geq1+\log N$.
  Given $N$, $\AlphabetSize$, and the packed representation
  $\PackedRepresentation{w}{\AlphabetSize}{\Text}$
  (\cref{def:packed-representation}) of any string
  $\Text\in\IntegerAlphabet^N$, we can in
  $\bigO(N/\log_{\AlphabetSize}N)$ time and using
  $\bigO(N/\log_{\AlphabetSize}N)$ words of space construct a data
  structure of size $\bigO(N/\log_{\AlphabetSize}N)$ words that supports
  the following query in $\bigO(1)$ time. Given $j\in[1\dd N]$
  and an integer $\tau\geq1$ satisfying
  $\tau\leq\tfrac16\log_{\AlphabetSize}N$, it determines whether
  $j\in\RTwo{\tau}{\Text}$ (\cref{def:periodic-positions}).
\end{proposition}
\begin{proof}
  Compute $h=\lfloor\tfrac12\log_{\AlphabetSize}N\rfloor$ in
  $\bigO(\log N)$ time by enumerating the powers of $\AlphabetSize$
  up to $N$. We multiply the current power $a$ by $\AlphabetSize$
  only if $a\leq\lfloor N/\AlphabetSize\rfloor$.
  If $h=0$, store only $N$; there are no permitted query
  values of $\tau$. Henceforth, assume that $h\geq1$, and denote
  $b_{\AlphabetSize}=\lceil\log\AlphabetSize\rceil$.

  \DSComponents
  The data structure consists of the following components:
  \begin{enumerate}
  \item The packed representation of $\Text$ and the integers $N$ and
    $b_{\AlphabetSize}$. They occupy
    $\bigO(1+N\log\AlphabetSize/w)=\bigO(N/\log_{\AlphabetSize}N)$ words.
  \item The shortest-period structure from \cref{pr:periods-of-short-strings}
    for strings of length at most $h$. It occupies $\bigO(\AlphabetSize^h)=\bigO(\sqrt N)$ words.
  \end{enumerate}
  Together, these components occupy $\bigO(N/\log_{\AlphabetSize}N)$ words.

  \DSQueries
  Given $(j,\tau)$, if $j>N-3\tau+2$, return that
  $j\notin\RTwo{\tau}{\Text}$. Otherwise, proceed in two steps:
  \begin{enumerate}
  \item Extract the packed representation of $X=\Text[j\dd j+3\tau-1)$
    using
    \cref{pr:packed-representation}\eqref{pr:packed-representation-substring}.
  \item Compute $p=\per{X}$ using the structure from
    \cref{pr:periods-of-short-strings}. Return that $j\in\RTwo{\tau}{\Text}$ if and only if
    $p\leq\tau/3$.
  \end{enumerate}
  The initial test excludes exactly the positions whose suffixes are
  shorter than $3\tau-1$. At every remaining position, the two steps
  test whether $\per{\Text[j\dd j+3\tau-1)}\leq\tau/3$.
  Correctness therefore follows from \cref{def:periodic-positions}.
  The extraction takes $\bigO(1+\tau\log\AlphabetSize/w)=\bigO(1)$ time,
  and the period query takes $\bigO(1)$ time.

  \DSConstruction
  Given $N$, $\AlphabetSize$, and $\PackedRepresentation{w}{\AlphabetSize}{\Text}$,
  construct the components in two steps:
  \begin{enumerate}
  \item Compute $b_{\AlphabetSize}=\lceil\log\AlphabetSize\rceil$
    in $\bigO(\log N)$ time. Copy the packed representation of $\Text$
    and store $N$ and $b_{\AlphabetSize}$. Copying takes
    $\bigO(1+N\log\AlphabetSize/w)=\bigO(N/\log_{\AlphabetSize}N)$
    time and words of space.
  \item Construct the structure from \cref{pr:periods-of-short-strings}.
    This takes
    $\bigO(\AlphabetSize^h h)=\bigO(\sqrt N\log N)
      =\bigO(N/\log_{\AlphabetSize}N)$ time and
    $\bigO(\AlphabetSize^h)=\bigO(\sqrt N)$ words of space.
  \end{enumerate}
  Thus, construction takes $\bigO(N/\log_{\AlphabetSize}N)$ time and
  words of space, including computing $h$ and $b_{\AlphabetSize}$.
  The packed operations use length bound $N$ and alphabet size
  $\AlphabetSize$, and the period structure uses size parameter $N$ and
  alphabet size $\AlphabetSize$. Their word-size assumptions follow from
  $w\geq1+\log N$ and $b_{\AlphabetSize}\leq\lceil\log N\rceil\leq w$.
  The period structure applies because $\AlphabetSize\leq N$ and
  $1\leq h\leq\lfloor\log_{\AlphabetSize}N\rfloor$.
  For every permitted query, $1\leq3\tau-1\leq\tfrac12\log_{\AlphabetSize}N$,
  and hence the integer $3\tau-1$ is at most $h$.
\end{proof}

%% file: tools/packed-sequence-to-string-mapping.tex
\subsection{Packed Sequence-to-String Mapping}
\label{sec:tools-sequence-to-string-mapping}

\begin{definition}[Sequence-to-string mapping]\label{def:seq-to-string}
  Let $\AlphabetSize \in \Z_{\geq 2}$.
  Let $W[1 \dd m]$ be a sequence of $m \geq \AlphabetSize$ strings of length $\ell \geq 1$
  over alphabet $\IntegerAlphabet$.
  For every $i \in [0 \dd m)$, let
  $s(i) = \increase{\AlphabetSize}{\AlphabetMap{m}{\AlphabetSize}{i}}$
  (\cref{def:alphabet-map,def:increase}), and denote
  $c=\AlphabetSize$. We define
  \[
    \SeqToString{\AlphabetSize}{W} :=
      \textstyle\bigodot_{i=1}^{m}
        \revstr{W[i]} \cdot c \cdot s(i-1)
    \in [0 \dd 2\AlphabetSize)^{*}.
  \]
\end{definition}

\begin{example}\label{ex:seq-to-string}
  Let $\AlphabetSize = 2$, $\ell = 2$, and $m = 3$, and consider the sequence
  $W$ with $W[1]=\zero\one$, $W[2]=\one\zero$, and $W[3]=\zero\zero$.
  Since $\lceil \log_2 3 \rceil = 2$, we have
  $\AlphabetMap{3}{2}{0} = \zero\zero$,
  $\AlphabetMap{3}{2}{1} = \zero\one$, and
  $\AlphabetMap{3}{2}{2} = \one\zero$.
  Hence
  $s(0) = \increase{2}{\zero\zero} = \two\two$,
  $s(1) = \increase{2}{\zero\one}  = \two\three$,
  $s(2) = \increase{2}{\one\zero}  = \three\two$, and $c=2$.
  Therefore,
  \[
    \SeqToString{2}{W}
      = \revstr{W[1]} \cdot c \cdot s(0)\cdot
        \revstr{W[2]} \cdot c \cdot s(1)\cdot
        \revstr{W[3]} \cdot c \cdot s(2)
      = \one\zero\,\two\,\two\two\,\zero\one\,\two\,\two\three\,
        \zero\zero\,\two\,\three\two.
  \]
  In particular, each block $\revstr{W[i]}$ uses symbols in $[0 \dd 2)$,
  whereas $c$ and each string $s(i)$ use symbols in $[2 \dd 4)$.
\end{example}

\begin{lemma}\label{lm:seq-to-string}
  Let $\AlphabetSize \in \Z_{\geq 2}$.
  Let $W[1 \dd m]$ be a sequence of $m \geq \AlphabetSize$ strings of length $\ell \geq 1$
  over alphabet $\IntegerAlphabet$.
  Denote $\Text = \SeqToString{\AlphabetSize}{W}$ (\cref{def:seq-to-string}),
  $k = \lceil \log_{\AlphabetSize} m \rceil$, $\gamma = \ell + k + 1$,
  and $c=\AlphabetSize$.
  Let $X \in \IntegerAlphabet^{+}$ satisfy $|X| \leq \ell$, denote
  $t = \PrefixRank{W}{m}{X}$, and, for every $r \in [1 \dd t]$, denote
  $i_r = \PrefixSelect{W}{r}{X}$ (\cref{def:prefix-rank-and-select}).
  Denote
  $b_X = \RangeBegTwo{\revstr{X} \cdot c}{\Text}$ and
  $e_X = \RangeEndTwo{\revstr{X} \cdot c}{\Text}$ (\cref{def:occ}). Then:
  \begin{enumerate}
  \item\label{lm:seq-to-string-range-size}
    It holds $e_X-b_X=t$.
  \item\label{lm:seq-to-string-locate}
    For every $r \in [1 \dd t]$, it holds
    $\SA{\Text}[b_X+r] = \gamma i_r-(k+|X|)$.
  \end{enumerate}
\end{lemma}
\begin{proof}
  Denote $Y=\revstr{X}$ and $\Pat=Y\cdot c$. For every
  $r\in[1\dd t]$, denote $q_r=\gamma i_r-(k+|X|)$.
  We prove the two claims in order:
  \begin{enumerate}

  \item First, we prove that $e_X-b_X=t$. To this end, we show that
    \[
      \OccTwo{\Pat}{\Text}=\{q_r:r\in[1\dd t]\}.
    \]
    For every $r\in[1\dd t]$, the string $X$ is a prefix of $W[i_r]$.
    Thus, $Y$ is a suffix of $\revstr{W[i_r]}$, and the construction
    of $\Text$ gives an occurrence of $\Pat=Y\cdot c$ starting at
    $q_r=\gamma i_r-(k+|X|)$.
    Conversely, let $q\in\OccTwo{\Pat}{\Text}$. Since $Y$ is nonempty
    and all its symbols belong to $\IntegerAlphabet$, whereas $c$
    and every symbol of every $s(i)$ belong to
    $[\AlphabetSize\dd2\AlphabetSize)$, the occurrence of $Y$ starting
    at $q$ lies entirely in a block $\revstr{W[i]}$ for some
    $i\in[1\dd m]$. The following symbol is $c=\AlphabetSize$,
    which does not occur inside this block. Therefore, $Y$ is a
    suffix of $\revstr{W[i]}$, or equivalently, $X$ is a prefix of
    $W[i]$. Hence $i=i_r$ for some $r\in[1\dd t]$, and
    $q=\gamma i_r-(k+|X|)=q_r$. This proves the set equality.
    The positions $q_r$ are distinct because the indices $i_r$ are
    distinct. By \cref{def:occ}, it follows that
    $e_X-b_X=|\OccTwo{\Pat}{\Text}|=t$.

  \item We now prove that
    $\SA{\Text}[b_X+r]=q_r$ for every $r\in[1\dd t]$.
    If $t=0$, there is nothing to prove. Otherwise,
    \cref{rm:occ} and the first step imply that the suffixes
    starting at $q_1,\ldots,q_t$ occupy suffix-array positions
    $b_X+1,\ldots,e_X$. It remains to determine their order.
    By \cref{def:prefix-rank-and-select}, we have
    $i_1<\cdots<i_t$. Alphabet mapping preserves lexicographic order,
    as does increasing every symbol by $\AlphabetSize$
    (\cref{ob:alphabet-map-properties}\eqref{ob:alphabet-map-lex-property} and \cref{def:increase}). Thus,
    $s(i_1-1)\prec\cdots\prec s(i_t-1)$.
    For every $r\in[1\dd t]$, the suffix
    $\Text[q_r\dd|\Text|]$ begins with $\Pat\cdot s(i_r-1)$.
    Since these strings $s(i_r-1)$ are distinct and all have length
    $k$, the comparison of any two such suffixes is determined
    within these strings. Hence the suffixes occur in the order
    $\Text[q_1\dd|\Text|]\prec\cdots\prec\Text[q_t\dd|\Text|]$,
    giving $\SA{\Text}[b_X+r]=q_r=\gamma i_r-(k+|X|)$ for every
    $r\in[1\dd t]$.\qedhere
  \end{enumerate}
\end{proof}

\begin{proposition}\label{pr:seq-to-str-range-beg-and-range-end}
  Let $N,\AlphabetSize,m\in\Z_{\geq2}$ satisfy
  $\AlphabetSize\leq m$, and let $\ell\in\Z_{\geq1}$ satisfy
  $\AlphabetSize^\ell\leq N$ and $m\ell\leq N$.
  Consider the word RAM model with word
  size $w$ satisfying $w\geq1+\log N$. Let
  $W[1\dd m]\in(\IntegerAlphabet^\ell)^m$, and denote
  $\Text=\SeqToString{\AlphabetSize}{W}$ and
  $c=\AlphabetSize$
  (\cref{def:seq-to-string}). Given $N$, $\AlphabetSize$, $m$, $\ell$, and the
  packed sequence representation
  $\PackedSeqRepresentation{w}{\AlphabetSize}{W}$
  (\cref{def:packed-sequence-representation}) of $W$, we can in
  $\bigO(m+\AlphabetSize^\ell)$ time construct a data structure using
  $\bigO(m+\AlphabetSize^\ell)$ words
  that, given $|X|$ and the packed representation
  $\PackedRepresentation{w}{\AlphabetSize}{X}$
  (\cref{def:packed-representation}) of any
  $X\in\IntegerAlphabet^{+}$ with $|X|\leq\ell$, returns the following
  values in $\bigO(1)$ time (see \cref{def:occ}):
  \begin{itemize}
  \item $\RangeBegTwo{\revstr{X}\cdot c}{\Text}$,
  \item $\RangeEndTwo{\revstr{X}\cdot c}{\Text}$.
  \end{itemize}
\end{proposition}
\begin{proof}
  Denote $N_0=\max\{m,\AlphabetSize^\ell\}$.
  For every $p\in[1\dd\ell]$, let
  $B_p[0\dd\AlphabetSize^p)$ and $E_p[0\dd\AlphabetSize^p)$ satisfy
  $B_p[\Val{\AlphabetSize}{X}]=\RangeBegTwo{\revstr{X}\cdot c}{\Text}$
  and $E_p[\Val{\AlphabetSize}{X}]=\RangeEndTwo{\revstr{X}\cdot c}{\Text}$
  for every $X\in\IntegerAlphabet^p$.

  \DSComponents
  The data structure consists of the following components:
  \begin{enumerate}
  \item The integer-value structure from \cref{th:val-encoding} with
    parameters $N_0$ and $\AlphabetSize$. It uses $\bigO(\sqrt{N_0})$ words.
  \item We store $B_p$ and $E_p$ for $p\in[1\dd\ell]$.
    Together, they occupy
    $\bigO(\sum_{p=1}^{\ell}\AlphabetSize^p)
      =\bigO(\AlphabetSize^\ell)$ words.
  \end{enumerate}
  The total size is $\bigO(m+\AlphabetSize^\ell)$ words.

  \DSQueries
  Given $|X|$ and the packed representation of $X$, answer the query
  in two steps:
  \begin{enumerate}
  \item Use the structure from \cref{th:val-encoding} to compute
    $x=\Val{\AlphabetSize}{X}$ in $\bigO(1)$ time.
  \item Return $B_{|X|}[x]$ and $E_{|X|}[x]$.
  \end{enumerate}
  Since $|X|\leq\ell\leq\lfloor\log_{\AlphabetSize}N_0\rfloor$,
  the conversion is supported, and the total query time is $\bigO(1)$.

  \DSConstruction
  We construct the components as follows:
  \begin{enumerate}

  \item First, compute the powers
    $A_{\rm pow}[p]:=\AlphabetSize^p$ for $p\in[0\dd\ell]$,
    $N_0$, and $b=\lceil\log\AlphabetSize\rceil$ in $\bigO(\log N_0)$ time.
    To keep all intermediate integers within one word, compute $b$
    using integer divisions. Construct the structure from
    \cref{th:val-encoding} in $\bigO(\sqrt{N_0})$ time.

  \item We now construct the arrays $B_p$ and $E_p$.
    For this purpose, let the auxiliary arrays
    $A_p[0\dd\AlphabetSize^p)$ and $H_p[0\dd\AlphabetSize^p)$ for
    $p\in[1\dd\ell]$ satisfy
    \[
      A_p[\Val{\AlphabetSize}{X}]=|\OccTwo{\revstr{X}}{\Text}|,
      \qquad
      H_p[\Val{\AlphabetSize}{X}]=\RangeBegTwo{\revstr{X}}{\Text}
      \quad\text{for }X\in\IntegerAlphabet^p.
    \]
    Thus, $A_p$ gives the number of suffixes beginning with each
    $\revstr{X}$, and $H_p$ gives the number of suffixes smaller than
    $\revstr{X}$ in lexicographic order. The array $H_0[0\dd1)$
    is defined by $H_0[0]=0$. We proceed in three steps:
    \begin{enumerate}

    \item First, we compute the prefix-frequency arrays
      $F_p[0\dd\AlphabetSize^p)$ for $p\in[0\dd\ell]$, defined by
      $F_p[\Val{\AlphabetSize}{X}]=\PrefixRank{W}{m}{X}$ for every
      $X\in\IntegerAlphabet^p$.
      Initialize every element of $F_\ell$ to zero.
      For every $i\in[1\dd m]$, proceed as follows:
      \begin{enumerate}
      \item Extract the length-$\ell$ substring starting at position
        $(i-1)\ell+1$ from the input representation using
        \cref{pr:packed-representation}\eqref{pr:packed-representation-substring}
        with input length $m\ell$.
        This gives $\PackedRepresentation{w}{\AlphabetSize}{W[i]}$.
      \item Use the structure from \cref{th:val-encoding} with length
        $\ell$ and this representation to compute
        $x=\Val{\AlphabetSize}{W[i]}$, and increment $F_\ell[x]$.
      \end{enumerate}
      For $p=\ell-1,\ell-2,\ldots,0$ and every
      $x\in[0\dd\AlphabetSize^p)$, set
      $F_p[x]:=\sum_{a=0}^{\AlphabetSize-1}F_{p+1}[\AlphabetSize x+a]$.
      This counts all strings beginning with a given length-$p$ prefix
      by summing over its one-symbol extensions.
      Each extraction and integer-value query takes $\bigO(1)$ time,
      since $\ell\leq\log_{\AlphabetSize}N_0$.
      The initialization and sums take $\bigO(\AlphabetSize^\ell)$ time,
      giving $\bigO(m+\AlphabetSize^\ell)$ time in total.

    \item Next, we compute $A_1,\ldots,A_\ell$, starting with the
      longest strings. Copy $F_\ell$ into $A_\ell$.
      For $p=\ell-1,\ell-2,\ldots,1$ and every
      $x\in[0\dd\AlphabetSize^p)$, set
      \[
        A_p[x]:=F_p[x]+\sum_{a=0}^{\AlphabetSize-1}
          A_{p+1}[a\cdot A_{\rm pow}[p]+x].
      \]
      To justify this computation, consider any $p\in[1\dd\ell]$ and
      $X\in\IntegerAlphabet^p$, and denote $Y=\revstr{X}$ and
      $x=\Val{\AlphabetSize}{X}$.
      Every occurrence of $Y$ lies inside a block $\revstr{W[i]}$,
      which is followed by $c$. Therefore, the suffixes beginning
      with $Y$ partition into those beginning with $Ya$, for
      $a\in[0\dd\AlphabetSize)$, and those beginning with $Yc$.
      The latter group has size $F_p[x]$ by
      \cref{lm:seq-to-string}\eqref{lm:seq-to-string-range-size}.
      Since $\revstr{Ya}=aX$ and
      $\Val{\AlphabetSize}{aX}=a\AlphabetSize^p+x$, the other groups
      have the sizes in the sum. When $p=\ell$, no occurrence of
      $Ya$ fits inside a block $\revstr{W[i]}$, giving
      $A_\ell[x]=F_\ell[x]$.
      The total number of additions is
      $\bigO(\sum_{p=1}^{\ell}\AlphabetSize^p)=\bigO(\AlphabetSize^\ell)$.

    \item Finally, we compute $H_0,\ldots,H_\ell$ and then the
      desired arrays $B_p$ and $E_p$.
      To this end, initialize $H_0[0]:=0$.
      For $p=0,1,\ldots,\ell-1$, process every
      $x\in[0\dd\AlphabetSize^p)$ as follows.
      Set $d:=H_p[x]$ and, for $a=0,1,\ldots,\AlphabetSize-1$,
      set $y:=a\cdot A_{\rm pow}[p]+x$, assign $H_{p+1}[y]:=d$,
      and then set $d:=d+A_{p+1}[y]$.
      To verify the computation, consider any $p\in[1\dd\ell]$ and
      $X\in\IntegerAlphabet^p$, and denote $Y=\revstr{X}$ and
      $x=\Val{\AlphabetSize}{X}$.
      In the partition from the preceding step, the groups of suffixes
      beginning with $Ya$ occur in ascending order of $a$, followed
      by the group beginning with $Yc$, since $c=\AlphabetSize$.
      Hence their starting offsets are obtained by adding their
      sizes to $H_p[x]$, exactly as in this computation.
      The initialization also gives the correct offsets for
      one-symbol strings: every suffix beginning with a symbol in
      $[0\dd\AlphabetSize)$ precedes every suffix beginning with
      a symbol in $[\AlphabetSize\dd2\AlphabetSize)$.
      For every $p\in[1\dd\ell]$ and
      $x\in[0\dd\AlphabetSize^p)$, now set
      \[
        B_p[x]:=H_p[x]+A_p[x]-F_p[x],
        \qquad E_p[x]:=H_p[x]+A_p[x].
      \]
      These assignments are correct because the suffixes beginning
      with $Yc$ are the last $F_p[x]$ suffixes among those beginning
      with $Y$. The same formulas apply when a group is empty.
      The scans and final assignments take
      $\bigO(\sum_{p=1}^{\ell}\AlphabetSize^p)=\bigO(\AlphabetSize^\ell)$ time.
    \end{enumerate}
  \end{enumerate}
  All constructed arrays together occupy $\bigO(\AlphabetSize^\ell)$ words.
  Their indices and the powers in $A_{\rm pow}$ are at most
  $\AlphabetSize^\ell\leq N_0\leq N$.
  The counts and offsets are at most $m\ell\leq N$, because exactly
  $m\ell$ suffixes begin with a symbol in $[0\dd\AlphabetSize)$, and
  all other suffixes are larger than every nonempty string over this alphabet.
  Thus, the input length and all computed integers fit in one word.
  The basic packed operations use length bound $N$ and the bit-width $b$,
  while the integer-value
  structure uses $N_0\leq N$ and supports strings of length up to
  $\lfloor\log_{\AlphabetSize}N_0\rfloor\geq\ell$.
  Their word-size assumptions follow from $w\geq1+\log N$.
  Including the $\bigO(\log N_0+\sqrt{N_0})$ setup cost, the complete
  construction takes $\bigO(m+\AlphabetSize^\ell)$ time and uses
  $\bigO(m+\AlphabetSize^\ell)$ words of space.
\end{proof}

\begin{proposition}\label{pr:seq-to-string-construction}
  Let $N,\AlphabetSize,\Seqlen\in\Z_{\geq2}$ satisfy
  $\AlphabetSize\leq\Seqlen$, and let $\ell\in\Z_{\geq1}$ satisfy
  $\AlphabetSize^\ell\leq N$. Denote
  $k=\lceil\log_{\AlphabetSize}\Seqlen\rceil$, and assume that
  $\Seqlen(\ell+k+1)\leq N$.
  Consider the word RAM model with word size $w$ satisfying
  $w\geq1+\log N$. Given $N$, $\AlphabetSize$, $\Seqlen$, $\ell$, and the packed
  sequence representation
  $\PackedSeqRepresentation{w}{\AlphabetSize}{W}$
  (\cref{def:packed-sequence-representation}) of a sequence
  $W[1\dd\Seqlen]\in(\IntegerAlphabet^\ell)^{\Seqlen}$, we can in
  $\bigO(\Seqlen+\AlphabetSize^\ell)$ time construct the packed representation
  (\cref{def:packed-representation}) of the string
  $S=\SeqToString{\AlphabetSize}{W}
  \in[0\dd2\AlphabetSize)^{\Seqlen(\ell+k+1)}$ (\cref{def:seq-to-string}).
\end{proposition}
\begin{proof}
  Denote $N_0=\max\{2\Seqlen,\AlphabetSize^\ell\}$,
  $\AlphabetSize'=2\AlphabetSize$, and $c=\AlphabetSize$.
  By \cref{def:seq-to-string},
  $S=\bigodot_{i=1}^{\Seqlen}(\revstr{W[i]}\cdot c\cdot s(i-1))$ and
  $|S|=\Seqlen(\ell+k+1)$.
  We construct the packed representation of $S$ as follows:
  \begin{enumerate}

  \item First, compute $\AlphabetSize^\ell$, $N_0$,
    $k=\lceil\log_{\AlphabetSize}\Seqlen\rceil$,
    $b=\lceil\log\AlphabetSize\rceil$,
    $b'=\lceil\log\AlphabetSize'\rceil$, and
    $b_m=\lceil\log\Seqlen\rceil$ in $\bigO(\log N_0)$ time.
    To keep all intermediate integers within one word, compute the
    logarithmic parameters using integer divisions.

  \item We now construct the three data structures used below:
    \begin{enumerate}

    \item Construct the data structure from \cref{th:alphabet-lifting}.

    \item Construct the string-reversal data structure from
      \cref{th:string-reversal}.

    \item Finally, construct the data structure from \cref{th:alphabet-map}.
      A query on a one-symbol string returns the packed representation
      of its length-$k$ base-$\AlphabetSize$ encoding.
    \end{enumerate}
    In total, this step takes $\bigO(\sqrt{N_0})$ time and uses
    $\bigO(\sqrt{N_0})$ words of space.

  \item\label{step:seq-to-string-compute-pieces}
    For every $i\in[1\dd\Seqlen]$, denote
    $U_{3i-2}=\revstr{W[i]}$, $U_{3i-1}=c$, and
    $U_{3i}=s(i-1)$. In this step, we
    construct the packed representation over alphabet
    $[0\dd\AlphabetSize')$ of every string in the sequence
    $(U_j)_{j\in[1\dd3\Seqlen]}$, and an array
    $A_{\rm len}[1\dd3\Seqlen]$ satisfying $A_{\rm len}[j]=|U_j|$.
    To this end, we first construct the packed representation
    $\PackedRepresentation{w}{\AlphabetSize'}{c}$ of the one-symbol
    string $c$ using
    \cref{pr:packed-representation}\eqref{pr:packed-representation-initialize}.
    For every $i\in[1\dd\Seqlen]$, we then proceed as follows:
    \begin{enumerate}

    \item First, we compute the packed representation of $U_{3i-2}$.
      To this end, extract the length-$\ell$ substring starting at position
      $(i-1)\ell+1$ from the input packed sequence representation using
      \cref{pr:packed-representation}\eqref{pr:packed-representation-substring}.
      This gives $\PackedRepresentation{w}{\AlphabetSize}{W[i]}$
      by \cref{def:packed-sequence-representation}.
      Query the structure from \cref{th:string-reversal} with this
      representation and length $\ell$ to obtain the packed representation
      of $\revstr{W[i]}$.
      Next, use the structure from \cref{th:alphabet-lifting} with offset $0$
      to obtain its packed representation over $[0\dd\AlphabetSize')$.
      Store the result as the packed representation of $U_{3i-2}$ and
      set $A_{\rm len}[3i-2]:=\ell$.

    \item Store a copy of
      $\PackedRepresentation{w}{\AlphabetSize'}{c}$ as the packed
      representation of $U_{3i-1}$ and set
      $A_{\rm len}[3i-1]:=1$.

    \item We now compute the packed representation of $U_{3i}=s(i-1)$ as follows.
      Initialize the packed representation of the one-symbol string
      $i-1$ over alphabet $[0\dd\Seqlen)$ using
      \cref{pr:packed-representation}\eqref{pr:packed-representation-initialize}.
      Query the structure from \cref{th:alphabet-map} with this representation
      and length $1$ to obtain the packed representation of
      $D_i=\AlphabetMap{\Seqlen}{\AlphabetSize}{i-1}$, which has length $k$.
      Next, query the structure from \cref{th:alphabet-lifting} with this
      representation, length $k$, and offset $\AlphabetSize$ to obtain the packed representation over
      $[0\dd\AlphabetSize')$ of
      $\increase{\AlphabetSize}{D_i}=s(i-1)$.
      Store this representation as the packed representation of $U_{3i}$
      and set $A_{\rm len}[3i]:=k$.
    \end{enumerate}
    Since $\ell\leq\log_{\AlphabetSize}N_0$,
    $k\leq\log_{\AlphabetSize}N_0+1\leq2\log_{\AlphabetSize}N_0$, and
    $\log\AlphabetSize'\leq2\log\AlphabetSize$, every extraction,
    reversal query, and alphabet-lifting query above takes $\bigO(1)$ time.
    Each alphabet-mapping query takes
    $\bigO(1+1/\log_{\Seqlen}N_0)=\bigO(1)$ time, since $\Seqlen\leq N_0$.
    Each constructed representation occupies $\bigO(1)$ words because
    $\max\{\ell,k\}\log\AlphabetSize'=\bigO(\log N_0)=\bigO(w)$.
    Thus, this step takes $\bigO(\Seqlen)$ time and uses
    $\bigO(\Seqlen)$ words of space.

  \item Finally, concatenate the packed representations of
    $U_1,\ldots,U_{3\Seqlen}$, in this order, using
    \cref{pr:packed-representation}\eqref{pr:packed-representation-concat}
    with their lengths in $A_{\rm len}$.
    Since $U_1\cdots U_{3\Seqlen}=S$, this constructs
    the packed representation of $S$. It takes
    $\bigO(\Seqlen+|S|\log\AlphabetSize'/w)=\bigO(\Seqlen)$ time because
    $(\ell+k+1)\log\AlphabetSize'=\bigO(\log N_0)$ and
    $w\geq1+\log N_0$.
  \end{enumerate}
  To verify the size and word-size assumptions, observe that
  $3\Seqlen\leq|S|=\Seqlen(\ell+k+1)\leq N$ because $\ell,k\geq1$.
  Consequently, $N_0\leq N$ and
  $2\AlphabetSize\leq2\Seqlen\leq N_0$.
  The basic packed operations use length bound $N$ and the bit-widths
  computed in the first step, and their alphabet and length bounds are
  valid, including the input length $\Seqlen\ell$, the $3\Seqlen$
  component strings, and their concatenation $S$.
  All three auxiliary structures use size parameter $N_0$, for which
  $w\geq1+\log N_0$. The lifting structure uses source alphabet size
  $\AlphabetSize$ and target alphabet size $\AlphabetSize'$, the reversal
  structure uses alphabet size $\AlphabetSize$, and the alphabet-mapping
  structure uses source alphabet size $\Seqlen$ and target alphabet size
  $\AlphabetSize$.
  The reversal and lifting queries satisfy their length bounds because
  $\ell\leq\log_{\AlphabetSize}N_0\leq N_0$ and $k\leq\Seqlen\leq N_0$.
  Each alphabet-mapping query converts one symbol into a string of length
  $k\leq N_0$. The powers used to compute $\AlphabetSize^\ell$ are at most
  $\AlphabetSize^\ell\leq N_0$, and every computed position is at most $|S|$.
  Since $N_0=\max\{2\Seqlen,\AlphabetSize^\ell\}$, the total construction
  time $\bigO(\log N_0+\sqrt{N_0}+\Seqlen)$ is
  $\bigO(\Seqlen+\AlphabetSize^\ell)$, and the construction uses
  $\bigO(\Seqlen+\AlphabetSize^\ell)$ words of space.
\end{proof}

%% file: tools/sparse-suffix-sorting.tex
\subsection{Sparse Suffix Sorting}
\label{sec:tools-sparse-suffix-sorting}

\begin{theorem}[{\cite{hierarchy}}]\label{th:sparse-suffix-sorting}
  Let $N,\AlphabetSize\in\Z_{\geq2}$ satisfy $\AlphabetSize\leq N$.
  Consider the word RAM model with word size $w\geq1+\log N$.
  Given $N$, $\AlphabetSize$, an integer $q\in[0\dd N+1]$, the packed representation
  $\PackedRepresentation{w}{\AlphabetSize}{\Text}$
  (\cref{def:packed-representation}) of a text
  $\Text\in\IntegerAlphabet^N$, and an array listing the elements
  of a set $Q\subseteq[1\dd N+1]$ of size $q$, we can
  deterministically compute the sequence $\LexSorted{Q}{\Text}$
  (\cref{def:lex-sorted}).
  The algorithm takes $\bigO(N/\log_{\AlphabetSize}N+q)$ time and
  uses $\bigO(N/\log_{\AlphabetSize}N+q)$ words of space.
\end{theorem}

%% file: tools/optimal-range-count-for-bounded-sum-instances.tex
\subsection{Optimal Range Counting and Selection on Bounded-Sum Instances}\label{sec:tools-optimal-range-count-on-bounded-sum-instances}

\begin{remark}
  Constant-time range selection on bounded-sum arrays, with linear
  preprocessing time and word space, was already obtained in~\cite{breaking}.
  We include selection in the following theorem so that both query types
  needed in the periodic suffix-array reduction are supported by one
  structure under a common word-size assumption.
\end{remark}

\begin{theorem}\label{th:range-count-for-small-sum}
  Let $c>0$ be a constant, and let $m,N\in\Z_{\geq2}$ satisfy $m\leq N$.
  In the word RAM model with word size $w\geq1+\log N$,
  given $N$, $m$, $m'$, and an array $A[1\dd m']$ of $m'\leq m$
  positive integers satisfying
  \[
    \sum_{i=1}^{m'}A[i]\leq N
    \qquad\text{and}\qquad
    \sum_{i=1}^{m'}A[i]\leq c\,m\log m,
  \]
  we can preprocess $A$ in $\bigO(m)$ time to obtain a data structure of
  size $\bigO(m)$ words that answers range counting and selection queries
  (\cref{def:range-count,def:range-select}) on $A$ in $\bigO(1)$ time.
\end{theorem}
\begin{proof}
  We assume that
  $m\geq\lceil4c^2+17\rceil$, since for smaller $m$ we can store
  $A$ and answer each query by scanning its $\bigO(1)$ integers.
  If $m'=0$, every counting query returns zero and every selection
  query returns $\infty$. If every integer in $A$ equals $1$, a counting
  query $(j,v)$ returns $j$ for $v\leq1$ and zero otherwise; a selection
  query $(r,v)$ returns $r$ if $v\leq1$ and $r\leq m'$, and $\infty$
  otherwise. Both cases require only constant-time queries and
  $\bigO(m)$-time preprocessing. Henceforth, assume that
  $a_{\max}:=\max_{j\in[1\dd m']}A[j]\geq2$, and denote
  $i_{\max}:=\lfloor\log_2a_{\max}\rfloor$,
  $s:=\sum_{j=1}^{m'}A[j]$, and $n_0:=2s-1$.
  The assumption on $m$ gives $\log m<\sqrt m$ and $2c<\sqrt m$,
  and hence $3\leq n_0<2cm\log m<m^2$.
  In particular, $i_{\max}=\bigO(\log m)$.
  Moreover, $n_0<2N$ implies $w>\log n_0$.

  We use the following bitvectors. Let $B[1\dd a_{\max}]$ satisfy
  $B[v]=\one$ if and only if $v$ is a power of two. For every
  $i\in[0\dd i_{\max}]$, let $B_i[1\dd m']$ satisfy
  $B_i[j]=\one$ if and only if $A[j]\geq2^i$, and denote
  $m_i:=\Rank{B_i}{m'}{\one}$. For each such $i$, let
  $P_i[1\dd m_i]$ list these
  positions in ascending order, that is,
  $P_i[t]=\Select{B_i}{t}{\one}$
  (\cref{def:rank-select}). For every
  $v\in[2^i\dd2^{i+1})$, the bitvector $C_v[1\dd m_i]$
  is defined by $C_v[t]=\one$ if and only if $A[P_i[t]]\geq v$.
  Denote their concatenation by
  \[
    C'_i=C_{2^i}C_{2^i+1}\cdots C_{2^{i+1}-1}
      \in\BinaryAlphabet^{m_i2^i}.
  \]
  Thus, $B_i$ selects the positions that can contribute to a query
  with threshold $v\in[2^i\dd2^{i+1})$, and $C_v$ specifies which
  of those positions satisfy $A[j]\geq v$.

  \DSComponents
  We store the following components:
  \begin{enumerate}
  \item The integers $m'$ and $a_{\max}$ and the packed representation of
    $B$, augmented with the rank/select structure from
    \cref{th:bin-rank-select}. They occupy
    $\bigO(1+a_{\max}/w)=\bigO(m)$ words.
  \item The integers $m_i$ and the packed representations of
    $B_i$, for $i\in[0\dd i_{\max}]$, augmented with rank/select
    structures. Together, they occupy
    $\bigO((i_{\max}+1)(1+m'/w))=\bigO(m)$ words.
  \item The packed representations of $C'_i$, for
    $i\in[0\dd i_{\max}]$, augmented with rank/select structures.
    To bound their total length, observe that $|C'_i|=m_i2^i$,
    where $m_i$ counts the positions $j$ satisfying $2^i\leq A[j]$.
    For each $j\in[1\dd m']$, this condition holds precisely for
    $0\leq i\leq\lfloor\log_2A[j]\rfloor$. Changing the order of
    summation therefore gives
    \[
      \sum_{i=0}^{i_{\max}}|C'_i|
        =\sum_{i=0}^{i_{\max}}m_i2^i
        =\sum_{j=1}^{m'}\sum_{i=0}^{\lfloor\log_2A[j]\rfloor}2^i
        <2\sum_{j=1}^{m'}A[j]=2s.
    \]
    Their total length is therefore at most $n_0$.
    Including the rank/select structures, these bitvectors occupy
    $\bigO(i_{\max}+1+n_0/w)=\bigO(m)$ words.
  \end{enumerate}
  The total space usage is therefore $\bigO(m)$ words.

  \DSQueries
  For range counting, given $j\in[0\dd m']$ and $v\in\Zn$, we answer
  $\TwoSidedRangeCount{A}{j}{v}$ in three steps:
  \begin{enumerate}
  \item If $v=0$, return $j$. If $v>a_{\max}$, return zero.
    Otherwise, compute $i:=\Rank{B}{v}{\one}-1$.
    Since $B$ marks the powers of two, $i=\lfloor\log_2v\rfloor$.
  \item Compute $j':=\Rank{B_i}{j}{\one}$. The first $j'$ positions
    in $P_i$ are exactly the positions in $[1\dd j]$ whose values
    in $A$ are at least $2^i$. Since $v\geq2^i$, the answer is
    $\Rank{C_v}{j'}{\one}$.
  \item Compute $d:=(v-2^i)m_i$ and return
    \[
      \Rank{C'_i}{d+j'}{\one}-\Rank{C'_i}{d}{\one}.
    \]
    This difference equals $\Rank{C_v}{j'}{\one}$ because
    $C_v=C'_i[d+1\dd d+m_i]$.
  \end{enumerate}
  Each step takes $\bigO(1)$ time.

  Consider a selection query $\RangeSelect{A}{r}{v}$ with $r\geq1$
  and $v\in\Zn$. First, use the counting algorithm above to compute
  $f:=\TwoSidedRangeCount{A}{m'}{v}$.
  If $r>f$, return $\infty$. If $v=0$, return $r$.
  Otherwise, $1\leq v\leq a_{\max}$, and we proceed as follows:
  \begin{enumerate}
  \item Compute $i:=\Rank{B}{v}{\one}-1$ and $d:=(v-2^i)m_i$.
    Thus, $C_v=C'_i[d+1\dd d+m_i]$.
  \item Compute
    $x:=\Select{C'_i}{\Rank{C'_i}{d}{\one}+r}{\one}-d$
    and return $\Select{B_i}{x}{\one}$.
  \end{enumerate}
  The value $\Rank{C'_i}{d}{\one}$ counts the ones preceding $C_v$ in $C'_i$.
  Since $r\leq f$, the selected one lies in $C_v$, and $x$ is the
  position of its $r$th one. Thus, $P_i[x]$ is the $r$th smallest
  position $j$ satisfying $A[j]\geq v$. By the definition of $P_i$,
  the final select query returns precisely this position.
  The selection query also takes $\bigO(1)$ time.

  \DSConstruction
  The bitvector initializations, character updates, copies, and
  concatenations below use \cref{pr:packed-representation}; a copy is
  obtained by extracting the entire string.
  We construct the three components in order.
  \begin{enumerate}

  \item First, we construct $B$ and its rank/select structure.
    Scan $A$ to compute $a_{\max}$ and $s$, and compute
    $n_0:=s+(s-1)$. Initialize the packed representation of
    $B:=\zero^{a_{\max}}$. Set $p:=1$ and $B[p]:=\one$.
    While $p\leq\lfloor a_{\max}/2\rfloor$, replace $p$ by $2p$
    and set $B[p]:=\one$.
    This takes $\bigO(m'+1+a_{\max}/w+\log a_{\max})=\bigO(m)$
    time. Apply \cref{th:bin-rank-select} to $B$.
    Its construction takes $\bigO(1+n_0/\log n_0)=\bigO(m)$ time:
    by $3\leq n_0=\bigO(m\log m)$ and the monotonicity of $x/\log x$
    for $x\geq3$, we have
    $n_0/\log n_0=\bigO(m\log m/\log(m\log m))=\bigO(m)$.
    Finally, compute $i_{\max}:=\Rank{B}{a_{\max}}{\one}-1$.

  \item Next, we construct $B_0,\ldots,B_{i_{\max}}$ using the
    values of $A$ in ascending order.
    \begin{enumerate}
    \item First, set $A_{\rm seq}[t]:=(A[t],t)$ for every
      $t\in[1\dd m']$, and sort this array lexicographically using
      radix sort. Both coordinates are smaller than $m^2$, and hence
      a constant number of passes with $\bigO(m)$ counters takes
      $\bigO(m)$ time. Denote the resulting pairs by
      $A_{\rm seq}[t]=(a_t,p_t)$, where
      $a_1\leq\cdots\leq a_{m'}$.
    \item For $i\in[0\dd i_{\max}+1]$, denote
      $d_i:=|\{t\in[1\dd m']:a_t<2^i\}|$.
      In particular, $d_0=0$ and $d_{i_{\max}+1}=m'$.
      Compute these integers in a scan of $A_{\rm seq}$: initialize
      $t:=0$ and $d_0:=0$, and for $i=1,\ldots,i_{\max}$, advance
      $t$ while $t<m'$ and $a_{t+1}<2^i$, and set $d_i:=t$.
      Finally, set $d_{i_{\max}+1}:=m'$.
      The scan takes $\bigO(m'+i_{\max}+1)$ time.
    \item Initialize a packed bitvector $D:=\one^{m'}$ and copy it to
      $B_0$. For $i=1,\ldots,i_{\max}$, set $D[p_t]:=\zero$ for
      every $t\in[d_{i-1}+1\dd d_i]$, and copy the resulting
      bitvector to $B_i$. After iteration $i$, $D[p_t]=\zero$ holds
      precisely for $t\leq d_i$, or equivalently, for $a_t<2^i$.
      This proves that the copies have the required contents.
      Each position is changed at most once, and each copy takes
      $\bigO(1+m'/w)$ time. The total time is
      $\bigO(m'+(i_{\max}+1)(1+m'/w))=\bigO(m)$.
      Augment each $B_i$ using \cref{th:bin-rank-select},
      and compute $m_i:=\Rank{B_i}{m'}{\one}$.
      This takes
      $\bigO((i_{\max}+1)(1+m/\log m))=\bigO(m)$ time.
    \end{enumerate}

  \item Finally, we construct the packed representations of $C'_i$
    and their rank/select structures. For every
    $i=0,\ldots,i_{\max}$, denote $q_i:=d_{i+1}-d_i$ and
    perform the following three substeps.
    \begin{enumerate}

    \item We first determine the intervals of thresholds $v$ for
      which the bitvectors $C_v$ are equal.
      Let $b_1<\cdots<b_{h-1}$ be the distinct values among
      $a_{d_i+1},\ldots,a_{d_{i+1}}$ that are smaller than
      $2^{i+1}-1$. Denote also $b_0:=2^i-1$ and
      $b_h:=2^{i+1}-1$. For every $r\in[1\dd h-1]$, let
      $u_r$ be the largest index $t$ with $a_t=b_r$, and denote
      $u_0:=d_i$. Thus, the pairs with first coordinate $b_r$
      occupy exactly $A_{\rm seq}[u_{r-1}+1\dd u_r]$.
      To keep all intermediate integers within one word, compute $2^{i+1}-1$
      as $2^i+(2^i-1)$. Compute $h$, the values $b_0,\ldots,b_h$,
      and the indices $u_0,\ldots,u_{h-1}$ by scanning
      $A_{\rm seq}[d_i+1\dd d_{i+1}]$ and grouping equal first
      coordinates. This takes $\bigO(1+q_i)$ time. In particular,
      $1\leq h\leq q_i+1$ and $h\leq2^i$.

    \item We now construct one packed block for each interval
      $[b_{r-1}+1\dd b_r]$ of thresholds.
      Initialize $D:=\one^{m_i}$ and allocate arrays
      $A_{\rm block}[1\dd h]$ and $A_{\rm len}[1\dd h]$.
      For $r=1,\ldots,h$, proceed as follows:
      \begin{enumerate}
      \item Compute $\delta_r:=b_r-b_{r-1}$ and the packed
        representation of $D^{\delta_r}$ using
        \cref{pr:packed-string-repetition}.
        Store this representation in $A_{\rm block}[r]$ and set
        $A_{\rm len}[r]:=\delta_rm_i$.
      \item If $r<h$, then for every
        $t\in[u_{r-1}+1\dd u_r]$, compute
        $p':=\Rank{B_i}{p_t}{\one}$ and set $D[p']:=\zero$.
      \end{enumerate}
      Before iteration $r$, we have $D[t]=\one$ if and only if
      $A[P_i[t]]>b_{r-1}$ for every $t\in[1\dd m_i]$.
      There is no value of $A$ strictly between
      $b_{r-1}$ and $b_r$. Hence
      $D=C_{b_{r-1}+1}=\cdots=C_{b_r}$, and the stored block is
      precisely $C_{b_{r-1}+1}\cdots C_{b_r}$.
      The updates remove exactly the positions with value $b_r$,
      establishing the same assertion for the next iteration.

    \item Concatenate the packed representations in
      $A_{\rm block}[1\dd h]$, using their lengths in
      $A_{\rm len}[1\dd h]$, to obtain the packed representation
      of $C'_i$. This takes $\bigO(h+|C'_i|/w)$ time.
    \end{enumerate}

    We next bound the time needed to construct these bitvectors.
    For a fixed $i$, the repetition in iteration $r$ takes
    $\bigO(1+\log\delta_r+\delta_rm_i/w)$ time.
    Since $\sum_{r=1}^h\delta_r=2^i$, concavity of the logarithm
    gives $\sum_{r=1}^h\log\delta_r\leq h\log(2^i/h)$.
    Including the scan, updates, and concatenation, the time for
    $C'_i$ is therefore
    \[
      \bigO\left(1+q_i+\frac{|C'_i|}{w}
        +h\log\frac{2^i}{h}\right).
    \]
    Denote by $h_i$ the number $h$ of blocks constructed for index
    $i$. We have $\sum_{i=0}^{i_{\max}}q_i=m'$ and $\sum_{i=0}^{i_{\max}}|C'_i|\leq n_0$.
    To bound the remaining sum
    $\sum_{i=0}^{i_{\max}} h_i\log(2^i/h_i)$, distinguish two cases:
    \begin{itemize}
    \item For $2^i<m$, the inequality
      $h_i\log(2^i/h_i)\leq h_i\cdot(2^i/h_i)=2^i$ gives a total of
      $\sum_{i\in[0\dd i_{\max}],\,2^i<m}2^i<2m$.
    \item For $2^i\geq m$, each of the $q_i$ corresponding
      values in $A$ is at least $2^i$, and hence
      $q_i2^i\leq\sum_{j=1}^{m'}A[j]=\bigO(m\log m)$.
      Thus, $q_i=\bigO(\log m)$ and
      $h_i\leq q_i+1=\bigO(\log m)$.
      There are $\bigO(\log m)$ such indices, and
      $\log(2^i/h_i)\leq i=\bigO(\log m)$. Their total is
      $\bigO(\log^3m)=\bigO(m)$.
    \end{itemize}
    Consequently, all the packed representations $C'_i$ are
    constructed in $\bigO(i_{\max}+1+m'+n_0/w+m)=\bigO(m)$ time.
    Their temporary arrays and packed blocks occupy
    $\bigO(\sum_{i=0}^{i_{\max}}(1+q_i+|C'_i|/w))=\bigO(m)$ words.

    It remains to augment the bitvectors with rank/select structures.
    Apply the batched construction from \cref{th:bin-rank-select}
    to $C'_0,\ldots,C'_{i_{\max}}$.
    Their total length is at most $n_0$, and $w>\log n_0$.
    The construction therefore takes
    \[
      \bigO(i_{\max}+1+n_0/\log n_0)
        \subseteq\bigO\left(\log m+
          \frac{m\log m}{\log(m\log m)}\right)
        =\bigO(m)
    \]
    time, using the monotonicity of $x/\log x$ for $x\geq3$.
  \end{enumerate}
  All packed operations use alphabet size $2$ and length bound $N$.
  They are applicable because $w>\log N$, $|B|,|B_i|\leq s\leq N$, and
  $|C'_i|=m_i2^i\leq s\leq N$. Each repetition also has a nonempty
  input, since $|D|=m_i\geq1$ follows from $2^i\leq a_{\max}$, and an
  output of length $\delta_rm_i\leq|C'_i|\leq N$.
  The rank/select constructions for $B$ and for
  $C'_0,\ldots,C'_{i_{\max}}$ use size parameter $n_0$, and those for
  $B_0,\ldots,B_{i_{\max}}$ use size parameter $m$.
  All powers of two computed above are at most $a_{\max}$.
  The remaining threshold endpoints are at most $n_0<2N$,
  and every offset into $C'_i$, including those used in queries, is at most
  $|C'_i|\leq s\leq N$. Thus, all these integers fit in one word.
  Since $n_0=\bigO(m\log m)$ and $w\geq\log m$, construction takes
  $\bigO(m)$ time and $\bigO(m)$ words of space.
\end{proof}

%% file: space-lower-bounds.tex
\section{Space Lower Bounds for String Queries over Integer Alphabets}\label{sec:space-lower-bounds}

\input{space-lower-bounds/space-lower-bound-suffix-array}

\input{space-lower-bounds/space-lower-bound-inverse-suffix-array}

\input{space-lower-bounds/space-lower-bound-select}

\input{space-lower-bounds/space-lower-bound-prefix-select}

\input{space-lower-bounds/space-lower-bound-special-rank}

\input{space-lower-bounds/space-lower-bound-prefix-special-rank}

%% file: space-lower-bounds/space-lower-bound-suffix-array.tex
\subsection{Suffix Array Queries}\label{sec:suffix-array-space-lower-bound}

\begin{proposition}\label{pr:symbol-counts}
  Let $\AlphabetSize,\Textlen \in \Z_{\geq 2}$ be such that $\AlphabetSize \leq \Textlen$.
  For every $\Text \in \IntegerAlphabet^{\Textlen}$, there exists
  a data structure using exactly $\lfloor2\min(\AlphabetSize \log \Textlen, \Textlen)\rfloor$ bits
  that, given any $c \in \IntegerAlphabet$, returns
  $|\OccTwo{c}{\Text}|$ (\cref{def:occ}), i.e., the number of occurrences
  of symbol $c$ in $\Text$.
\end{proposition}
\begin{proof}
  We use one of the following two representations.
  \begin{itemize}
  \item If $\AlphabetSize\log\Textlen<\Textlen$, let
    $C[0\dd\AlphabetSize)$ denote the array defined by
    $C[c]=|\OccTwo{c}{\Text}|$ for every $c\in\IntegerAlphabet$.
    Each count belongs to $[0\dd\Textlen]$ and can be stored in
    $\lfloor\log\Textlen\rfloor+1$ bits. We store $C$ as a
    fixed-width array, using
    $\AlphabetSize(\lfloor\log\Textlen\rfloor+1)
      \leq2\AlphabetSize\log\Textlen
      =2\min(\AlphabetSize\log\Textlen,\Textlen)$ bits.
    A query for $c$ returns $C[c]$.
  \item If $\AlphabetSize\log\Textlen\geq\Textlen$, denote
    $p_a:=a+\sum_{c=0}^{a}|\OccTwo{c}{\Text}|$ for every
    $a\in\IntegerAlphabet$.
    The sequence $(p_a)_{a\in\IntegerAlphabet}$ is strictly
    increasing, with $p_0\geq0$ and
    $p_{\AlphabetSize-1}=\Textlen+\AlphabetSize-1<2\Textlen$.
    Let $B[0\dd2\Textlen)$ denote the bitvector whose ones are
    exactly at the positions $p_a$ for $a\in\IntegerAlphabet$.
    We store $B$ in plain form, using
    $2\Textlen=2\min(\AlphabetSize\log\Textlen,\Textlen)$ bits.
    Given $c\in\IntegerAlphabet$, scan $B$ to locate its
    $(c+1)$st one at position $p_c$ and, if $c>0$, the preceding
    one at position $p_{c-1}$.
    For $c=0$, return $p_0=|\OccTwo{0}{\Text}|$.
    Otherwise, return
    $p_c-p_{c-1}-1=|\OccTwo{c}{\Text}|$, where the equality
    follows from the definition of $p_a$.
  \end{itemize}
  In both cases, the representation is an integer number of bits that
  is at most $2\min(\AlphabetSize\log\Textlen,\Textlen)$. We pad it with
  zeros to exactly $\lfloor2\min(\AlphabetSize\log\Textlen,\Textlen)\rfloor$
  bits. The queries read only the unpadded part, whose length is
  determined by $\AlphabetSize$ and $\Textlen$.
\end{proof}

\begin{theorem}\label{th:suffix-array-space-lower-bound}
  Let $\epsilon \in (0,1)$ be a constant.
  Let $\AlphabetSize,\Textlen \in \Z_{\geq 2}$ be such that
  $\AlphabetSize \leq \Textlen$ and
  $\Textlen > \tfrac{36}{\epsilon^2} \cdot 2^{6/\epsilon}$.
  There is no data structure that, for every
  $\Text \in \IntegerAlphabet^{\Textlen}$,
  uses at most $(1-\epsilon) \Textlen \log \AlphabetSize$
  bits of space, and answers suffix array queries on $\Text$
  (that, given any $i \in [1 \dd \Textlen]$,
  return $\SA{\Text}[i]$; see \cref{def:suffix-array}).
\end{theorem}
\begin{proof}
  Suppose that such a data structure exists. For each
  $\Text\in\IntegerAlphabet^{\Textlen}$, consider the encoding obtained
  by concatenating the symbol-count representation from
  \cref{pr:symbol-counts} with the assumed representation for suffix
  array queries. The symbol-count representation has a fixed length
  determined by $\AlphabetSize$ and $\Textlen$.
  Knowing this length allows us to separate the two representations.
  The encoding uses at most
  $(1-\epsilon)\Textlen\log\AlphabetSize
    +2\min(\AlphabetSize\log\Textlen,\Textlen)$~bits.

  We first show how to reconstruct $\Text$ from this encoding. For
  every $c\in\IntegerAlphabet$, exactly $|\OccTwo{c}{\Text}|$ suffixes
  begin with $c$, and all suffixes beginning with smaller symbols precede
  them in lexicographic order. Consequently, for every
  $r\in[1\dd\Textlen]$ and $c\in\IntegerAlphabet$, it holds
  \[
    \Text[\SA{\Text}[r]]=c
      \quad\Longleftrightarrow\quad
      \sum_{a=0}^{c-1}|\OccTwo{a}{\Text}|<r
      \leq\sum_{a=0}^{c}|\OccTwo{a}{\Text}|,
  \]
  where the empty sum for $c=0$ is zero. For every
  $r\in[1\dd\Textlen]$, we use the symbol counts to find the
  unique symbol $c$ satisfying the inequalities above. The equivalence
  then gives $\Text[\SA{\Text}[r]]=c$.
  We also query the suffix-array representation to obtain the position
  $j=\SA{\Text}[r]$, and assign $c$ to position $j$ in the
  reconstructed text. Since $\SA{\Text}$ is a permutation of
  $[1\dd\Textlen]$, repeating this for $r=1,\ldots,\Textlen$
  recovers every symbol of $\Text$.

  It remains to show that the encoding uses at most
  $S:=\Textlen\log\AlphabetSize-1$ bits. We first establish
  $3\min(\AlphabetSize\log\Textlen,\Textlen)
    \leq\epsilon\Textlen\log\AlphabetSize$ by considering two cases:
  \begin{itemize}
  \item If $\AlphabetSize>2^{3/\epsilon}$, then
    $3\min(\AlphabetSize\log\Textlen,\Textlen)
      \leq3\Textlen<\epsilon\Textlen\log\AlphabetSize$.
  \item If $\AlphabetSize\leq2^{3/\epsilon}$, then the assumption
    $\Textlen>\tfrac{36}{\epsilon^2}\cdot2^{6/\epsilon}$ implies
    $\Textlen>\tfrac{36}{\epsilon^2}\AlphabetSize^2$, and hence
    $6\AlphabetSize\sqrt{\Textlen}<\epsilon\Textlen$.
    Since $\log\Textlen=2\log\sqrt{\Textlen}<2\sqrt{\Textlen}$, we obtain
    \[
      3\min(\AlphabetSize\log\Textlen,\Textlen)
        \leq3\AlphabetSize\log\Textlen
        <6\AlphabetSize\sqrt{\Textlen}
        <\epsilon\Textlen
        \leq\epsilon\Textlen\log\AlphabetSize.
    \]
  \end{itemize}
  Since $\min(\AlphabetSize\log\Textlen,\Textlen)\geq1$, this gives
  $2\min(\AlphabetSize\log\Textlen,\Textlen)+1
    \leq3\min(\AlphabetSize\log\Textlen,\Textlen)
    \leq\epsilon\Textlen\log\AlphabetSize$.
  Thus, the encoding uses at most
  $(1-\epsilon)\Textlen\log\AlphabetSize
    +2\min(\AlphabetSize\log\Textlen,\Textlen)
    \leq\Textlen\log\AlphabetSize-1=S$ bits.

  The number of bit strings of length at most $S$ is
  \[
    \sum_{i=0}^{\lfloor S\rfloor}2^i
      <2^{S+1}=\AlphabetSize^{\Textlen}.
  \]
  There are $\AlphabetSize^{\Textlen}$ possible texts, and therefore
  two distinct texts must have the same encoding. However, reconstructing
  the text from this encoding would give the same result in both cases,
  a contradiction.
\end{proof}

\begin{remark}\label{rm:suffix-array-space-lower-bound}
  The lower bound for $\Textlen$ in \cref{th:suffix-array-space-lower-bound}
  was chosen to simplify the analysis (requiring only that it depends solely on $\epsilon$)
  and has not been optimized.
\end{remark}

\begin{theorem}\label{th:suffix-array-asymptotic-space-lower-bound}
  Let $\AlphabetSize, \Textlen \in \Z_{\geq 2}$ satisfy
  $\AlphabetSize \leq \Textlen$.
  Any data structure supporting suffix array queries
  (\cref{def:suffix-array}) for every text
  $\Text \in \IntegerAlphabet^{\Textlen}$ has worst-case space usage
  $\Omega(\Textlen \log \AlphabetSize)$ bits.
\end{theorem}
\begin{proof}
  It suffices to choose any constant $\epsilon \in (0,1)$ and apply
  \cref{th:suffix-array-space-lower-bound}. Observe that the bound
  holds for all $\AlphabetSize$ and $\Textlen$, with the exception
  of only a constant number of pairs $(\AlphabetSize,\Textlen)$.
  For example, choosing $\epsilon = 0.99$ (yielding a lower bound of
  $\lfloor 0.01 \Textlen \log \AlphabetSize \rfloor + 1 = \Omega(\Textlen \log \AlphabetSize)$ bits)
  excludes only integer pairs $(\AlphabetSize,\Textlen)$ where
  $2 \leq \AlphabetSize \leq \Textlen \leq 2451$.
  These remaining pairs $(\AlphabetSize,\Textlen)$ form a finite
  set, and excluding finitely many parameter pairs does not affect
  the asymptotic statement. More precisely, for any fixed pair
  $(\AlphabetSize,\Textlen)$, any data structure answering suffix
  array queries on strings $\Text \in [0 \dd \AlphabetSize)^{\Textlen}$
  needs to store at least a single bit
  (since otherwise, it would always return the same answer,
  which is not correct since there always exist two strings in
  $[0 \dd \AlphabetSize)^{\Textlen}$ with distinct suffix arrays).
  We can therefore obtain a universal constant $\delta$ that works
  for \emph{all} valid pairs $(\AlphabetSize,\Textlen)$ by setting
  $\delta = \min(\delta_0,\delta_1)$, where $\delta_0 = 0.01$
  and $\delta_1 = \tfrac{1}{2}\min\{1/(\Textlen \log \AlphabetSize) : \AlphabetSize,\Textlen \in \Z_{\geq 2}\text{ and }
  \AlphabetSize \leq \Textlen \leq 2451\}$. For such a constant $\delta > 0$,
  it holds that, for \emph{every} pair $(\AlphabetSize,\Textlen)$ satisfying $2 \leq \AlphabetSize \leq \Textlen$, every data structure answering
  suffix array queries for strings
  $\Text \in [0 \dd \AlphabetSize)^{\Textlen}$ must use at least
  $\lfloor \delta \Textlen \log \AlphabetSize \rfloor + 1 = \Omega(\Textlen \log \AlphabetSize)$ bits of space.
\end{proof}

\begin{theorem}\label{th:suffix-array-bounded-length-space-lower-bound}
  Let $\AlphabetSize, \Textlen \in \Z_{\geq 2}$ satisfy
  $\AlphabetSize \leq \Textlen$.
  Any data structure supporting suffix array queries
  (\cref{def:suffix-array}) for every text
  $\Text \in \IntegerAlphabet^{*}$ satisfying
  $\AlphabetSize \leq |\Text| \leq \Textlen$ has worst-case space usage
  $\Omega(\Textlen \log \AlphabetSize)$ bits.
\end{theorem}
\begin{proof}
  Any data structure satisfying the premise must in particular answer suffix
  array queries on every text $\Text \in \IntegerAlphabet^{\Textlen}$.
  Thus, the claim follows by \cref{th:suffix-array-asymptotic-space-lower-bound}.
\end{proof}

%% file: space-lower-bounds/space-lower-bound-inverse-suffix-array.tex
\subsection{Inverse Suffix Array Queries}\label{sec:inverse-suffix-array-space-lower-bound}

\begin{theorem}\label{th:inverse-suffix-array-space-lower-bound}
  Let $\epsilon \in (0,1)$ be a constant.
  Let $\AlphabetSize,\Textlen \in \Z_{\geq 2}$ be such that
  $\AlphabetSize \leq \Textlen$ and
  $\Textlen > \tfrac{36}{\epsilon^2} \cdot 2^{6/\epsilon}$.
  There is no data structure that, for every
  $\Text \in \IntegerAlphabet^{\Textlen}$,
  uses at most $(1-\epsilon) \Textlen \log \AlphabetSize$
  bits of space, and answers inverse suffix array queries on $\Text$
  (that, given any $j \in [1 \dd \Textlen]$,
  return $\ISA{\Text}[j]$; see \cref{def:inverse-suffix-array}).
\end{theorem}
\begin{proof}
  Suppose that such a data structure exists. We can use the same
  representation to answer suffix array queries as follows. Given
  $r\in[1\dd\Textlen]$, query $\ISA{\Text}[j]$ for
  $j=1,\ldots,\Textlen$ until finding the unique position $j$
  satisfying $\ISA{\Text}[j]=r$. Since $\ISA{\Text}$ is the inverse
  permutation of $\SA{\Text}$, this position is $\SA{\Text}[r]$,
  which we return. This requires no additional stored data and would
  contradict \cref{th:suffix-array-space-lower-bound}.
\end{proof}

\begin{theorem}\label{th:inverse-suffix-array-asymptotic-space-lower-bound}
  Let $\AlphabetSize, \Textlen \in \Z_{\geq 2}$ satisfy
  $\AlphabetSize \leq \Textlen$.
  Any data structure supporting inverse suffix array queries
  (\cref{def:inverse-suffix-array}) for every text
  $\Text \in \IntegerAlphabet^{\Textlen}$ has worst-case space usage
  $\Omega(\Textlen \log \AlphabetSize)$ bits.
\end{theorem}
\begin{proof}
  A data structure supporting inverse suffix array queries for every
  $\Text\in\IntegerAlphabet^{\Textlen}$ also supports suffix array queries
  for every such text without additional stored data. Given
  $r\in[1\dd\Textlen]$, query $\ISA{\Text}[j]$ for $j=1,\ldots,\Textlen$
  until finding the unique position $j$ satisfying $\ISA{\Text}[j]=r$, and
  return $j=\SA{\Text}[r]$. Thus, the claim follows by
  \cref{th:suffix-array-asymptotic-space-lower-bound}.
\end{proof}

\begin{theorem}\label{th:inverse-suffix-array-bounded-length-space-lower-bound}
  Let $\AlphabetSize, \Textlen \in \Z_{\geq 2}$ satisfy
  $\AlphabetSize \leq \Textlen$.
  Any data structure supporting inverse suffix array queries
  (\cref{def:inverse-suffix-array}) for every text
  $\Text \in \IntegerAlphabet^{*}$ satisfying
  $\AlphabetSize \leq |\Text| \leq \Textlen$ has worst-case space usage
  $\Omega(\Textlen \log \AlphabetSize)$ bits.
\end{theorem}
\begin{proof}
  Any data structure satisfying the premise must in particular answer
  inverse suffix array queries on every text
  $\Text\in\IntegerAlphabet^{\Textlen}$. Thus, the claim follows by
  \cref{th:inverse-suffix-array-asymptotic-space-lower-bound}.
\end{proof}

%% file: space-lower-bounds/space-lower-bound-select.tex
\subsection{Select Queries}\label{sec:select-space-lower-bound}

\begin{theorem}\label{th:select-space-lower-bound}
  Let $\epsilon \in (0,1)$ be a constant.
  Let $\AlphabetSize,\Textlen \in \Z_{\geq 2}$ be such that
  $\AlphabetSize \leq \Textlen$ and
  $\Textlen > \tfrac{1}{\epsilon}$.
  There is no data structure that, for every
  $\Text \in \IntegerAlphabet^{\Textlen}$,
  uses at most $(1-\epsilon) \Textlen \log \AlphabetSize$
  bits of space, and answers select queries (\cref{def:rank-select}) on $\Text$.
\end{theorem}
\begin{proof}
  Suppose that the claim does not hold, i.e., there exists a data structure,
  denoted $D^{\rm S}_{\AlphabetSize,\Textlen}$, that for every
  string $\Text \in [0 \dd \AlphabetSize)^{\Textlen}$,
  uses at most $b := (1-\epsilon)\Textlen \log \AlphabetSize$ bits of space
  and answers select queries on $\Text$. We will prove that using this
  data structure, we can reconstruct $\Text$, leading to a contradiction.

  Consider any string $\Text \in [0 \dd \AlphabetSize)^{\Textlen}$.
  Given any $i \in [1 \dd \Textlen]$, we can determine $\Text[i]$
  using $D^{\rm S}_{\AlphabetSize,\Textlen}(\Text)$ by trying every
  integer $r \in [1 \dd \Textlen]$ and every symbol
  $c \in [0 \dd \AlphabetSize)$, and computing $\Select{\Text}{r}{c}$.
  Upon finding the pair $(r,c)$ such that $\Select{\Text}{r}{c} = i$,
  it follows by \cref{def:rank-select} that $\Text[i] = c$.
  Such a pair exists for every $i$: the symbol $\Text[i]$ has its
  $r$th occurrence at position $i$ for some $r \in [1 \dd \Textlen]$.
  Repeating this for every $i \in [1 \dd \Textlen]$ reconstructs
  the entire string $\Text$.

  To obtain a contradiction, we now compare the number of possible
  stored bit strings with the number of texts.
  Every data structure using at most $b$ bits
  corresponds to a bit string of length at most $b$.
  The number of such bit strings is
  $\sum_{k=0}^{\lfloor b \rfloor} 2^k < 2^{b+1}$.
  On the other hand, the number of distinct strings $\Text \in [0 \dd \AlphabetSize)^{\Textlen}$
  is $\AlphabetSize^{\Textlen}$.
  To show that there are fewer possible stored bit strings than texts,
  it suffices to prove that $b+1<\Textlen\log\AlphabetSize$.
  Since $\AlphabetSize\geq2$ and $\Textlen>1/\epsilon$, we have
  $\epsilon\Textlen\log\AlphabetSize\geq\epsilon\Textlen>1$.
  Recalling that $b=(1-\epsilon)\Textlen\log\AlphabetSize$, we
  therefore obtain
  $b+1=(1-\epsilon)\Textlen\log\AlphabetSize+1<\Textlen\log\AlphabetSize$.
  Consequently,
  \[
    2^{b+1}<2^{\Textlen\log\AlphabetSize}=\AlphabetSize^{\Textlen}.
  \]
  The number of distinct bit strings (possible states of the data structure)
  is strictly less than the number of distinct strings that must be
  represented. By the pigeonhole principle, there must exist at least two
  distinct strings $\Text, \Text'$ that map to the same stored bit
  string. Since the reconstruction of the string
  is determined solely by the stored bit string,
  the reconstruction procedure would return the same result for both,
  contradicting the fact that $\Text \neq \Text'$.
\end{proof}

\begin{theorem}\label{th:select-asymptotic-space-lower-bound}
  Let $\AlphabetSize, \Textlen \in \Z_{\geq 2}$ satisfy
  $\AlphabetSize \leq \Textlen$.
  Any data structure supporting select queries (\cref{def:rank-select})
  for every string $\Text \in \IntegerAlphabet^{\Textlen}$ has worst-case
  space usage $\Omega(\Textlen \log \AlphabetSize)$ bits.
\end{theorem}
\begin{proof}
  It suffices to choose any constant $\epsilon \in (0,1)$ and apply
  \cref{th:select-space-lower-bound}. Observe that the bound
  holds for all $\AlphabetSize$ and $\Textlen$, with the exception
  of only a constant number of pairs $(\AlphabetSize,\Textlen)$.
  For example, choosing $\epsilon = 0.5$ (yielding a lower bound of
  $0.5 \Textlen \log \AlphabetSize = \Omega(\Textlen \log \AlphabetSize)$ bits)
  excludes only the integer pair $(\AlphabetSize,\Textlen) = (2, 2)$.
  Excluding finitely many parameter pairs does not affect the asymptotic statement.
  Indeed, for each excluded pair, there exist two strings
  with different answers, and hence every such data structure stores at
  least one bit.
\end{proof}

\begin{theorem}\label{th:select-bounded-length-space-lower-bound}
  Let $\AlphabetSize, \Textlen \in \Z_{\geq 2}$ satisfy
  $\AlphabetSize \leq \Textlen$.
  Any data structure supporting select queries (\cref{def:rank-select})
  for every string $\Text \in \IntegerAlphabet^{*}$ satisfying
  $\AlphabetSize \leq |\Text| \leq \Textlen$ has worst-case space usage
  $\Omega(\Textlen \log \AlphabetSize)$ bits.
\end{theorem}
\begin{proof}
  Any data structure satisfying the premise must in particular answer select
  queries on every string $\Text \in \IntegerAlphabet^{\Textlen}$.
  Thus, the claim follows by \cref{th:select-asymptotic-space-lower-bound}.
\end{proof}

%% file: space-lower-bounds/space-lower-bound-prefix-select.tex
\subsection{Prefix Select Queries}\label{sec:prefix-select-space-lower-bound}

\begin{lemma}\label{lm:reduce-select-to-prefix-select}
  Let $\AlphabetSize,\Seqlen \in \Z_{\geq 2}$ be such that $\AlphabetSize \leq \Seqlen$, and let $\ell \in \Z_{\geq 1}$
  and $s\in\Z_{\geq0}$.
  Suppose that there exists a data structure, denoted $D^{\rm PS}_{\AlphabetSize,\Seqlen,\ell}$, that for every
  sequence $W[1 \dd \Seqlen]$ of $\Seqlen$ strings over alphabet $[0 \dd \AlphabetSize)$, each of length exactly $\ell$,
  answers prefix select queries (\cref{def:prefix-rank-and-select}) on $W$
  for every rank in $[1 \dd \Seqlen]$ and uses at most $s$ bits of space.
  Then there exists a data structure, denoted $D^{\rm S}_{\AlphabetSize^{\ell},\Seqlen}$, that for every
  string $A \in [0 \dd \AlphabetSize^\ell)^{\Seqlen}$, answers select queries (\cref{def:rank-select}) on $A$ and uses at most
  $s$ bits of space.
\end{lemma}
\begin{proof}

  Consider any string $A \in [0 \dd \AlphabetSize^\ell)^{\Seqlen}$.
  We define a sequence $W_A[1 \dd \Seqlen]$ of $\Seqlen$ strings over alphabet
  $[0 \dd \AlphabetSize)$ by setting
  $W_A[i] := \AlphabetMap{\AlphabetSize^{\ell}}{\AlphabetSize}{A[i]}$
  (\cref{def:alphabet-map}) for every $i \in [1 \dd \Seqlen]$.
  By \cref{def:alphabet-map}, each string
  $\AlphabetMap{\AlphabetSize^{\ell}}{\AlphabetSize}{A[i]}$ has length
  exactly $\lceil \log_{\AlphabetSize} (\AlphabetSize^{\ell}) \rceil = \ell$.
  Thus, the sequence $W_A$ is a valid input for the data structure
  $D^{\rm PS}_{\AlphabetSize,\Seqlen,\ell}$.

  We define the data structure
  $D^{\rm S}_{\AlphabetSize^{\ell},\Seqlen}(A)$ for the string $A$
  to consist of a single component: the data structure
  $D^{\rm PS}_{\AlphabetSize,\Seqlen,\ell}(W_A)$ for the sequence $W_A$.
  By assumption, this uses at most $s$ bits of space.

  Let $a \in [0 \dd \AlphabetSize^{\ell})$ and $r \in \Z_{\geq 1}$.
  The select query $\Select{A}{r}{a}$ is then implemented as follows.
  If $r>\Seqlen$, return $\infty$. This is correct because $A$ has length
  $\Seqlen$. In the remainder, assume $r\in[1\dd\Seqlen]$.
  Since the function $\AlphabetMap{\AlphabetSize^{\ell}}{\AlphabetSize}{\cdot}$ (\cref{def:alphabet-map}) is a bijection,
  for every $j \in [1 \dd \Seqlen]$ we have $A[j] = a$ if and only if $W_A[j] = \AlphabetMap{\AlphabetSize^{\ell}}{\AlphabetSize}{a}$.
  Consequently, letting $X_a = \AlphabetMap{\AlphabetSize^{\ell}}{\AlphabetSize}{a}$,
  it holds $\Select{A}{r}{a} = \PrefixSelect{W_A}{r}{X_a}$.
  Thus, using $D^{\rm PS}_{\AlphabetSize,\Seqlen,\ell}(W_A)$,
  we can compute $\Select{A}{r}{a}$ by
  returning $\PrefixSelect{W_A}{r}{X_a}$.
\end{proof}

\begin{theorem}\label{th:prefix-select-space-lower-bound}
  Let $\epsilon \in (0,1)$ and $c \in (0,1]$ be constants.
  Let $\AlphabetSize,\Seqlen \in \Z_{\geq 2}$ be such that $\AlphabetSize \leq \Seqlen$ and $\Seqlen > \tfrac{1}{\epsilon}$.
  Let also $\ell := \lfloor c \log_{\AlphabetSize} \Seqlen \rfloor$ and assume that $\ell \geq 1$.
  There is no data structure that, for every sequence $W[1 \dd \Seqlen]$ of $\Seqlen$ strings
  over alphabet $[0 \dd \AlphabetSize)$, each of length exactly $\ell$,
  uses at most $(1-\epsilon)\Seqlen\ell\log \AlphabetSize$ bits of space
  and answers prefix select queries (\cref{def:prefix-rank-and-select}) on
  $W$ for every rank in $[1 \dd \Seqlen]$.
\end{theorem}
\begin{proof}

  Suppose that the claim does not hold. Then there exists a data structure,
  denoted $D^{\rm PS}_{\AlphabetSize,\Seqlen,\ell}$, that for every sequence $W[1 \dd \Seqlen]$ of $\Seqlen$ strings
  over alphabet $[0 \dd \AlphabetSize)$, each of length exactly $\ell := \lfloor c \log_{\AlphabetSize} \Seqlen \rfloor$,
    answers prefix select queries on $W$ for every rank in
    $[1 \dd \Seqlen]$ and uses at most
  $(1-\epsilon)\Seqlen\ell\log \AlphabetSize$ bits of space.

  By \cref{lm:reduce-select-to-prefix-select}, this implies the existence of a data structure
  $D^{\rm S}_{\AlphabetSize^\ell,\Seqlen}$ that, for every string $A \in [0 \dd \AlphabetSize^\ell)^{\Seqlen}$,
  answers select queries on $A$ and uses at most
  $(1-\epsilon)\Seqlen\ell\log \AlphabetSize$ bits of space.
  By $c \in (0,1]$ and $\ell = \lfloor c\log_{\AlphabetSize} \Seqlen \rfloor$, we have
  $\AlphabetSize^\ell
    =       \AlphabetSize^{\lfloor c\log_{\AlphabetSize} \Seqlen \rfloor}
    \leq    \AlphabetSize^{\log_{\AlphabetSize} \Seqlen}
    =       \Seqlen$.
  On the other hand, $\ell \geq 1$ implies that $\AlphabetSize^{\ell} \geq \AlphabetSize \geq 2$.
  Thus, we have $2 \leq \AlphabetSize^{\ell} \leq \Seqlen$.
  Therefore \cref{th:select-space-lower-bound} applies with alphabet size $\AlphabetSize^{\ell}$
  and string length $\Seqlen$, and implies that no such data structure can use at most
  $(1-\epsilon) \Seqlen\log(\AlphabetSize^\ell)$ bits of space. Since
  $(1-\epsilon) \Seqlen\log(\AlphabetSize^\ell) =
  (1-\epsilon) \Seqlen\ell\log \AlphabetSize$, we thus obtain a contradiction.
\end{proof}

\begin{theorem}\label{th:prefix-select-asymptotic-space-lower-bound}
  Let $c \in (0,1]$ be a constant. Let
  $\AlphabetSize, \Seqlen \in \Z_{\geq 2}$ satisfy
  $\AlphabetSize \leq \Seqlen$. Let
  $\ell := \lfloor c \log_{\AlphabetSize} \Seqlen \rfloor$, and assume
  $\ell \geq 1$.
  Any data structure supporting prefix select queries with ranks in
  $[1 \dd \Seqlen]$
  (\cref{def:prefix-rank-and-select}) for every sequence
  $W[1 \dd \Seqlen]$ of $\Seqlen$ strings over alphabet
  $\IntegerAlphabet$, each of length exactly $\ell$, has worst-case space
  usage $\Omega(\Seqlen\ell \log \AlphabetSize)$ bits.
\end{theorem}
\begin{proof}
  It suffices to choose any constant $\epsilon \in (0,1)$ and apply \cref{th:prefix-select-space-lower-bound}.
  The bound then holds for all admissible pairs $(\AlphabetSize,\Seqlen)$, with the exception of only finitely many such pairs.
  Excluding finitely many parameter pairs does not affect the asymptotic statement.
  Indeed, for each excluded pair, there exist two sequences
  with different answers, and hence every such data structure stores at
  least one bit.
\end{proof}

\begin{theorem}\label{th:prefix-select-bounded-total-length-space-lower-bound}
  Let $c \in (0,1]$ be a constant, and let
  $\AlphabetSize, N \in \Z_{\geq 2}$. Assume that there exists
  $\Seqlen \in \Z_{\geq 2}$ such that, for
  $\ell := \lfloor c\log_{\AlphabetSize} \Seqlen \rfloor$, it holds
  $\ell \geq 1$ and $\Seqlen\ell \leq N$.
  Any data structure supporting prefix select queries with ranks in
  $[1 \dd \Seqlen]$
  (\cref{def:prefix-rank-and-select}) for every such $\Seqlen$ and every
  sequence $W[1 \dd \Seqlen]$ of $\Seqlen$ strings over alphabet
  $\IntegerAlphabet$, each of length exactly $\ell$, has worst-case space
  usage $\Omega(N \log \AlphabetSize)$ bits.
\end{theorem}
\begin{proof}
  Let $\Seqlen^*$ be the largest positive integer such that, for
  $\ell^* := \lfloor c\log_{\AlphabetSize} \Seqlen^* \rfloor$, we have
  $\ell^* \geq 1$ and $\Seqlen^*\ell^* \leq N$.
  Since the data structure supports queries for every admissible value of $\Seqlen$,
  it must in particular support queries for $\Seqlen=\Seqlen^*$.
  Since $\ell^*\geq1$ and $c\leq1$, we have
  $\AlphabetSize\leq\AlphabetSize^{\ell^*/c}\leq\Seqlen^*$.
  Therefore, by \cref{th:prefix-select-asymptotic-space-lower-bound}, it must use
  at least $\Omega(\Seqlen^*\ell^*\log \AlphabetSize)$ bits of space.
  It remains to show that $\Seqlen^*\ell^* = \Omega(N)$.
  Let $\ell' := \lfloor c\log_{\AlphabetSize} (\Seqlen^*+1) \rfloor$.
  Since $\ell'\geq\ell^*\geq1$, the maximality of $\Seqlen^*$ gives
  $(\Seqlen^*+1)\ell' > N$.
  On the other hand, since $c \in (0,1]$, we have
  $c\log_{\AlphabetSize}(\Seqlen^*+1)
    <    c\log_{\AlphabetSize}\Seqlen^* + c
    \leq c\log_{\AlphabetSize}\Seqlen^* + 1$,
  and therefore $\ell' \leq \ell^*+1$.
  We also have $\Seqlen^*+1 \leq 2\Seqlen^*$ and $\ell^*+1 \leq 2\ell^*$ (since $\ell^* \geq 1$).
  Consequently, $N < (\Seqlen^*+1)\ell' \leq (\Seqlen^*+1)(\ell^*+1) \leq 4\Seqlen^*\ell^*$.
  Thus $\Seqlen^*\ell^* > N/4$, i.e., $\Seqlen^*\ell^* = \Omega(N)$.
  We therefore obtain $\Omega(\Seqlen^*\ell^*\log \AlphabetSize) = \Omega(N\log \AlphabetSize)$.
\end{proof}

%% file: space-lower-bounds/space-lower-bound-special-rank.tex
\subsection{Special Rank Queries}\label{sec:special-rank-space-lower-bound}

\begin{lemma}\label{lm:binary-special-rank-family}
  Let $\Textlen \in \Z_{\geq 2}$.
  There exists a family $\mathcal{T}_{\Textlen} \subseteq \BinaryAlphabet^{\Textlen}$ of
  size $|\mathcal{T}_{\Textlen}| = 2^{\lfloor \Textlen/2 \rfloor - 1}$
  such that, for any two distinct strings
  $S,S' \in \mathcal{T}_{\Textlen}$, the sequences (see \cref{def:rank-select})
  \begin{itemize}
  \item $(\SpecialRank{S}{1},\SpecialRank{S}{2},\dots,\SpecialRank{S}{\Textlen})$ and
  \item $(\SpecialRank{S'}{1},\SpecialRank{S'}{2},\dots,\SpecialRank{S'}{\Textlen})$
  \end{itemize}
  are distinct.
\end{lemma}
\begin{proof}
  Denote $m=\lfloor \Textlen/2 \rfloor$. We prove the claim in three steps:
  \begin{enumerate}

  \item\label{step:binary-special-rank-family-construction}
    First, we construct a family
    $\mathcal{T}_{\Textlen}\subseteq\BinaryAlphabet^{\Textlen}$. For every
    finite sequence $(a_1,\dots,a_r)$ of $r \geq 1$ positive integers
    satisfying $\sum_{i=1}^{r}a_i=m$, let
    \[
      S^{(a_1,\dots,a_r)}=
        \begin{cases}
          \zero^{a_1}\one^{a_1}\zero^{a_2}\one^{a_2}\cdots
            \zero^{a_r}\one^{a_r}
              & \text{if } \Textlen=2m,\\
          \zero^{a_1}\one^{a_1}\zero^{a_2}\one^{a_2}\cdots
            \zero^{a_r}\one^{a_r}\zero
              & \text{if } \Textlen=2m+1.
        \end{cases}
    \]
    Let $\mathcal{T}_{\Textlen}$ be the family of all such strings. In the
    first case, every string has length $2m=\Textlen$; in the second case,
    every string has length $2m+1=\Textlen$. Thus,
    $\mathcal{T}_{\Textlen}\subseteq\BinaryAlphabet^{\Textlen}$.

  \item\label{step:binary-special-rank-family-size}
    Next, we prove
    $|\mathcal{T}_{\Textlen}|=2^{\lfloor\Textlen/2\rfloor-1}$. In every
    string $S^{(a_1,\dots,a_r)}$ constructed in
    Step~\ref{step:binary-special-rank-family-construction}, the maximal
    zero-blocks in $S^{(a_1,\dots,a_r)}[1\dd2m]$ have lengths
    $a_1,\dots,a_r$ from left to right. Hence, distinct sequences
    $(a_1,\dots,a_r)$ yield distinct strings. It remains to count the finite
    sequences of positive integers summing to $m$. Consider a sequence of
    $m$ ones and the $m-1$ gaps between consecutive ones. Choosing any subset
    of these gaps, placing separators in the chosen gaps, and summing every
    maximal group of ones gives a sequence of positive integers summing to
    $m$. This is a bijection between the subsets of the $m-1$ gaps and the
    sequences under consideration. Consequently,
    $|\mathcal{T}_{\Textlen}|=2^{m-1}
      =2^{\lfloor\Textlen/2\rfloor-1}$.

  \item Finally, we prove that every $S\in\mathcal{T}_{\Textlen}$ is uniquely
    determined by its sequence of special-rank answers. For every
    $i\in[1\dd\Textlen]$, denote
    $R_S[i]=\SpecialRank{S}{i}
      =|\{j\in[1\dd i]:S[j]=S[i]\}|$. By the construction in
    Step~\ref{step:binary-special-rank-family-construction}, the prefix
    $S[1\dd2m]$ is a concatenation of blocks
    $\zero^{a_j}\one^{a_j}$. Hence, every prefix of $S[1\dd2m]$ contains at
    least as many zeros as ones. For every $k\in[1\dd m]$, the $k$th
    occurrence of $\zero$ therefore precedes the $k$th occurrence of
    $\one$. It follows that every value $k\in[1\dd m]$ occurs exactly twice
    in $R_S[1\dd2m]$: first at the position of the $k$th zero and then at
    the position of the $k$th one. Thus, for every $i\in[1\dd2m]$, the
    sequence $R_S$ determines $S[i]$ through
    \[
      S[i]=
      \begin{cases}
        \zero,
          & \text{if $R_S[i]$ is the first occurrence of its value in
            $R_S[1\dd i]$},\\
        \one, & \text{otherwise.}
      \end{cases}
    \]
    If $\Textlen=2m$, this determines the whole string. If
    $\Textlen=2m+1$, it determines the first $2m$ symbols, while the last
    symbol is $\zero$ for every string in $\mathcal{T}_{\Textlen}$. Hence,
    the sequence $(R_S[1],\dots,R_S[\Textlen])$ determines $S$, and the
    special-rank sequences of any two distinct strings in
    $\mathcal{T}_{\Textlen}$ are distinct. Together with
    Steps~\ref{step:binary-special-rank-family-construction}
    and~\ref{step:binary-special-rank-family-size}, this proves the claim.\qedhere
  \end{enumerate}
\end{proof}

\begin{lemma}\label{lm:tableau-special-rank-family}
  Let $\Textlen, \AlphabetSize \in \Z_{\geq 2}$ be such that $\AlphabetSize \leq \Textlen$.
  Consider any $a,b \in \Z_{\geq 1}$ such that $1 \leq b \leq a$, $b \leq \AlphabetSize$, and $ab \leq \Textlen$.
  Then there exists a family $\mathcal{T}_{\Textlen,\AlphabetSize,a,b} \subseteq [0 \dd \AlphabetSize)^{\Textlen}$ of size
  $|\mathcal{T}_{\Textlen,\AlphabetSize,a,b}| \geq \big(\frac{b}{2e}\big)^{ab}$
  such that, for any two distinct strings
  $S,S' \in \mathcal{T}_{\Textlen,\AlphabetSize,a,b}$, the sequences (see \cref{def:rank-select})
  \begin{itemize}
  \item $(\SpecialRank{S}{1},\SpecialRank{S}{2},\dots,\SpecialRank{S}{\Textlen})$ and
  \item $(\SpecialRank{S'}{1},\SpecialRank{S'}{2},\dots,\SpecialRank{S'}{\Textlen})$
  \end{itemize}
  are distinct.
\end{lemma}
\begin{proof}
  We prove the claim in three steps:
  \begin{enumerate}

  \item\label{step:tableau-special-rank-family-construction} First, we construct a family
  $\mathcal{T}_{\Textlen,\AlphabetSize,a,b} \subseteq [0 \dd \AlphabetSize)^{\Textlen}$.
  Let $\mathcal{M}_{a,b}$ denote the set of fillings of a $b \times a$ matrix storing the set of numbers
  $1,2,\dots,ab$ such that numbers strictly increase left-to-right in each row and
  strictly increase top-to-bottom in each column.
  For $M \in \mathcal{M}_{a,b}$ and $t \in [1 \dd ab]$, let $\Row{M}{t} \in [1 \dd b]$ and
  $\Col{M}{t} \in [1 \dd a]$ denote the row and column (respectively) containing number $t$.
  For every $M \in \mathcal{M}_{a,b}$, we define a string $X_{M} \in [0 \dd b)^{ab}$ by
  $X_{M}[t] := \Row{M}{t} - 1$ for $t \in [1 \dd ab]$, and then let
  $S_{M} := X_{M} \cdot \zero^{\Textlen-ab} \in [0 \dd \AlphabetSize)^{\Textlen}$.
  Let $\mathcal{T}_{\Textlen,\AlphabetSize,a,b} := \{S_{M} : M \in \mathcal{M}_{a,b}\}$.
  Since $b \leq \AlphabetSize$, we indeed have
  $\mathcal{T}_{\Textlen,\AlphabetSize,a,b} \subseteq [0 \dd \AlphabetSize)^{\Textlen}$.

  \item\label{step:tableau-special-rank-family-size} Next, we prove
  $|\mathcal{T}_{\Textlen,\AlphabetSize,a,b}| \geq (b/(2e))^{ab}$.
  For the family constructed in Step~\ref{step:tableau-special-rank-family-construction},
  it suffices to lower-bound $|\mathcal{M}_{a,b}|$, because the map $M \mapsto S_{M}$ is injective. Indeed, for every
  $t \in [1 \dd ab]$, we have $\Row{M}{t} = S_{M}[t] + 1$, and then
  $\Col{M}{t} = 1 + |\{u \in [1 \dd t-1] : S_{M}[u] = S_{M}[t]\}|$.
  Hence $S_{M}[1 \dd ab]$ determines both $\Row{M}{t}$ and $\Col{M}{t}$ for every
  $t \in [1 \dd ab]$, and therefore determines $M$ uniquely. Thus
  $|\mathcal{T}_{\Textlen,\AlphabetSize,a,b}| = |\mathcal{M}_{a,b}|$.
  A filling in $\mathcal{M}_{a,b}$ is known as a \emph{Young tableau}.
  By the hook length formula \cite[Theorem~1]{FrameRT54} (see also~\cite{MoralesPP18}), the number of such
  fillings is exactly
  \[
    |\mathcal{M}_{a,b}| =
      (ab)! \Big/ \left(
        \prod_{i=1}^{b}\prod_{j=1}^{a}\bigl((a-j)+(b-i)+1\bigr)
      \right).
  \]
  Since $(a-j)+(b-i)+1\leq a+b$ for every $i\in[1\dd b]$ and
  $j\in[1\dd a]$, the denominator and numerator satisfy, respectively,
  \[
    \prod_{i=1}^{b}\prod_{j=1}^{a}\bigl((a-j)+(b-i)+1\bigr) \leq (a+b)^{ab}
    \qquad\text{and}\qquad
    (ab)! \geq \bigl(ab/e\bigr)^{ab},
  \]
  and hence $|\mathcal{M}_{a,b}| \geq (ab/(e(a+b)))^{ab}$. Therefore
  $|\mathcal{T}_{\Textlen,\AlphabetSize,a,b}| \geq (ab/(e(a+b)))^{ab}$.
  Since $a \geq b$, we have $a+b \leq 2a$, which gives $ab/(a+b) \geq b/2$. Therefore
  $|\mathcal{T}_{\Textlen,\AlphabetSize,a,b}| \geq (b/(2e))^{ab}$.

  \item Finally, we prove that distinct strings in $\mathcal{T}_{\Textlen,\AlphabetSize,a,b}$
  have distinct sequences of special-rank answers.
  Let $M \in \mathcal{M}_{a,b}$. For every $i \in [1 \dd \Textlen]$, let
  $R_{S_{M}}[i] := \SpecialRank{S_{M}}{i}$.
  If $t \in [1 \dd ab]$, then among numbers at most $t$ in row $\Row{M}{t}$ there are
  exactly $\Col{M}{t}$ values, because rows increase left-to-right. Therefore
  $X_{M}[t] = \Row{M}{t}-1$ is the $\Col{M}{t}$th occurrence of that symbol in the prefix
  $X_{M}[1 \dd t]$, and hence also in the prefix $S_{M}[1 \dd t]$. Thus
  $R_{S_{M}}[t] = \Col{M}{t}$ for every $t \in [1 \dd ab]$.
  It follows that $R_{S_{M}}[1 \dd ab]$ determines $\Col{M}{t}$ for every
  $t \in [1 \dd ab]$, and then $\Row{M}{t}$ is the number of occurrences of the value
  $R_{S_{M}}[t]$ in the prefix $R_{S_{M}}[1 \dd t]$, because entries in each fixed column
  increase top-to-bottom. Hence $R_{S_{M}}[1 \dd ab]$ determines $M$ uniquely.
  Thus the mapping $M \mapsto (R_{S_{M}}[1], \dots, R_{S_{M}}[\Textlen])$ is injective, and
  hence any two distinct strings $S, S' \in \mathcal{T}_{\Textlen,\AlphabetSize,a,b}$ yield
  distinct sequences of special-rank answers. Together with
  Steps~\ref{step:tableau-special-rank-family-construction}
  and~\ref{step:tableau-special-rank-family-size}, this proves the claim.\qedhere
  \end{enumerate}
\end{proof}

\begin{theorem}\label{th:special-rank-space-lower-bound}
  Let $\Textlen,\AlphabetSize \in \Z_{\geq 2}$ be such that $\AlphabetSize \leq \Textlen$ and $\Textlen \geq 2^{16}$.
  There is no data structure that, for every string $S \in [0 \dd \AlphabetSize)^{\Textlen}$,
  uses at most $\frac{1}{32}\Textlen\log \AlphabetSize - 1$ bits of space
  and answers special rank queries (\cref{def:rank-select}) on $S$.
\end{theorem}
\begin{proof}

  Suppose that the claim does not hold. Then there exists a data structure,
  denoted $D^{\rm SR}_{\AlphabetSize,\Textlen}$, that for every string $S \in [0\dd \AlphabetSize)^{\Textlen}$
  answers special rank queries on $S$ and uses at most
  $\frac{1}{32} \Textlen\log \AlphabetSize - 1$ bits of space.
  We derive a contradiction by showing that there are more than
  $\AlphabetSize^{\Textlen/32}$ distinct special-rank answer
  tables that must be represented.

  We distinguish three cases:
  \begin{itemize}

  \item \emph{Case 1: $\AlphabetSize \leq 32$.}
    By \cref{lm:binary-special-rank-family}, there exists a family
    $\mathcal{T}_{\Textlen} \subseteq \BinaryAlphabet^{\Textlen} \subseteq [0\dd \AlphabetSize)^{\Textlen}$
    of size $2^{\lfloor \Textlen/2 \rfloor - 1}$
    such that distinct strings in the family have distinct special-rank answer tables.
    Since $\AlphabetSize \leq 32$, we have $\AlphabetSize^{\Textlen/32} \leq 32^{\Textlen/32} = 2^{5\Textlen/32}$.
    We also have $\lfloor \Textlen/2 \rfloor - 1 \geq \Textlen/2 - 2 > 5\Textlen/32$ for every $\Textlen \geq 2^{16}$.
    Therefore $|\mathcal{T}_{\Textlen}| = 2^{\lfloor \Textlen/2 \rfloor - 1} > \AlphabetSize^{\Textlen/32}$.

  \item \emph{Case 2: $32 < \AlphabetSize \leq \lfloor \sqrt{\Textlen} \rfloor$.}
    Let $b := \AlphabetSize$ and $a := \lfloor \Textlen/\AlphabetSize \rfloor$. Then $b \leq a$, $b \leq \AlphabetSize$,
    and $ab \leq \Textlen$. Also, $ab = b\lfloor \Textlen/b \rfloor \geq \Textlen-b \geq \Textlen/2$, since
    $b \leq \sqrt{\Textlen}$ and $\Textlen \geq 4$ imply $\sqrt{\Textlen} \leq \Textlen/2$.
    By \cref{lm:tableau-special-rank-family}, there exists a family
    $\mathcal{T}_{\Textlen,\AlphabetSize,a,b} \subseteq [0 \dd \AlphabetSize)^{\Textlen}$ such that distinct strings
    in the family have distinct special-rank answer tables and
    $|\mathcal{T}_{\Textlen,\AlphabetSize,a,b}| \geq (b/(2e))^{ab}$.
    Since $b=\AlphabetSize>2e$ and $ab\geq\Textlen/2$, it follows that
    \[
      |\mathcal{T}_{\Textlen,\AlphabetSize,a,b}| \geq \left(\frac{\AlphabetSize}{2e}\right)^{\Textlen/2}.
    \]
    To prove that this is greater than $\AlphabetSize^{\Textlen/32}$, it suffices to show that
    $(\AlphabetSize/(2e))^{16} > \AlphabetSize$, or equivalently $\AlphabetSize^{15} > (2e)^{16}$.
    This inequality already holds for $\AlphabetSize = 33$, and the left-hand side increases
    with $\AlphabetSize$. Thus it holds for every $\AlphabetSize \geq 33$, and therefore
    $|\mathcal{T}_{\Textlen,\AlphabetSize,a,b}| > \AlphabetSize^{\Textlen/32}$.

  \item \emph{Case 3: $\AlphabetSize > \lfloor \sqrt{\Textlen} \rfloor$.}
    Let $b := \lfloor \sqrt{\Textlen} \rfloor$ and $a := \lfloor \Textlen/b \rfloor$. Then
    $b \leq a$, $b \leq \AlphabetSize$, and $ab \leq \Textlen$. As in Case~2, we have
    $ab = b\lfloor \Textlen/b \rfloor \geq \Textlen-b \geq \Textlen/2$.
    By \cref{lm:tableau-special-rank-family}, there exists a family
    $\mathcal{T}_{\Textlen,\AlphabetSize,a,b} \subseteq [0 \dd \AlphabetSize)^{\Textlen}$ such that distinct strings
    in the family have distinct special-rank answer tables and
    $|\mathcal{T}_{\Textlen,\AlphabetSize,a,b}| \geq (b/(2e))^{ab}$.
    Since $\Textlen \geq 4$, we have $b = \lfloor \sqrt{\Textlen} \rfloor \geq \sqrt{\Textlen}/2 \geq \sqrt{\AlphabetSize}/2$.
    Since $\Textlen \geq 2^{16}$, we also have $b\geq2^8=256>2e$.
    Therefore
    \[
      |\mathcal{T}_{\Textlen,\AlphabetSize,a,b}|
        \geq \left(\frac{b}{2e}\right)^{ab}
        \geq \left(\frac{\sqrt{\AlphabetSize}}{4e}\right)^{\Textlen/2}.
    \]
    To prove that this is greater than $\AlphabetSize^{\Textlen/32}$, it suffices to show that
    $(\sqrt{\AlphabetSize}/(4e))^{16} > \AlphabetSize$, or equivalently $\AlphabetSize^7 > (4e)^{16}$.
    Moreover, $\AlphabetSize > b \geq 256$.
    The inequality $256^7 > (4e)^{16}$ is true, and the left-hand side increases with $\AlphabetSize$.
    Therefore $\AlphabetSize^7 > (4e)^{16}$ for every $\AlphabetSize > 256$, and hence
    $|\mathcal{T}_{\Textlen,\AlphabetSize,a,b}| > \AlphabetSize^{\Textlen/32}$.
  \end{itemize}

  In all three cases, the number of distinct answer tables that must be represented is
  greater than $\AlphabetSize^{\Textlen/32}$.
  On the other hand, every data structure using at most $\frac{1}{32} \Textlen\log \AlphabetSize - 1$
  bits of space corresponds to a bit string of length at most
  $\frac{1}{32} \Textlen\log \AlphabetSize - 1$.
  The number of such bit strings is
  \[
    \sum_{k=0}^{\lfloor (1/32)\Textlen\log \AlphabetSize - 1 \rfloor} 2^k
      < 2^{(1/32)\Textlen\log \AlphabetSize}
      = \AlphabetSize^{\Textlen/32}.
  \]
  Thus the number of possible stored bit strings is strictly smaller than the number
  of distinct answer tables that must be represented.
  By the pigeonhole principle, there must exist two inputs that map to the same stored
  bit string but have different answers to some special rank query, which is impossible.
  This contradiction completes the proof.
\end{proof}

\begin{remark}\label{rm:special-rank-space-lower-bound}
  The thresholds $1/32$ and $2^{16}$ in \cref{th:special-rank-space-lower-bound}
  were chosen only to simplify the case analysis and have not been optimized.
\end{remark}

\begin{theorem}\label{th:special-rank-asymptotic-space-lower-bound}
  Let $\AlphabetSize, \Textlen \in \Z_{\geq 2}$ satisfy $\AlphabetSize \leq \Textlen$.
  Any data structure supporting special rank queries (\cref{def:rank-select})
  for every string $\Text \in [0 \dd \AlphabetSize)^{\Textlen}$
  has worst-case space usage $\Omega(\Textlen \log \AlphabetSize)$ bits.
\end{theorem}
\begin{proof}
  By \cref{th:special-rank-space-lower-bound}, for every pair of integers
  $(\AlphabetSize,\Textlen)$ satisfying $2 \leq \AlphabetSize \leq \Textlen$ and $\Textlen \geq 2^{16}$, every data structure
  answering special rank queries on strings of length exactly $\Textlen$ over alphabet
  $[0 \dd \AlphabetSize)$ must use more than $\frac{1}{32} \Textlen\log \AlphabetSize - 1$ bits of space.
  Since $\frac{1}{32} \Textlen\log \AlphabetSize - 1 = \Omega(\Textlen\log \AlphabetSize)$, this yields an
  $\Omega(\Textlen\log \AlphabetSize)$ lower bound for all such pairs $(\AlphabetSize,\Textlen)$.
  The remaining pairs $(\AlphabetSize,\Textlen)$ with $2 \leq \AlphabetSize \leq \Textlen < 2^{16}$ form a finite set,
  and excluding finitely many parameter pairs does not affect the asymptotic statement.
  Indeed, for each excluded pair, there exist two strings
  with different answers, and hence every such data structure stores at
  least one bit.
  Therefore every data structure answering special rank queries on strings of length
  exactly $\Textlen$ over alphabet $[0 \dd \AlphabetSize)$ must use at least $\Omega(\Textlen\log \AlphabetSize)$ bits.
\end{proof}

\begin{theorem}\label{th:special-rank-bounded-length-space-lower-bound}
  Let $\AlphabetSize, \Textlen \in \Z_{\geq 2}$ satisfy
  $\AlphabetSize \leq \Textlen$.
  Any data structure supporting special rank queries
  (\cref{def:rank-select}) for every string
  $\Text \in \IntegerAlphabet^{*}$ satisfying
  $\AlphabetSize \leq |\Text| \leq \Textlen$ has worst-case space usage
  $\Omega(\Textlen \log \AlphabetSize)$ bits.
\end{theorem}
\begin{proof}
  Any data structure satisfying the premise must in particular answer special
  rank queries on every string $\Text \in \IntegerAlphabet^{\Textlen}$.
  Thus, the claim follows by \cref{th:special-rank-asymptotic-space-lower-bound}.
\end{proof}

%% file: space-lower-bounds/space-lower-bound-prefix-special-rank.tex
\subsection{Prefix Special Rank Queries}\label{sec:prefix-special-rank-space-lower-bound}

\begin{lemma}\label{lm:reduce-special-rank-to-prefix-special-rank}
  Let $\AlphabetSize,\Seqlen \in \Z_{\geq 2}$ be such that $\AlphabetSize \leq \Seqlen$, and let $\ell \in \Z_{\geq 1}$
  and $s\in\Z_{\geq0}$.
  Suppose that there exists a data structure, denoted $D^{\rm PSR}_{\AlphabetSize,\Seqlen,\ell}$, that for every
  sequence $W[1 \dd \Seqlen]$ of $\Seqlen$ strings over alphabet $[0 \dd \AlphabetSize)$, each of length exactly $\ell$,
  answers prefix special rank queries (\cref{def:prefix-rank-and-select}) on $W$ and uses at most $s$ bits of space.
  Then there exists a data structure, denoted $D^{\rm SR}_{\AlphabetSize^{\ell},\Seqlen}$, that for every
  string $A \in [0 \dd \AlphabetSize^\ell)^{\Seqlen}$, answers special rank queries (\cref{def:rank-select}) on $A$ and uses at most
  $s$ bits of space.
\end{lemma}
\begin{proof}

  Consider any string $A \in [0 \dd \AlphabetSize^\ell)^{\Seqlen}$.
  We define a sequence $W_A[1 \dd \Seqlen]$ of $\Seqlen$ strings over alphabet
  $[0 \dd \AlphabetSize)$ by setting
  $W_A[i] := \AlphabetMap{\AlphabetSize^{\ell}}{\AlphabetSize}{A[i]}$
  (\cref{def:alphabet-map}) for every $i \in [1 \dd \Seqlen]$.
  By \cref{def:alphabet-map}, each string
  $\AlphabetMap{\AlphabetSize^{\ell}}{\AlphabetSize}{A[i]}$ has length
  exactly $\lceil \log_{\AlphabetSize} (\AlphabetSize^{\ell}) \rceil = \ell$.
  Thus, the sequence $W_A$ is a valid input for the data structure
  $D^{\rm PSR}_{\AlphabetSize,\Seqlen,\ell}$.

  We define the data structure
  $D^{\rm SR}_{\AlphabetSize^{\ell},\Seqlen}(A)$ for the string $A$
  to consist of a single component: the data structure
  $D^{\rm PSR}_{\AlphabetSize,\Seqlen,\ell}(W_A)$ for the sequence $W_A$.
  By assumption, this uses at most $s$ bits of space.

  Let $i \in [1 \dd \Seqlen]$. To answer the special rank query $\SpecialRank{A}{i}$,
  note that $\SpecialRank{A}{i} = |\{j \in [1 \dd i] : A[j] = A[i]\}|$.
  Since the function $\AlphabetMap{\AlphabetSize^{\ell}}{\AlphabetSize}{\cdot}$ (\cref{def:alphabet-map}) is a bijection,
  for every $j \in [1 \dd i]$ we have
  $A[j] = A[i]$ if and only if $W_A[j] = W_A[i]$.
  Also, since $W_A[i]$ has length exactly $\ell$, its prefix of length $\ell$
  is the whole string. Therefore
  \begin{align*}
    \PrefixSpecialRank{W_A}{i}{\ell}
      &= |\{j \in [1 \dd i] : W_A[i]\text{ is a prefix of }W_A[j]\}|\\
      &= |\{j \in [1 \dd i] : W_A[j] = W_A[i]\}| = \SpecialRank{A}{i}.
  \end{align*}
  Thus, using $D^{\rm PSR}_{\AlphabetSize,\Seqlen,\ell}(W_A)$,
  we can compute $\SpecialRank{A}{i}$ by
  returning $\PrefixSpecialRank{W_A}{i}{\ell}$.
\end{proof}

\begin{theorem}\label{th:prefix-special-rank-space-lower-bound}
  Let $c \in (0,1]$ be a constant.
  Let $\AlphabetSize,\Seqlen \in \Z_{\geq 2}$ be such that $\AlphabetSize \leq \Seqlen$ and $\Seqlen \geq 2^{16}$.
  Let also $\ell := \lfloor c \log_{\AlphabetSize} \Seqlen \rfloor$ and assume that $\ell \geq 1$.
  There is no data structure that, for every sequence $W[1 \dd \Seqlen]$ of $\Seqlen$ strings
  over alphabet $[0 \dd \AlphabetSize)$, each of length exactly $\ell$,
  uses at most $\frac{1}{32} \Seqlen\ell\log \AlphabetSize - 1$ bits of space
  and answers prefix special rank queries (\cref{def:prefix-rank-and-select}) on $W$.
\end{theorem}
\begin{proof}

  Suppose that the claim does not hold. Then there exists a data structure,
  denoted $D^{\rm PSR}_{\AlphabetSize,\Seqlen,\ell}$, that for every sequence $W[1 \dd \Seqlen]$ of $\Seqlen$ strings
  over alphabet $[0 \dd \AlphabetSize)$, each of length exactly $\ell := \lfloor c \log_{\AlphabetSize} \Seqlen \rfloor$,
  answers prefix special rank queries on $W$ and uses at most
  $\frac{1}{32} \Seqlen\ell\log \AlphabetSize - 1$ bits of space.

  By \cref{lm:reduce-special-rank-to-prefix-special-rank}, this implies the existence of a data structure
  $D^{\rm SR}_{\AlphabetSize^\ell,\Seqlen}$ that, for every string $A \in [0 \dd \AlphabetSize^\ell)^{\Seqlen}$,
  answers special rank queries on $A$ and uses at most
  $\frac{1}{32} \Seqlen\ell\log \AlphabetSize - 1$ bits of space.
  By $c \in (0,1]$ and $\ell = \lfloor c\log_{\AlphabetSize} \Seqlen \rfloor$, we have
  $\AlphabetSize^\ell
    =       \AlphabetSize^{\lfloor c\log_{\AlphabetSize} \Seqlen \rfloor}
    \leq    \AlphabetSize^{\log_{\AlphabetSize} \Seqlen}
    =       \Seqlen$.
  On the other hand, $\ell \geq 1$ implies that $\AlphabetSize^{\ell} \geq \AlphabetSize \geq 2$.
  Thus, we have $2 \leq \AlphabetSize^{\ell} \leq \Seqlen$.
  Therefore \cref{th:special-rank-space-lower-bound} applies with alphabet size $\AlphabetSize^{\ell}$
  and string length $\Seqlen$, and implies that no such data structure can use at most
  $\frac{1}{32} \Seqlen\log(\AlphabetSize^\ell) - 1$ bits of space. Since
  $\frac{1}{32} \Seqlen\log(\AlphabetSize^\ell) - 1 =
  \frac{1}{32} \Seqlen\ell\log \AlphabetSize - 1$, we thus obtain
  a contradiction.
\end{proof}

\begin{remark}\label{rm:prefix-special-rank-space-lower-bound}
  The thresholds $1/32$ and $2^{16}$ in \cref{th:prefix-special-rank-space-lower-bound}
  were chosen only to simplify the analysis (see
  \cref{th:special-rank-space-lower-bound}) and have not been optimized.
\end{remark}

\begin{theorem}\label{th:prefix-special-rank-asymptotic-space-lower-bound}
  Let $c \in (0,1]$ be a constant. Let
  $\AlphabetSize, \Seqlen \in \Z_{\geq 2}$ satisfy
  $\AlphabetSize \leq \Seqlen$. Let
  $\ell := \lfloor c \log_{\AlphabetSize} \Seqlen \rfloor$, and assume
  $\ell \geq 1$.
  Any data structure supporting prefix special rank queries
  (\cref{def:prefix-rank-and-select}) for every sequence
  $W[1 \dd \Seqlen]$ of $\Seqlen$ strings over alphabet
  $\IntegerAlphabet$, each of length exactly $\ell$, has worst-case space
  usage $\Omega(\Seqlen\ell \log \AlphabetSize)$ bits.
\end{theorem}
\begin{proof}
  By \cref{th:prefix-special-rank-space-lower-bound}, for every pair of integers
  $(\AlphabetSize,\Seqlen)$ satisfying $2 \leq \AlphabetSize \leq \Seqlen$ and $\Seqlen \geq 2^{16}$, every data structure
  answering prefix special rank queries on sequences of $\Seqlen$ strings over alphabet
  $[0\dd \AlphabetSize)$, each of length exactly $\ell = \lfloor c\log_{\AlphabetSize} \Seqlen \rfloor$ (where $\ell \geq 1$),
  must use more than $\frac{1}{32} \Seqlen\ell\log \AlphabetSize - 1$ bits of space.
  Since $\frac{1}{32} \Seqlen\ell\log \AlphabetSize - 1 = \Omega(\Seqlen\ell\log \AlphabetSize)$, this yields an
  $\Omega(\Seqlen\ell\log \AlphabetSize)$ lower bound for all such pairs $(\AlphabetSize,\Seqlen)$.
  The remaining pairs $(\AlphabetSize,\Seqlen)$ with $2 \leq \AlphabetSize \leq \Seqlen < 2^{16}$ form a finite set,
  and excluding finitely many parameter pairs does not affect the asymptotic statement.
  Indeed, for each excluded pair, there exist two sequences
  with different answers, and hence every such data structure stores at
  least one bit.
  Therefore every data structure answering prefix special rank queries in this setting
  must use at least $\Omega(\Seqlen\ell\log \AlphabetSize)$ bits of space.
\end{proof}

\begin{theorem}\label{th:prefix-special-rank-bounded-total-length-space-lower-bound}
  Let $c \in (0,1]$ be a constant, and let
  $\AlphabetSize, N \in \Z_{\geq 2}$. Assume that there exists
  $\Seqlen \in \Z_{\geq 2}$ such that, for
  $\ell := \lfloor c\log_{\AlphabetSize} \Seqlen \rfloor$, it holds
  $\ell \geq 1$ and $\Seqlen\ell \leq N$.
  Any data structure supporting prefix special rank queries
  (\cref{def:prefix-rank-and-select}) for every such $\Seqlen$ and every
  sequence $W[1 \dd \Seqlen]$ of $\Seqlen$ strings over alphabet
  $\IntegerAlphabet$, each of length exactly $\ell$, has worst-case space
  usage $\Omega(N \log \AlphabetSize)$ bits.
\end{theorem}
\begin{proof}
  Let $\Seqlen^*$ be the largest positive integer such that, for
  $\ell^* := \lfloor c\log_{\AlphabetSize} \Seqlen^* \rfloor$, we have
  $\ell^* \geq 1$ and $\Seqlen^*\ell^* \leq N$.
  Since the data structure supports queries for every admissible value of $\Seqlen$,
  it must in particular support queries for $\Seqlen=\Seqlen^*$.
  Since $\ell^*\geq1$ and $c\leq1$, we have
  $\AlphabetSize\leq\AlphabetSize^{\ell^*/c}\leq\Seqlen^*$.
  Therefore, by \cref{th:prefix-special-rank-asymptotic-space-lower-bound}, it must use
  at least $\Omega(\Seqlen^*\ell^*\log \AlphabetSize)$ bits of space.
  It remains to show that $\Seqlen^*\ell^* = \Omega(N)$.
  Let $\ell' := \lfloor c\log_{\AlphabetSize} (\Seqlen^*+1) \rfloor$.
  Since $\ell'\geq\ell^*\geq1$, the maximality of $\Seqlen^*$ gives
  $(\Seqlen^*+1)\ell' > N$.
  Since $c \in (0,1]$, we have
  $c\log_{\AlphabetSize}(\Seqlen^*+1)
    <    c\log_{\AlphabetSize}\Seqlen^* + c
    \leq c\log_{\AlphabetSize}\Seqlen^* + 1$,
  and therefore $\ell' \leq \ell^*+1$.
  Consequently, $N < (\Seqlen^*+1)(\ell^*+1) \leq 4\Seqlen^*\ell^*$.
  We therefore obtain $\Omega(\Seqlen^*\ell^*\log \AlphabetSize) = \Omega(N\log \AlphabetSize)$.
\end{proof}

%% file: prefix-select.tex
\section{Perfect Equivalence of Prefix Select and
  Suffix Array Queries for Every Alphabet Size}\label{sec:equiv-prefix-select-and-sa}

\input{prefix-select/problem-def}

\input{prefix-select/alphabet-reduction-suffix-array}

\input{prefix-select/alphabet-reduction-prefix-select}

\input{prefix-select/reduce-prefix-select-to-suffix-array}

\input{prefix-select/reduce-suffix-array-to-prefix-select}
\input{prefix-select/lower-bound}

\input{prefix-select/summary}

%% file: prefix-select/problem-def.tex
\subsection{Problem Definitions}\label{sec:prefix-select-and-sa-problem-def}
\vspace{-1.5ex}

\begin{framed}
  \noindent
  \probname{Indexing for Prefix Select Queries over Alphabet $[0 \dd \AlphabetSize)$}
  \begin{bfdescription}
  \item[Input:]
    Let $\AlphabetSize, \Seqlen \in \Z_{\geq 2}$ be such that
    $\AlphabetSize \leq \Seqlen$. Denote
    $\ell = \lfloor \log_{\AlphabetSize} \Seqlen \rfloor$.
    Consider the word RAM model with word size $w > \log \Seqlen$.
    The input consists of the integer $\Seqlen$ and the packed sequence representation
    $\PackedSeqRepresentation{w}{\AlphabetSize}{W}$
    (\cref{def:packed-sequence-representation}) of a sequence
    $W[1 \dd \Seqlen] \in (\IntegerAlphabet^\ell)^{\Seqlen}$.
  \item[Output:]
    A data structure that, given the length $|X|$ and the packed representation
    $\PackedRepresentation{w}{\AlphabetSize}{X}$ of any string
    $X \in \IntegerAlphabet^{\leq \ell}$, and any integer
    $r \in [1 \dd \Seqlen]$, returns
    $\PrefixSelect{W}{r}{X}$ (\cref{def:prefix-rank-and-select}),
    i.e., the $r$th smallest element of the set
    $\{j \in [1 \dd \Seqlen] : X\text{ is a prefix of }W[j]\}$
    (if $r \leq \PrefixRank{W}{\Seqlen}{X}$) or $\infty$ (otherwise).
  \end{bfdescription}
\end{framed}

\begin{framed}
  \noindent
  \probname{Indexing for Suffix Array Queries over Alphabet $[0 \dd \AlphabetSize)$}
  \begin{bfdescription}
  \item[Input:]
    Let $\AlphabetSize, \Textlen \in \Z_{\geq 2}$ be such that
    $\AlphabetSize \leq \Textlen$. Consider the word RAM model with word size
    $w > \log \Textlen$. The input consists of the length $\Textlen$ and the packed representation
    $\PackedRepresentation{w}{\AlphabetSize}{\Text}$
    (\cref{def:packed-representation}) of a string
    $\Text\in\IntegerAlphabet^{\Textlen}$.
  \item[Output:]
    A data structure that, given
    any $i \in [1 \dd \Textlen]$, returns $\SA{\Text}[i]$ (\cref{def:suffix-array}),
    i.e., the starting position of the lexicographically $i$th smallest suffix
    of $\Text$.
  \end{bfdescription}
\end{framed}
\vspace{2ex}

%% file: prefix-select/alphabet-reduction-suffix-array.tex
\subsection{General Alphabet Reduction for Suffix Array Queries}\label{sec:general-suffix-array-alphabet-reduction}

\begin{theorem}\label{th:general-suffix-array-alphabet-reduction}
  Let $N, \AlphabetSize_1, \AlphabetSize_2 \in \Z_{\geq 2}$ be such that $\AlphabetSize_2 \leq \AlphabetSize_1 \leq N$, and denote
  $k=\lceil\log_{\AlphabetSize_2}\AlphabetSize_1\rceil$.
  In the word RAM model with word size $w \geq 1+\log N$, given $N$, $\AlphabetSize_1$, $\AlphabetSize_2$, $|\Text|$, and the packed representation
  $\PackedRepresentation{w}{\AlphabetSize_1}{\Text}$ (\cref{def:packed-representation}) of any string $\Text \in [0 \dd \AlphabetSize_1)^+$ satisfying $(2k+3)|\Text|\leq N$,
  we can in $\bigO(\sqrt{N} + |\Text| / \log_{\AlphabetSize_1} N)$ time compute integers $\alpha$, $\beta$, $\gamma$, and $\mu$, and the
  packed representation
  $\PackedRepresentation{w}{\AlphabetSize_2}{\Text'}$
  of a string $\Text' \in [0 \dd \AlphabetSize_2)^{*}$
  such that $|\Text'| = (2k+3)|\Text|$ and, for every $i \in [1 \dd |\Text|]$,
  it holds
  $
    \SA{\Text}[i] = \frac{1}{\gamma}(\SA{\Text'}[\alpha + i] - \beta) + \mu
  $.
\end{theorem}
\begin{proof}
  The algorithm consists of two steps.
  \begin{enumerate}
  \item First, we compute the packed representation
    $\PackedRepresentation{w}{\AlphabetSize_2}{\Text'}$ of
    $\Text'=\ExtAlphabetMap{\AlphabetSize_1}{\AlphabetSize_2}{\Text}$
    (\cref{def:ext-alphabet-map}). To this end, we construct the data
    structure from \cref{th:ext-alphabet-map} in $\bigO(\sqrt{N})$ time and
    apply it to $\Text$ in $\bigO(1+|\Text|/\log_{\AlphabetSize_1}N)$ time.
  \item Next, to keep the intermediate integers within one word,
    compute $k$ by counting the successive integer divisions of $\AlphabetSize_1-1$ by $\AlphabetSize_2$ needed to reach zero.
    Set $\gamma=2k+3$, $\alpha=|\Text|\gamma-|\Text|$, $\beta=1$, and $\mu=1$.
    This takes $\bigO(\log\AlphabetSize_1)\subseteq\bigO(\log N)$ time.
  \end{enumerate}
  The mapping structure uses size parameter $N$, source alphabet size
  $\AlphabetSize_1$, and target alphabet size $\AlphabetSize_2$, and its
  query is valid because $(2k+3)|\Text|\leq N$.
  This bound also ensures that all returned parameters fit in one word.
  The total running time is
  $\bigO(\sqrt{N}+|\Text|/\log_{\AlphabetSize_1}N)$,
  since $\log N=\bigO(\sqrt{N})$.
  By \cref{def:ext-alphabet-map}, the string $\Text'$ has length
  $|\Text|\gamma$, as required, and $\alpha=|\Text'|-|\Text|$.
  The suffix-array identity follows from
  \cref{lm:general-suffix-array-alphabet-reduction}.
\end{proof}

\begin{theorem}\label{th:general-suffix-array-alphabet-reduction-2}
  Let $N, \AlphabetSize_1, \AlphabetSize_2 \in \Z_{\geq 2}$ be such that $\AlphabetSize_2 \leq \AlphabetSize_1 \leq N$, and denote
  $k=\lceil\log_{\AlphabetSize_2}\AlphabetSize_1\rceil$.
  In the word RAM model with word size $w \geq 1+\log N$, given $N$, $\AlphabetSize_1$, $\AlphabetSize_2$, $|\Text|$, and the packed representation
  $\PackedRepresentation{w}{\AlphabetSize_1}{\Text}$ (\cref{def:packed-representation}) of any string $\Text \in [0 \dd \AlphabetSize_1)^+$ satisfying $(2k+3)|\Text|\leq N$,
  we can in $\bigO(N / \log_{\AlphabetSize_2} N)$ time compute integers $\alpha$, $\beta$, $\gamma$, and $\mu$, and the
  packed representation
  $\PackedRepresentation{w}{\AlphabetSize_2}{\Text'}$
  of a string $\Text' \in [0 \dd \AlphabetSize_2)^{*}$
  such that $|\Text'| = (2k+3)|\Text|$ and, for every $i \in [1 \dd |\Text|]$,
  it holds
  $
    \SA{\Text}[i] = \frac{1}{\gamma}(\SA{\Text'}[\alpha + i] - \beta) + \mu
  $.
\end{theorem}
\begin{proof}
  Apply \cref{th:general-suffix-array-alphabet-reduction} with the supplied
  parameters. Since
  $|\Text|\log\AlphabetSize_1\leq k|\Text|\log\AlphabetSize_2
    \leq N\log\AlphabetSize_2$, we have
  $|\Text|/\log_{\AlphabetSize_1}N\leq N/\log_{\AlphabetSize_2}N$.
  The setup time is also covered by this bound, since
  $\sqrt N=\bigO(N/\log_{\AlphabetSize_2}N)$.
\end{proof}

%% file: prefix-select/alphabet-reduction-prefix-select.tex
\subsection{General Alphabet Reduction for Prefix Select Queries}\label{sec:general-prefix-select-alphabet-reduction}

\begin{theorem}\label{th:general-prefix-select-alphabet-reduction}
  Let $\AlphabetSize_1, \AlphabetSize_2, N \in \Z_{\geq 2}$ be such that $\AlphabetSize_1 \geq \AlphabetSize_2$.
  Consider the word RAM model with word size $w$ satisfying
  $1+\log N \leq w = \bigO(\log N)$.
  Assume that there exists a data structure answering prefix select
  queries (\cref{def:prefix-rank-and-select})
  that, given $N$, $\AlphabetSize_2$, $m_2$, and the packed sequence representation
  $\PackedSeqRepresentation{w}{\AlphabetSize_2}{W_2}$
  (\cref{def:packed-sequence-representation}) of a sequence
  $W_2[1 \dd m_2]$ of $m_2 \geq \AlphabetSize_2$ nonempty strings
  of common length $\ell_2 = \lfloor \log_{\AlphabetSize_2} m_2 \rfloor$
  over alphabet $[0 \dd \AlphabetSize_2)$,
  where $m_2 \cdot \ell_2 \leq N$,
  returns $\PrefixSelect{W_2}{r}{X}$ given the length $|X|$ and the packed
  representation $\PackedRepresentation{w}{\AlphabetSize_2}{X}$ of any string
  $X\in[0\dd\AlphabetSize_2)^{\leq\ell_2}$ and any integer $r\in[1\dd m_2]$,
  and achieves the following complexities:
  \begin{itemize}
  \item space usage $S(\AlphabetSize_2,N)$ bits,
  \item preprocessing time $P_t(\AlphabetSize_2,N)$,
  \item preprocessing space $P_s(\AlphabetSize_2,N)$ bits,
  \item query time $Q(\AlphabetSize_2,N)$.
  \end{itemize}
  Then, letting $N' = \lfloor N / (4k) \rfloor$
  (where $k = \lceil \log_{\AlphabetSize_2} \AlphabetSize_1 \rceil$),
  there exists a data structure answering prefix select queries
  that, given $N$, $\AlphabetSize_1$, $\AlphabetSize_2$, $m_1$, and the packed sequence representation
  $\PackedSeqRepresentation{w}{\AlphabetSize_1}{W_1}$
  (\cref{def:packed-sequence-representation}) of a sequence
  $W_1[1 \dd m_1]$ of $m_1 \geq \AlphabetSize_1$ nonempty strings
  of common length $\ell_1 = \lfloor \tfrac{1}{2} \log_{\AlphabetSize_1} m_1 \rfloor$
  over alphabet $[0 \dd \AlphabetSize_1)$,
  where $m_1 \cdot \ell_1 \leq N'$,
  returns $\PrefixSelect{W_1}{r}{X}$ given the length $|X|$ and the packed
  representation $\PackedRepresentation{w}{\AlphabetSize_1}{X}$ of any string
  $X\in[0\dd\AlphabetSize_1)^{\leq\ell_1}$ and any integer $r\in[1\dd m_1]$,
  and achieves the following complexities:
  \begin{itemize}
  \item space usage $\bigO(N \log \AlphabetSize_2 + S(\AlphabetSize_2,N))$ bits,
  \item preprocessing time $\bigO(N / \log_{\AlphabetSize_2} N + P_t(\AlphabetSize_2,N))$,
  \item preprocessing space $\bigO(N \log \AlphabetSize_2 + P_s(\AlphabetSize_2,N))$ bits,
  \item query time $\bigO(1 + Q(\AlphabetSize_2,N))$.
  \end{itemize}
\end{theorem}
\begin{proof}
  Consider a sequence $W_1[1\dd m_1]$ as in the claim, with
  $\ell_1=\lfloor\tfrac12\log_{\AlphabetSize_1}m_1\rfloor\geq1$
  and $m_1\ell_1\leq N'$.
  Denote $m_2=m_1$ and
  $\ell_2=\lfloor\log_{\AlphabetSize_2}m_1\rfloor$.
  Since $\log_{\AlphabetSize_2}\AlphabetSize_1\geq1$, we have
  $k\leq2\log_{\AlphabetSize_2}\AlphabetSize_1$. Consequently,
  $k\ell_1\leq\log_{\AlphabetSize_2}m_1$.
  Since $k\ell_1$ is an integer, we obtain
  $k\ell_1\leq\lfloor\log_{\AlphabetSize_2}m_1\rfloor=\ell_2$. In the other direction,
  \[
    \ell_2\leq\log_{\AlphabetSize_2}m_1
      =(\log_{\AlphabetSize_1}m_1)\cdot
        \log_{\AlphabetSize_2}\AlphabetSize_1
      <2(\ell_1+1)k\leq4k\ell_1,
  \]
  where the last inequality uses $\ell_1\geq1$.
  Let $W_2[1\dd m_2]$ denote the sequence satisfying
  \[
    W_2[i]=\AlphabetMap{\AlphabetSize_1}{\AlphabetSize_2}{W_1[i]}
      \cdot\zero^{\ell_2-k\ell_1}
      \qquad\text{for every }i\in[1\dd m_1]
  \]
  (\cref{def:alphabet-map}). Each string $W_2[i]$ has length $\ell_2$,
  and $m_2\geq\AlphabetSize_2$. Moreover,
  $m_2\ell_2\leq4km_1\ell_1\leq4kN'\leq N$.
  Thus, $W_2$ satisfies the assumptions of the prefix-select structure
  from the claim.

  For any $X\in[0\dd\AlphabetSize_1)^{\leq\ell_1}$, denote
  $R=\AlphabetMap{\AlphabetSize_1}{\AlphabetSize_2}{X}$.
  By \cref{ob:alphabet-map-properties}\eqref{ob:alphabet-map-prefix-property},
  for every $i\in[1\dd m_1]$,
  $X$ is a prefix of $W_1[i]$ if and only if $R$ is a prefix of
  $\AlphabetMap{\AlphabetSize_1}{\AlphabetSize_2}{W_1[i]}$.
  Since $|R|\leq k\ell_1$, appending zeros does not change this condition.
  By \cref{ob:prefix-rank-and-select-padding}, for every $r\geq1$,
  \[
    \PrefixSelect{W_1}{r}{X}=\PrefixSelect{W_2}{r}{R}.
  \]

  \DSComponents
  The data structure consists of the following components:
  \begin{enumerate}
  \item The structure from \cref{th:alphabet-map} with size parameter
    $N$ and alphabet sizes $\AlphabetSize_1$ and $\AlphabetSize_2$,
    together with the integer $k$. It occupies $\bigO(\sqrt N)$ words.
  \item The prefix-select structure from the claim constructed for $W_2$.
    It occupies $S(\AlphabetSize_2,N)$ bits.
  \end{enumerate}
  Since $w=\bigO(\log N)$, the total space is
  $\bigO(\sqrt N\log N+S(\AlphabetSize_2,N))
    \subseteq\bigO(N\log\AlphabetSize_2+S(\AlphabetSize_2,N))$ bits.

  \DSQueries
  Given $|X|$, the packed representation
  $\PackedRepresentation{w}{\AlphabetSize_1}{X}$ of
  $X\in[0\dd\AlphabetSize_1)^{\leq\ell_1}$, and $r\in[1\dd m_1]$,
  we compute $\PrefixSelect{W_1}{r}{X}$ in two steps:
  \begin{enumerate}
  \item First, apply the structure from \cref{th:alphabet-map} to
    $|X|$ and $\PackedRepresentation{w}{\AlphabetSize_1}{X}$ to obtain
    the packed representation $\PackedRepresentation{w}{\AlphabetSize_2}{R}$.
    This query is valid because $k|X|\leq k\ell_1\leq\ell_2\leq N$.
    It takes $\bigO(1+|X|/\log_{\AlphabetSize_1}N)=\bigO(1)$ time
    because $|X|\leq\ell_1\leq\log_{\AlphabetSize_1}N$.
  \item Query the prefix-select structure for $W_2$ with $r$,
    $|R|=k|X|$, and $\PackedRepresentation{w}{\AlphabetSize_2}{R}$,
    and return its answer. The query is valid because $|R|\leq k\ell_1\leq\ell_2$,
    and it takes $Q(\AlphabetSize_2,N)$ time.
  \end{enumerate}
  The identity above proves correctness, and the total query time is
  $\bigO(1+Q(\AlphabetSize_2,N))$.

  \DSConstruction
  Given $N$, $\AlphabetSize_1$, $\AlphabetSize_2$, $m_1$, and
  $\PackedSeqRepresentation{w}{\AlphabetSize_1}{W_1}$,
  we construct the two components as follows:
  \begin{enumerate}
  \item First, compute $k$, $\ell_1$, $\ell_2$, and
    $b_2=\lceil\log\AlphabetSize_2\rceil$ in $\bigO(\log N)$ time,
    and store $k$. To keep all intermediate integers within one word,
    we compute these parameters using integer divisions. Construct the structure from \cref{th:alphabet-map}
    with the parameters specified above in $\bigO(\sqrt N)$ time.
    Its assumptions hold because
    $w\geq1+\log N$ and
    $\AlphabetSize_2\leq\AlphabetSize_1\leq m_1\leq m_1\ell_1\leq N'\leq N$,
    where the second and third inequalities follow from
    $m_1\geq\AlphabetSize_1$ and $\ell_1\geq1$.
  \item Next, we construct the prefix-select structure for $W_2$.
    Denote by $M[1\dd m_1]$ and $Z[1\dd m_1]$ the sequences satisfying
    $M[i]=\AlphabetMap{\AlphabetSize_1}{\AlphabetSize_2}{W_1[i]}$ and
    $Z[i]=\zero^{\ell_2-k\ell_1}$ for every $i\in[1\dd m_1]$.
    Since $W_2[i]=M[i]Z[i]$, we proceed as follows:
    \begin{enumerate}
    \item To obtain $\PackedSeqRepresentation{w}{\AlphabetSize_2}{M}$,
      apply the structure from \cref{th:alphabet-map} to the concatenation
      represented by $\PackedSeqRepresentation{w}{\AlphabetSize_1}{W_1}$,
      with length $m_1\ell_1$. This takes
      $\bigO(1+m_1\ell_1/\log_{\AlphabetSize_1}N)$ time.
      Since $\log\AlphabetSize_1\leq k\log\AlphabetSize_2$ and
      $km_1\ell_1\leq kN'\leq N/4$, we have
      \[
        \frac{m_1\ell_1}{\log_{\AlphabetSize_1}N}
          \leq\frac{km_1\ell_1}{\log_{\AlphabetSize_2}N}
          \leq\frac{N}{4\log_{\AlphabetSize_2}N}.
      \]
    \item Initialize the packed representation of
      $\zero^{m_1(\ell_2-k\ell_1)}$ over alphabet $[0\dd\AlphabetSize_2)$
      using
      \cref{pr:packed-representation}\eqref{pr:packed-representation-initialize}.
      This is $\PackedSeqRepresentation{w}{\AlphabetSize_2}{Z}$.
      Its construction takes
      $\bigO(1+m_1(\ell_2-k\ell_1)\log\AlphabetSize_2/w)
        =\bigO(N/\log_{\AlphabetSize_2}N)$ time.
    \item We now construct $\PackedSeqRepresentation{w}{\AlphabetSize_2}{W_2}$
      by applying \cref{th:packed-pointwise-concatenation-partition-offline}
      to $M$ and $Z$. The running time is
      $\bigO(\sqrt N+m_1\ell_2/\log_{\AlphabetSize_2}N)
        =\bigO(N/\log_{\AlphabetSize_2}N)$.
    \item Finally, construct the prefix-select structure from the claim
      using $m_2$ and $\PackedSeqRepresentation{w}{\AlphabetSize_2}{W_2}$.
      This takes $P_t(\AlphabetSize_2,N)$ time and uses
      $P_s(\AlphabetSize_2,N)$ bits of space.
    \end{enumerate}
  \end{enumerate}
  The initialization uses length bound $N$. The pointwise concatenation
  and the prefix-select structure from the claim use size parameter $N$
  and alphabet size $\AlphabetSize_2$, and the concatenation's length
  conditions hold because $1\leq\ell_2\leq N/m_1$ and $m_1\ell_2\leq N$.
  Since $\log_{\AlphabetSize_2}N\leq\log N$, it holds
  $\sqrt N+\log N=\bigO(N/\log_{\AlphabetSize_2}N)$.
  The work preceding the last substep therefore takes
  $\bigO(N/\log_{\AlphabetSize_2}N)$ time and uses at most this many
  words of auxiliary space. The input representation occupies
  $\bigO(w+m_1\ell_1\log\AlphabetSize_1)=\bigO(N\log\AlphabetSize_2)$ bits,
  by $\log\AlphabetSize_1\leq k\log\AlphabetSize_2$ and
  $km_1\ell_1\leq N/4$. Using $w=\bigO(\log N)$, the total preprocessing
  time is $\bigO(N/\log_{\AlphabetSize_2}N+P_t(\AlphabetSize_2,N))$,
  and the preprocessing space is
  $\bigO(N\log\AlphabetSize_2+P_s(\AlphabetSize_2,N))$ bits.
\end{proof}

%% file: prefix-select/reduce-prefix-select-to-suffix-array.tex
\subsection{Reducing Prefix Select to Suffix Array Queries}\label{sec:reduce-prefix-select-to-sa}

\begin{lemma}\label{lm:reduce-prefix-select-to-sa}
  Let $\AlphabetSize\in\Z_{\geq2}$, and let $W[1\dd m]$ be a sequence
  of $m\geq\AlphabetSize$ strings of length $\ell\geq1$ over alphabet
  $\IntegerAlphabet$. Denote
  $\Text=\SeqToString{\AlphabetSize}{W}$
  (\cref{def:seq-to-string}), $k=\lceil\log_{\AlphabetSize}m\rceil$,
  $\gamma=\ell+k+1$, and $c=\AlphabetSize$.
  For every nonempty $X\in\IntegerAlphabet^{\leq\ell}$, denote
  $b_X=\RangeBegTwo{\revstr{X}\cdot c}{\Text}$ and
  $e_X=\RangeEndTwo{\revstr{X}\cdot c}{\Text}$ (\cref{def:occ}).
  For every $r\in[1\dd m]$, it holds
  \[
    \PrefixSelect{W}{r}{X}
      =\begin{cases}
        \frac{1}{\gamma}\cdot(\SA{\Text}[b_X+r]+k+|X|)
          & \text{if $r\leq e_X-b_X$},\\[1ex]
        \infty & \text{otherwise.}
      \end{cases}
  \]
\end{lemma}
\begin{proof}
  By \cref{lm:seq-to-string}\eqref{lm:seq-to-string-range-size},
  it holds $e_X-b_X=\PrefixRank{W}{m}{X}$.
  If $r\leq e_X-b_X$, then
  $\SA{\Text}[b_X+r]=\gamma\cdot\PrefixSelect{W}{r}{X}-(k+|X|)$
  by \cref{lm:seq-to-string}\eqref{lm:seq-to-string-locate}.
  Rearranging this equality gives the first case of the formula.
  Otherwise, $r>\PrefixRank{W}{m}{X}$, and the answer is $\infty$
  by \cref{def:prefix-rank-and-select}.
\end{proof}

\begin{theorem}\label{th:reduce-prefix-select-to-sa-for-all-alphabet-sizes}
  Let $\AlphabetSize, N \in \Z_{\geq 2}$ be such that $\AlphabetSize \leq N$.
  Consider the word RAM model with word size $w$ satisfying
  $1+\log N \leq w = \bigO(\log N)$.
  Assume that there exists a data structure for
  \probname{Indexing}
  \probname{for}
  \probname{Suffix}
  \probname{Array}
  \probname{Queries}
  \probname{over}
  \probname{Alphabet}
  $[0 \dd \AlphabetSize)$
  that, given $N$, $\AlphabetSize$, and a valid input of at most $N$ symbols over alphabet
  $\IntegerAlphabet$,\footnote{\label{footnote:sa-input-2} That is, the length $\Textlen$ and the
  packed representation
  $\PackedRepresentation{w}{\AlphabetSize}{\Text}$ of $\Text \in \IntegerAlphabet^{\Textlen}$ with
  $\AlphabetSize \leq \Textlen \leq N$
  (\cref{sec:prefix-select-and-sa-problem-def}).}
  achieves:
  \begin{itemize}
  \item space usage $S(\AlphabetSize,N)$ bits,
  \item preprocessing time $P_t(\AlphabetSize,N)$,
  \item preprocessing space $P_s(\AlphabetSize,N)$ bits,
  \item query time $Q(\AlphabetSize,N)$.
  \end{itemize}
  Then, there exists a data structure for
  \probname{Indexing}
  \probname{for}
  \probname{Prefix}
  \probname{Select}
  \probname{Queries}
  \probname{over}
  \probname{Alphabet}
  $[0 \dd \AlphabetSize)$
  that, given $N$, $\AlphabetSize$, and a valid input of at most $N$ symbols over alphabet
  $\IntegerAlphabet$,\footnote{\label{footnote:prefix-select-input-2} That is, the integer $\Seqlen$ and the
  packed sequence representation
  $\PackedSeqRepresentation{w}{\AlphabetSize}{W}$ of a sequence
  $W[1 \dd \Seqlen]$ of $\Seqlen \geq \AlphabetSize$ strings of length
  $\ell = \lfloor \log_{\AlphabetSize} \Seqlen \rfloor$, where
  $\Seqlen \ell \leq N$; see \cref{sec:prefix-select-and-sa-problem-def}.}
  achieves:
  \begin{itemize}
  \item space usage $\bigO(N \log \AlphabetSize + S(\AlphabetSize,N))$ bits,
  \item preprocessing time $\bigO(N / \log_{\AlphabetSize} N + P_t(\AlphabetSize,N))$,
  \item preprocessing space $\bigO(N \log \AlphabetSize + P_s(\AlphabetSize,N))$ bits,
  \item query time $\bigO(1 + Q(\AlphabetSize,N))$.
  \end{itemize}
\end{theorem}
\begin{proof}
  Consider a sequence $W[1\dd\Seqlen]$ of
  $\Seqlen\geq\AlphabetSize$ strings of length
  $\ell=\lfloor\log_{\AlphabetSize}\Seqlen\rfloor\geq1$ over
  $\IntegerAlphabet$, with $\Seqlen\ell\leq N$.
  Our main construction divides $W$ into a constant number of consecutive
  blocks and constructs a separate text for each block. This keeps every
  text supplied to the suffix-array structure within the length bound $N$.
  To answer a prefix-select query, we first count the strings having the
  queried prefix in each block. These counts determine the block containing
  the requested string, after which one suffix-array query gives its position.

  First, suppose that $\Seqlen<64\AlphabetSize$. Then
  $\ell<1+\log_{\AlphabetSize}64\leq7$, and hence $\ell\leq6$.
  For every $p\in[1\dd\ell]$ and $x\in[0\dd\AlphabetSize^p)$,
  we store an array $L_{p,x}$ listing in ascending order all indices
  $i\in[1\dd\Seqlen]$ such that
  $\Val{\AlphabetSize}{W[i][1\dd p]}=x$, together with its length.
  To construct these arrays, first compute $\ell$ and
  $b=\lceil\log\AlphabetSize\rceil$, and store them.
  For each $p\in[1\dd\ell]$, proceed as follows:
  \begin{enumerate}
  \item Initialize an array $C[0\dd\AlphabetSize^p)$ to zero and scan
    $i=1,\ldots,\Seqlen$, incrementing $C[x]$ for
    $x=\Val{\AlphabetSize}{W[i][1\dd p]}$.
  \item Allocate $L_{p,x}[1\dd C[x]]$ for every $x\in[0\dd\AlphabetSize^p)$
    and store its length. Reset $C$ to zero and scan $i=1,\ldots,\Seqlen$
    again. At each index $i$, compute $x=\Val{\AlphabetSize}{W[i][1\dd p]}$,
    increment $C[x]$, and set $L_{p,x}[C[x]]:=i$.
  \end{enumerate}
  Computing $x$ in these steps requires at most six symbol accesses using
  \cref{pr:packed-representation}\eqref{pr:packed-representation-access}
  with length bound $N$, and
  $\bigO(1)$ arithmetic operations. Since $\AlphabetSize^p\leq\Seqlen$
  and $\ell\leq6$, the construction takes $\bigO(\Seqlen)$ time and
  uses $\bigO(\Seqlen)$ words. Given a query with rank $r$ and string $X$,
  return $r$ if $X=\emptystring$. Otherwise, compute
  $x=\Val{\AlphabetSize}{X}$ from the supplied packed representation,
  using the same access operation with input length $|X|$ and length
  bound $N$. Return $L_{|X|,x}[r]$ if
  $r\leq|L_{|X|,x}|$, or $\infty$ otherwise. The query takes $\bigO(1)$
  time. The bounds at the end of the proof also apply to this construction.

  Henceforth, assume that $\Seqlen\geq64\AlphabetSize$.
  Let $W_1,\ldots,W_{64}$ denote the consecutive blocks of $W$ in which
  the first $\Seqlen\bmod64$ blocks contain
  $\lfloor\Seqlen/64\rfloor+1$ strings each, and the remaining blocks
  contain $\lfloor\Seqlen/64\rfloor$ strings each.
  For every $t\in[1\dd64]$, let $q_t$ denote the number of strings
  in $W_t$, and let $a_t=\sum_{u=1}^{t}q_u$, with $a_0=0$.
  In particular, $W_t[j]=W[a_{t-1}+j]$ for every $j\in[1\dd q_t]$.
  The block sizes satisfy $\AlphabetSize\leq q_t\leq\Seqlen/32$ and
  $\Seqlen\leq128q_t$ for every $t\in[1\dd64]$.
  Denote $c=\AlphabetSize$, $k_t=\lceil\log_{\AlphabetSize}q_t\rceil$,
  and $Y_t=\SeqToString{\AlphabetSize}{W_t}$ (\cref{def:seq-to-string}).

  For each $t\in[1\dd64]$, let $\alpha_t,\beta_t,\gamma_t,\mu_t$
  and $\Text_t\in\IntegerAlphabet^*$ denote the integers and string
  obtained from \cref{th:general-suffix-array-alphabet-reduction-2}
  with input $Y_t$, size parameter $7|Y_t|$, and alphabet sizes
  $2\AlphabetSize$ and $\AlphabetSize$. Since
  $\AlphabetSize<2\AlphabetSize\leq\AlphabetSize^2$, we have
  $\lceil\log_{\AlphabetSize}(2\AlphabetSize)\rceil=2$, and hence the length
  condition of the reduction holds with equality and gives
  $|\Text_t|=7|Y_t|$. Since $k_t\leq\ell+1$,
  these strings satisfy
  \[
    2\AlphabetSize\leq|Y_t|=q_t(\ell+k_t+1),
    \qquad
    |\Text_t|=7|Y_t|\leq28q_t\ell
      \leq\tfrac78\Seqlen\ell\leq N.
  \]
  These bounds ensure that the alphabet reduction satisfies its alphabet
  and word-size assumptions and that each $\Text_t$ is a valid input to
  the suffix-array data structure from the premise.

  \DSComponents
  The data structure consists of the following components:
  \begin{enumerate}
  \item The integers $\Seqlen$ and $\ell$, the boundaries
    $a_0,\ldots,a_{64}$, and the integers
    $q_t,k_t,\alpha_t,\beta_t,\gamma_t,\mu_t$ for every $t\in[1\dd64]$.
    They occupy $\bigO(1)$ words, or $\bigO(\log N)$ bits.
  \item For every $t\in[1\dd64]$, the structure from
    \cref{pr:seq-to-str-range-beg-and-range-end} constructed for $W_t$.
    These structures occupy $\bigO(\Seqlen)$ words in total, or
    $\bigO(\Seqlen\log N)$ bits.
  \item For every $t\in[1\dd64]$, an instance of the suffix-array
    data structure from the premise constructed for $\Text_t$.
    Together, these instances use $\bigO(S(\AlphabetSize,N))$ bits.
  \end{enumerate}
  The total space is $\bigO(\Seqlen\log N+S(\AlphabetSize,N))$ bits.

  \DSQueries
  Given $|X|$, the packed representation
  $\PackedRepresentation{w}{\AlphabetSize}{X}$ of
  $X\in\IntegerAlphabet^{\leq\ell}$, and $r\in[1\dd\Seqlen]$,
  return $r$ if $|X|=0$. Otherwise, proceed as follows:
  \begin{enumerate}[beginpenalty=10000]
  \item First, we count the strings having prefix $X$ in each block.
    Initialize $d_0:=0$. For $t=1,\ldots,64$, compute
    $b_t=\RangeBegTwo{\revstr{X}\cdot c}{Y_t}$ and
    $e_t=\RangeEndTwo{\revstr{X}\cdot c}{Y_t}$ using the structure from
    \cref{pr:seq-to-str-range-beg-and-range-end} for $W_t$, and set
    $d_t:=d_{t-1}+e_t-b_t$. The value $d_t$ counts the strings having
    prefix $X$ in the first $t$ blocks. There are $64$ queries, each
    taking $\bigO(1)$ time.
  \item If $r>d_{64}$, return $\infty$. Otherwise, scan
    $d_1,\ldots,d_{64}$ to find the smallest $h\in[1\dd64]$ such that
    $d_h\geq r$, and compute $r':=r-d_{h-1}$. Thus, the requested
    string is the $r'$th string having prefix $X$ in $W_h$.
    This step takes $\bigO(1)$ time.
  \item We now compute its position using one suffix-array query.
    Obtain $j:=\SA{\Text_h}[\alpha_h+b_h+r']$ from the structure for
    $\Text_h$, compute $v:=(j-\beta_h)/\gamma_h+\mu_h$, and return
    $a_{h-1}+(v+k_h+|X|)/(\ell+k_h+1)$.
    This step takes $\bigO(1+Q(\AlphabetSize,N))$ time.
  \end{enumerate}
  To prove correctness, first note that
  \cref{lm:seq-to-string}\eqref{lm:seq-to-string-range-size} gives
  $e_t-b_t=\PrefixRank{W_t}{q_t}{X}$ for every $t\in[1\dd64]$.
  Consequently, $d_{64}=\PrefixRank{W}{\Seqlen}{X}$, which justifies
  returning $\infty$ when $r>d_{64}$. Otherwise, the choice of $h$
  gives $d_{h-1}<r\leq d_h$, and hence
  $1\leq r'\leq e_h-b_h$ and $1\leq b_h+r'\leq|Y_h|$.
  By \cref{th:general-suffix-array-alphabet-reduction-2}, we have
  $v=\SA{Y_h}[b_h+r']$. Applying \cref{lm:reduce-prefix-select-to-sa}
  to $W_h$ shows that $(v+k_h+|X|)/(\ell+k_h+1)$ equals
  $\PrefixSelect{W_h}{r'}{X}$ and is an integer.
  Since the blocks occur consecutively in $W$, we obtain
  \[
    \PrefixSelect{W}{r}{X}
      =a_{h-1}+\PrefixSelect{W_h}{r'}{X}
      =a_{h-1}+\frac{v+k_h+|X|}{\ell+k_h+1},
  \]
  as required. For $X=\emptystring$, the answer is $r$ by
  \cref{def:prefix-rank-and-select}.
  The total query time is $\bigO(1+Q(\AlphabetSize,N))$.

  \DSConstruction
  We construct the three components as follows:
  \begin{enumerate}
  \item First, we compute the integers and packed texts needed for the
    suffix-array structures.
    Compute $\ell$ and $b=\lceil\log\AlphabetSize\rceil$, and store
    $\Seqlen$ and $\ell$. Compute and store the block sizes $q_t$
    and the integers $k_t$ for $t\in[1\dd64]$.
    Set $a_0:=0$, and compute and store $a_t:=a_{t-1}+q_t$
    for $t=1,\ldots,64$.
    Then, for every $t\in[1\dd64]$, proceed as follows:
    \begin{enumerate}
    \item Extract $\PackedSeqRepresentation{w}{\AlphabetSize}{W_t}$, the
      length-$q_t\ell$ substring of the input representation starting at
      position $a_{t-1}\ell+1$, using
      \cref{pr:packed-representation}\eqref{pr:packed-representation-substring}.
    \item Compute $\PackedRepresentation{w}{2\AlphabetSize}{Y_t}$ using
      \cref{pr:seq-to-string-construction}.
    \item Apply \cref{th:general-suffix-array-alphabet-reduction-2}
      to $Y_t$ to compute
      $\PackedRepresentation{w}{\AlphabetSize}{\Text_t}$ and
      $\alpha_t,\beta_t,\gamma_t,\mu_t$.
      Store these integers, and retain the representations of $W_t$
      and $\Text_t$ for the next two steps.
    \end{enumerate}
  \item For every $t\in[1\dd64]$, construct the structure from
    \cref{pr:seq-to-str-range-beg-and-range-end} for $W_t$.
  \item Finally, for each $t\in[1\dd64]$, construct the suffix-array
    data structure from the premise for $\Text_t$.
    Each construction takes $P_t(\AlphabetSize,N)$ time and uses
    $P_s(\AlphabetSize,N)$ bits of space.
  \end{enumerate}
  Both sequence-to-string tools and the suffix-array structures from
  the premise are applied with size parameter $N$ and alphabet size
  $\AlphabetSize$.
  The assumptions of the sequence-to-string tools hold because $\AlphabetSize\leq q_t\leq\Seqlen\leq N$,
  $\AlphabetSize^\ell\leq\Seqlen\leq N$,
  $q_t\ell\leq q_t(\ell+k_t+1)=|Y_t|\leq N$, and
  $w\geq1+\log N$.
  Each invocation takes $\bigO(q_t+\AlphabetSize^\ell)=\bigO(\Seqlen)$ time
  and uses $\bigO(\Seqlen)$ words.
  To keep all operands within one word, the integer logarithms used above
  are computed by successive integer divisions.
  To avoid computing the product $64\AlphabetSize$, we test
  $\Seqlen<64\AlphabetSize$ by comparing $\lfloor\Seqlen/64\rfloor$
  with $\AlphabetSize$. If $N<64$, then $\Seqlen\leq N<64\AlphabetSize$,
  and the case $\Seqlen<64\AlphabetSize$ applies without this test.
  Otherwise, the constant $64\leq N$ fits in one word.
  The block boundaries and prefix counts are at most $\Seqlen\leq N$.
  The extractions use length bound $N$, and all string positions used
  above are within the corresponding strings, whose lengths are at most $N$.
  Since $q_t=\Theta(\Seqlen)$, $|Y_t|\leq4q_t\ell$, and
  $\ell\log\AlphabetSize\leq\log\Seqlen$, each alphabet reduction
  takes $\bigO(7|Y_t|\log\AlphabetSize/\log(7|Y_t|))
    =\bigO(q_t)$ time.
  Extracting all blocks also takes $\bigO(\Seqlen)$ time, since the
  input occupies $\bigO(\Seqlen)$ words. The first two steps therefore
  take $\bigO(\Seqlen)$ time and use $\bigO(\Seqlen)$ words.
  There are only $64$ suffix-array structures to construct, and each
  occupies at most its preprocessing space after construction.
  Consequently, the total preprocessing time is
  $\bigO(\Seqlen+P_t(\AlphabetSize,N))$, and the preprocessing space,
  including the structures already constructed, is
  $\bigO(\Seqlen\log N+P_s(\AlphabetSize,N))$ bits.

  Finally, the bounds in both cases imply the claimed complexities.
  Indeed, $\log\Seqlen<(\ell+1)\log\AlphabetSize
    \leq2\ell\log\AlphabetSize$, and
  $\log(N/\Seqlen)\leq N/\Seqlen$. Hence
  \[
    \Seqlen\log N
      =\Seqlen\log\Seqlen+\Seqlen\log(N/\Seqlen)
      \leq2\Seqlen\ell\log\AlphabetSize+N
      \leq3N\log\AlphabetSize.
  \]
  Thus, $\Seqlen=\bigO(N/\log_{\AlphabetSize}N)$ and
  $\Seqlen\log N=\bigO(N\log\AlphabetSize)$, as required.
\end{proof}

\begin{theorem}\label{th:reduce-prefix-select-to-sa-for-all-alphabet-sizes-clean}
  Let $\AlphabetSize, N \in \Z_{\geq 2}$ be such that $\AlphabetSize \leq N$.
  Consider the word RAM model with word size $w$ satisfying
  $1+\log N \leq w = \bigO(\log N)$.
  Assume that there exists a data structure for
  \probname{Indexing}
  \probname{for}
  \probname{Suffix}
  \probname{Array}
  \probname{Queries}
  \probname{over}
  \probname{Alphabet}
  $[0 \dd \AlphabetSize)$
  that, given $N$, $\AlphabetSize$, and a valid input of at most $N$ symbols over alphabet $\IntegerAlphabet$,\footref{footnote:sa-input-2} achieves:
  \begin{itemize}
  \item space usage $S(\AlphabetSize,N)$ bits,
  \item preprocessing time $P_t(\AlphabetSize,N)$,
  \item preprocessing space $P_s(\AlphabetSize,N)$ bits,
  \item query time $Q(\AlphabetSize,N)$.
  \end{itemize}
  Then, there exists a data structure for
  \probname{Indexing}
  \probname{for}
  \probname{Prefix}
  \probname{Select}
  \probname{Queries}
  \probname{over}
  \probname{Alphabet}
  $[0 \dd \AlphabetSize)$
  that, given $N$, $\AlphabetSize$, and a valid input of at most $N$ symbols over alphabet
  $\IntegerAlphabet$,\footref{footnote:prefix-select-input-2} achieves:
  \begin{itemize}
  \item space usage $\bigO(S(\AlphabetSize,N))$ bits,
  \item preprocessing time $\bigO(P_t(\AlphabetSize,N))$,
  \item preprocessing space $\bigO(P_s(\AlphabetSize,N))$ bits,
  \item query time $\bigO(Q(\AlphabetSize,N))$.
  \end{itemize}
\end{theorem}
\begin{proof}
  The claim follows by \cref{th:reduce-prefix-select-to-sa-for-all-alphabet-sizes}, except we observe that
  the final complexities can be simplified as follows:
  \begin{itemize}

  \item By \cref{th:suffix-array-bounded-length-space-lower-bound},
    it holds $S(\AlphabetSize,N) = \Omega(N \log \AlphabetSize)$.
    Consequently, we can simplify the space bound of the index in
    \cref{th:reduce-prefix-select-to-sa-for-all-alphabet-sizes}
    to $\bigO(S(\AlphabetSize,N))$ bits.

  \item Since every suffix array query returns an integer, it holds
    $Q(\AlphabetSize,N) = \Omega(1)$. Consequently, we can
    simplify the query time of the index in
    \cref{th:reduce-prefix-select-to-sa-for-all-alphabet-sizes}
    to $\bigO(Q(\AlphabetSize,N))$.

  \item Recall that the index in
    \cref{th:reduce-prefix-select-to-sa-for-all-alphabet-sizes}
    takes $\bigO(N / \log_{\AlphabetSize} N + P_t(\AlphabetSize,N))$ time
    to construct.
    For every $\AlphabetSize,N \in \Z_{\geq 2}$ satisfying
    $\AlphabetSize \leq N$,
    \cref{th:suffix-array-bounded-length-space-lower-bound} states that the
    suffix array data structure in the premise of this theorem has worst-case
    space usage $\Omega(N \log \AlphabetSize)$ bits. Consequently, for every
    such pair $(\AlphabetSize,N)$, there exists a valid input text for which
    preprocessing produces an output data structure occupying
    $\Omega(N \log \AlphabetSize)$ bits. Writing this output to memory takes
    $\Omega((N \log \AlphabetSize)/w)
    = \Omega(N / \log_{\AlphabetSize} N)$ time, and storing it requires
    $\Omega(N \log \AlphabetSize)$ bits of preprocessing space
    (\cref{rm:space}). Hence,
    $P_t(\AlphabetSize,N) = \Omega(N / \log_{\AlphabetSize} N)$ and
    $P_s(\AlphabetSize,N) = \Omega(N \log \AlphabetSize)$.
    Therefore the construction time bound in
    \cref{th:reduce-prefix-select-to-sa-for-all-alphabet-sizes}
    can be simplified to
    $\bigO(P_t(\AlphabetSize,N))$.

  \item Lastly, recall that the space complexity of the index construction in
    \cref{th:reduce-prefix-select-to-sa-for-all-alphabet-sizes}
    is $\bigO(N \log \AlphabetSize + P_s(\AlphabetSize,N))$ bits.
    Since $P_s(\AlphabetSize,N) = \Omega(N \log \AlphabetSize)$, this bound
    can be simplified to $\bigO(P_s(\AlphabetSize,N))$.
    \qedhere
  \end{itemize}
\end{proof}

%% file: prefix-select/reduce-suffix-array-to-prefix-select.tex
\subsection{Reducing Suffix Array to Prefix Select Queries}\label{sec:reduce-sa-to-prefix-select}

\subsubsection{The Nonperiodic Positions}\label{sec:reduce-sa-to-prefix-select-nonperiodic}

\begin{proposition}[{\cite{breaking}}]\label{pr:reduce-sa-to-prefix-select-nonperiodic}
  Let $\AlphabetSize, N \in \Z_{\geq 2}$ be such that $\AlphabetSize < N^{1/7}$
  and let $\tau = \lfloor \tfrac{1}{7}\log_{\AlphabetSize} N \rfloor \geq 1$.
  Consider the word RAM model with word size $w$ satisfying
  $1+\log N \leq w = \bigO(\log N)$.
  Assume that there exists a data structure that, given $N$, $\AlphabetSize$, $m$, and the packed sequence representation
  $\PackedSeqRepresentation{w}{\AlphabetSize}{W}$
  (\cref{def:packed-sequence-representation}) of any sequence
  $W[1 \dd m]$ of $m \geq \AlphabetSize$
  nonempty strings of common length
  $\ell = \lfloor \tfrac{1}{2} \log_{\AlphabetSize} m \rfloor$
  over alphabet $\IntegerAlphabet$, where $m \cdot \ell \leq \lfloor N/8\rfloor$, returns $\PrefixSelect{W}{r}{X}$ (see \cref{def:prefix-rank-and-select})
  given the length $|X|$ and the packed representation
  $\PackedRepresentation{w}{\AlphabetSize}{X}$ of any string
  $X\in\IntegerAlphabet^{\leq\ell}$ and any integer $r\in[1\dd m]$, and
  achieves the following complexities:
  \begin{itemize}
  \item space usage $S(\AlphabetSize,N)$ bits,
  \item preprocessing time $P_t(\AlphabetSize,N)$,
  \item preprocessing space $P_s(\AlphabetSize,N)$ bits,
  \item query time $Q(\AlphabetSize,N)$.
  \end{itemize}
  Then, there exists a data structure that, given $N$, $\AlphabetSize$, and the packed representation
  $\PackedRepresentation{w}{\AlphabetSize}{\Text}$
  (\cref{def:packed-representation}) of any
  string $\Text \in \IntegerAlphabet^{N}$ such that
  $\Text[N]$ does not occur in $\Text[1 \dd N)$,\footnote{Although~\cite{breaking}
    assumes that the final symbol is unique and smallest, its constructions
    for answering SA and ISA queries require only uniqueness.}
  answers queries that, given any $i \in [1 \dd N]$ satisfying
  $\SA{\Text}[i] \in [1 \dd N] \setminus \RTwo{\tau}{\Text}$
  (\cref{def:suffix-array,def:periodic-positions}),
  return $\SA{\Text}[i]$, and achieves the following complexities:
  \begin{itemize}
  \item space usage $\bigO(N \log \AlphabetSize + S(\AlphabetSize,N))$ bits,
  \item preprocessing time $\bigO(N / \log_{\AlphabetSize} N + P_t(\AlphabetSize,N))$,
  \item preprocessing space $\bigO(N \log \AlphabetSize + P_s(\AlphabetSize,N))$ bits,
  \item query time $\bigO(1 + Q(\AlphabetSize,N))$.
  \end{itemize}
\end{proposition}
\begin{proof}
  In the full version of~\cite{breaking}, this reduction is part of a
  joint reduction for suffix array and inverse suffix array queries, and
  it is presented together with a concrete prefix-select data structure.
  Here, we state it for an arbitrary prefix-select data structure as in
  the claim, which we apply to a constant number of equal-size blocks of
  strings, and
  we adapt the construction to word sizes as small as $1+\log N$.
  See \cref{pr:nonperiodic-sa} in Appendix~\ref{sec:appendix-sa}
  for the complete construction and proof.
\end{proof}

\subsubsection{The Periodic Positions}\label{sec:reduce-sa-to-prefix-select-periodic}

\begin{proposition}[{\cite{breaking}}]\label{pr:reduce-small-sum-range-count-to-periodic-sa-queries}
  Let $\AlphabetSize,N\in\Z_{\geq2}$ satisfy
  $\AlphabetSize<N^{1/7}$, and let
  $\tau=\lfloor\tfrac17\log_{\AlphabetSize}N\rfloor\geq1$.
  Consider the word RAM model with word size
  $w\geq1+\log N$.
  Assume that there exists a data structure answering range counting
  queries (\cref{def:range-count}) that, given $N$, integers $m\in[2\dd N]$
  and $m'\in[0\dd m]$, and an array $A[1\dd m']$ of positive integers satisfying
  $\sum_{i=1}^{m'}A[i]\leq N$ and
  $\sum_{i=1}^{m'}A[i]\leq m\log m$,
  achieves the following complexities:
  \begin{itemize}
  \item space usage $S(m,N)$ words,
  \item preprocessing time $P_t(m,N)$,
  \item preprocessing space $P_s(m,N)$ words,
  \item query time $Q(m,N)$.
  \end{itemize}
  Then, there exists $m=\Theta(N/\log_{\AlphabetSize}N)$ with $m\leq N$
  and a data structure that, given $N$, $\AlphabetSize$, and the packed representation
  $\PackedRepresentation{w}{\AlphabetSize}{\Text}$
  (\cref{def:packed-representation}) of any text
  $\Text\in\IntegerAlphabet^N$, answers queries that, given
  $i\in[1\dd N]$ satisfying $\SA{\Text}[i]\in\RTwo{\tau}{\Text}$
  (\cref{def:periodic-positions}), return
  $\SA{\Text}[i]$ (\cref{def:suffix-array}), and achieves:
  \begin{itemize}
  \item space usage $\bigO(N/\log_{\AlphabetSize}N+S(m,N))$ words,
  \item preprocessing time $\bigO(N/\log_{\AlphabetSize}N+P_t(m,N))$,
  \item preprocessing space $\bigO(N/\log_{\AlphabetSize}N+P_s(m,N))$ words,
  \item query time $\bigO(1+Q(m,N))$.
  \end{itemize}
\end{proposition}
\begin{proof}
  In~\cite{breaking}, this reduction is presented together with a
  concrete data structure for range counting and selection on arrays
  with bounded sum. Here, we state it for an arbitrary range-counting
  data structure as in the claim, answer the range-selection queries
  using \cref{th:range-count-for-small-sum}, and adapt the construction
  to word sizes as small as $1+\log N$. See \cref{pr:periodic-sa} in
  Appendix~\ref{sec:appendix-periodic-sa} for the complete construction
  and proof.
\end{proof}

\begin{proposition}\label{pr:periodic-pos-optimal-sa}
  Let $\AlphabetSize, N \in \Z_{\geq 2}$ be such that $\AlphabetSize < N^{1/7}$
  and let $\tau = \lfloor \tfrac{1}{7}\log_{\AlphabetSize} N \rfloor \geq 1$.
  Consider the word RAM model
  with word size $w\geq1+\log N$.
  Given $N$, $\AlphabetSize$, and the packed representation
  $\PackedRepresentation{w}{\AlphabetSize}{\Text}$ (\cref{def:packed-representation})
  of any string $\Text \in \IntegerAlphabet^{N}$,
  we can in $\bigO(N / \log_{\AlphabetSize} N)$ time and using
  $\bigO(N/\log_{\AlphabetSize}N)$ words of space construct a data structure
  of size $\bigO(N / \log_{\AlphabetSize} N)$ words that, given
  any $i \in [1 \dd N]$ satisfying $\SA{\Text}[i] \in \RTwo{\tau}{\Text}$ (\cref{def:periodic-positions}),
  returns $\SA{\Text}[i]$ (\cref{def:suffix-array}) in $\bigO(1)$ time.
\end{proposition}
\begin{proof}
  By \cref{th:range-count-for-small-sum} with $c=1$, the range-counting
  structure required by \cref{pr:reduce-small-sum-range-count-to-periodic-sa-queries} exists with
  $S(m,N)=P_t(m,N)=P_s(m,N)=\bigO(m)$ and $Q(m,N)=\bigO(1)$.
  Applying that proposition gives $m=\Theta(N/\log_{\AlphabetSize}N)$
  and a data structure with the following complexities:
  \begin{itemize}
  \item space usage
    $\bigO(N/\log_{\AlphabetSize}N+m)=\bigO(N/\log_{\AlphabetSize}N)$ words,
  \item preprocessing time
    $\bigO(N/\log_{\AlphabetSize}N+m)=\bigO(N/\log_{\AlphabetSize}N)$,
  \item preprocessing space
    $\bigO(N/\log_{\AlphabetSize}N+m)=\bigO(N/\log_{\AlphabetSize}N)$ words,
  \item query time $\bigO(1+Q(m,N))=\bigO(1)$.
    \qedhere
  \end{itemize}
\end{proof}

\subsubsection{The Final Reduction}\label{sec:reduce-sa-to-prefix-select-final-reduction}

\begin{proposition}\label{pr:reduce-sa-to-prefix-select-for-nearly-all-alphabet-sizes-with-dollar}
  Let $\AlphabetSize, N \in \Z_{\geq 2}$ be such that $\AlphabetSize < N^{1/7}$.
  Consider the word RAM model with word size $w$ satisfying
  $1+\log N \leq w = \bigO(\log N)$.
  Assume that there exists a data structure that, given $N$, $\AlphabetSize$, $m$, and the packed sequence representation
  $\PackedSeqRepresentation{w}{\AlphabetSize}{W}$
  (\cref{def:packed-sequence-representation}) of a sequence
  $W[1 \dd m]$ of $m \geq \AlphabetSize$ nonempty strings
  of common length $\ell = \lfloor \tfrac{1}{2} \log_{\AlphabetSize} m \rfloor$
  over alphabet $\IntegerAlphabet$, where $m \cdot \ell \leq \lfloor N/8\rfloor$, returns $\PrefixSelect{W}{r}{X}$ (see \cref{def:prefix-rank-and-select})
  given the length $|X|$ and the packed representation
  $\PackedRepresentation{w}{\AlphabetSize}{X}$ of any string
  $X\in\IntegerAlphabet^{\leq\ell}$ and any integer $r\in[1\dd m]$, and
  achieves the following complexities:
  \begin{itemize}
  \item space usage $S(\AlphabetSize,N)$ bits,
  \item preprocessing time $P_t(\AlphabetSize,N)$,
  \item preprocessing space $P_s(\AlphabetSize,N)$ bits,
  \item query time $Q(\AlphabetSize,N)$.
  \end{itemize}
  Then, there exists a data structure answering suffix array queries
  (\cref{def:suffix-array}) that, given $N$, $\AlphabetSize$, and the packed representation $\PackedRepresentation{w}{\AlphabetSize}{\Text}$ of any text $\Text \in [0 \dd \AlphabetSize)^{N}$
  such that $\Text[N]$ does not appear in $\Text[1 \dd N)$, achieves:
  \begin{itemize}
  \item space usage $\bigO(N \log \AlphabetSize + S(\AlphabetSize,N))$ bits,
  \item preprocessing time $\bigO(N / \log_{\AlphabetSize} N + P_t(\AlphabetSize,N))$,
  \item preprocessing space $\bigO(N \log \AlphabetSize + P_s(\AlphabetSize,N))$ bits,
  \item query time $\bigO(1 + Q(\AlphabetSize,N))$.
  \end{itemize}
\end{proposition}
\begin{proof}

  Let $\Text \in \IntegerAlphabet^{N}$ be any text such that $\Text[N]$ does not occur in $\Text[1 \dd N)$.
  Let $\tau = \lfloor \tfrac{1}{7}\log_{\AlphabetSize} N \rfloor \geq 1$.
  In particular, $\tau\leq\tfrac16\log_{\AlphabetSize}N$.

  \DSComponents
  The data structure consists of the following components:
  \begin{enumerate}

  \item The integer $\tau$ and the data structure from \cref{pr:sa-core}.
    It uses $\bigO(N/\log N)$ words, i.e., $\bigO(N)$ bits of space,
    because $w=\bigO(\log N)$.

  \item The data structure from \cref{pr:reduce-sa-to-prefix-select-nonperiodic}.
    It uses $\bigO(N \log \AlphabetSize + S(\AlphabetSize,N))$ bits of space.

  \item The data structure from \cref{pr:periodic-pos-optimal-sa}.
    It uses $\bigO(N/\log_{\AlphabetSize}N)$ words, i.e.,
    $\bigO(N\log\AlphabetSize)$ bits of space.
  \end{enumerate}

  In total, the data structure needs $\bigO(N \log \AlphabetSize + S(\AlphabetSize,N))$ bits of space.

  \DSQueries
  Let $i \in [1 \dd N]$. Given the index $i$, we compute $\SA{\Text}[i]$ as follows:
  \begin{enumerate}

  \item Query the structure from \cref{pr:sa-core} with $(i,\tau)$
    to determine in $\bigO(1)$ time whether
    $\SA{\Text}[i]\in\RTwo{\tau}{\Text}$ (\cref{def:periodic-positions}).

  \item We consider two cases:
    \begin{itemize}

      \item If $\SA{\Text}[i] \in \RTwo{\tau}{\Text}$, then we
      compute $\SA{\Text}[i]$ using \cref{pr:periodic-pos-optimal-sa}
      in $\bigO(1)$ time.

    \item Otherwise (i.e., if $\SA{\Text}[i] \in [1 \dd N] \setminus \RTwo{\tau}{\Text}$),
      we compute $\SA{\Text}[i]$ using the
      data structure from \cref{pr:reduce-sa-to-prefix-select-nonperiodic}
      in $\bigO(1 + Q(\AlphabetSize,N))$ time.
    \end{itemize}
  \end{enumerate}

  In total, the query takes $\bigO(1 + Q(\AlphabetSize,N))$ time.

  \DSConstruction
  Given $N$, $\AlphabetSize$, and the packed representation
  $\PackedRepresentation{w}{\AlphabetSize}{\Text}$ of $\Text$, we construct the
  components of the above data structure as follows:
  \begin{enumerate}

  \item Compute and store $\tau$ in $\bigO(\log N)$ time.
    Construct the structure from \cref{pr:sa-core} with
    $\PackedRepresentation{w}{\AlphabetSize}{\Text}$. This takes
    $\bigO(N/\log_{\AlphabetSize}N)$ time.

  \item The second component is constructed in
    $\bigO(N / \log_{\AlphabetSize} N + P_t(\AlphabetSize,N))$ time and using
    $\bigO(N \log \AlphabetSize + P_s(\AlphabetSize,N))$ bits of space using \cref{pr:reduce-sa-to-prefix-select-nonperiodic}.

  \item The third component is constructed in
    $\bigO(N / \log_{\AlphabetSize} N)$ time using \cref{pr:periodic-pos-optimal-sa}.
  \end{enumerate}

  All three constructions use size parameter $N$ and alphabet size
  $\AlphabetSize$.
  In total, the construction takes
  $\bigO(N / \log_{\AlphabetSize} N + P_t(\AlphabetSize,N))$ time and uses
  $\bigO(N \log \AlphabetSize + P_s(\AlphabetSize,N))$ bits of space.
\end{proof}

\begin{theorem}\label{th:reduce-sa-to-prefix-select-for-nearly-all-alphabet-sizes}
  Let $\AlphabetSize, N \in \Z_{\geq 2}$ be such that $\AlphabetSize < N^{1/17}$.
  Consider the word RAM model with word size $w$ satisfying
  $1+\log N \leq w = \bigO(\log N)$.
  Assume that there exists a data structure for
  \probname{Indexing}
  \probname{for}
  \probname{Prefix}
  \probname{Select}
  \probname{Queries}
  \probname{over}
  \probname{Alphabet}
  $[0 \dd \AlphabetSize)$
  that, given $N$, $\AlphabetSize$, and a valid input of at most $N$ symbols over alphabet
  $\IntegerAlphabet$,\footnote{\label{footnote:prefix-select-input} That is, the integer $\Seqlen$ and the
  packed sequence representation
  $\PackedSeqRepresentation{w}{\AlphabetSize}{W}$
  (\cref{def:packed-sequence-representation}) of a sequence
  $W[1 \dd \Seqlen]$ of $\Seqlen \geq \AlphabetSize$ strings
  of length $\ell = \lfloor \log_{\AlphabetSize} \Seqlen \rfloor$, where $\Seqlen \ell \leq N$;
  see \cref{sec:prefix-select-and-sa-problem-def}.} achieves:
  \begin{itemize}
  \item space usage $S(\AlphabetSize,N)$ bits,
  \item preprocessing time $P_t(\AlphabetSize,N)$,
  \item preprocessing space $P_s(\AlphabetSize,N)$ bits,
  \item query time $Q(\AlphabetSize,N)$.
  \end{itemize}
  Then, there exists a data structure for
  \probname{Indexing}
  \probname{for}
  \probname{Suffix}
  \probname{Array}
  \probname{Queries}
  \probname{over}
  \probname{Alphabet}
  $[0 \dd \AlphabetSize)$
  that, given $N$, $\AlphabetSize$, and a valid input of at most $N$ symbols over alphabet
  $\IntegerAlphabet$,\footnote{\label{footnote:sa-input} That is, the length $\Textlen$ and the
  packed representation $\PackedRepresentation{w}{\AlphabetSize}{\Text}$ of $\Text \in \IntegerAlphabet^{\Textlen}$ with $\AlphabetSize \leq \Textlen \leq N$;
  see \cref{sec:prefix-select-and-sa-problem-def}.} achieves:
  \begin{itemize}
  \item space usage $\bigO(N \log \AlphabetSize + S(\AlphabetSize,N))$ bits,
  \item preprocessing time $\bigO(N / \log_{\AlphabetSize} N + P_t(\AlphabetSize,N))$,
  \item preprocessing space $\bigO(N \log \AlphabetSize + P_s(\AlphabetSize,N))$ bits,
  \item query time $\bigO(1 + Q(\AlphabetSize,N))$.
  \end{itemize}
\end{theorem}
\begin{proof}
  Denote $\hat{\AlphabetSize}=\AlphabetSize+2$.
  Let $D$ denote the prefix-select structure from the claim.
  Applying \cref{th:general-prefix-select-alphabet-reduction}
  to $D$ with source alphabet size $\hat{\AlphabetSize}$ and target
  alphabet size $\AlphabetSize$ yields a structure $D'$ over
  alphabet $[0\dd\hat{\AlphabetSize})$.
  Since $\AlphabetSize<\hat{\AlphabetSize}\leq\AlphabetSize^2$,
  we have $k=\lceil\log_{\AlphabetSize}\hat{\AlphabetSize}\rceil=2$.
  The structure $D'$ answers prefix select queries on sequences
  $W'[1\dd m']$ of
  $m'\geq\hat{\AlphabetSize}$ nonempty strings of common length
  $\ell'=\lfloor\tfrac12\log_{\hat{\AlphabetSize}}m'\rfloor$
  over alphabet $[0\dd\hat{\AlphabetSize})$, where
  $m'\ell'\leq\lfloor N/(4k)\rfloor=\lfloor N/8\rfloor$.
  Given a rank $r\in[1\dd m']$, the length $|X|$, and the packed
  representation of $X\in[0\dd\hat{\AlphabetSize})^{\leq\ell'}$,
  it returns $\PrefixSelect{W'}{r}{X}$.
  Its construction obtains the target alphabet size as
  $\AlphabetSize=\hat{\AlphabetSize}-2$.
  It achieves the following complexities:
  \begin{itemize}
  \item space usage
    $\bigO(N\log\AlphabetSize+S(\AlphabetSize,N))$ bits,
  \item preprocessing time
    $\bigO(N/\log_{\AlphabetSize}N+P_t(\AlphabetSize,N))$,
  \item preprocessing space
    $\bigO(N\log\AlphabetSize+P_s(\AlphabetSize,N))$ bits,
  \item query time $\bigO(1+Q(\AlphabetSize,N))$.
  \end{itemize}

  Consider a text $\Text\in\IntegerAlphabet^{\Textlen}$ with
  $\AlphabetSize\leq\Textlen\leq N$, and denote
  $t=\Text[\Textlen]$ and $\delta=N-\Textlen$.
  To make the final symbol unique without increasing the text length,
  we use the map $f:[0\dd\AlphabetSize)\to[1\dd\hat{\AlphabetSize})$
  defined by
  \[
    f(a)=
    \begin{cases}
      a+1 & \text{if }a<t,\\
      a+2 & \text{if }a\geq t.
    \end{cases}
  \]
  Let $U\in[1\dd\hat{\AlphabetSize})^{\Textlen}$ denote the string
  satisfying $U[i]=f(\Text[i])$ for every $i\in[1\dd\Textlen)$ and
  $U[\Textlen]=t+1$, and let
  $\Text_{\rm aux}=\zero^\delta U\in[0\dd\hat{\AlphabetSize})^N$.
  The last symbol $\Text_{\rm aux}[N]=t+1$ is unique because
  $t+1\ne\zero$ and $f(a)\ne t+1$ for every $a\in[0\dd\AlphabetSize)$.
  Moreover, $\hat{\AlphabetSize}^7\leq\AlphabetSize^{14}<N$.
  Thus, $D'$ provides the prefix select queries required by
  \cref{pr:reduce-sa-to-prefix-select-for-nearly-all-alphabet-sizes-with-dollar}
  for $\Text_{\rm aux}$, with alphabet size $\hat{\AlphabetSize}$ and
  size parameter $N$.

  \DSComponents
  We store $\delta$ and the data structure from
  \cref{pr:reduce-sa-to-prefix-select-for-nearly-all-alphabet-sizes-with-dollar}
  for $\Text_{\rm aux}$, using $D'$ for prefix select queries.
  The bounds for the two reductions, together with
  $\log\hat{\AlphabetSize}=\Theta(\log\AlphabetSize)$, give a total
  space usage of $\bigO(N\log\AlphabetSize+S(\AlphabetSize,N))$ bits.

  \DSQueries
  We first prove that $\SA{U}[i]=\SA{\Text}[i]$ for every
  $i\in[1\dd\Textlen]$. Consider two distinct suffixes of $\Text$.
  \begin{itemize}
  \item If neither suffix is a prefix of the other, their order is
    determined by the first unequal symbols. The map $f$ is strictly
    increasing, and replacing $f(t)=t+2$ at the final position by $t+1$
    preserves its order relative to $f(a)$ for every $a\ne t$.
    The corresponding suffixes of $U$ therefore have the same order.
  \item If one suffix is a proper prefix of the other, the final symbol
    $t$ of the shorter suffix is compared with a nonfinal occurrence of
    $t$ in the longer suffix. In $U$, these symbols become $t+1$ and
    $t+2$, respectively, while all preceding symbols are mapped
    identically. Hence, the shorter suffix remains lexicographically
    smaller.
  \end{itemize}
  Since all symbols of $U$ are positive, the $\delta$ suffixes of
  $\Text_{\rm aux}$ starting in $[1\dd\delta]$ precede every suffix
  starting in $[\delta+1\dd N]$. The latter suffixes are the suffixes
  of $U$, with their starting positions increased by $\delta$.
  Consequently, for every $i\in[1\dd\Textlen]$, it holds
  \[
    \SA{\Text_{\rm aux}}[\delta+i]
      =\delta+\SA{U}[i]=\delta+\SA{\Text}[i].
  \]
  Given $i$, we query the stored structure at rank $\delta+i$ and
  subtract $\delta$ from the returned position. The query takes
  $\bigO(1+Q(\AlphabetSize,N))$ time.

  \DSConstruction
  Given $N$, $\AlphabetSize$, $\Textlen$, and
  $\PackedRepresentation{w}{\AlphabetSize}{\Text}$,
  the construction proceeds in two steps:
  \begin{enumerate}

  \item First, we construct
    $\PackedRepresentation{w}{\hat{\AlphabetSize}}{\Text_{\rm aux}}$
    as follows:
    \begin{enumerate}

    \item Compute and store $\delta=N-\Textlen$.
      Compute $b=\lceil\log\AlphabetSize\rceil$ and
      $\hat b=\lceil\log\hat{\AlphabetSize}\rceil$ in $\bigO(\log N)$
      time, and retrieve $t=\Text[\Textlen]$ using
      \cref{pr:packed-representation}\eqref{pr:packed-representation-access}.
      Construct the structure from \cref{th:alphabet-lifting}.
      Query it with $\Textlen$ and
      $\PackedRepresentation{w}{\AlphabetSize}{\Text}$ to obtain
      $\PackedRepresentation{w}{\hat{\AlphabetSize}}{\Text}$.

    \item Using
      \cref{pr:packed-representation}\eqref{pr:packed-representation-initialize},
      construct an array $A[0\dd\hat{\AlphabetSize})$ containing the
      packed representations of the one-symbol strings
      $f(0),\ldots,f(\AlphabetSize-1),\zero,\zero$, in this order,
      over alphabet $[0\dd\hat{\AlphabetSize})$.
      Apply \cref{th:packed-morphism} to $A$ with $d=1$.
      Query the resulting structure with $\Textlen$ and
      $\PackedRepresentation{w}{\hat{\AlphabetSize}}{\Text}$.
      In the returned representation, replace the last symbol by $t+1$
      using \cref{pr:packed-representation}\eqref{pr:packed-representation-update}
      to obtain $\PackedRepresentation{w}{\hat{\AlphabetSize}}{U}$.

    \item Initialize the packed representation of $\zero^\delta$ over
      $[0\dd\hat{\AlphabetSize})$ using
      \cref{pr:packed-representation}\eqref{pr:packed-representation-initialize}.
      Concatenate it with $\PackedRepresentation{w}{\hat{\AlphabetSize}}{U}$
      using \cref{pr:packed-representation}\eqref{pr:packed-representation-concat}
      to obtain
      $\PackedRepresentation{w}{\hat{\AlphabetSize}}{\Text_{\rm aux}}$.
    \end{enumerate}

  \item Apply
    \cref{pr:reduce-sa-to-prefix-select-for-nearly-all-alphabet-sizes-with-dollar}
    to $\Text_{\rm aux}$, using $D'$ for prefix select queries.
    Store the resulting suffix-array structure.
  \end{enumerate}

  The packed operations use length bound $N$ and alphabet sizes at most
  $\hat{\AlphabetSize}\leq N$, which satisfy their word-size assumptions.
  The lifting structure uses size parameter $N$, source alphabet size
  $\AlphabetSize$, and target alphabet size $\hat{\AlphabetSize}$, and
  the morphism structure uses size parameter $N$ and alphabet size
  $\hat{\AlphabetSize}$.
  Since $\hat{\AlphabetSize}<N^{1/7}$ and
  $\log\hat{\AlphabetSize}=\Theta(\log\AlphabetSize)$, the first step takes
  $\bigO(\sqrt N+N/\log_{\hat{\AlphabetSize}}N)
    =\bigO(N/\log_{\AlphabetSize}N)$ time.
  Since $w=\bigO(\log N)$, these operations use
  $\bigO(N\log\AlphabetSize)$ bits of space.
  The bounds for $D'$ and
  \cref{pr:reduce-sa-to-prefix-select-for-nearly-all-alphabet-sizes-with-dollar},
  together with $\log\hat{\AlphabetSize}=\Theta(\log\AlphabetSize)$,
  imply that the second step takes
  $\bigO(N/\log_{\AlphabetSize}N+P_t(\AlphabetSize,N))$ time and uses
  $\bigO(N\log\AlphabetSize+P_s(\AlphabetSize,N))$ bits of space.
  These are also the total construction bounds.
\end{proof}

\begin{theorem}\label{th:reduce-sa-to-prefix-select-for-all-alphabet-sizes}
  Let $\AlphabetSize, N \in \Z_{\geq 2}$ be such that $\AlphabetSize \leq N$.
  Consider the word RAM model with word size $w$ satisfying
  $1+\log N \leq w = \bigO(\log N)$.
  Assume that there exists a data structure for
  \probname{Indexing}
  \probname{for}
  \probname{Prefix}
  \probname{Select}
  \probname{Queries}
  \probname{over}
  \probname{Alphabet}
  $[0 \dd \AlphabetSize)$
  that, given $N$, $\AlphabetSize$, and a valid input of at most $N$ symbols over alphabet
  $\IntegerAlphabet$,\footref{footnote:prefix-select-input}
  achieves:
  \begin{itemize}
  \item space usage $S(\AlphabetSize,N)$ bits,
  \item preprocessing time $P_t(\AlphabetSize,N)$,
  \item preprocessing space $P_s(\AlphabetSize,N)$ bits,
  \item query time $Q(\AlphabetSize,N)$.
  \end{itemize}
  Then, there exists a data structure for
  \probname{Indexing}
  \probname{for}
  \probname{Suffix}
  \probname{Array}
  \probname{Queries}
  \probname{over}
  \probname{Alphabet}
  $[0 \dd \AlphabetSize)$
  that, given $N$, $\AlphabetSize$, and a valid input of at most $N$ symbols over alphabet
  $\IntegerAlphabet$,\footref{footnote:sa-input}
  achieves:
  \begin{itemize}
  \item space usage $\bigO(N \log \AlphabetSize + S(\AlphabetSize,N))$ bits,
  \item preprocessing time $\bigO(N / \log_{\AlphabetSize} N + P_t(\AlphabetSize,N))$,
  \item preprocessing space $\bigO(N \log \AlphabetSize + P_s(\AlphabetSize,N))$ bits,
  \item query time $\bigO(1 + Q(\AlphabetSize,N))$.
  \end{itemize}
\end{theorem}
\begin{proof}
  The case $\AlphabetSize<N^{1/17}$ follows from
  \cref{th:reduce-sa-to-prefix-select-for-nearly-all-alphabet-sizes}.
  Assume that $\AlphabetSize\geq N^{1/17}$.
  Since $\AlphabetSize\leq N$, we have
  $\log\AlphabetSize=\Theta(\log N)$.
  Consider a string $\Text\in\IntegerAlphabet^+$ with
  $\AlphabetSize\leq|\Text|\leq N$.

  \DSComponents
  We store the suffix array $\SA{\Text}$ in $|\Text|$ words.
  It occupies $\bigO(Nw)=\bigO(N\log\AlphabetSize)$ bits.

  \DSQueries
  Given $i\in[1\dd|\Text|]$, we return $\SA{\Text}[i]$ in $\bigO(1)$ time.

  \DSConstruction
  We first compute $b=\lceil\log\AlphabetSize\rceil$ in $\bigO(\log N)$ time.
  Using \cref{pr:packed-representation}\eqref{pr:packed-representation-access},
  we construct an array $A[1\dd|\Text|]$ with $A[i]=\Text[i]$ for every
  $i\in[1\dd|\Text|]$ in $\bigO(|\Text|)$ time.
  We then compute $\SA{\Text}$ from $A$ in $\bigO(|\Text|)$ time using
  \cref{th:sa-and-isa-construction}.

  The construction takes $\bigO(N)=\bigO(N/\log_{\AlphabetSize}N)$ time
  and uses $\bigO(N)$ words, or $\bigO(N\log\AlphabetSize)$ bits of space.
  These bounds imply the claimed complexities.
\end{proof}

\begin{theorem}\label{th:reduce-sa-to-prefix-select-for-all-alphabet-sizes-clean}
  Let $\AlphabetSize, N \in \Z_{\geq 2}$ be such that $\AlphabetSize \leq N$.
  Consider the word RAM model with word size $w$ satisfying
  $1+\log N \leq w = \bigO(\log N)$.
  Assume that there exists a data structure for
  \probname{Indexing}
  \probname{for}
  \probname{Prefix}
  \probname{Select}
  \probname{Queries}
  \probname{over}
  \probname{Alphabet}
  $[0 \dd \AlphabetSize)$
  that, given $N$, $\AlphabetSize$, and a valid input of at most $N$ symbols over alphabet
  $\IntegerAlphabet$,\footref{footnote:prefix-select-input}
  achieves:
  \begin{itemize}
  \item space usage $S(\AlphabetSize,N)$ bits,
  \item preprocessing time $P_t(\AlphabetSize,N)$,
  \item preprocessing space $P_s(\AlphabetSize,N)$ bits,
  \item query time $Q(\AlphabetSize,N)$.
  \end{itemize}
  Then, there exists a data structure for
  \probname{Indexing}
  \probname{for}
  \probname{Suffix}
  \probname{Array}
  \probname{Queries}
  \probname{over}
  \probname{Alphabet}
  $[0 \dd \AlphabetSize)$
  that, given $N$, $\AlphabetSize$, and a valid input of at most $N$ symbols over alphabet
  $\IntegerAlphabet$,\footref{footnote:sa-input}
  achieves:
  \begin{itemize}
  \item space usage $\bigO(S(\AlphabetSize,N))$ bits,
  \item preprocessing time $\bigO(P_t(\AlphabetSize,N))$,
  \item preprocessing space $\bigO(P_s(\AlphabetSize,N))$ bits,
  \item query time $\bigO(Q(\AlphabetSize,N))$.
  \end{itemize}
\end{theorem}
\begin{proof}
  We apply \cref{th:reduce-sa-to-prefix-select-for-all-alphabet-sizes}
  and show that its additive terms can be removed.
  By \cref{th:prefix-select-bounded-total-length-space-lower-bound},
  applied with $c=1$, the prefix-select structure from the claim requires
  $\Omega(N\log\AlphabetSize)$ bits for some valid input.
  The theorem is applicable: the structure from the claim supports every
  sequence considered there, because
  $\lfloor\log_{\AlphabetSize}\Seqlen\rfloor\geq1$ implies
  $\Seqlen\geq\AlphabetSize$, and the existence assumption of the
  theorem holds because $\Seqlen=\AlphabetSize$ gives
  $\ell=\lfloor\log_{\AlphabetSize}\Seqlen\rfloor=1$ and
  $\Seqlen\ell=\AlphabetSize\leq N$.
  In particular, $S(\AlphabetSize,N)=\Omega(N\log\AlphabetSize)$.

  Writing this structure to memory takes
  $\Omega((N\log\AlphabetSize)/w)=\Omega(N/\log_{\AlphabetSize}N)$ time,
  and storing it requires $\Omega(N\log\AlphabetSize)$ bits of
  preprocessing space (\cref{rm:space}).
  Hence $P_t(\AlphabetSize,N)=\Omega(N/\log_{\AlphabetSize}N)$ and
  $P_s(\AlphabetSize,N)=\Omega(N\log\AlphabetSize)$.
  Together with $Q(\AlphabetSize,N)=\Omega(1)$, these lower bounds
  give the claimed complexities.
\end{proof}

%% file: prefix-select/lower-bound.tex
\subsection{Cell-Probe Lower Bounds for Prefix Select and Suffix Array Queries over Binary Strings}

\newcommand{\PSQBin}{\probname{Indexing} \probname{for} \probname{Prefix} \probname{Select} \probname{Queries} \probname{over} \probname{Alphabet} $\{0,1\}$\xspace}
\newcommand{\SABin}{\probname{Indexing}
    \probname{for}
    \probname{Suffix}
    \probname{Array}
    \probname{Queries}
    \probname{over}
    \probname{Alphabet}
    $\{0,1\}$\xspace}
\newcommand{\BI}{\probname{Ball} \probname{Inheritance}\xspace}

In this subsection, we derive a cell-probe lower bound for \PSQBin and \SABin.
We do so by reducing the \BI problem, introduced by
Chan, Larsen, and Pătrașcu \cite{ChanLP11} and analyzed from the lower-bound perspective by Gr{\o}nlund and
Larsen~\cite{GronlundL16}.

\begin{framed}
  \noindent
  \BI~\cite{ChanLP11,GronlundL16}
  \begin{bfdescription}
  \item[Input:]
    A perfect binary tree $T$ with $n$ leaves, where $n$ is a power of two,
    and an ordered list $L$ of $n$ \emph{balls}. The balls are assigned
    bijectively to the leaves of $T$. We say that a ball \emph{reaches}
    the leaf assigned to it. For every node $v$ of $T$, let
    $L_v$ denote the list of balls reaching a leaf in the subtree rooted at
    $v$, ordered as in $L$.
  \item[Output:]
    A data structure that, given a query $(v, i)$, where $v$ is a node of
    $T$ and $i\in [1\dd |L_v|]$, returns the index of the leaf reached by the
    $i$th ball in $L_v$.
  \end{bfdescription}
\end{framed}

\begin{theorem}[{\cite{GronlundL16}}]\label{th:lb-bi}
  Let $n$ be a sufficiently large power of two.
  In the cell probe model\footnote{Gr{\o}nlund and Larsen~\cite{GronlundL16} state this lower bound
  in the word RAM model, but, as they note (see, e.g., the introduction to
  Section~2 in~\cite{GronlundL16}), it also holds in the cell probe model.}
  with $\Theta(\log n)$-bit words,
  every deterministic data structure solving \BI and using at most
  $S(n)=\Omega(n\log n)$ bits of space must have worst-case query time
  \[
    Q(n) \ge \Omega\left(
      \frac{\log \log n}{\log((S(n)/(n\log n))\log \log n)}
    \right).
  \]
\end{theorem}

\medskip
\begin{lemma}\label{lm:prefix-select-to-balls}
  Let $m \in \Z_{\geq 2}$ be a power of two, and consider the cell probe
  model with words of size $\Theta(\log m)$.
  Suppose that \PSQBin admits a data structure that, given a sequence
  $W[1 \dd m]$ of $m$ binary strings of length $\log m$, uses $S$ bits of
  space and answers queries in time $Q$.
  Then \BI admits a data structure that, given a tree $T$ of $m$ leaves and a
  list $L$ of $m$ balls, uses $S+\Oh(m\log m)$ bits of space and answers
  queries in time $Q+\Oh(1)$. If the prefix-select data structure is
  deterministic, then so is the resulting ball-inheritance data structure.
\end{lemma}
\begin{proof}
  Label every left edge of $T$ by $0$ and every right edge by $1$.
  For each node $v$ of $T$, let $I_v$ denote the \emph{identifier} of $v$,
  defined as the sequence of edge labels on the unique root-to-$v$ path.
  Let us index the leaves of $T$ from left to right by $0,1,\dots,m-1$.
  Since $T$ is perfect, the identifier of the leaf numbered $i$ is
  exactly the length-$\log m$ binary representation of $i$ (with leading zeros
  if needed). We construct a sequence $W[1 \dd m]$ by letting $W[j]$ be the
  identifier of the leaf reached by the ball $L[j]$.

  \DSComponents
  The data structure consists of the following components:
  \begin{enumerate}
  \item The data structure for prefix select queries on $W$. It uses $S$ bits
    of space.
  \item The sequence $W[1 \dd m]$. It uses $m\log m$ bits of space.
  \item The identifiers $I_v$ for all nodes $v$ of $T$. They use
    $\Oh(m\log m)$ bits of space, since the perfect binary tree has
    $\Oh(m)$ nodes and every identifier has length at most $\log m$.
  \end{enumerate}
  In total, the data structure uses $S+\Oh(m\log m)$ bits of space.

  \DSQueries
  Given a query $(v,i)$, we return the requested leaf index as follows:
  \begin{enumerate}
  \item Retrieve the identifier $I_v$.
  \item Using the prefix-select data structure, compute
    $p:=\PrefixSelect{W}{i}{I_v}$.
  \item Return the integer represented by $W[p]$.
  \end{enumerate}
  To prove correctness, observe that, for every node $v$, the list $L_v$ is
  precisely the subsequence of $L$ consisting of balls whose leaves have
  identifiers with prefix $I_v$. Consequently, $p$ is the position in $L$ of the $i$th ball in
  $L_v$, and $W[p]$ is the identifier of the leaf reached by this ball. The
  returned integer is therefore the requested leaf index. The query takes
  $Q+\Oh(1)$ time. Moreover, if the prefix-select data structure is
  deterministic, then so is the resulting data structure.
\end{proof}

\begin{corollary}\label{cor:lb-prefix-select}
  Let $N$ be a sufficiently large positive integer.
  In the cell probe model with words of size $\Theta(\log N)$,
  every deterministic data structure solving \PSQBin for all valid inputs of
  at most $N$ symbols and using at most $S(N)=\Omega(N)$ bits of space
  must have worst-case query time
  \[
    Q(N) \ge \Omega\left(
      \frac{\log \log N}{\log((S(N)/N)\log \log N)}
    \right).
  \]
\end{corollary}
\begin{proof}
  Let $m$ be the largest power of two such that $m\log m \le N$.
  Then $m = \Theta(N/\log N)$.

  By assumption, every valid input of at most $N$ symbols for \PSQBin can be
  indexed using $S(N)$ bits of space and query time $Q(N)$.
  In particular, this holds for sequences of $m$ binary strings of length
  $\log m$, since such an input contains exactly $m\log m \le N$ symbols.
  By \cref{lm:prefix-select-to-balls}, every size-$m$ instance of \BI can
  therefore be solved using $S(N)+\Oh(m\log m)$ bits of space and query time
  $Q(N)+\Oh(1)$.
  Since $m\log m \le N$ and $S(N)=\Omega(N)$, the additive $\Oh(m\log m)$ term is absorbed, and the \BI structure uses $\Oh(S(N))$ bits.

  Applying \cref{th:lb-bi}, we obtain
  \[
    Q(N)+\Oh(1) \ge \Omega\left(
      \frac{\log \log m}{\log((S(N)/(m \log m))\log \log m)}
    \right).
  \]
  Here, we replaced the space bound $\Oh(S(N))$ by $S(N)$. This changes
  the denominator only by an additive constant, which is absorbed because
  $S(N)=\Omega(N)$ and $m\log m\leq N$ give
  $(S(N)/(m\log m))\log\log m=\Omega(\log\log m)$.
  A correct data structure has $Q(N)\geq 1$, because
  $\PrefixSelect{W}{1}{\zero}$ is $1$ for $W=(\zero,\one)$ and is $2$ for
  $W=(\one,\zero)$. Hence, $Q(N)+\Oh(1)=\Oh(Q(N))$.
  Since $m = \Theta(N/\log N)$, we have
  $\log\log m = \Theta(\log\log N)$ and $m\log m = \Theta(N)$. For the
  same reason as above, the latter changes the denominator only by an
  additive constant. Therefore,
  \[
    Q(N) \ge \Omega\left(
      \frac{\log \log N}{\log((S(N)/N)\log \log N)}
    \right). \qedhere
  \]
\end{proof}

\begin{theorem}\label{th:lb-suffix-array}
  Let $N$ be a sufficiently large positive integer.
  In the cell probe model with words of size $\Theta(\log N)$,
  every deterministic data structure solving \SABin for binary strings of
  length at most $N$ and using at most $S(N)=\Omega(N)$ bits of space
  must have worst-case query time
  \[
    Q(N) \ge \Omega\left(
      \frac{\log \log N}{\log((S(N)/N)\log \log N)}
    \right).
  \]
\end{theorem}
\begin{proof}
  Consider any suffix-array data structure satisfying the assumptions in the
  claim. We use the construction from the proof of
  \cref{th:reduce-prefix-select-to-sa-for-all-alphabet-sizes} with
  $\AlphabetSize=2$ and word size $w=\lceil1+\log N\rceil$, but account for
  its space usage and number of probes directly. This is necessary because
  that theorem is stated in the word RAM model, whereas the suffix-array data
  structure considered here is an arbitrary cell-probe data structure.
  We prove the claim in four steps:
  \begin{enumerate}

  \item\label{step:lb-suffix-array-space}
    First, we prove that the resulting prefix-select data structure uses
    $\bigO(S(N))$ bits of space. Consider a valid input $W[1\dd m]$ for
    \PSQBin containing at most $N$ symbols. If $m<128$, the construction
    stores its occurrence lists and their lengths in $\bigO(m)$ words.
    Otherwise, it constructs $64$ binary texts $\Text_t$, each of length
    at most $N$, and stores the suffix-array structure for each text.
    Together, these structures occupy at most $64S(N)$ bits. The block
    boundaries, conversion parameters, and structures from
    \cref{pr:seq-to-str-range-beg-and-range-end} occupy
    $\bigO(m)$ logical words, each containing $\bigO(\log N)$ bits.
    The space analysis in the proof of
    \cref{th:reduce-prefix-select-to-sa-for-all-alphabet-sizes} gives
    $m\log N=\bigO(N)$ for $\AlphabetSize=2$. In either case, the
    auxiliary space is therefore $\bigO(N)$ bits, and the total space
    is $\bigO(S(N))$ because $S(N)=\Omega(N)$.
    The static cell-probe model places no restriction on the preprocessing
    computation. The time needed to construct these components is therefore
    irrelevant.

  \item\label{step:lb-suffix-array-query-additive}
    Next, we prove that the resulting prefix-select data structure answers
    queries using $Q(N)+\bigO(1)$ probes. Consider a query with rank $r$
    and string $X$. An empty query string is handled directly, and for
    $m<128$ a query uses $\bigO(1)$ accesses to the stored
    arrays. Otherwise, the query algorithm from
    \cref{th:reduce-prefix-select-to-sa-for-all-alphabet-sizes} uses the
    range structures to count the strings having prefix $X$ in each block.
    If fewer than $r$ strings have this prefix, it returns $\infty$.
    Otherwise, the cumulative counts determine the block containing the
    $r$th such string and its rank within that block. One suffix-array query
    then yields the answer, using the stored conversion parameters.
    There are only $64$ range structures, and each query to one of them
    accesses $\bigO(1)$ logical words of $\bigO(\log N)$ bits. Each
    logical word occupies $\bigO(1)$ cells because the cell size is
    $\Theta(\log N)$ bits. Accessing the block boundaries and conversion
    parameters also takes $\bigO(1)$ probes. Each suffix-array structure
    is stored in cells of its original size, and its query takes at most
    $Q(N)$ probes. The remaining arithmetic is free in this model,
    giving $Q(N)+\bigO(1)$ probes in total.

  \item\label{step:lb-suffix-array-query}
    We now prove that the bound in
    Step~\ref{step:lb-suffix-array-query-additive} is $\bigO(Q(N))$. Any
    correct suffix-array data structure for all binary strings of length at
    most $N$ has worst-case query time $Q(N)\geq1$. Indeed, without probing
    its stored representation, the query algorithm could not distinguish the
    strings $01$ and $10$, whose first suffix-array entries are different.
    Consequently, $Q(N)+\bigO(1)=\bigO(Q(N))$.

  \item Finally, we derive the claimed lower bound on $Q(N)$. By
    Steps~\ref{step:lb-suffix-array-space}--\ref{step:lb-suffix-array-query},
    we have obtained a deterministic cell-probe data structure for prefix
    select queries over a binary alphabet on inputs of at most $N$ symbols
    with $\bigO(S(N))$ bits of space and $\bigO(Q(N))$ query time.
    Let $c_s,c_q\geq1$ be constants such that this data structure uses
    at most $c_sS(N)$ bits of space and at most $c_qQ(N)$ probes.
    Applying \cref{cor:lb-prefix-select} with input-size parameter $N$
    yields
    \[
      c_qQ(N) \geq \Omega\left(
        \frac{\log\log N}
        {\log((c_s S(N)/N)\log\log N)}
      \right).
    \]
    The denominator equals
    $\log((S(N)/N)\log\log N)+\log c_s$.
    Since $S(N)=\Omega(N)$, its first term is
    $\Omega(\log\log\log N)$, and adding the constant $\log c_s$
    does not change its asymptotic value. Dividing by the constant $c_q$
    gives the claimed lower bound.\qedhere
  \end{enumerate}
\end{proof}

%% file: prefix-select/summary.tex
\subsection{Summary}\label{sec:prefix-select-and-sa-summary}

\begin{theorem}[Perfect equivalence of prefix select and suffix array queries for every alphabet size]
  \label{th:equivalence-of-prefix-select-and-sa-summary}
  Let $\AlphabetSize, N \in \Z_{\geq 2}$ be such that $\AlphabetSize \leq N$.
  In the word RAM model with word size $w$ satisfying
  $1+\log N \leq w = \bigO(\log N)$, the problems of
  \probname{Indexing}
  \probname{for}
  \probname{Prefix}
  \probname{Select}
  \probname{Queries}
  \probname{over}
  \probname{Alphabet}
  $[0 \dd \AlphabetSize)$ and
  \probname{Indexing}
  \probname{for}
  \probname{Suffix}
  \probname{Array}
  \probname{Queries}
  \probname{over}
  \probname{Alphabet}
  $[0 \dd \AlphabetSize)$
  are asymptotically equivalent:
  If there exists a data structure for one of these two problems that, given
  $N$, $\AlphabetSize$, and a valid input of at most $N$ symbols over alphabet
  $\IntegerAlphabet$,\footnote{For the problem of
  \probname{Indexing}
  \probname{for}
  \probname{Prefix}
  \probname{Select}
  \probname{Queries}
  \probname{over}
  \probname{Alphabet}
  $[0 \dd \AlphabetSize)$,
  a valid input consists of the integer $m$ and the packed sequence representation
  $\PackedSeqRepresentation{w}{\AlphabetSize}{W}$
  (\cref{def:packed-sequence-representation})
  of a sequence $W[1 \dd m]$ of $m \geq \AlphabetSize$ nonempty strings of common length
  $\ell = \lfloor \log_{\AlphabetSize} m \rfloor$ such
  that $m \cdot \ell \leq N$. Given any $r \in [1 \dd m]$, the length $|X|$, and the packed representation
  $\PackedRepresentation{w}{\AlphabetSize}{X}$ of any string $X \in [0 \dd \AlphabetSize)^{\leq \ell}$,
  the query returns $\PrefixSelect{W}{r}{X}$ (\cref{def:prefix-rank-and-select}),
  i.e., the $r$th smallest element of the set
  $\{j \in [1 \dd m] : X\text{ is a prefix of }W[j]\}$
  (if $r \leq \PrefixRank{W}{m}{X}$) or $\infty$ (otherwise).
  In
  \probname{Indexing}
  \probname{for}
  \probname{Suffix}
  \probname{Array}
  \probname{Queries}
  \probname{over}
  \probname{Alphabet}
  $[0 \dd \AlphabetSize)$,
  a valid input consists of the length $\Textlen$ and the packed representation
  $\PackedRepresentation{w}{\AlphabetSize}{\Text}$ of a string $\Text \in \IntegerAlphabet^{\Textlen}$ such that
  $\AlphabetSize \leq \Textlen \leq N$. Given any $i \in [1 \dd \Textlen]$, the query returns the value $\SA{\Text}[i]$ (\cref{def:suffix-array}), i.e.,
  the starting position of the lexicographically $i$th smallest suffix of $\Text$.}
  achieves:
  \begin{itemize}
  \item space usage of $S(\AlphabetSize,N)$ bits,
  \item preprocessing time $P_t(\AlphabetSize,N)$,
  \item preprocessing space $P_s(\AlphabetSize,N)$ bits,
  \item query time $Q(\AlphabetSize,N)$,
  \end{itemize}
  then there exists a data structure for the other problem that, given
  $N$, $\AlphabetSize$, and a valid input of at most $N$ symbols over alphabet $\IntegerAlphabet$, achieves:
  \begin{itemize}
  \item space usage of $\bigO(S(\AlphabetSize,N))$ bits,
  \item preprocessing time $\bigO(P_t(\AlphabetSize,N))$,
  \item preprocessing space $\bigO(P_s(\AlphabetSize,N))$ bits,
  \item query time $\bigO(Q(\AlphabetSize,N))$.
  \end{itemize}
\end{theorem}
\begin{proof}
  The result follows by combining
  \cref{th:reduce-prefix-select-to-sa-for-all-alphabet-sizes-clean} and
  \cref{th:reduce-sa-to-prefix-select-for-all-alphabet-sizes-clean}.
\end{proof}

%% file: prefix-special-rank.tex
\section{Perfect Equivalence of Prefix Special Rank and
  Inverse Suffix Array Queries for Every Alphabet Size}\label{sec:equiv-prefix-special-rank-and-isa}

\input{prefix-special-rank/problem-def}

\input{prefix-special-rank/alphabet-reduction-inverse-suffix-array}

\input{prefix-special-rank/alphabet-reduction-prefix-special-rank}

\input{prefix-special-rank/reduce-prefix-special-rank-to-inverse-suffix-array}

\input{prefix-special-rank/reduce-inverse-suffix-array-to-prefix-special-rank}

\input{prefix-special-rank/summary}

%% file: prefix-special-rank/problem-def.tex
\subsection{Problem Definitions}\label{sec:prefix-special-rank-and-isa-problem-def}
\vspace{-1.5ex}

\begin{framed}
  \noindent
  \probname{Indexing for Prefix Special Rank Queries over Alphabet $[0 \dd \AlphabetSize)$}
  \begin{bfdescription}
  \item[Input:]
    Let $\AlphabetSize, \Seqlen \in \Z_{\geq 2}$ be such that
    $\AlphabetSize \leq \Seqlen$. Denote
    $\ell = \lfloor \log_{\AlphabetSize} \Seqlen \rfloor$.
    Consider the word RAM model with word size $w > \log \Seqlen$.
    The input consists of the integer $\Seqlen$ and the packed sequence representation
    $\PackedSeqRepresentation{w}{\AlphabetSize}{W}$
    (\cref{def:packed-sequence-representation}) of a sequence
    $W[1 \dd \Seqlen] \in (\IntegerAlphabet^\ell)^{\Seqlen}$.
  \item[Output:]
    A data structure that, given any index $i \in [1 \dd \Seqlen]$ and any length $p \in [0 \dd \ell]$,
    returns $\PrefixSpecialRank{W}{i}{p}$ (\cref{def:prefix-rank-and-select}),
    i.e., the number of strings in $W[1 \dd i]$ having $X = W[i][1 \dd p]$ as a prefix.
  \end{bfdescription}
\end{framed}

\begin{framed}
  \noindent
  \probname{Indexing for Inverse Suffix Array Queries over Alphabet $[0 \dd \AlphabetSize)$}
  \begin{bfdescription}
  \item[Input:]
    Let $\AlphabetSize, \Textlen \in \Z_{\geq 2}$ be such that
    $\AlphabetSize \leq \Textlen$. Consider the word RAM model with word size
    $w > \log \Textlen$. The input consists of the length $\Textlen$ and the packed representation
    $\PackedRepresentation{w}{\AlphabetSize}{\Text}$
    (\cref{def:packed-representation}) of a string
    $\Text\in\IntegerAlphabet^{\Textlen}$.
  \item[Output:]
    A data structure that, given
    any $j \in [1 \dd \Textlen]$, returns $\ISA{\Text}[j]$ (\cref{def:inverse-suffix-array}),
    i.e., the position of the index $j$ in the suffix array of $\Text$.
  \end{bfdescription}
\end{framed}
\vspace{2ex}

%% file: prefix-special-rank/alphabet-reduction-inverse-suffix-array.tex
\subsection{General Alphabet Reduction for Inverse Suffix Array Queries}\label{sec:general-inverse-suffix-array-alphabet-reduction}

\begin{lemma}\label{lm:general-inverse-suffix-array-alphabet-reduction}
  Let $\AlphabetSize_1, \AlphabetSize_2 \in \Z_{\geq 2}$ be such that $\AlphabetSize_1 \geq \AlphabetSize_2$.
  Let $\Text_1 \in [0 \dd \AlphabetSize_1)^{+}$. For every $j \in [1 \dd |\Text_1|]$, it holds
  \[
    \ISA{\Text_1}[j] = \ISA{\Text_2}[(j - 1) \cdot \delta + 1] - \Delta,
  \]
  where
  $\Text_2 = \ExtAlphabetMap{\AlphabetSize_1}{\AlphabetSize_2}{\Text_1} \in [0 \dd \AlphabetSize_2)^{+}$
  (\cref{def:ext-alphabet-map}),
  $\Delta = |\Text_2| - |\Text_1|$,
  $k = \lceil \log_{\AlphabetSize_2} \AlphabetSize_1 \rceil$, and
  $\delta = 2k + 3$.
\end{lemma}
\begin{proof}
  Let $j\in[1\dd|\Text_1|]$ and denote $i=\ISA{\Text_1}[j]$.
  Then $\SA{\Text_1}[i]=j$, and rearranging the identity in
  \cref{lm:general-suffix-array-alphabet-reduction} yields
  \[
    \SA{\Text_2}[\Delta+i]=(j-1)\cdot\delta+1.
  \]
  Since $\SA{\Text_2}$ and $\ISA{\Text_2}$ are inverse permutations,
  it follows that $\ISA{\Text_2}[(j-1)\cdot\delta+1]=\Delta+i$.
  Substituting $i=\ISA{\Text_1}[j]$ and subtracting $\Delta$ proves the claim.
\end{proof}

\begin{theorem}\label{th:general-inverse-suffix-array-alphabet-reduction}
  Let $N, \AlphabetSize_1, \AlphabetSize_2 \in \Z_{\geq 2}$ be such that $\AlphabetSize_2 \leq \AlphabetSize_1 \leq N$, and denote
  $k=\lceil\log_{\AlphabetSize_2}\AlphabetSize_1\rceil$.
  In the word RAM model with word size $w \geq 1+\log N$, given $N$, $\AlphabetSize_1$, $\AlphabetSize_2$, $|\Text|$, and the packed representation
  $\PackedRepresentation{w}{\AlphabetSize_1}{\Text}$ (\cref{def:packed-representation}) of any string $\Text \in [0 \dd \AlphabetSize_1)^+$ satisfying $(2k+3)|\Text|\leq N$,
  we can in $\bigO(\sqrt{N} + |\Text| / \log_{\AlphabetSize_1} N)$ time compute integers $\alpha$, $\beta$, $\gamma$, and $\mu$, and the
  packed representation
  $\PackedRepresentation{w}{\AlphabetSize_2}{\Text'}$
  of a string $\Text' \in [0 \dd \AlphabetSize_2)^{*}$
  such that $|\Text'| = (2k+3)|\Text|$ and, for every $j \in [1 \dd |\Text|]$,
  it holds
  $
    \ISA{\Text}[j] = \ISA{\Text'}[(j - \alpha) \cdot \beta + \gamma] - \mu
  $.
\end{theorem}
\begin{proof}
  We proceed in two steps.
  \begin{enumerate}
  \item We first compute the packed representation
    $\PackedRepresentation{w}{\AlphabetSize_2}{\Text'}$ of
    $\Text'=\ExtAlphabetMap{\AlphabetSize_1}{\AlphabetSize_2}{\Text}$
    (\cref{def:ext-alphabet-map}). For this purpose, construct the data
    structure from \cref{th:ext-alphabet-map} in $\bigO(\sqrt{N})$ time and
    apply it to $\Text$ in $\bigO(1+|\Text|/\log_{\AlphabetSize_1}N)$ time.
  \item Next, to keep the intermediate integers within one word,
    compute $k$ by counting the successive integer divisions of
    $\AlphabetSize_1-1$ by $\AlphabetSize_2$ needed to reach zero.
    Set $\beta=2k+3$, $\alpha=\gamma=1$, and $\mu=|\Text|\beta-|\Text|$.
    This takes $\bigO(\log\AlphabetSize_1)\subseteq\bigO(\log N)$ time.
  \end{enumerate}
  The mapping structure uses size parameter $N$, source alphabet size
  $\AlphabetSize_1$, and target alphabet size $\AlphabetSize_2$.
  Since $(2k+3)|\Text|\leq N$, its query satisfies the length
  bound, and all returned parameters fit in one word.
  By \cref{def:ext-alphabet-map}, it holds $|\Text'|=|\Text|\beta$,
  which is the required length, and $\mu=|\Text'|-|\Text|$.
  With these parameters, \cref{lm:general-inverse-suffix-array-alphabet-reduction}
  gives the claimed inverse-suffix-array identity.
  The two steps take
  $\bigO(\sqrt{N}+|\Text|/\log_{\AlphabetSize_1}N)$
  time in total, since $\log N=\bigO(\sqrt{N})$.
\end{proof}

\begin{theorem}\label{th:general-inverse-suffix-array-alphabet-reduction-2}
  Let $N, \AlphabetSize_1, \AlphabetSize_2 \in \Z_{\geq 2}$ be such that $\AlphabetSize_2 \leq \AlphabetSize_1 \leq N$, and denote
  $k=\lceil\log_{\AlphabetSize_2}\AlphabetSize_1\rceil$.
  In the word RAM model with word size $w \geq 1+\log N$, given $N$, $\AlphabetSize_1$, $\AlphabetSize_2$, $|\Text|$, and the packed representation
  $\PackedRepresentation{w}{\AlphabetSize_1}{\Text}$ (\cref{def:packed-representation}) of any string $\Text \in [0 \dd \AlphabetSize_1)^+$ satisfying $(2k+3)|\Text|\leq N$,
  we can in $\bigO(N / \log_{\AlphabetSize_2} N)$ time compute integers $\alpha$, $\beta$, $\gamma$, and $\mu$, and the
  packed representation
  $\PackedRepresentation{w}{\AlphabetSize_2}{\Text'}$
  of a string $\Text' \in [0 \dd \AlphabetSize_2)^{*}$
  such that $|\Text'| = (2k+3)|\Text|$ and, for every $j \in [1 \dd |\Text|]$,
  it holds
  $
    \ISA{\Text}[j] =  \ISA{\Text'}[(j - \alpha) \cdot \beta + \gamma] - \mu
  $.
\end{theorem}
\begin{proof}
  Use \cref{th:general-inverse-suffix-array-alphabet-reduction} with the
  supplied parameters. The inequality
  $|\Text|\log\AlphabetSize_1\leq k|\Text|\log\AlphabetSize_2
    \leq N\log\AlphabetSize_2$ gives
  $|\Text|/\log_{\AlphabetSize_1}N\leq N/\log_{\AlphabetSize_2}N$.
  Moreover, $\sqrt N=\bigO(N/\log_{\AlphabetSize_2}N)$, which also
  covers the setup time.
\end{proof}

%% file: prefix-special-rank/alphabet-reduction-prefix-special-rank.tex
\subsection{General Alphabet Reduction for Prefix Special Rank Queries}\label{sec:general-prefix-special-rank-alphabet-reduction}

\begin{theorem}\label{th:general-prefix-special-rank-alphabet-reduction}
  Let $\AlphabetSize_1, \AlphabetSize_2, N \in \Z_{\geq 2}$ be such that $\AlphabetSize_1 \geq \AlphabetSize_2$.
  Consider the word RAM model with word size $w$ satisfying
  $1+\log N \leq w = \bigO(\log N)$.
  Assume that there exists a data structure answering prefix special rank
  queries (\cref{def:prefix-rank-and-select})
  that, given $N$, $\AlphabetSize_2$, $m_2$, and the packed sequence representation
  $\PackedSeqRepresentation{w}{\AlphabetSize_2}{W_2}$
  (\cref{def:packed-sequence-representation}) of a sequence
  $W_2[1 \dd m_2]$ of $m_2 \geq \AlphabetSize_2$ nonempty strings
  of common length $\ell_2 = \lfloor \log_{\AlphabetSize_2} m_2 \rfloor$
  over alphabet $[0 \dd \AlphabetSize_2)$,
  where $m_2 \cdot \ell_2 \leq N$,
  achieves the following complexities:
  \begin{itemize}
  \item space usage $S(\AlphabetSize_2,N)$ bits,
  \item preprocessing time $P_t(\AlphabetSize_2,N)$,
  \item preprocessing space $P_s(\AlphabetSize_2,N)$ bits,
  \item query time $Q(\AlphabetSize_2,N)$.
  \end{itemize}
  Then, letting $N' = \lfloor N / (4k) \rfloor$
  (where $k = \lceil \log_{\AlphabetSize_2} \AlphabetSize_1 \rceil$),
  there exists a data structure answering prefix special rank queries
  that, given $N$, $\AlphabetSize_1$, $\AlphabetSize_2$, $m_1$, and the packed sequence representation
  $\PackedSeqRepresentation{w}{\AlphabetSize_1}{W_1}$
  (\cref{def:packed-sequence-representation}) of a sequence
  $W_1[1 \dd m_1]$ of $m_1 \geq \AlphabetSize_1$ nonempty strings
  of common length $\ell_1 = \lfloor \tfrac{1}{2} \log_{\AlphabetSize_1} m_1 \rfloor$
  over alphabet $[0 \dd \AlphabetSize_1)$,
  where $m_1 \cdot \ell_1 \leq N'$,
  achieves the following complexities:
  \begin{itemize}[beginpenalty=10000,midpenalty=10000]
  \item space usage $\bigO(N \log \AlphabetSize_2 + S(\AlphabetSize_2,N))$ bits,
  \item preprocessing time $\bigO(N / \log_{\AlphabetSize_2} N + P_t(\AlphabetSize_2,N))$,
  \item preprocessing space $\bigO(N \log \AlphabetSize_2 + P_s(\AlphabetSize_2,N))$ bits,
  \item query time $\bigO(1 + Q(\AlphabetSize_2,N))$.
  \end{itemize}
\end{theorem}
\begin{proof}
  Consider a sequence $W_1[1\dd m_1]$ as in the claim, with
  $\ell_1=\lfloor\tfrac12\log_{\AlphabetSize_1}m_1\rfloor\geq1$
  and $m_1\ell_1\leq N'$.
  Denote $m_2=m_1$ and
  $\ell_2=\lfloor\log_{\AlphabetSize_2}m_1\rfloor$.
  Since $\log_{\AlphabetSize_2}\AlphabetSize_1\geq1$, we have
  $k\leq2\log_{\AlphabetSize_2}\AlphabetSize_1$. Consequently,
  $k\ell_1\leq\log_{\AlphabetSize_2}m_1$.
  Since $k\ell_1$ is an integer, we obtain
  $k\ell_1\leq\lfloor\log_{\AlphabetSize_2}m_1\rfloor=\ell_2$. Furthermore,
  \[
    \ell_2\leq\log_{\AlphabetSize_2}m_1
      =(\log_{\AlphabetSize_1}m_1)\cdot
        \log_{\AlphabetSize_2}\AlphabetSize_1
      <2(\ell_1+1)k\leq4k\ell_1,
  \]
  where the last inequality uses $\ell_1\geq1$.
  Let $W_2[1\dd m_2]$ denote the sequence satisfying
  \[
    W_2[i]=\AlphabetMap{\AlphabetSize_1}{\AlphabetSize_2}{W_1[i]}
      \cdot\zero^{\ell_2-k\ell_1}
      \qquad\text{for every }i\in[1\dd m_1]
  \]
  (\cref{def:alphabet-map}). Each string $W_2[i]$ has length $\ell_2$,
  and $m_2\geq\AlphabetSize_2$. Moreover,
  $m_2\ell_2\leq4km_1\ell_1\leq4kN'\leq N$.
  Thus, $W_2$ satisfies the assumptions of the prefix-special-rank
  structure from the claim.

  Consider any $i\in[1\dd m_1]$ and $p\in[0\dd\ell_1]$.
  For every $j\in[1\dd m_1]$, the definition of $W_2$ gives
  $W_2[j][1\dd kp]
    =\AlphabetMap{\AlphabetSize_1}{\AlphabetSize_2}{W_1[j][1\dd p]}$.
  By \cref{ob:alphabet-map-properties}\eqref{ob:alphabet-map-prefix-property},
  the equality $W_1[j][1\dd p]=W_1[i][1\dd p]$ holds if and only if
  $W_2[j][1\dd kp]=W_2[i][1\dd kp]$.
  Counting these indices $j\in[1\dd i]$ gives
  \[
    \PrefixSpecialRank{W_1}{i}{p}=\PrefixSpecialRank{W_2}{i}{kp}.
  \]

  \DSComponents
  We store the integer $k$ and the prefix-special-rank structure from
  the claim constructed for $W_2$. They occupy
  $\bigO(\log N+S(\AlphabetSize_2,N))
    \subseteq\bigO(N\log\AlphabetSize_2+S(\AlphabetSize_2,N))$ bits.

  \DSQueries
  Given $i\in[1\dd m_1]$ and $p\in[0\dd\ell_1]$, compute $kp$ and
  return $\PrefixSpecialRank{W_2}{i}{kp}$ using the stored structure.
  The query is valid because $kp\leq k\ell_1\leq\ell_2$, and the
  identity above proves correctness. The total query time is
  $\bigO(1+Q(\AlphabetSize_2,N))$.

  \DSConstruction
  Given $N$, $\AlphabetSize_1$, $\AlphabetSize_2$, $m_1$, and
  $\PackedSeqRepresentation{w}{\AlphabetSize_1}{W_1}$,
  we construct the data structure in three steps:
  \begin{enumerate}
  \item First, compute $k$, $\ell_1$, $\ell_2$, and
    $b_2=\lceil\log\AlphabetSize_2\rceil$ in $\bigO(\log N)$ time,
    and store $k$. To keep all intermediate integers within one word,
    we compute these parameters using integer divisions. Construct a temporary structure from
    \cref{th:alphabet-map} in $\bigO(\sqrt N)$ time. Its assumptions hold because
    $w\geq1+\log N$ and
    $\AlphabetSize_2\leq\AlphabetSize_1\leq m_1\leq m_1\ell_1\leq N'\leq N$,
    where the second and third inequalities follow from
    $m_1\geq\AlphabetSize_1$ and $\ell_1\geq1$.
  \item We now construct
    $\PackedSeqRepresentation{w}{\AlphabetSize_2}{W_2}$.
    For this purpose, let $M[1\dd m_1]$ and $Z[1\dd m_1]$ denote the sequences defined by
    $M[i]=\AlphabetMap{\AlphabetSize_1}{\AlphabetSize_2}{W_1[i]}$ and
    $Z[i]=\zero^{\ell_2-k\ell_1}$ for every $i\in[1\dd m_1]$.
    Since $W_2[i]=M[i]Z[i]$, we proceed as follows:
    \begin{enumerate}
    \item Apply the temporary structure from \cref{th:alphabet-map}
      to the length-$m_1\ell_1$ concatenation represented by
      $\PackedSeqRepresentation{w}{\AlphabetSize_1}{W_1}$.
      The query returns $\PackedSeqRepresentation{w}{\AlphabetSize_2}{M}$
      in $\bigO(1+m_1\ell_1/\log_{\AlphabetSize_1}N)$ time.
      The bounds $\log\AlphabetSize_1\leq k\log\AlphabetSize_2$ and
      $km_1\ell_1\leq kN'\leq N/4$ imply
      \[
        \frac{m_1\ell_1}{\log_{\AlphabetSize_1}N}
          \leq\frac{km_1\ell_1}{\log_{\AlphabetSize_2}N}
          \leq\frac{N}{4\log_{\AlphabetSize_2}N}.
      \]
    \item To construct $\PackedSeqRepresentation{w}{\AlphabetSize_2}{Z}$,
      initialize the packed representation of $\zero^{m_1(\ell_2-k\ell_1)}$
      over alphabet $[0\dd\AlphabetSize_2)$ using
      \cref{pr:packed-representation}\eqref{pr:packed-representation-initialize}.
      This takes
      $\bigO(1+m_1(\ell_2-k\ell_1)\log\AlphabetSize_2/w)
        =\bigO(N/\log_{\AlphabetSize_2}N)$ time.
    \item Finally, apply the pointwise concatenation algorithm from
      \cref{th:packed-pointwise-concatenation-partition-offline} to $M$
      and $Z$. This produces $\PackedSeqRepresentation{w}{\AlphabetSize_2}{W_2}$
      in $\bigO(\sqrt N+m_1\ell_2/\log_{\AlphabetSize_2}N)
        =\bigO(N/\log_{\AlphabetSize_2}N)$ time.
    \end{enumerate}
  \item Finally, construct the prefix-special-rank structure from the
    claim using $m_2$ and $\PackedSeqRepresentation{w}{\AlphabetSize_2}{W_2}$.
    This takes $P_t(\AlphabetSize_2,N)$ time and uses
    $P_s(\AlphabetSize_2,N)$ bits of space.
  \end{enumerate}
  The alphabet-mapping structure uses size parameter $N$, source
  alphabet size $\AlphabetSize_1$, and target alphabet size
  $\AlphabetSize_2$. The initialization uses length bound $N$. The
  pointwise concatenation and the prefix-special-rank structure from the
  claim use size parameter $N$ and alphabet size $\AlphabetSize_2$, and
  the concatenation's length conditions hold because
  $1\leq\ell_2\leq N/m_1$ and $m_1\ell_2\leq N$.
  The inequality $\log_{\AlphabetSize_2}N\leq\log N$ implies
  $\sqrt N+\log N=\bigO(N/\log_{\AlphabetSize_2}N)$.
  Thus, the first two steps take $\bigO(N/\log_{\AlphabetSize_2}N)$ time.
  Their running time also bounds their auxiliary space in words.
  Since $\log\AlphabetSize_1\leq k\log\AlphabetSize_2$ and
  $km_1\ell_1\leq N/4$, the input representation requires
  $\bigO(w+m_1\ell_1\log\AlphabetSize_1)=\bigO(N\log\AlphabetSize_2)$ bits.
  With $w=\bigO(\log N)$, the preprocessing
  takes $\bigO(N/\log_{\AlphabetSize_2}N+P_t(\AlphabetSize_2,N))$ time and uses
  $\bigO(N\log\AlphabetSize_2+P_s(\AlphabetSize_2,N))$ bits.
\end{proof}

%% file: prefix-special-rank/reduce-prefix-special-rank-to-inverse-suffix-array.tex
\subsection{Reducing Prefix Special Rank to Inverse Suffix Array Queries}\label{sec:reduce-prefix-special-rank-to-isa}

\begin{lemma}\label{lm:reduce-prefix-special-rank-to-isa}
  Let $\AlphabetSize \in \Z_{\geq 2}$.
  Let $W[1 \dd m]$ be a sequence of $m \geq \AlphabetSize$ strings of length $\ell \geq 1$ over
  alphabet $\IntegerAlphabet$. Denote
  $\Text = \SeqToString{\AlphabetSize}{W} \in [0 \dd 2\AlphabetSize)^{*}$
  (\cref{def:seq-to-string}). For every $i \in [1 \dd m]$ and every
  $p \in [1 \dd \ell]$, it holds
  \[
    \PrefixSpecialRank{W}{i}{p}=\ISA{\Text}[q]-b_X,
  \]
  where
  $k=\lceil\log_{\AlphabetSize}m\rceil$,
  $c=\AlphabetSize$,
  $q=(i-1)(\ell+k+1)+\ell-p+1$,
  $X=W[i][1\dd p]$, and
  $b_X=\RangeBegTwo{\revstr{X}\cdot c}{\Text}$
  (\cref{def:occ}).
\end{lemma}
\begin{proof}
  Denote $r=\PrefixSpecialRank{W}{i}{p}$. By
  \cref{def:prefix-rank-and-select}, it holds
  $r\in[1\dd\PrefixRank{W}{m}{X}]$ and
  $\PrefixSelect{W}{r}{X}=i$. Applying
  \cref{lm:seq-to-string}\eqref{lm:seq-to-string-locate} yields
  $\SA{\Text}[b_X+r]=i(\ell+k+1)-(k+p)
    =(i-1)(\ell+k+1)+\ell-p+1=q$.
  Consequently, $\ISA{\Text}[q]=b_X+r$, which proves the claim.
\end{proof}

\begin{theorem}\label{th:reduce-prefix-special-rank-to-isa-for-all-alphabet-sizes}
  Let $\AlphabetSize, N \in \Z_{\geq 2}$ be such that $\AlphabetSize \leq N$.
  Consider the word RAM model with word size $w$ satisfying
  $1+\log N \leq w = \bigO(\log N)$.
  Assume that there exists a data structure for
  \probname{Indexing}
  \probname{for}
  \probname{Inverse}
  \probname{Suffix}
  \probname{Array}
  \probname{Queries}
  \probname{over}
  \probname{Alphabet}
  $[0 \dd \AlphabetSize)$
  that, given $N$, $\AlphabetSize$, and a valid input of at most $N$ symbols over alphabet
  $\IntegerAlphabet$,\footnote{\label{footnote:isa-input-2} That is, the length $\Textlen$ and the
  packed representation
  $\PackedRepresentation{w}{\AlphabetSize}{\Text}$ of a string $\Text \in \IntegerAlphabet^{\Textlen}$ such that
  $\AlphabetSize \leq \Textlen \leq N$; see \cref{sec:prefix-special-rank-and-isa-problem-def}.}
  achieves:
  \begin{itemize}
  \item space usage $S(\AlphabetSize,N)$ bits,
  \item preprocessing time $P_t(\AlphabetSize,N)$,
  \item preprocessing space $P_s(\AlphabetSize,N)$ bits,
  \item query time $Q(\AlphabetSize,N)$.
  \end{itemize}
  Then, there exists a data structure for
  \probname{Indexing}
  \probname{for}
  \probname{Prefix}
  \probname{Special}
  \probname{Rank}
  \probname{Queries}
  \probname{over}
  \probname{Alphabet}
  $[0 \dd \AlphabetSize)$
  that, given $N$, $\AlphabetSize$, and a valid input of at most $N$ symbols over alphabet
  $\IntegerAlphabet$,\footnote{\label{footnote:prefix-special-rank-input-2} That is, the integer $\Seqlen$ and the
  packed sequence representation
  $\PackedSeqRepresentation{w}{\AlphabetSize}{W}$
  (\cref{def:packed-sequence-representation}) of a sequence
  $W[1 \dd \Seqlen]$ of $\Seqlen \geq \AlphabetSize$ strings of length
  $\ell = \lfloor \log_{\AlphabetSize} \Seqlen \rfloor$, where
  $\Seqlen \ell \leq N$; see \cref{sec:prefix-special-rank-and-isa-problem-def}.}
  achieves:
  \begin{itemize}
  \item space usage $\bigO(N \log \AlphabetSize + S(\AlphabetSize,N))$ bits,
  \item preprocessing time $\bigO(N / \log_{\AlphabetSize} N + P_t(\AlphabetSize,N))$,
  \item preprocessing space $\bigO(N \log \AlphabetSize + P_s(\AlphabetSize,N))$ bits,
  \item query time $\bigO(1 + Q(\AlphabetSize,N))$.
  \end{itemize}
\end{theorem}
\begin{proof}
  Consider a sequence $W[1\dd\Seqlen]$ of
  $\Seqlen\geq\AlphabetSize$ strings of length
  $\ell=\lfloor\log_{\AlphabetSize}\Seqlen\rfloor\geq1$ over
  $\IntegerAlphabet$, with $\Seqlen\ell\leq N$.
  To keep each text supplied to the inverse-suffix-array structure within
  the length bound $N$, our main construction applies the sequence-to-string
  mapping separately to a constant number of blocks of $W$. A prefix-special-rank
  query is answered by counting the matching strings in the preceding
  blocks and using one inverse-suffix-array query in the block containing
  the queried string.

  First, suppose that $\Seqlen<64\AlphabetSize$. In this case,
  $\ell<1+\log_{\AlphabetSize}64\leq7$, and hence $\ell\leq6$.
  We can therefore store all answers in arrays
  $D_p[1\dd\Seqlen]$, where $p\in[1\dd\ell]$ and
  $D_p[i]=\PrefixSpecialRank{W}{i}{p}$.
  To construct them, compute $\ell$ and $b=\lceil\log\AlphabetSize\rceil$.
  For each $p\in[1\dd\ell]$, initialize an array
  $C[0\dd\AlphabetSize^p)$ to zero and scan $i=1,\ldots,\Seqlen$.
  At each index $i$, compute $x=\Val{\AlphabetSize}{W[i][1\dd p]}$
  from the at most six symbols of this prefix in the input representation,
  accessed using
  \cref{pr:packed-representation}\eqref{pr:packed-representation-access}
  with length bound $N$.
  Next, increment $C[x]$ and set $D_p[i]:=C[x]$.
  Since $\AlphabetSize^p\leq\Seqlen$, this takes $\bigO(\Seqlen)$
  time and words. A query returns $i$ if $p=0$, and $D_p[i]$ otherwise.
  The bounds below also apply to this construction.

  Henceforth, assume that $\Seqlen\geq64\AlphabetSize$.
  Let $W_1,\ldots,W_{64}$ denote the consecutive blocks of $W$ in which
  the first $\Seqlen\bmod64$ blocks contain
  $\lfloor\Seqlen/64\rfloor+1$ strings each, and the remaining blocks
  contain $\lfloor\Seqlen/64\rfloor$ strings each.
  For every $t\in[1\dd64]$, let $q_t$ denote the number of strings
  in $W_t$, and let $a_t=\sum_{u=1}^{t}q_u$, with $a_0=0$.
  In particular, $W_t[j]=W[a_{t-1}+j]$ for every $j\in[1\dd q_t]$.
  These block sizes satisfy $\AlphabetSize\leq q_t\leq\Seqlen/32$ and
  $\Seqlen\leq128q_t$ for every $t\in[1\dd64]$.
  Denote $c=\AlphabetSize$, $b=\lceil\log\AlphabetSize\rceil$,
  $k_t=\lceil\log_{\AlphabetSize}q_t\rceil$, and
  $Y_t=\SeqToString{\AlphabetSize}{W_t}$
  (\cref{def:seq-to-string}).

  For $t\in[1\dd64]$, let $\alpha_t,\beta_t,\gamma_t,\mu_t$
  and $\Text_t\in\IntegerAlphabet^*$ denote the integers and string
  obtained from \cref{th:general-inverse-suffix-array-alphabet-reduction-2}
  with input $Y_t$, size parameter $7|Y_t|$, and alphabet sizes
  $2\AlphabetSize$ and $\AlphabetSize$. Since
  $\AlphabetSize<2\AlphabetSize\leq\AlphabetSize^2$, we have
  $\lceil\log_{\AlphabetSize}(2\AlphabetSize)\rceil=2$, and hence the length
  condition of the reduction holds with equality and gives
  $|\Text_t|=7|Y_t|$. Since $k_t\leq\ell+1$,
  these strings satisfy
  \[
    2\AlphabetSize\leq|Y_t|=q_t(\ell+k_t+1),
    \qquad
    |\Text_t|=7|Y_t|\leq28q_t\ell
      \leq\tfrac78\Seqlen\ell\leq N.
  \]
  In particular, the alphabet reduction satisfies its alphabet and
  word-size assumptions, and each $\Text_t$ is a valid input to the
  inverse-suffix-array data structure from the premise.

  \DSComponents
  The data structure consists of the following components:
  \begin{enumerate}
  \item The packed sequence representation
    $\PackedSeqRepresentation{w}{\AlphabetSize}{W}$, the integers
    $\Seqlen$, $\ell$, and $b$, the boundaries $a_0,\ldots,a_{64}$,
    and the integers $q_t,k_t,\alpha_t,\beta_t,\gamma_t,\mu_t$ for
    every $t\in[1\dd64]$.
    Together, they occupy $\bigO(\Seqlen\log N)$ bits.
  \item For every $t\in[1\dd64]$, the structure from
    \cref{pr:seq-to-str-range-beg-and-range-end} constructed for $W_t$.
    These structures occupy $\bigO(\Seqlen)$ words in total, or
    $\bigO(\Seqlen\log N)$ bits.
  \item For every $t\in[1\dd64]$, an instance of the inverse-suffix-array
    data structure from the premise constructed for $\Text_t$.
    Together, these instances use $\bigO(S(\AlphabetSize,N))$ bits.
  \end{enumerate}
  The total space is
  $\bigO(\Seqlen\log N+S(\AlphabetSize,N))$ bits.

  \DSQueries
  Given $i\in[1\dd\Seqlen]$ and $p\in[0\dd\ell]$, return $i$
  if $p=0$. Otherwise, proceed as follows:
  \begin{enumerate}[beginpenalty=10000]
  \item First, scan the stored boundaries to determine the index
    $h\in[1\dd64]$ satisfying
    $a_{h-1}<i\leq a_h$, and compute $u:=i-a_{h-1}$.
    Extract the packed representation
    $\PackedRepresentation{w}{\AlphabetSize}{X}$ of $X=W[i][1\dd p]$
    from the input representation, starting at position $(i-1)\ell+1$,
    using \cref{pr:packed-representation}\eqref{pr:packed-representation-substring}.
    This takes $\bigO(1)$ time because
    $p\log\AlphabetSize\leq\ell\log\AlphabetSize\leq\log\Seqlen\leq w$.
  \item Next, for each $t\in[1\dd h]$, compute
    $b_t=\RangeBegTwo{\revstr{X}\cdot c}{Y_t}$ and
    $e_t=\RangeEndTwo{\revstr{X}\cdot c}{Y_t}$ using the structure from
    \cref{pr:seq-to-str-range-beg-and-range-end} for $W_t$.
    Compute $d:=\sum_{t=1}^{h-1}(e_t-b_t)$, the number of strings
    having prefix $X$ in the blocks preceding $W_h$.
    There are at most $64$ queries, each taking $\bigO(1)$ time.
  \item We now compute the answer within $W_h$ using one
    inverse-suffix-array query.
    Compute $v:=(u-1)(\ell+k_h+1)+\ell-p+1$ and
    $j:=(v-\alpha_h)\cdot\beta_h+\gamma_h$.
    Obtain $r:=\ISA{\Text_h}[j]$ from the structure for $\Text_h$,
    and return $d+(r-\mu_h-b_h)$.
    This step takes $\bigO(1+Q(\AlphabetSize,N))$ time.
  \end{enumerate}
  To verify correctness, first note that
  \cref{lm:seq-to-string}\eqref{lm:seq-to-string-range-size} gives
  $e_t-b_t=\PrefixRank{W_t}{q_t}{X}$ for every $t\in[1\dd h]$.
  Moreover, $v\in[1\dd|Y_h|]$, and
  \cref{th:general-inverse-suffix-array-alphabet-reduction-2} gives
  $r-\mu_h=\ISA{Y_h}[v]$.
  By \cref{lm:reduce-prefix-special-rank-to-isa}, it follows that
  $r-\mu_h-b_h=\PrefixSpecialRank{W_h}{u}{p}$.
  Since $W_h[u]=W[i]$ and $i=a_{h-1}+u$, we obtain
  \[
    \PrefixSpecialRank{W}{i}{p}
      =\sum_{t=1}^{h-1}\PrefixRank{W_t}{q_t}{X}
        +\PrefixSpecialRank{W_h}{u}{p}
      =d+(r-\mu_h-b_h).
  \]
  For $p=0$, correctness follows from \cref{def:prefix-rank-and-select}.
  The total query time is $\bigO(1+Q(\AlphabetSize,N))$.

  \DSConstruction
  We construct the three components as follows:
  \begin{enumerate}
  \item First, we compute the integers and packed texts needed for the
    inverse-suffix-array structures.
    Compute $\ell$ and $b$, copy the packed sequence representation of $W$,
    and store $\Seqlen$, $\ell$, and $b$.
    Compute and store $q_t$ and $k_t$ for $t\in[1\dd64]$.
    Initialize $a_0:=0$, and compute and store $a_t:=a_{t-1}+q_t$
    for $t=1,\ldots,64$.
    Then, for every $t\in[1\dd64]$, proceed as follows:
    \begin{enumerate}
    \item Extract $\PackedSeqRepresentation{w}{\AlphabetSize}{W_t}$, the
      length-$q_t\ell$ substring of the input representation starting at
      position $a_{t-1}\ell+1$, using
      \cref{pr:packed-representation}\eqref{pr:packed-representation-substring}.
    \item Compute $\PackedRepresentation{w}{2\AlphabetSize}{Y_t}$ using
      \cref{pr:seq-to-string-construction}.
    \item Apply
      \cref{th:general-inverse-suffix-array-alphabet-reduction-2}
      to $Y_t$ to compute
      $\PackedRepresentation{w}{\AlphabetSize}{\Text_t}$ and
      $\alpha_t,\beta_t,\gamma_t,\mu_t$.
      Store these integers, and retain the representations of $W_t$
      and $\Text_t$ for the next two steps.
    \end{enumerate}
  \item For each $t\in[1\dd64]$, construct the structure from
    \cref{pr:seq-to-str-range-beg-and-range-end} for $W_t$.
  \item Finally, for every $t\in[1\dd64]$, construct the
    inverse-suffix-array data structure from the premise for $\Text_t$.
    Each construction takes $P_t(\AlphabetSize,N)$ time and uses
    $P_s(\AlphabetSize,N)$ bits of space.
  \end{enumerate}
  Both sequence-to-string tools and the inverse-suffix-array structures
  from the premise are applied with size parameter $N$ and alphabet
  size $\AlphabetSize$.
  The assumptions of the sequence-to-string tools hold because $\AlphabetSize\leq q_t\leq\Seqlen\leq N$,
  $\AlphabetSize^\ell\leq\Seqlen\leq N$,
  $q_t\ell\leq q_t(\ell+k_t+1)=|Y_t|\leq N$, and
  $w\geq1+\log N$.
  Each invocation takes $\bigO(q_t+\AlphabetSize^\ell)=\bigO(\Seqlen)$ time
  and uses $\bigO(\Seqlen)$ words.
  To keep all operands within one word, the integer logarithms used above
  are computed by successive integer divisions.
  To avoid computing the product $64\AlphabetSize$, we test
  $\Seqlen<64\AlphabetSize$ by comparing $\lfloor\Seqlen/64\rfloor$
  with $\AlphabetSize$. If $N<64$, then $\Seqlen\leq N<64\AlphabetSize$,
  and the case $\Seqlen<64\AlphabetSize$ applies without this test.
  Otherwise, the constant $64\leq N$ fits in one word.
  The block boundaries and prefix counts are at most $\Seqlen\leq N$.
  The extractions use length bound $N$, and all string positions used
  above are within the corresponding
  strings, whose lengths are at most $N$.
  Since $q_t=\Theta(\Seqlen)$, $|Y_t|\leq4q_t\ell$, and
  $\ell\log\AlphabetSize\leq\log\Seqlen$, each alphabet reduction
  takes $\bigO(7|Y_t|\log\AlphabetSize/\log(7|Y_t|))
    =\bigO(q_t)$ time.
  Copying the input and extracting all blocks also take
  $\bigO(\Seqlen)$ time, since the input occupies $\bigO(\Seqlen)$ words.
  The first two steps therefore take $\bigO(\Seqlen)$ time and use
  $\bigO(\Seqlen)$ words.
  There are only $64$ inverse-suffix-array structures to construct,
  and each occupies at most its preprocessing space after construction.
  Consequently, the total preprocessing time is
  $\bigO(\Seqlen+P_t(\AlphabetSize,N))$, and the preprocessing space,
  including the structures already constructed, is
  $\bigO(\Seqlen\log N+P_s(\AlphabetSize,N))$ bits.

  Finally, the bounds in both cases imply the claimed complexities.
  Indeed, $\log\Seqlen<(\ell+1)\log\AlphabetSize
    \leq2\ell\log\AlphabetSize$, and
  $\log(N/\Seqlen)\leq N/\Seqlen$. Hence
  \[
    \Seqlen\log N
      =\Seqlen\log\Seqlen+\Seqlen\log(N/\Seqlen)
      \leq2\Seqlen\ell\log\AlphabetSize+N
      \leq3N\log\AlphabetSize.
  \]
  Thus, $\Seqlen=\bigO(N/\log_{\AlphabetSize}N)$ and
  $\Seqlen\log N=\bigO(N\log\AlphabetSize)$, as required.
\end{proof}

\begin{theorem}\label{th:reduce-prefix-special-rank-to-isa-for-all-alphabet-sizes-clean}
  Let $\AlphabetSize, N \in \Z_{\geq 2}$ be such that $\AlphabetSize \leq N$.
  Consider the word RAM model with word size $w$ satisfying
  $1+\log N \leq w = \bigO(\log N)$.
  Assume that there exists a data structure for
  \probname{Indexing}
  \probname{for}
  \probname{Inverse}
  \probname{Suffix}
  \probname{Array}
  \probname{Queries}
  \probname{over}
  \probname{Alphabet}
  $[0 \dd \AlphabetSize)$
  that, given $N$, $\AlphabetSize$, and a valid input of at most $N$ symbols over alphabet $\IntegerAlphabet$,\footref{footnote:isa-input-2} achieves:
  \begin{itemize}
  \item space usage $S(\AlphabetSize,N)$ bits,
  \item preprocessing time $P_t(\AlphabetSize,N)$,
  \item preprocessing space $P_s(\AlphabetSize,N)$ bits,
  \item query time $Q(\AlphabetSize,N)$.
  \end{itemize}
  Then, there exists a data structure for
  \probname{Indexing}
  \probname{for}
  \probname{Prefix}
  \probname{Special}
  \probname{Rank}
  \probname{Queries}
  \probname{over}
  \probname{Alphabet}
  $[0 \dd \AlphabetSize)$
  that, given $N$, $\AlphabetSize$, and a valid input of at most $N$ symbols over alphabet
  $\IntegerAlphabet$,\footref{footnote:prefix-special-rank-input-2} achieves:
  \begin{itemize}
  \item space usage $\bigO(S(\AlphabetSize,N))$ bits,
  \item preprocessing time $\bigO(P_t(\AlphabetSize,N))$,
  \item preprocessing space $\bigO(P_s(\AlphabetSize,N))$ bits,
  \item query time $\bigO(Q(\AlphabetSize,N))$.
  \end{itemize}
\end{theorem}
\begin{proof}
  By \cref{th:reduce-prefix-special-rank-to-isa-for-all-alphabet-sizes}, it
  remains to show that $N\log\AlphabetSize$,
  $N/\log_{\AlphabetSize}N$, $N\log\AlphabetSize$, and $1$ are dominated by
  $S(\AlphabetSize,N)$, $P_t(\AlphabetSize,N)$,
  $P_s(\AlphabetSize,N)$, and $Q(\AlphabetSize,N)$, respectively.
  For every $\AlphabetSize,N\in\Z_{\geq 2}$ satisfying
  $\AlphabetSize\leq N$,
  \cref{th:inverse-suffix-array-bounded-length-space-lower-bound} states that the
  inverse suffix array data structure in the premise of this theorem has
  worst-case space usage $\Omega(N\log\AlphabetSize)$ bits. Consequently, for
  every such pair $(\AlphabetSize,N)$, there exists a valid input text for
  which preprocessing produces an output data structure occupying
  $\Omega(N\log\AlphabetSize)$ bits. Writing this output to memory takes
  $\Omega((N\log\AlphabetSize)/w)
  =\Omega(N/\log_{\AlphabetSize}N)$ time, and storing it requires
  $\Omega(N\log\AlphabetSize)$ bits of preprocessing space
  (\cref{rm:space}). Hence,
  $S(\AlphabetSize,N)=\Omega(N\log\AlphabetSize)$,
  $P_t(\AlphabetSize,N)=\Omega(N/\log_{\AlphabetSize}N)$, and
  $P_s(\AlphabetSize,N)=\Omega(N\log\AlphabetSize)$. Together with
  $Q(\AlphabetSize,N)=\Omega(1)$, these inequalities prove the claim.
\end{proof}

%% file: prefix-special-rank/reduce-inverse-suffix-array-to-prefix-special-rank.tex
\subsection{Reducing Inverse Suffix Array to Prefix Special Rank Queries}\label{sec:reduce-isa-to-prefix-special-rank}

\subsubsection{The Nonperiodic Positions}\label{sec:reduce-isa-to-prefix-special-rank-nonperiodic}

\begin{proposition}[{\cite{breaking}}]\label{pr:reduce-isa-to-prefix-special-rank-nonperiodic}
  Let $\AlphabetSize, N \in \Z_{\geq 2}$ be such that $\AlphabetSize < N^{1/7}$
  and let $\tau = \lfloor \tfrac{1}{7}\log_{\AlphabetSize} N \rfloor \geq 1$.
  Consider the word RAM model with word size $w$ satisfying
  $1+\log N \leq w = \bigO(\log N)$.
  Assume that there exists a data structure answering prefix special rank
  queries that, given $N$, $\AlphabetSize$, $m$, and the packed sequence representation
  $\PackedSeqRepresentation{w}{\AlphabetSize}{W}$
  (\cref{def:packed-sequence-representation}) of any sequence
  $W[1 \dd m]$ of $m \geq \AlphabetSize$ nonempty strings
  of common length $\ell = \lfloor \tfrac{1}{2} \log_{\AlphabetSize} m \rfloor$
  over alphabet $\IntegerAlphabet$, where $m\ell\leq\lfloor N/8\rfloor$,
  achieves the following complexities:
  \begin{itemize}
  \item space usage $S(\AlphabetSize,N)$ bits,
  \item preprocessing time $P_t(\AlphabetSize,N)$,
  \item preprocessing space $P_s(\AlphabetSize,N)$ bits,
  \item query time $Q(\AlphabetSize,N)$.
  \end{itemize}
  Then, there exists a data structure that, given $N$, $\AlphabetSize$, and the packed representation
  $\PackedRepresentation{w}{\AlphabetSize}{\Text}$ (\cref{def:packed-representation})
  of any string $\Text \in \IntegerAlphabet^{N}$ such that $\Text[N]$ does not occur in $\Text[1 \dd N)$,
  answers queries that, given any $j \in [1 \dd N] \setminus \RTwo{\tau}{\Text}$ (\cref{def:periodic-positions}),
  return $\ISA{\Text}[j]$, and achieves the following complexities:
  \begin{itemize}
  \item space usage $\bigO(N \log \AlphabetSize + S(\AlphabetSize,N))$ bits,
  \item preprocessing time $\bigO(N / \log_{\AlphabetSize} N + P_t(\AlphabetSize,N))$,
  \item preprocessing space $\bigO(N \log \AlphabetSize + P_s(\AlphabetSize,N))$ bits,
  \item query time $\bigO(1 + Q(\AlphabetSize,N))$.
  \end{itemize}
\end{proposition}
\begin{proof}
  In the full version of~\cite{breaking}, this reduction is part of a
  joint reduction for suffix array and inverse suffix array queries, and
  it is presented together with a concrete prefix-rank data structure.
  Here, we state it for an arbitrary prefix-special-rank data structure
  as in the claim, which we apply to a constant number of equal-size
  blocks of strings, and we adapt the construction to word sizes as small as
  $1+\log N$. See \cref{pr:nonperiodic-isa} in
  Appendix~\ref{sec:appendix-isa} for the complete construction and
  proof.
\end{proof}

\subsubsection{The Periodic Positions}\label{sec:reduce-isa-to-prefix-special-rank-periodic}

\begin{proposition}[{\cite{breaking}}]\label{pr:reduce-small-sum-range-count-to-periodic-isa-queries}
  Let $\AlphabetSize,N\in\Z_{\geq2}$ satisfy
  $\AlphabetSize<N^{1/7}$, and let
  $\tau=\lfloor\tfrac17\log_{\AlphabetSize}N\rfloor\geq1$.
  Consider the word RAM model with word size
  $w\geq1+\log N$.
  Assume that there exists a data structure answering range counting
  queries (\cref{def:range-count}) that, given $N$, integers $m\in[2\dd N]$
  and $m'\in[0\dd m]$, and an array $A[1\dd m']$ of positive integers satisfying
  $\sum_{i=1}^{m'}A[i]\leq N$ and
  $\sum_{i=1}^{m'}A[i]\leq m\log m$,
  achieves the following complexities:
  \begin{itemize}
  \item space usage $S(m,N)$ words,
  \item preprocessing time $P_t(m,N)$,
  \item preprocessing space $P_s(m,N)$ words,
  \item query time $Q(m,N)$.
  \end{itemize}
  Then, there exists $m=\Theta(N/\log_{\AlphabetSize}N)$ with $m\leq N$
  and a data structure that, given $N$, $\AlphabetSize$, and the packed representation
  $\PackedRepresentation{w}{\AlphabetSize}{\Text}$
  (\cref{def:packed-representation}) of any text
  $\Text\in\IntegerAlphabet^N$, answers queries that, given
  $j\in\RTwo{\tau}{\Text}$ (\cref{def:periodic-positions}), return
  $\ISA{\Text}[j]$ (\cref{def:inverse-suffix-array}), and achieves:
  \begin{itemize}
  \item space usage $\bigO(N/\log_{\AlphabetSize}N+S(m,N))$ words,
  \item preprocessing time $\bigO(N/\log_{\AlphabetSize}N+P_t(m,N))$,
  \item preprocessing space $\bigO(N/\log_{\AlphabetSize}N+P_s(m,N))$ words,
  \item query time $\bigO(1+Q(m,N))$.
  \end{itemize}
\end{proposition}
\begin{proof}
  In~\cite{breaking}, this reduction is presented together with a
  concrete range-counting data structure for arrays with bounded sum.
  Here, we state it for an arbitrary range-counting data structure as in
  the claim, and we adapt the construction to word sizes as small as
  $1+\log N$. See \cref{pr:periodic-isa} in
  Appendix~\ref{sec:appendix-periodic-isa} for the complete construction
  and proof.
\end{proof}

\begin{proposition}\label{pr:periodic-pos-optimal-isa}
  Let $\AlphabetSize, N \in \Z_{\geq 2}$ be such that $\AlphabetSize < N^{1/7}$
  and let $\tau = \lfloor \tfrac{1}{7}\log_{\AlphabetSize} N \rfloor \geq 1$.
  Consider the word RAM model
  with word size $w\geq1+\log N$.
  Given $N$, $\AlphabetSize$, and the packed representation
  $\PackedRepresentation{w}{\AlphabetSize}{\Text}$ (\cref{def:packed-representation})
  of any string $\Text \in \IntegerAlphabet^{N}$,
  we can in $\bigO(N / \log_{\AlphabetSize} N)$ time and using
  $\bigO(N/\log_{\AlphabetSize}N)$ words of space construct a data structure
  of size $\bigO(N / \log_{\AlphabetSize} N)$ words that, given
  any $j \in \RTwo{\tau}{\Text}$ (\cref{def:periodic-positions}),
  returns $\ISA{\Text}[j]$ (\cref{def:inverse-suffix-array}) in $\bigO(1)$ time.
\end{proposition}
\begin{proof}
  By \cref{th:range-count-for-small-sum} with $c=1$, the range-counting
  structure required by \cref{pr:reduce-small-sum-range-count-to-periodic-isa-queries} exists with
  $S(m,N)=P_t(m,N)=P_s(m,N)=\bigO(m)$ and $Q(m,N)=\bigO(1)$.
  Applying that proposition gives $m=\Theta(N/\log_{\AlphabetSize}N)$
  and a data structure with the following complexities:
  \begin{itemize}
  \item space usage
    $\bigO(N/\log_{\AlphabetSize}N+m)=\bigO(N/\log_{\AlphabetSize}N)$ words,
  \item preprocessing time
    $\bigO(N/\log_{\AlphabetSize}N+m)=\bigO(N/\log_{\AlphabetSize}N)$,
  \item preprocessing space
    $\bigO(N/\log_{\AlphabetSize}N+m)=\bigO(N/\log_{\AlphabetSize}N)$ words,
  \item query time $\bigO(1+Q(m,N))=\bigO(1)$.
    \qedhere
  \end{itemize}
\end{proof}

\subsubsection{The Final Reduction}\label{sec:reduce-isa-to-prefix-special-rank-final-reduction}

\begin{proposition}\label{pr:reduce-isa-to-prefix-special-rank-for-nearly-all-alphabet-sizes-with-dollar}
  Let $\AlphabetSize, N \in \Z_{\geq 2}$ be such that $\AlphabetSize < N^{1/7}$.
  Consider the word RAM model with word size $w$ satisfying
  $1+\log N \leq w = \bigO(\log N)$.
  Assume that there exists a data structure answering prefix special rank
  queries that, given $N$, $\AlphabetSize$, $m$, and the packed sequence representation
  $\PackedSeqRepresentation{w}{\AlphabetSize}{W}$
  (\cref{def:packed-sequence-representation}) of a sequence
  $W[1 \dd m]$ of $m \geq \AlphabetSize$ nonempty strings
  of common length $\ell = \lfloor \tfrac{1}{2} \log_{\AlphabetSize} m \rfloor$
  over alphabet $\IntegerAlphabet$, where $m\ell\leq\lfloor N/8\rfloor$,
  achieves the following complexities:
  \begin{itemize}
  \item space usage $S(\AlphabetSize,N)$ bits,
  \item preprocessing time $P_t(\AlphabetSize,N)$,
  \item preprocessing space $P_s(\AlphabetSize,N)$ bits,
  \item query time $Q(\AlphabetSize,N)$.
  \end{itemize}
  Then, there exists a data structure answering inverse suffix array queries
  (\cref{def:inverse-suffix-array}) that, given $N$, $\AlphabetSize$, and the packed representation $\PackedRepresentation{w}{\AlphabetSize}{\Text}$ of any text $\Text \in [0 \dd \AlphabetSize)^{N}$
  such that $\Text[N]$ does not appear in $\Text[1 \dd N)$, achieves:
  \begin{itemize}
  \item space usage $\bigO(N \log \AlphabetSize + S(\AlphabetSize,N))$ bits,
  \item preprocessing time $\bigO(N / \log_{\AlphabetSize} N + P_t(\AlphabetSize,N))$,
  \item preprocessing space $\bigO(N \log \AlphabetSize + P_s(\AlphabetSize,N))$ bits,
  \item query time $\bigO(1 + Q(\AlphabetSize,N))$.
  \end{itemize}
\end{proposition}
\begin{proof}

  Let $\Text \in \IntegerAlphabet^{N}$ be any text such that $\Text[N]$ does not occur in $\Text[1 \dd N)$.
  Let $\tau = \lfloor \tfrac{1}{7}\log_{\AlphabetSize} N \rfloor \geq 1$.
  In particular, $\tau\leq\tfrac16\log_{\AlphabetSize}N$.

  \DSComponents
  The data structure consists of the following components:
  \begin{enumerate}

  \item The integer $\tau$ and the data structure from \cref{pr:isa-core}.

  \item The data structure from \cref{pr:reduce-isa-to-prefix-special-rank-nonperiodic}.
    It uses $\bigO(N \log \AlphabetSize + S(\AlphabetSize,N))$ bits of space.

  \item The data structure from \cref{pr:periodic-pos-optimal-isa}.
  \end{enumerate}

  By \cref{pr:isa-core,pr:periodic-pos-optimal-isa}, the first and third
  components each use $\bigO(N/\log_{\AlphabetSize}N)$ words, i.e.,
  $\bigO(N\log\AlphabetSize)$ bits of space, because $w=\bigO(\log N)$.
  Thus, the
  data structure uses
  $\bigO(N \log \AlphabetSize + S(\AlphabetSize,N))$ bits of space in total.

  \DSQueries
  Let $j \in [1 \dd N]$. Given the index $j$, we compute $\ISA{\Text}[j]$ as follows:
  \begin{enumerate}

  \item Query the structure from \cref{pr:isa-core} with $(j,\tau)$
    to determine in $\bigO(1)$ time whether
    $j\in\RTwo{\tau}{\Text}$ (\cref{def:periodic-positions}).

  \item We consider two cases:
    \begin{itemize}

      \item If $j \in \RTwo{\tau}{\Text}$, then we
      compute $\ISA{\Text}[j]$ using \cref{pr:periodic-pos-optimal-isa}
      in $\bigO(1)$ time.

    \item Otherwise (i.e., if $j \in [1 \dd N] \setminus \RTwo{\tau}{\Text}$),
      we compute $\ISA{\Text}[j]$ using the
      data structure from \cref{pr:reduce-isa-to-prefix-special-rank-nonperiodic}
      in $\bigO(1 + Q(\AlphabetSize,N))$ time.
    \end{itemize}
  \end{enumerate}

  In total, the query takes $\bigO(1 + Q(\AlphabetSize,N))$ time.

  \DSConstruction
  Given $N$, $\AlphabetSize$, and the packed representation
  $\PackedRepresentation{w}{\AlphabetSize}{\Text}$ of $\Text$, we construct the
  components of the above data structure as follows:
  \begin{enumerate}

  \item Compute and store $\tau$ in $\bigO(\log N)$ time.
    Construct the structure from \cref{pr:isa-core} with
    $\PackedRepresentation{w}{\AlphabetSize}{\Text}$. This takes
    $\bigO(N/\log_{\AlphabetSize}N)$ time.

  \item The second component is constructed in
    $\bigO(N / \log_{\AlphabetSize} N + P_t(\AlphabetSize,N))$ time and using
    $\bigO(N \log \AlphabetSize + P_s(\AlphabetSize,N))$ bits of space using \cref{pr:reduce-isa-to-prefix-special-rank-nonperiodic}.

  \item The third component is constructed in
    $\bigO(N / \log_{\AlphabetSize} N)$ time using \cref{pr:periodic-pos-optimal-isa}.
  \end{enumerate}

  All three constructions use size parameter $N$ and alphabet size
  $\AlphabetSize$.
  In total, the construction takes
  $\bigO(N / \log_{\AlphabetSize} N + P_t(\AlphabetSize,N))$ time and uses
  $\bigO(N \log \AlphabetSize + P_s(\AlphabetSize,N))$ bits of space.
\end{proof}

\begin{theorem}\label{th:reduce-isa-to-prefix-special-rank-for-nearly-all-alphabet-sizes}
  Let $\AlphabetSize, N \in \Z_{\geq 2}$ be such that $\AlphabetSize < N^{1/17}$.
  Consider the word RAM model with word size $w$ satisfying
  $1+\log N \leq w = \bigO(\log N)$.
  Assume that there exists a data structure for
  \probname{Indexing}
  \probname{for}
  \probname{Prefix}
  \probname{Special}
  \probname{Rank}
  \probname{Queries}
  \probname{over}
  \probname{Alphabet}
  $[0 \dd \AlphabetSize)$
  that, given $N$, $\AlphabetSize$, and a valid input of at most $N$ symbols over alphabet
  $\IntegerAlphabet$,\footnote{\label{footnote:prefix-special-rank-input} That is, the integer $\Seqlen$ and the
  packed sequence representation
  $\PackedSeqRepresentation{w}{\AlphabetSize}{W}$
  (\cref{def:packed-sequence-representation}) of a sequence
  $W[1 \dd \Seqlen]$ of $\Seqlen \geq \AlphabetSize$ strings
  of length $\ell = \lfloor \log_{\AlphabetSize} \Seqlen \rfloor$, where $\Seqlen \ell \leq N$;
  see \cref{sec:prefix-special-rank-and-isa-problem-def}.} achieves:
  \begin{itemize}
  \item space usage $S(\AlphabetSize,N)$ bits,
  \item preprocessing time $P_t(\AlphabetSize,N)$,
  \item preprocessing space $P_s(\AlphabetSize,N)$ bits,
  \item query time $Q(\AlphabetSize,N)$.
  \end{itemize}
  Then, there exists a data structure for
  \probname{Indexing}
  \probname{for}
  \probname{Inverse}
  \probname{Suffix}
  \probname{Array}
  \probname{Queries}
  \probname{over}
  \probname{Alphabet}
  $[0 \dd \AlphabetSize)$
  that, given $N$, $\AlphabetSize$, and a valid input of at most $N$ symbols over alphabet
  $\IntegerAlphabet$,\footnote{\label{footnote:isa-input} That is, the length $\Textlen$ and the
  packed representation $\PackedRepresentation{w}{\AlphabetSize}{\Text}$ of $\Text \in \IntegerAlphabet^{\Textlen}$ with $\AlphabetSize \leq \Textlen \leq N$;
  see \cref{sec:prefix-special-rank-and-isa-problem-def}.} achieves:
  \begin{itemize}
  \item space usage $\bigO(N \log \AlphabetSize + S(\AlphabetSize,N))$ bits,
  \item preprocessing time $\bigO(N / \log_{\AlphabetSize} N + P_t(\AlphabetSize,N))$,
  \item preprocessing space $\bigO(N \log \AlphabetSize + P_s(\AlphabetSize,N))$ bits,
  \item query time $\bigO(1 + Q(\AlphabetSize,N))$.
  \end{itemize}
\end{theorem}
\begin{proof}
  Denote $\hat{\AlphabetSize}=\AlphabetSize+2$.
  Let $D$ denote the prefix-special-rank structure from the claim.
  Applying \cref{th:general-prefix-special-rank-alphabet-reduction}
  to $D$ with source alphabet size $\hat{\AlphabetSize}$ and target
  alphabet size $\AlphabetSize$ yields a structure $D'$ over
  alphabet $[0\dd\hat{\AlphabetSize})$.
  Since $\AlphabetSize<\hat{\AlphabetSize}\leq\AlphabetSize^2$,
  we have $k=\lceil\log_{\AlphabetSize}\hat{\AlphabetSize}\rceil=2$.
  The structure $D'$ answers prefix special rank queries on sequences
  $W'[1\dd m']$ of
  $m'\geq\hat{\AlphabetSize}$ nonempty strings of common length
  $\ell'=\lfloor\tfrac12\log_{\hat{\AlphabetSize}}m'\rfloor$
  over alphabet $[0\dd\hat{\AlphabetSize})$, where
  $m'\ell'\leq\lfloor N/(4k)\rfloor=\lfloor N/8\rfloor$.
  Given any $i'\in[1\dd m']$ and $p'\in[0\dd\ell']$, it returns
  $\PrefixSpecialRank{W'}{i'}{p'}$.
  Its construction obtains the target alphabet size as
  $\AlphabetSize=\hat{\AlphabetSize}-2$.
  It achieves the following complexities:
  \begin{itemize}
  \item space usage
    $\bigO(N\log\AlphabetSize+S(\AlphabetSize,N))$ bits,
  \item preprocessing time
    $\bigO(N/\log_{\AlphabetSize}N+P_t(\AlphabetSize,N))$,
  \item preprocessing space
    $\bigO(N\log\AlphabetSize+P_s(\AlphabetSize,N))$ bits,
  \item query time $\bigO(1+Q(\AlphabetSize,N))$.
  \end{itemize}

  Consider a text $\Text\in\IntegerAlphabet^{\Textlen}$ with
  $\AlphabetSize\leq\Textlen\leq N$, and denote
  $t=\Text[\Textlen]$ and $\delta=N-\Textlen$.
  To make the final symbol unique without increasing the text length,
  we use the map $f:[0\dd\AlphabetSize)\to[1\dd\hat{\AlphabetSize})$
  defined by
  \[
    f(a)=
    \begin{cases}
      a+1 & \text{if }a<t,\\
      a+2 & \text{if }a\geq t.
    \end{cases}
  \]
  Let $U\in[1\dd\hat{\AlphabetSize})^{\Textlen}$ denote the string
  satisfying $U[i]=f(\Text[i])$ for every $i\in[1\dd\Textlen)$ and
  $U[\Textlen]=t+1$, and let
  $\Text_{\rm aux}=\zero^\delta U\in[0\dd\hat{\AlphabetSize})^N$.
  The last symbol $\Text_{\rm aux}[N]=t+1$ is unique because
  $t+1\ne\zero$ and $f(a)\ne t+1$ for every $a\in[0\dd\AlphabetSize)$.
  Moreover, $\hat{\AlphabetSize}^7\leq\AlphabetSize^{14}<N$.
  Thus, $D'$ provides the prefix special rank queries required by
  \cref{pr:reduce-isa-to-prefix-special-rank-for-nearly-all-alphabet-sizes-with-dollar}
  for $\Text_{\rm aux}$, with alphabet size $\hat{\AlphabetSize}$ and
  size parameter $N$.

  \DSComponents
  We store the integer $\delta$ together with the data structure from
  \cref{pr:reduce-isa-to-prefix-special-rank-for-nearly-all-alphabet-sizes-with-dollar}
  for $\Text_{\rm aux}$, using $D'$ for prefix special rank queries.
  The bounds for the two reductions, together with
  $\log\hat{\AlphabetSize}=\Theta(\log\AlphabetSize)$, give a total
  space usage of $\bigO(N\log\AlphabetSize+S(\AlphabetSize,N))$ bits.

  \DSQueries
  We first show that $\ISA{U}[j]=\ISA{\Text}[j]$ for every
  $j\in[1\dd\Textlen]$. Consider two distinct suffixes of $\Text$.
  \begin{itemize}
  \item If neither suffix is a prefix of the other, their order is
    determined by the first unequal symbols. The map $f$ is strictly
    increasing, and replacing $f(t)=t+2$ at the final position by $t+1$
    does not change its order relative to $f(a)$ for any $a\ne t$.
    Hence, the corresponding suffixes of $U$ have the same order.
  \item If one suffix is a proper prefix of the other, its final symbol
    $t$ is compared with a nonfinal occurrence of $t$ in the longer suffix.
    These symbols become $t+1$ and $t+2$, respectively, whereas all
    preceding symbols are mapped identically. Thus, the shorter suffix
    remains lexicographically smaller.
  \end{itemize}
  All symbols of $U$ are positive, and hence the $\delta$ suffixes of
  $\Text_{\rm aux}$ starting in $[1\dd\delta]$ precede all suffixes
  starting in $[\delta+1\dd N]$. Consequently, for every
  $j\in[1\dd\Textlen]$, it holds
  \[
    \ISA{\Text_{\rm aux}}[\delta+j]
      =\delta+\ISA{U}[j]=\delta+\ISA{\Text}[j].
  \]
  Given $j$, we return $\ISA{\Text_{\rm aux}}[\delta+j]-\delta$ using
  the stored structure in $\bigO(1+Q(\AlphabetSize,N))$ time.

  \DSConstruction
  Given $N$, $\AlphabetSize$, $\Textlen$, and
  $\PackedRepresentation{w}{\AlphabetSize}{\Text}$,
  the construction proceeds in two steps:
  \begin{enumerate}

  \item First, we construct
    $\PackedRepresentation{w}{\hat{\AlphabetSize}}{\Text_{\rm aux}}$
    as follows:
    \begin{enumerate}

    \item Compute and store $\delta=N-\Textlen$.
      Compute $b=\lceil\log\AlphabetSize\rceil$ and
      $\hat b=\lceil\log\hat{\AlphabetSize}\rceil$ in $\bigO(\log N)$
      time, and retrieve $t=\Text[\Textlen]$ using
      \cref{pr:packed-representation}\eqref{pr:packed-representation-access}.
      Construct the structure from \cref{th:alphabet-lifting}.
      Query it with $\Textlen$ and
      $\PackedRepresentation{w}{\AlphabetSize}{\Text}$ to obtain
      $\PackedRepresentation{w}{\hat{\AlphabetSize}}{\Text}$.

    \item Using
      \cref{pr:packed-representation}\eqref{pr:packed-representation-initialize},
      construct an array $A[0\dd\hat{\AlphabetSize})$ containing the
      packed representations of the one-symbol strings
      $f(0),\ldots,f(\AlphabetSize-1),\zero,\zero$, in this order,
      over alphabet $[0\dd\hat{\AlphabetSize})$.
      Apply \cref{th:packed-morphism} to $A$ with $d=1$.
      Query the resulting structure with $\Textlen$ and
      $\PackedRepresentation{w}{\hat{\AlphabetSize}}{\Text}$.
      In the returned representation, replace the last symbol by $t+1$
      using \cref{pr:packed-representation}\eqref{pr:packed-representation-update}
      to obtain $\PackedRepresentation{w}{\hat{\AlphabetSize}}{U}$.

    \item Initialize the packed representation of $\zero^\delta$ over
      $[0\dd\hat{\AlphabetSize})$ using
      \cref{pr:packed-representation}\eqref{pr:packed-representation-initialize}.
      Concatenate it with $\PackedRepresentation{w}{\hat{\AlphabetSize}}{U}$
      using \cref{pr:packed-representation}\eqref{pr:packed-representation-concat}
      to obtain
      $\PackedRepresentation{w}{\hat{\AlphabetSize}}{\Text_{\rm aux}}$.
    \end{enumerate}

  \item Apply
    \cref{pr:reduce-isa-to-prefix-special-rank-for-nearly-all-alphabet-sizes-with-dollar}
    to $\Text_{\rm aux}$, using $D'$ for prefix special rank queries.
    Store the resulting inverse-suffix-array structure.
  \end{enumerate}

  The packed operations use length bound $N$ and alphabet sizes at most
  $\hat{\AlphabetSize}\leq N$, which satisfy their word-size assumptions.
  The lifting structure uses size parameter $N$, source alphabet size
  $\AlphabetSize$, and target alphabet size $\hat{\AlphabetSize}$, and
  the morphism structure uses size parameter $N$ and alphabet size
  $\hat{\AlphabetSize}$.
  Since $\hat{\AlphabetSize}<N^{1/7}$ and
  $\log\hat{\AlphabetSize}=\Theta(\log\AlphabetSize)$, the first step takes
  $\bigO(\sqrt N+N/\log_{\hat{\AlphabetSize}}N)
    =\bigO(N/\log_{\AlphabetSize}N)$ time.
  Since $w=\bigO(\log N)$, these operations use
  $\bigO(N\log\AlphabetSize)$ bits of space.
  The bounds for $D'$ and
  \cref{pr:reduce-isa-to-prefix-special-rank-for-nearly-all-alphabet-sizes-with-dollar},
  together with $\log\hat{\AlphabetSize}=\Theta(\log\AlphabetSize)$,
  imply that the second step takes
  $\bigO(N/\log_{\AlphabetSize}N+P_t(\AlphabetSize,N))$ time and uses
  $\bigO(N\log\AlphabetSize+P_s(\AlphabetSize,N))$ bits of space.
  These are also the total construction bounds.
\end{proof}

\begin{theorem}\label{th:reduce-isa-to-prefix-special-rank-for-all-alphabet-sizes}
  Let $\AlphabetSize, N \in \Z_{\geq 2}$ be such that $\AlphabetSize \leq N$.
  Consider the word RAM model with word size $w$ satisfying
  $1+\log N \leq w = \bigO(\log N)$.
  Assume that there exists a data structure for
  \probname{Indexing}
  \probname{for}
  \probname{Prefix}
  \probname{Special}
  \probname{Rank}
  \probname{Queries}
  \probname{over}
  \probname{Alphabet}
  $[0 \dd \AlphabetSize)$
  that, given $N$, $\AlphabetSize$, and a valid input of at most $N$ symbols over alphabet
  $\IntegerAlphabet$,\footref{footnote:prefix-special-rank-input}
  achieves:
  \begin{itemize}
  \item space usage $S(\AlphabetSize,N)$ bits,
  \item preprocessing time $P_t(\AlphabetSize,N)$,
  \item preprocessing space $P_s(\AlphabetSize,N)$ bits,
  \item query time $Q(\AlphabetSize,N)$.
  \end{itemize}
  Then, there exists a data structure for
  \probname{Indexing}
  \probname{for}
  \probname{Inverse}
  \probname{Suffix}
  \probname{Array}
  \probname{Queries}
  \probname{over}
  \probname{Alphabet}
  $[0 \dd \AlphabetSize)$
  that, given $N$, $\AlphabetSize$, and a valid input of at most $N$ symbols over alphabet
  $\IntegerAlphabet$,\footref{footnote:isa-input}
  achieves:
  \begin{itemize}
  \item space usage $\bigO(N \log \AlphabetSize + S(\AlphabetSize,N))$ bits,
  \item preprocessing time $\bigO(N / \log_{\AlphabetSize} N + P_t(\AlphabetSize,N))$,
  \item preprocessing space $\bigO(N \log \AlphabetSize + P_s(\AlphabetSize,N))$ bits,
  \item query time $\bigO(1 + Q(\AlphabetSize,N))$.
  \end{itemize}
\end{theorem}
\begin{proof}
  For $\AlphabetSize<N^{1/17}$, the claim follows from
  \cref{th:reduce-isa-to-prefix-special-rank-for-nearly-all-alphabet-sizes}.
  Assume now that $\AlphabetSize\geq N^{1/17}$.
  Since $\AlphabetSize\leq N$, we have $\log\AlphabetSize=\Theta(\log N)$.
  Consider a string $\Text\in\IntegerAlphabet^+$ with
  $\AlphabetSize\leq|\Text|\leq N$.

  \DSComponents
  We store the array $\ISA{\Text}$ in $|\Text|$ words.
  It occupies $\bigO(Nw)=\bigO(N\log\AlphabetSize)$ bits.

  \DSQueries
  Given $j\in[1\dd|\Text|]$, we return $\ISA{\Text}[j]$ in $\bigO(1)$ time.

  \DSConstruction
  First, we compute $b=\lceil\log\AlphabetSize\rceil$ in $\bigO(\log N)$ time.
  We then use \cref{pr:packed-representation}\eqref{pr:packed-representation-access}
  to construct an array $A[1\dd|\Text|]$ with $A[i]=\Text[i]$ for every
  $i\in[1\dd|\Text|]$ in $\bigO(|\Text|)$ time.
  From this array, \cref{th:sa-and-isa-construction} computes
  $\ISA{\Text}$ in $\bigO(|\Text|)$ time.

  In total, construction takes $\bigO(N)=\bigO(N/\log_{\AlphabetSize}N)$ time
  and uses $\bigO(N)$ words, or $\bigO(N\log\AlphabetSize)$ bits of space.
  Together with the space and query bounds above, this proves the claim.
\end{proof}

\begin{theorem}\label{th:reduce-isa-to-prefix-special-rank-for-all-alphabet-sizes-clean}
  Let $\AlphabetSize, N \in \Z_{\geq 2}$ be such that $\AlphabetSize \leq N$.
  Consider the word RAM model with word size $w$ satisfying
  $1+\log N \leq w = \bigO(\log N)$.
  Assume that there exists a data structure for
  \probname{Indexing}
  \probname{for}
  \probname{Prefix}
  \probname{Special}
  \probname{Rank}
  \probname{Queries}
  \probname{over}
  \probname{Alphabet}
  $[0 \dd \AlphabetSize)$
  that, given $N$, $\AlphabetSize$, and a valid input of at most $N$ symbols over alphabet
  $\IntegerAlphabet$,\footref{footnote:prefix-special-rank-input}
  achieves:
  \begin{itemize}
  \item space usage $S(\AlphabetSize,N)$ bits,
  \item preprocessing time $P_t(\AlphabetSize,N)$,
  \item preprocessing space $P_s(\AlphabetSize,N)$ bits,
  \item query time $Q(\AlphabetSize,N)$.
  \end{itemize}
  Then, there exists a data structure for
  \probname{Indexing}
  \probname{for}
  \probname{Inverse}
  \probname{Suffix}
  \probname{Array}
  \probname{Queries}
  \probname{over}
  \probname{Alphabet}
  $[0 \dd \AlphabetSize)$
  that, given $N$, $\AlphabetSize$, and a valid input of at most $N$ symbols over alphabet
  $\IntegerAlphabet$,\footref{footnote:isa-input}
  achieves:
  \begin{itemize}
  \item space usage $\bigO(S(\AlphabetSize,N))$ bits,
  \item preprocessing time $\bigO(P_t(\AlphabetSize,N))$,
  \item preprocessing space $\bigO(P_s(\AlphabetSize,N))$ bits,
  \item query time $\bigO(Q(\AlphabetSize,N))$.
  \end{itemize}
\end{theorem}
\begin{proof}
  It suffices to remove the additive terms in
  \cref{th:reduce-isa-to-prefix-special-rank-for-all-alphabet-sizes}.
  We use \cref{th:prefix-special-rank-bounded-total-length-space-lower-bound}
  with $c=1$. The theorem applies: the structure from the claim supports every
  sequence considered there, because
  $\lfloor\log_{\AlphabetSize}\Seqlen\rfloor\geq1$ implies
  $\Seqlen\geq\AlphabetSize$, and the existence assumption of the
  theorem holds because $\Seqlen=\AlphabetSize$ gives
  $\ell=\lfloor\log_{\AlphabetSize}\Seqlen\rfloor=1$ and
  $\Seqlen\ell=\AlphabetSize\leq N$.
  It follows that the prefix-special-rank structure from the claim requires
  $\Omega(N\log\AlphabetSize)$ bits on some valid input, and hence
  $S(\AlphabetSize,N)=\Omega(N\log\AlphabetSize)$.

  On this input, writing the output structure takes
  $\Omega((N\log\AlphabetSize)/w)=\Omega(N/\log_{\AlphabetSize}N)$ time,
  and its storage requires $\Omega(N\log\AlphabetSize)$ bits of
  preprocessing space (\cref{rm:space}).
  Therefore $P_t(\AlphabetSize,N)=\Omega(N/\log_{\AlphabetSize}N)$ and
  $P_s(\AlphabetSize,N)=\Omega(N\log\AlphabetSize)$.
  These bounds, together with $Q(\AlphabetSize,N)=\Omega(1)$,
  imply the claim.
\end{proof}

%% file: prefix-special-rank/summary.tex
\Needspace{7\baselineskip}
\subsection{Summary}\label{sec:prefix-special-rank-and-isa-summary}

\begin{theorem}[Perfect equivalence of prefix special rank and inverse suffix array queries for every alphabet size]
  \label{th:equivalence-of-prefix-special-rank-and-isa-summary}
  Let $\AlphabetSize, N \in \Z_{\geq 2}$ be such that $\AlphabetSize \leq N$.
  In the word RAM model with word size $w$ satisfying
  $1+\log N \leq w = \bigO(\log N)$, the problems of
  \probname{Indexing}
  \probname{for}
  \probname{Prefix}
  \probname{Special}
  \probname{Rank}
  \probname{Queries}
  \probname{over}
  \probname{Alphabet}
  $[0 \dd \AlphabetSize)$ and
  \probname{Indexing}
  \probname{for}
  \probname{Inverse}
  \probname{Suffix}
  \probname{Array}
  \probname{Queries}
  \probname{over}
  \probname{Alphabet}
  $[0 \dd \AlphabetSize)$
  are asymptotically equivalent:
  If there exists a data structure for one of these two problems that, given
  $N$, $\AlphabetSize$, and a valid input of at most $N$ symbols over alphabet
  $\IntegerAlphabet$,\footnote{For the problem of
  \probname{Indexing}
  \probname{for}
  \probname{Prefix}
  \probname{Special}
  \probname{Rank}
  \probname{Queries}
  \probname{over}
  \probname{Alphabet}
  $[0 \dd \AlphabetSize)$,
    a valid input consists of the integer $m$ and the packed sequence representation
    $\PackedSeqRepresentation{w}{\AlphabetSize}{W}$
    (\cref{def:packed-sequence-representation})
    of a sequence $W[1 \dd m]$ of $m \geq \AlphabetSize$ nonempty strings of common length
  $\ell = \lfloor \log_{\AlphabetSize} m \rfloor$ such
  that $m \cdot \ell \leq N$. Given any $i \in [1 \dd m]$ and any $p \in [0 \dd \ell]$, the query
  returns the value $\PrefixSpecialRank{W}{i}{p}$ (\cref{def:prefix-rank-and-select}), i.e., the
  number of strings in $W[1 \dd i]$ having $X = W[i][1 \dd p]$ as a prefix.
  In
  \probname{Indexing}
  \probname{for}
  \probname{Inverse}
  \probname{Suffix}
  \probname{Array}
  \probname{Queries}
  \probname{over}
  \probname{Alphabet}
  $[0 \dd \AlphabetSize)$,
  a valid input consists of the length $\Textlen$ and the packed representation
  $\PackedRepresentation{w}{\AlphabetSize}{\Text}$ of a string $\Text \in \IntegerAlphabet^{\Textlen}$ such that
  $\AlphabetSize \leq \Textlen \leq N$. Given any $j \in [1 \dd \Textlen]$, the query returns the value $\ISA{\Text}[j]$ (\cref{def:inverse-suffix-array}), i.e.,
  the position of the value $j$ in the suffix array of $\Text$.} achieves:
  \begin{itemize}
  \item space usage of $S(\AlphabetSize,N)$ bits,
  \item preprocessing time $P_t(\AlphabetSize,N)$,
  \item preprocessing space $P_s(\AlphabetSize,N)$ bits,
  \item query time $Q(\AlphabetSize,N)$,
  \end{itemize}
  then there exists a data structure for the other problem that, given
  $N$, $\AlphabetSize$, and a valid input of at most $N$ symbols over alphabet $\IntegerAlphabet$, achieves:
  \begin{itemize}
  \item space usage of $\bigO(S(\AlphabetSize,N))$ bits,
  \item preprocessing time $\bigO(P_t(\AlphabetSize,N))$,
  \item preprocessing space $\bigO(P_s(\AlphabetSize,N))$ bits,
  \item query time $\bigO(Q(\AlphabetSize,N))$.
  \end{itemize}
\end{theorem}
\begin{proof}
  The result follows by combining
  \cref{th:reduce-prefix-special-rank-to-isa-for-all-alphabet-sizes-clean} and
  \cref{th:reduce-isa-to-prefix-special-rank-for-all-alphabet-sizes-clean}.
\end{proof}

%% file: acknowledgements.tex
\section*{Declarations}

We used OpenAI's GPT 6 Astra and GPT 6.1 Sol, and Anthropic's
Opus 5.5 for proofreading, checking proofs and citations,
and preparing the LaTeX sources.
The authors assume responsibility for all content.

For Appendix~\ref{app:stronger-lower-bound}, we also used OpenAI's
GPT-5.4 through the Chat and Codex interfaces to explore stronger bounds
and simplify the proofs. The final arguments combine our ideas with
AI-generated suggestions and were revised and verified by the authors.

%% file: appendix.tex
\appendix

\section{Black-Box Formulations of Earlier Reductions}

The reductions in~\cite{breaking} from suffix array and inverse suffix
array queries use prefix queries for nonperiodic positions and bounded-sum
range queries for periodic positions. They were presented together with
concrete data structures answering these queries. In this appendix, we
revisit these constructions with two goals. First, we state the reductions
in black-box form, so that they apply to any data structures satisfying
the corresponding query interfaces. Second, we adapt their construction
algorithms to work with word sizes as small as $1+\log N$, where $N$ is
the length of the text.

\subsection{Nonperiodic Positions}\label{sec:appendix-nonperiodic}

\input{appendix/preliminaries}
\input{appendix/suffix-array}
\input{appendix/inverse-suffix-array}

\subsection{Periodic Positions}\label{sec:appendix-periodic}

\input{appendix/periodic-preliminaries}
\input{appendix/periodic-suffix-array}
\input{appendix/periodic-inverse-suffix-array}

\section{Stronger Lower Bounds for (Inverse) Suffix Array Size}\label{app:stronger-lower-bound}

\input{appendix/stronger-lower-bound}

%% file: appendix/preliminaries.tex
\subsubsection{Preliminaries}\label{sec:appendix-prelim}

\begin{definition}[Successor]\label{def:succ}
  Consider any totally ordered set $\mathcal{U}$, and let
  $A \subseteq \mathcal{U}$ be a finite nonempty subset of $\mathcal{U}$.
  For every $x \in \mathcal{U}$ satisfying $x \leq \max A$,
  we denote $\Successor{A}{x} = \min\{x' \in A : x \leq x'\}$.
\end{definition}

\begin{definition}[Distinguishing prefix of a suffix~\cite{sss,breaking}]\label{def:dist-prefixes}
  Let $\Textlen\in\Z_{\geq2}$ and $\Text \in \Sigma^{\Textlen}$.
  For $\tau \in [1 \dd \lfloor \tfrac{\Textlen}{2} \rfloor]$, let
  $\SSS$ be a $\tau$-synchronizing set of $\Text$ (\cref{def:sss}).
  For every $j \in [1 \dd \Textlen - 3\tau + 2] \setminus \RTwo{\tau}{\Text}$,
  we denote (see \cref{def:periodic-positions,def:succ})
  \[
    \DistPrefixPos{j}{\tau}{\Text}{\SSS} := \Text[j \dd \Successor{\SSS}{j} + 2\tau).
  \]
  We then let
  \[
    \DistPrefixes{\tau}{\Text}{\SSS}
      := \{\DistPrefixPos{j}{\tau}{\Text}{\SSS} : j \in [1 \dd \Textlen -
          3\tau + 2] \setminus \RTwo{\tau}{\Text}\}.
  \]
\end{definition}

\begin{remark}\label{rm:dist-prefixes}
  Note that $\Successor{\SSS}{j}$ in \cref{def:dist-prefixes}
  is well-defined for every $j \in [1 \dd \Textlen - 3\tau + 2]
  \setminus \RTwo{\tau}{\Text}$, because by
  \cref{def:sss}\eqref{def:sss-density}, for such $j$
  we have $[j \dd j + \tau) \cap \SSS \neq \emptyset$.
\end{remark}

\begin{lemma}[{\cite{sss,breaking}; see also~\cite{sublinearlz}}]\label{lm:dist-prefixes}
  Let $\Textlen\in\Z_{\geq2}$ and $\Text \in \Sigma^{\Textlen}$.
  For $\tau \in [1 \dd \lfloor \tfrac{\Textlen}{2} \rfloor]$,
  let $\SSS$ be a $\tau$-synchronizing set of $\Text$ (\cref{def:sss}). Then:
  \begin{enumerate}
  \item\label{lm:dist-prefixes-it-1}
    It holds $\DistPrefixes{\tau}{\Text}{\SSS} \subseteq \Sigma^{\leq 3\tau - 1}$
    (\cref{def:dist-prefixes}).
  \item\label{lm:dist-prefixes-it-2}
    $\DistPrefixes{\tau}{\Text}{\SSS}$ is prefix-free, i.e.,
    for $D, D' \in \DistPrefixes{\tau}{\Text}{\SSS}$,
    $D \neq D'$ implies that $D$ is not a prefix~of~$D'$.
  \end{enumerate}
\end{lemma}

\begin{lemma}[{\cite{breaking}; see also~\cite{sublinearlz}}]\label{lm:dist-prefix-existence}
  Let $\Textlen\in\Z_{\geq2}$ and $\Text \in \Sigma^{\Textlen}$.
  For $\tau \in [1 \dd \lfloor \tfrac{\Textlen}{2} \rfloor]$, let
  $\SSS$ be a $\tau$-synchronizing set of $\Text$ (\cref{def:sss}).
  Let $m\in\Z_{\geq1}$ and $\Pat \in \Sigma^{m}$ be a $\tau$-nonperiodic pattern (\cref{def:periodic-pattern}) such
  that $m \geq 3\tau - 1$ and $\OccTwo{\Pat}{\Text} \neq \emptyset$.
  Then, there exists a unique
  $D \in \DistPrefixes{\tau}{\Text}{\SSS}$ (\cref{def:dist-prefixes})
  that is a prefix of $\Pat$.
\end{lemma}

\begin{remark}\label{rm:dist-prefixes-arbitrary-text}
  Although~\cite{sublinearlz} assumes that the final symbol of $\Text$
  occurs only once, its proofs of the claims in
  \cref{lm:dist-prefixes,lm:dist-prefix-existence} do not use this assumption.
  These lemmas therefore apply to arbitrary texts.
\end{remark}

\begin{lemma}[{\cite{hierarchy}}]\label{lm:nonperiodic-occ-cyclic}
  Let $\Textlen\in\Z_{\geq2}$, $m\in\Z_{\geq1}$, $\Text \in \Sigma^{\Textlen}$,
  $\tau \in [1 \dd \lfloor \tfrac{\Textlen}{2} \rfloor]$, and let
  $\SSS$ be a $\tau$-synchronizing set of $\Text$.
  Let $D \in \DistPrefixes{\tau}{\Text}{\SSS}$ (\cref{def:dist-prefixes}),
  and let $\Pat \in \Sigma^{m}$ be a $\tau$-nonperiodic pattern
  (\cref{def:periodic-pattern}) such that no nonempty suffix of $\Text$ is a proper prefix of $\Pat$,
  and $D$ is a prefix of $\Pat$. Denote $\deltatext = |D| - 2\tau$,
  and let $A$ and $B$ be such that $\Pat = A B$ and $|A| = \deltatext$.
  Then, it holds
  \begin{align*}
    \OccTwo{\Pat}{\Text}
      &= \{s  - \deltatext : s \in \SSS,\ A\text{ is a suffix of }\Textinf[s - m \dd s),\text{ and }\\
      &\hspace{3.55cm}B\text{ is a prefix of }\Textinf[s \dd s + m)\}.
  \end{align*}
\end{lemma}

\begin{proposition}[{\cite{breaking}; see also~\cite{hierarchy}}]
\label{pr:dist-prefix-lookup}
  Let $N,\AlphabetSize\in\Z_{\geq2}$ satisfy $\AlphabetSize\leq N$.
  Consider the word RAM model with word size $w\geq1+\log N$.
  Given $N$, $\AlphabetSize$, an integer $\tau\geq1$ satisfying
  $\tau\leq\tfrac16\log_{\AlphabetSize}N$, the packed representation
  $\PackedRepresentation{w}{\AlphabetSize}{\Text}$
  (\cref{def:packed-representation}) of a text
  $\Text\in\IntegerAlphabet^N$, and a
  $\tau$-synchronizing set $\SSS$ of $\Text$ (\cref{def:sss}) specified by its
  size and an array listing its elements,
  we can in $\bigO(N/\log_{\AlphabetSize}N+|\SSS|)$ time
  and using $\bigO(N/\log_{\AlphabetSize}N+|\SSS|)$ words
  of space construct a data structure of size $\bigO(\sqrt N)$
  words that supports the following query in $\bigO(1)$ time.
  Given $|X|$ and the packed representation
  $\PackedRepresentation{w}{\AlphabetSize}{X}$ of any string
  $X\in\IntegerAlphabet^{\leq3\tau-1}$, it returns $|D|$ and
  $\PackedRepresentation{w}{\AlphabetSize}{D}$, where $D$ is the
  unique string in $\DistPrefixes{\tau}{\Text}{\SSS}$
  (\cref{def:dist-prefixes}) that is a prefix of $X$, if such a
  string exists. Otherwise, it reports that no such prefix exists.
\end{proposition}

%% file: appendix/suffix-array.tex
\subsubsection{Suffix Array Queries}
\label{sec:appendix-sa}

\begin{lemma}[Detailed version adapted
    from~\cite{breaking}]\label{lm:reduce-suffix-array-to-prefix-select}
  Let $\Textlen\in\Z_{\geq2}$, $\Text \in \Sigma^{\Textlen}$,
  $\tau \in [1 \dd \lfloor \tfrac{\Textlen}{2} \rfloor]$, and assume that
  $\Text[\Textlen]$ does not occur in $\Text[1 \dd \Textlen)$.
  Let $\SSS$ be a $\tau$-synchronizing set of $\Text$ (\cref{def:sss}).
  Denote $n_{\SSS} = |\SSS|$ and $(s_i)_{i \in [1 \dd n_{\SSS}]} = \LexSorted{\SSS}{\Text}$
  (\cref{def:lex-sorted}). Let $A_{\SSS}[1 \dd n_{\SSS}]$ and $A_{\rm str}[1 \dd n_{\SSS}]$
  be defined by
  \begin{itemize}
  \item $A_{\SSS}[i] = s_i$,
  \item $A_{\rm str}[i] = \revstr{X_i}$,
    where $X_i = \Textinf[s_i - \tau \dd s_i + 2\tau)$.
  \end{itemize}
  Let $i \in [1 \dd \Textlen]$ be such that
  $\SA{\Text}[i] \in [1 \dd \Textlen - 3\tau + 2] \setminus \RTwo{\tau}{\Text}$.
  Let $D$ be the unique string in $\DistPrefixes{\tau}{\Text}{\SSS}$
  (\cref{def:dist-prefixes}) that is a prefix of $\Text[\SA{\Text}[i] \dd \Textlen]$
  (such $D$ exists by applying \cref{lm:dist-prefix-existence}).
  Denote $b = \RangeBegTwo{D}{\Text}$ (\cref{def:occ}) and $\deltatext = |D| - 2\tau$.
  Then, it holds $1 \leq i -b \leq \PrefixRank{A_{\rm str}}{n_{\SSS}}{\revstr{D}}$
  (see \cref{def:prefix-rank-and-select}),
  and furthermore (see \cref{def:suffix-array,def:prefix-rank-and-select}):
  \[
    \SA{\Text}[i] = A_{\SSS}[\PrefixSelect{A_{\rm str}}{i-b}{\revstr{D}}] - \deltatext.
  \]
\end{lemma}
\begin{proof}
  Denote $p=\SA{\Text}[i]$. Since
  $p\in[1\dd\Textlen-3\tau+2]\setminus\RTwo{\tau}{\Text}$,
  the suffix $\Text[p\dd\Textlen]$ is a $\tau$-nonperiodic pattern of
  length at least $3\tau-1$. It occurs in $\Text$, and hence
  \cref{lm:dist-prefix-existence} gives the unique distinguishing prefix
  $D$ used in the claim. By \cref{def:dist-prefixes,def:sss}, it holds
  $2\tau\leq|D|\leq3\tau-1$, and therefore $0\leq\deltatext<\tau$.
  Let $A=D[1\dd\deltatext]$ and $B=D[\deltatext+1\dd|D|]$,
  so that $D=AB$ and $|B|=2\tau$.

  Denote $k=\PrefixRank{A_{\rm str}}{n_{\SSS}}{\revstr{D}}$ and,
  for every $t\in[1\dd k]$, let
  $a_t=\PrefixSelect{A_{\rm str}}{t}{\revstr{D}}$.
  We prove the claim in two steps.
  \begin{enumerate}
  \item\label{step:nonperiodic-sa-identity-occurrences} First, we show that
    \[
      \OccTwo{D}{\Text}
        =\{A_{\SSS}[a_t]-\deltatext:t\in[1\dd k]\}.
    \]
    We establish this equality using \cref{lm:nonperiodic-occ-cyclic}.
    \begin{enumerate}
    \item We first verify that this lemma applies with pattern $D$.
      If $|D|<3\tau-1$, then $D$ is $\tau$-nonperiodic by
      \cref{def:periodic-pattern}. If $|D|=3\tau-1$, then
      $D=\Text[p\dd p+3\tau-1)$ and
      $\per{D}>\tfrac{1}{3}\tau$, since $p\notin\RTwo{\tau}{\Text}$.
      Thus, $D$ is $\tau$-nonperiodic in either case. Moreover, no proper
      prefix of $D$ contains $\Text[\Textlen]$, because $D$ is a substring
      of $\Text$ and this symbol occurs only at position $\Textlen$ in $\Text$.
      Every nonempty suffix of $\Text$ contains $\Text[\Textlen]$.
      Consequently, no such suffix is a proper prefix of $D$.
    \item We now relate the two conditions in \cref{lm:nonperiodic-occ-cyclic}
      to prefix matches in $A_{\rm str}$.
      Let $q\in[1\dd n_{\SSS}]$ and $s=A_{\SSS}[q]$.
      Since $|D|=\deltatext+2\tau$ and $\deltatext<\tau$, the
      length-$|D|$ prefix of
      $A_{\rm str}[q]=\revstr{\Textinf[s-\tau\dd s+2\tau)}$
      is $\revstr{\Textinf[s-\deltatext\dd s+2\tau)}$.
      Hence, $\revstr{D}$ is a prefix of $A_{\rm str}[q]$ if and only if
      $D=\Textinf[s-\deltatext\dd s+2\tau)$.
      With $m=|D|$, the latter equality holds if and only if $A$ is a
      suffix of $\Textinf[s-m\dd s)$ and $B$ is a prefix of
      $\Textinf[s\dd s+m)$, because $|A|=\deltatext$ and $|B|=2\tau$.
      By the definition of prefix select, the indices $q$ satisfying
      these conditions are precisely $a_1,\ldots,a_k$.
      Applying \cref{lm:nonperiodic-occ-cyclic} with pattern $D$ and split
      $D=AB$ therefore yields the required equality.
    \end{enumerate}
  \item Next, we prove that, for every $t\in[1\dd k]$, it holds
    \[
      \SA{\Text}[b+t]=A_{\SSS}[a_t]-\deltatext.
    \]
    The positions $A_{\SSS}[a_t]$ are distinct. By Step~\ref{step:nonperiodic-sa-identity-occurrences},
    $|\OccTwo{D}{\Text}|=k\geq1$, since $D$ occurs at $p$.
    By \cref{rm:occ}, the
    suffixes beginning with $D$ have precisely the ranks
    $b+1,\ldots,b+k$. For every $t\in[1\dd k]$, Step~\ref{step:nonperiodic-sa-identity-occurrences} also
    gives $A_{\SSS}[a_t]-\deltatext\in\OccTwo{D}{\Text}$.
    Since $A$ is the length-$\deltatext$ prefix of $D$, it follows that
    $\Text[A_{\SSS}[a_t]-\deltatext\dd\Textlen]
      =A\cdot\Text[A_{\SSS}[a_t]\dd\Textlen]$.
    The definition of prefix select gives $a_1<\cdots<a_k$, and
    $A_{\SSS}$ lists the starting positions of suffixes in ascending
    lexicographic order. Prepending the same string $A$ preserves this
    order. Thus, the suffixes starting at
    $A_{\SSS}[a_1]-\deltatext,\ldots,A_{\SSS}[a_k]-\deltatext$
    are exactly the suffixes beginning with $D$, in lexicographic order.
    This proves the identity above. Finally, $D$ is a prefix of
    $\Text[\SA{\Text}[i]\dd\Textlen]$, which implies
    $i\in[b+1\dd b+k]$. Hence, $1\leq i-b\leq k$, and substituting
    $t=i-b$ and the definition of $a_t$ into the identity proves the
    suffix-array formula in the claim.\qedhere
  \end{enumerate}
\end{proof}

\begin{proposition}\label{pr:nonperiodic-sa}
  Let $\AlphabetSize, N \in \Z_{\geq 2}$ be such that $\AlphabetSize < N^{1/7}$
  and let $\tau = \lfloor \tfrac{1}{7}\log_{\AlphabetSize} N \rfloor \geq 1$.
  Consider the word RAM model with word size $w$ satisfying
  $1+\log N \leq w = \bigO(\log N)$.
  Assume that there exists a data structure that, given $N$, $\AlphabetSize$, $m$, and the packed sequence representation
  $\PackedSeqRepresentation{w}{\AlphabetSize}{W}$
  (\cref{def:packed-sequence-representation}) of any sequence
  $W[1 \dd m]$ of $m \geq \AlphabetSize$
  nonempty strings of common length
  $\ell = \lfloor \tfrac{1}{2} \log_{\AlphabetSize} m \rfloor$
  over alphabet $\IntegerAlphabet$, where $m \cdot \ell \leq \lfloor N/8\rfloor$, returns $\PrefixSelect{W}{r}{X}$ (see \cref{def:prefix-rank-and-select})
  given the length $|X|$ and the packed representation
  $\PackedRepresentation{w}{\AlphabetSize}{X}$ of any string
  $X\in\IntegerAlphabet^{\leq\ell}$ and any integer $r\in[1\dd m]$, and
  achieves the following complexities:
  \begin{itemize}
  \item space usage $S(\AlphabetSize,N)$ bits,
  \item preprocessing time $P_t(\AlphabetSize,N)$,
  \item preprocessing space $P_s(\AlphabetSize,N)$ bits,
  \item query time $Q(\AlphabetSize,N)$.
  \end{itemize}
  Then, there exists a data structure that, given $N$, $\AlphabetSize$, and the packed representation
  $\PackedRepresentation{w}{\AlphabetSize}{\Text}$
  (\cref{def:packed-representation}) of any
  string $\Text \in \IntegerAlphabet^{N}$ such that
  $\Text[N]$ does not occur in $\Text[1 \dd N)$,
  answers queries that, given any $i \in [1 \dd N]$ satisfying
  $\SA{\Text}[i] \in [1 \dd N] \setminus \RTwo{\tau}{\Text}$
  (\cref{def:suffix-array,def:periodic-positions}),
  return $\SA{\Text}[i]$, and achieves the following complexities:
  \begin{itemize}
  \item space usage $\bigO(N \log \AlphabetSize + S(\AlphabetSize,N))$ bits,
  \item preprocessing time $\bigO(N / \log_{\AlphabetSize} N + P_t(\AlphabetSize,N))$,
  \item preprocessing space $\bigO(N \log \AlphabetSize + P_s(\AlphabetSize,N))$ bits,
  \item query time $\bigO(1 + Q(\AlphabetSize,N))$.
  \end{itemize}
\end{proposition}
\begin{proof}

  Let $N_0\geq2$ be an absolute constant such that
  $\lfloor 7x/(64\log x)\rfloor\geq x^{6/7}$ for every integer
  $x\geq N_0$.
  If $N<N_0$, construct and store the suffix array directly in
  $\bigO(1)$ time and words. Otherwise, denote
  $m=\lfloor N/(64\tau)\rfloor$ and
  $\ell=\lfloor\tfrac12\log_{\AlphabetSize}m\rfloor$.
  Since $64\tau\leq\tfrac{64}{7}\log_{\AlphabetSize}N\leq\tfrac{64}{7}\log N$, we have $m\geq N^{6/7}$, and therefore
  $3\tau\leq\tfrac37\log_{\AlphabetSize}N
    \leq\tfrac12\log_{\AlphabetSize}m$.
  Since $3\tau$ is an integer, this gives $3\tau\leq\ell$.

  Let $\SSS$ be any $\tau$-synchronizing set of $\Text$ (\cref{def:sss}) satisfying
  $|\SSS| = \bigO(N/\tau)$. Such $\SSS$ exists by \cref{th:sss-existence-and-construction}.
  Denote $n_{\SSS}=|\SSS|$ and
  $(s^{\rm lex}_i)_{i\in[1\dd n_{\SSS}]}=\LexSorted{\SSS}{\Text}$
  (\cref{def:lex-sorted}).
  Let $A_{\SSS}[1 \dd n_{\SSS}]$ denote an array defined such that,
  for every $i \in [1 \dd n_{\SSS}]$, it holds $A_{\SSS}[i] = s^{\rm lex}_i$.
  Let $A_{\rm str}[1 \dd n_{\SSS}]$ be an array defined such that,
  for any $i \in [1 \dd n_{\SSS}]$, it holds $A_{\rm str}[i] = \revstr{X_i}$,
  where $X_i = \Textinf[s^{\rm lex}_i - \tau \dd s^{\rm lex}_i + 2\tau)$.

  To obtain input sequences with the parameters required by the claim,
  we pad the context strings to length $\ell$ and divide their sequence
  into blocks of $m$ strings.
  Denote $m' := m \cdot \lceil \tfrac{\max(n_{\SSS}, m)}{m} \rceil$ and let
  $W[1 \dd m']$ denote an array defined such that:
  \begin{itemize}
  \item for $i \in [1 \dd n_{\SSS}]$,
    $W[i]=A_{\rm str}[i]\cdot\zero^{\ell-3\tau}$;
  \item for every $i \in (n_{\SSS} \dd m']$, it holds $W[i] := \zero^{\ell}$.
  \end{itemize}
  Denote $n_{\rm blocks} := m' / m$. For any $i \in [1 \dd n_{\rm blocks}]$,
  let $W_i[1 \dd m]$ be an array defined such that,
  for every $j \in [1 \dd m]$, it holds $W_i[j] = W[(i-1)m + j]$.
  In other words, we split the array $W$ into blocks of size $m$, and
  the strings in the $i$th block are stored in $W_i[1 \dd m]$.

  For every $i\in[1\dd n_{\rm blocks}]$ and $p\in[0\dd\ell]$,
  let $L_{i,p}[0\dd\AlphabetSize^p)$ be an array such that, for every
  $X\in\IntegerAlphabet^p$, it holds
  $L_{i,p}[\Val{\AlphabetSize}{X}]
    =\big|\{j\in[1\dd m]:X\text{ is a prefix of }W_i[j]\}\big|$
  (see \cref{def:val}).

  \DSComponents
  The data structure consists of the following components:
  \begin{enumerate}

  \item The data structure from \cref{th:val-encoding} with
    parameters $N$ and $\AlphabetSize$.
    It uses $\bigO(\sqrt N)$ words of space.

  \item The data structure from \cref{th:string-reversal} for reversing
    packed strings of length at most $N$ over alphabet $\IntegerAlphabet$.
    It uses $\bigO(\sqrt N)$ words of space.

  \item The packed representation
    $\PackedRepresentation{w}{\AlphabetSize}{\Text}$
    (\cref{def:packed-representation}) of $\Text$.
    It needs $\bigO(N \log \AlphabetSize)$ bits.

  \item The structure from \cref{pr:range-beg-and-end} for text $\Text$.
    It uses $\bigO(\sqrt N)$ words of space.

  \item The structure from \cref{pr:rank-to-pref} for text $\Text$.
    It also uses $\bigO(N \log \AlphabetSize)$ bits of space.

  \item The array $A_{\SSS}[1\dd n_{\SSS}]$ in plain form.
    It uses $\bigO(1+n_{\SSS})$ words of space.

  \item The distinguishing-prefix lookup structure from
    \cref{pr:dist-prefix-lookup} for $\Text$ and $\SSS$, denoted
    $\mathcal D_{\rm pref}$. It uses $\bigO(\sqrt N)$ words.

  \item For every $i\in[1\dd n_{\rm blocks}]$, the prefix-select
    structure from the claim for $W_i[1\dd m]$.
    These structures use $\bigO(n_{\rm blocks}\cdot S(\AlphabetSize,N))$
    bits in total.

  \item The arrays $L_{i,0},\ldots,L_{i,\ell}$ for all
    $i\in[1\dd n_{\rm blocks}]$, stored in plain form.
    They use $\bigO(n_{\rm blocks}\cdot\AlphabetSize^\ell)$ words,
    since $\sum_{p=0}^{\ell}\AlphabetSize^p=\bigO(\AlphabetSize^\ell)$.
  \end{enumerate}

  Since $m=\Theta(N/\log_{\AlphabetSize}N)$ and
  $n_{\SSS}=\bigO(N/\tau)=\bigO(m)$, we have
  $m'=\bigO(m+n_{\SSS})=\bigO(m)$ and
  $n_{\rm blocks}=m'/m=\bigO(1)$.
  Moreover, $\AlphabetSize^\ell\leq\sqrt m\leq\sqrt N$ and
  $\sqrt N=\bigO(N/\log_{\AlphabetSize}N)$.
  Using $w=\bigO(\log N)$ to convert words to bits, the total
  space is therefore $\bigO(N\log\AlphabetSize+S(\AlphabetSize,N))$ bits.

  \DSQueries
  Let $i\in[1\dd N]$ satisfy
  $\SA{\Text}[i]\notin\RTwo{\tau}{\Text}$.
  We compute $\SA{\Text}[i]$ in four steps.
  \begin{enumerate}
  \item\label{step:nonperiodic-sa-query-prefix}
    We first obtain a prefix of the suffix starting at
    $\SA{\Text}[i]$. Query the structure from \cref{pr:rank-to-pref}
    with $(i,3\tau-1)$ to compute $\ell_X=|X|$ and the packed
    representation $\PackedRepresentation{w}{\AlphabetSize}{X}$,
    where $X=\Text[\SA{\Text}[i]\dd
      \min(\SA{\Text}[i]+3\tau-1,N+1))$.
    If $\ell_X<3\tau-1$, return $N+1-\ell_X$.
  \item\label{step:nonperiodic-sa-query-dist}
    Next, we compute the distinguishing prefix $D$ and the
    parameters of the prefix-select query. Apply $\mathcal D_{\rm pref}$
    to $\ell_X$ and $\PackedRepresentation{w}{\AlphabetSize}{X}$,
    obtaining $\ell_D=|D|$ and
    $\PackedRepresentation{w}{\AlphabetSize}{D}$.
    Set $\deltatext=\ell_D-2\tau$ and query the structure from
    \cref{pr:range-beg-and-end} with $\ell_D$ and
    $\PackedRepresentation{w}{\AlphabetSize}{D}$ to compute
    $b=\RangeBegTwo{D}{\Text}$. Set $r=i-b$.
    Using the structures from \cref{th:string-reversal,th:val-encoding},
    compute the packed representation
    $\PackedRepresentation{w}{\AlphabetSize}{\revstr{D}}$ of
    $\revstr{D}$ and its integer value $d'=\Val{\AlphabetSize}{\revstr{D}}$.
  \item We now compute
    $\delta_{\rm lex}=\PrefixSelect{A_{\rm str}}{r}{\revstr{D}}$
    using a prefix-select query on one of the blocks $W_q$, as follows:
    \begin{enumerate}
    \item\label{step:nonperiodic-sa-query-block}
      First, we locate the block containing the $r$th string of $W$
      with prefix $\revstr{D}$. Initialize $q=1$ and $r_{\rm base}=0$.
      While $r_{\rm base}+L_{q,\ell_D}[d']<r$, add
      $L_{q,\ell_D}[d']$ to $r_{\rm base}$ and increase $q$ by one.
      Set $r_{\rm local}=r-r_{\rm base}$.
    \item\label{step:nonperiodic-sa-query-select}
      We then query the prefix-select structure for $W_q$ with
      $r_{\rm local}$, $\ell_D$, and
      $\PackedRepresentation{w}{\AlphabetSize}{\revstr{D}}$,
      obtaining $p'=\PrefixSelect{W_q}{r_{\rm local}}{\revstr{D}}$.
      Set $\delta_{\rm lex}=(q-1)m+p'$.
    \end{enumerate}
  \item\label{step:nonperiodic-sa-query-return}
    Finally, return $A_{\SSS}[\delta_{\rm lex}]-\deltatext$.
  \end{enumerate}

  We now show that the returned value is $\SA{\Text}[i]$.
  If $\ell_X<3\tau-1$, then $X$ is the entire suffix starting at
  $\SA{\Text}[i]$, and Step~\ref{step:nonperiodic-sa-query-prefix}
  returns $N+1-\ell_X=\SA{\Text}[i]$. Otherwise,
  $\SA{\Text}[i]\in[1\dd N-3\tau+2]\setminus\RTwo{\tau}{\Text}$.
  Hence, $X$ occurs in $\Text$ and is $\tau$-nonperiodic
  (\cref{def:periodic-positions,def:periodic-pattern}).
  By \cref{lm:dist-prefix-existence}, $X$ has a unique prefix
  $D\in\DistPrefixes{\tau}{\Text}{\SSS}$, which
  Step~\ref{step:nonperiodic-sa-query-dist} obtains from
  $\mathcal D_{\rm pref}$. Since every string in this
  set has length at most $3\tau-1$, the same $D$ is the unique
  distinguishing prefix of $\Text[\SA{\Text}[i]\dd N]$.
  By \cref{lm:reduce-suffix-array-to-prefix-select}, it holds
  $1\leq r\leq\PrefixRank{A_{\rm str}}{n_{\SSS}}{\revstr{D}}$.
  For every $t\in[1\dd n_{\SSS}]$, the strings $W[t]$ and
  $A_{\rm str}[t]$ have the same prefix of length $3\tau$.
  Thus, $\revstr{D}$ is a prefix of one if and only if it is a
  prefix of the other. The bound on $r$ ensures that the $r$th
  matching string in $W$ has index at most $n_{\SSS}$, irrespective
  of whether the appended strings $\zero^\ell$ have this prefix.
  Consequently,
  $\PrefixSelect{A_{\rm str}}{r}{\revstr{D}}
    =\PrefixSelect{W}{r}{\revstr{D}}$.
  Since $L_{q,\ell_D}[d']$ counts the strings in $W_q$ having
  prefix $\revstr{D}$, the loop in
  Step~\ref{step:nonperiodic-sa-query-block} stops with
  $q\leq n_{\rm blocks}$
  and $r_{\rm base}=\sum_{t=1}^{q-1}L_{t,\ell_D}[d']$. The stopping
  condition and the choice of the first such block imply
  $1\leq r_{\rm local}\leq L_{q,\ell_D}[d']\leq m$.
  Together with $\ell_D\leq3\tau-1\leq\ell$, this shows that the query
  in Step~\ref{step:nonperiodic-sa-query-select} is valid.
  By the definition of the blocks,
  \[
    \delta_{\rm lex}
      =\PrefixSelect{W}{r}{\revstr{D}}
      =\PrefixSelect{A_{\rm str}}{r}{\revstr{D}}.
  \]
  By \cref{lm:reduce-suffix-array-to-prefix-select}, the value
  returned in Step~\ref{step:nonperiodic-sa-query-return} is
  $\SA{\Text}[i]$.

  The requested suffix prefix and the strings passed to the range
  structure have length at most $3\tau-1<\tfrac12\log_{\AlphabetSize}N$,
  as required by \cref{pr:rank-to-pref,pr:range-beg-and-end}.
  The distinguishing-prefix lookup, reversal, and integer encoding
  also take $\bigO(1)$ time on these strings.
  Since $n_{\rm blocks}=\bigO(1)$, locating the block takes constant
  time. Together with the single prefix-select query, this gives
  $\bigO(1+Q(\AlphabetSize,N))$ query time.

  \DSConstruction
  We construct the components from the packed representation of $\Text$
  in four steps.
  \begin{enumerate}

  \item\label{step:nonperiodic-sa-construct-support} First, prepare the auxiliary structures.
    To obtain $\tau$, $m$, and $\ell$, observe that
    $\tau$ and $\ell$ are the largest integers $t$ satisfying
    $\AlphabetSize^{7t}\leq N$ and $\AlphabetSize^{2t}\leq m$,
    respectively. We thus compute $\tau$ by repeatedly dividing $N$ by
    $\AlphabetSize^7$, set $m=\lfloor N/(64\tau)\rfloor$, and compute
    $\ell$ by repeatedly dividing $m$ by $\AlphabetSize^2$, in
    $\bigO(\log N)$ time in total.
    Construct the encoder
    from \cref{th:val-encoding} and the reversal structure from
    \cref{th:string-reversal}. Store a copy of the packed text, and construct
    the structures from \cref{pr:range-beg-and-end,pr:rank-to-pref}
    for $\Text$.

  \item\label{step:nonperiodic-sa-construct-synchronizing} We next construct $A_{\SSS}$ and $\mathcal D_{\rm pref}$.
    Using \cref{th:sss-packed-construction}, compute an array
    $A[1\dd n_{\SSS}]$ listing a $\tau$-synchronizing set $\SSS$ of $\Text$
    of size $\bigO(N/\tau)$. Sort its entries using \cref{th:sss-lex-sort},
    obtaining $A_{\SSS}$, and construct $\mathcal D_{\rm pref}$ for $\SSS$
    using \cref{pr:dist-prefix-lookup}.

  \item We now construct the prefix-select structures for
    $W_1,\ldots,W_{n_{\rm blocks}}$.
    \begin{enumerate}

    \item\label{step:nonperiodic-sa-construct-sequences} First, we construct the packed sequence representations of
      $W_1,\ldots,W_{n_{\rm blocks}}$ as follows. Compute
      $\lceil\log\AlphabetSize\rceil$ and use
      \cref{pr:packed-representation} to initialize the packed
      representations of $\zero^{\ell-3\tau}$ and $\zero^\ell$.
      For every $j\in[1\dd n_{\SSS}]$, denote $s:=A_{\SSS}[j]$.
      Since $s\in\SSS\subseteq[1\dd N-2\tau+1]$, we construct the
      packed representation of $X_j=\Textinf[s-\tau\dd s+2\tau)$
      using the substring and concatenation operations from
      \cref{pr:packed-representation}:
      \begin{itemize}
      \item If $s>\tau$, extract the packed representation of
        $\Text[s-\tau\dd s+2\tau)$ from that of $\Text$.
      \item If $s\leq\tau$, extract the packed representations of
        $\Text[N+s-\tau\dd N]$ and $\Text[1\dd s+2\tau)$ from that of
        $\Text$, and concatenate them in this order. Their lengths are
        $\tau-s+1$ and $s+2\tau-1$, respectively, and their concatenation
        is exactly $X_j$.
      \end{itemize}
      Reverse the packed representation of $X_j$ using the structure from
      \cref{th:string-reversal}
      constructed above, and concatenate the result with the packed
      representation of $\zero^{\ell-3\tau}$ to obtain that of $W[j]$.
      For every $j\in(n_{\SSS}\dd m']$, use the packed representation
      of $\zero^\ell$ for $W[j]$.
      Next, for every $i\in[1\dd n_{\rm blocks}]$, apply the concatenation
      operation to the packed representations of
      $W[(i-1)m+1],\ldots,W[im]$, each of length $\ell$, obtaining
      $\PackedSeqRepresentation{w}{\AlphabetSize}{W_i}$.

    \item\label{step:nonperiodic-sa-construct-prefix-select} For each $i\in[1\dd n_{\rm blocks}]$, construct the
      prefix-select structure from the claim using $m$
      and a copy of $\PackedSeqRepresentation{w}{\AlphabetSize}{W_i}$. Keep the
      original representation.
    \end{enumerate}

  \item\label{step:nonperiodic-sa-construct-prefix-counts} Finally, compute the arrays $L_{i,p}$. Apply
    \cref{th:packed-prefix-frequencies} to each
    $\PackedSeqRepresentation{w}{\AlphabetSize}{W_i}$, kept in
    Step~\ref{step:nonperiodic-sa-construct-prefix-select}. For every $p\in[0\dd\ell]$, the returned
    array $F_p$ is precisely $L_{i,p}$ by their definitions.
  \end{enumerate}

  We first verify the remaining parameter requirements used in
  the construction. The structures from
  \cref{pr:range-beg-and-end,pr:rank-to-pref,pr:dist-prefix-lookup},
  the algorithms from
  \cref{th:sss-packed-construction,th:sss-lex-sort,th:packed-prefix-frequencies},
  and the prefix-select structures from the claim use size parameter
  $N$ and alphabet size $\AlphabetSize$.
  Moreover, $1\leq\tau\leq\tfrac17\log_{\AlphabetSize}N$. This gives
  $\tau\leq\lfloor N/2\rfloor$ and $\tau\leq\tfrac16\log_{\AlphabetSize}N$, and
  \cref{th:sss-packed-construction} applies with $\mu=\tfrac17$.
  The bound $\ell\geq3\tau\geq1$ gives
  $m\geq\AlphabetSize^2\geq\AlphabetSize$.
  Since $\ell\leq\tfrac12\log_{\AlphabetSize}N$ and
  $\log_{\AlphabetSize}N<7(\tau+1)\leq14\tau$, the definition of $m$
  gives $m\ell<(N/(64\tau))\cdot7\tau<N/8$, and hence
  $m\ell\leq\lfloor N/8\rfloor$.
  These bounds verify the assumptions on the prefix-select structures.
  They also justify using length bound $N$ in all packed operations:
  each context has length $3\tau\leq\ell$ and each concatenated
  block has length $m\ell<N/8$.
  The sequence indices fit in one word because
  $m'\leq n_{\SSS}+m\leq17N/16<2N\leq2^w$.
  Finally, $\AlphabetSize^\ell\leq m$, $m\ell\leq N$, and
  $w\geq1+\log N$ verify the hypotheses of
  \cref{th:packed-prefix-frequencies}.

  We now bound the construction time and space.
  Steps~\ref{step:nonperiodic-sa-construct-support}--\ref{step:nonperiodic-sa-construct-synchronizing} take $\bigO(N/\log_{\AlphabetSize}N)$ time and use
  $\bigO(N/\log_{\AlphabetSize}N)$ words, by the cited results
  and the bounds $\sqrt N=\bigO(N/\log_{\AlphabetSize}N)$ and
  $n_{\SSS}=\bigO(N/\tau)=\bigO(N/\log_{\AlphabetSize}N)$.
  In Step~\ref{step:nonperiodic-sa-construct-sequences}, computing $\lceil\log\AlphabetSize\rceil$ takes
  $\bigO(\log N)=\bigO(m)$ time. Each $W[j]$ takes constant time because
  $3\tau\leq\ell\leq\log_{\AlphabetSize}N$.
  Concatenating one block takes
  $\bigO(m+m\ell\log\AlphabetSize/w)=\bigO(m)$ time.
  Since $m'=\bigO(m)$, Step~\ref{step:nonperiodic-sa-construct-sequences} takes $\bigO(m)$ time and words.
  Copying these representations in Step~\ref{step:nonperiodic-sa-construct-prefix-select} also takes $\bigO(m)$
  time and words in total. The $\bigO(1)$ preprocessing calls take
  $\bigO(P_t(\AlphabetSize,N))$ time and
  $\bigO(P_s(\AlphabetSize,N))$ bits of space.
  In Step~\ref{step:nonperiodic-sa-construct-prefix-counts}, \cref{th:packed-prefix-frequencies} takes
  $\bigO(m\ell\log\AlphabetSize/\log m)=\bigO(m)$ time and words
  per block, because $\ell\leq\tfrac12\log_{\AlphabetSize}m$.
  There are only $\bigO(1)$ blocks.
  Converting the auxiliary word-space bounds to bits using
  $w=\bigO(\log N)$ gives the claimed
  $\bigO(N/\log_{\AlphabetSize}N+P_t(\AlphabetSize,N))$ time and
  $\bigO(N\log\AlphabetSize+P_s(\AlphabetSize,N))$ bits of working space.
\end{proof}

%% file: appendix/inverse-suffix-array.tex
\subsubsection{Inverse Suffix Array Queries}
\label{sec:appendix-isa}

\begin{lemma}[Detailed version adapted
    from~\cite{breaking}]\label{lm:reduce-inverse-suffix-array-to-prefix-special-rank}
  Let $\Textlen\in\Z_{\geq2}$, $\Text \in \Sigma^{\Textlen}$,
  $\tau \in [1 \dd \lfloor \tfrac{\Textlen}{2} \rfloor]$, and assume that
  $\Text[\Textlen]$ does not occur in $\Text[1 \dd \Textlen)$.
  Let $\SSS$ be a $\tau$-synchronizing set of $\Text$ (\cref{def:sss}).
  Denote $n_{\SSS} = |\SSS|$ and $(s_i)_{i \in [1 \dd n_{\SSS}]} = \LexSorted{\SSS}{\Text}$
  (\cref{def:lex-sorted}). Let $A_{\SSS}[1 \dd n_{\SSS}]$ and $A_{\rm str}[1 \dd n_{\SSS}]$
  be defined by
  \begin{itemize}
  \item $A_{\SSS}[i] = s_i$,
  \item $A_{\rm str}[i] = \revstr{X_i}$,
    where $X_i = \Textinf[s_i - \tau \dd s_i + 2\tau)$.
  \end{itemize}
  Let $j \in [1 \dd \Textlen - 3\tau + 2] \setminus \RTwo{\tau}{\Text}$
  (see \cref{def:periodic-positions}). Denote $s = \Successor{\SSS}{j}$ (\cref{def:succ}),
  $\deltatext = s - j$, and $D = \Text[j \dd s + 2\tau)$.
  Then, letting $i \in [1 \dd n_{\SSS}]$ be such that $s_i = s$,
  it holds (see \cref{def:inverse-suffix-array,def:occ,def:prefix-rank-and-select})
  \[
    \ISA{\Text}[j] =
      \RangeBegTwo{D}{\Text} + \PrefixSpecialRank{A_{\rm str}}{i}{\deltatext + 2\tau}.
  \]
\end{lemma}
\begin{proof}
  Denote $r=\ISA{\Text}[j]$ and $b=\RangeBegTwo{D}{\Text}$.
  We prove the claim in two steps.
  \begin{enumerate}
  \item First, we show that $D$ is the distinguishing prefix of
    $\Text[j\dd\Textlen]$ and that $\revstr{D}$ is the
    length-$|D|$ prefix of $A_{\rm str}[i]$.
    The density condition in \cref{def:sss}\eqref{def:sss-density}
    gives $s\in[j\dd j+\tau)$, and hence
    $0\leq\deltatext<\tau$.
    By \cref{def:dist-prefixes}, the string
    $D=\Text[j\dd\Successor{\SSS}{j}+2\tau)$ belongs to
    $\DistPrefixes{\tau}{\Text}{\SSS}$. It is a prefix of
    $\Text[j\dd\Textlen]=\Text[\SA{\Text}[r]\dd\Textlen]$,
    and thus is the unique distinguishing prefix used in
    \cref{lm:reduce-suffix-array-to-prefix-select}.
    Moreover, $|D|=\deltatext+2\tau$ and
    $D=\Text[s-\deltatext\dd s+2\tau)$.
    Since $0\leq\deltatext<\tau$, the string $D$ is a suffix of
    $\Textinf[s-\tau\dd s+2\tau)$. Reversing these strings gives
    $A_{\rm str}[i][1\dd|D|]=\revstr{D}$.
  \item Next, we apply \cref{lm:reduce-suffix-array-to-prefix-select}
    with suffix rank $r$, obtaining
    \[
      A_{\SSS}[\PrefixSelect{A_{\rm str}}{r-b}{\revstr{D}}]
        =j+\deltatext=s=A_{\SSS}[i].
    \]
    Since $A_{\SSS}$ contains distinct positions, it follows that
    $\PrefixSelect{A_{\rm str}}{r-b}{\revstr{D}}=i$.
    By the definition of prefix select, exactly $r-b$ of the strings
    $A_{\rm str}[1],\ldots,A_{\rm str}[i]$ have prefix $\revstr{D}$.
    Since $\revstr{D}$ is the length-$|D|$ prefix of $A_{\rm str}[i]$,
    we obtain
    \[
      r-b=\PrefixRank{A_{\rm str}}{i}{\revstr{D}}
        =\PrefixSpecialRank{A_{\rm str}}{i}{|D|}.
    \]
    Substituting the definitions of $r$ and $b$, and the equality
    $|D|=\deltatext+2\tau$, proves the claim.\qedhere
  \end{enumerate}
\end{proof}

\begin{proposition}\label{pr:nonperiodic-isa}
  Let $\AlphabetSize, N \in \Z_{\geq 2}$ be such that $\AlphabetSize < N^{1/7}$
  and let $\tau = \lfloor \tfrac{1}{7}\log_{\AlphabetSize} N \rfloor \geq 1$.
  Consider the word RAM model with word size $w$ satisfying
  $1+\log N \leq w = \bigO(\log N)$.
  Assume that there exists a data structure answering prefix special rank
  queries (see \cref{def:prefix-rank-and-select}) that, given $N$, $\AlphabetSize$, $m$, and the packed
  sequence representation $\PackedSeqRepresentation{w}{\AlphabetSize}{W}$
  (\cref{def:packed-sequence-representation}) of any sequence
  $W[1 \dd m]$ of $m \geq \AlphabetSize$
  nonempty strings of common length
  $\ell = \lfloor \tfrac{1}{2} \log_{\AlphabetSize} m \rfloor$
  over alphabet $\IntegerAlphabet$, where $m\ell\leq\lfloor N/8\rfloor$,
  achieves the following complexities:
  \begin{itemize}
  \item space usage $S(\AlphabetSize,N)$ bits,
  \item preprocessing time $P_t(\AlphabetSize,N)$,
  \item preprocessing space $P_s(\AlphabetSize,N)$ bits,
  \item query time $Q(\AlphabetSize,N)$.
  \end{itemize}
  Then, there exists a data structure that, given $N$, $\AlphabetSize$, and the packed representation
  $\PackedRepresentation{w}{\AlphabetSize}{\Text}$
  (\cref{def:packed-representation}) of any
  string $\Text \in \IntegerAlphabet^{N}$ such that
  $\Text[N]$ does not occur in $\Text[1 \dd N)$,
  answers queries that, given any $j \in [1 \dd N] \setminus \RTwo{\tau}{\Text}$ (\cref{def:periodic-positions}),
  return $\ISA{\Text}[j]$ (\cref{def:inverse-suffix-array}),
  and achieves the following complexities:
  \begin{itemize}
  \item space usage $\bigO(N \log \AlphabetSize + S(\AlphabetSize,N))$ bits,
  \item preprocessing time $\bigO(N / \log_{\AlphabetSize} N + P_t(\AlphabetSize,N))$,
  \item preprocessing space $\bigO(N \log \AlphabetSize + P_s(\AlphabetSize,N))$ bits,
  \item query time $\bigO(1 + Q(\AlphabetSize,N))$.
  \end{itemize}
\end{proposition}
\begin{proof}

  Let $N_0\geq2$ be an absolute constant satisfying
  $\lfloor 7x/(64\log x)\rfloor\geq x^{6/7}$ for every integer
  $x\geq N_0$.
  For $N<N_0$, construct and store the inverse suffix array directly
  in $\bigO(1)$ time and words. Henceforth, assume that $N\geq N_0$,
  and denote $m=\lfloor N/(64\tau)\rfloor$ and
  $\ell=\lfloor\tfrac12\log_{\AlphabetSize}m\rfloor$.
  Since $64\tau\leq\tfrac{64}{7}\log_{\AlphabetSize}N\leq\tfrac{64}{7}\log N$, we have $m\geq N^{6/7}$ and
  $3\tau\leq\tfrac37\log_{\AlphabetSize}N
    \leq\tfrac12\log_{\AlphabetSize}m$.
  Taking the floor on the right gives $3\tau\leq\ell$.

  Let $\SSS$ be any $\tau$-synchronizing set of $\Text$ (\cref{def:sss}) satisfying
  $|\SSS| = \bigO(N/\tau)$. Such $\SSS$ exists by \cref{th:sss-existence-and-construction}.
  Denote $n_{\SSS}=|\SSS|$. Let
  $(s^{\rm text}_i)_{i\in[1\dd n_{\SSS}]}$ list $\SSS$ in increasing
  text order, and let
  $(s^{\rm lex}_i)_{i\in[1\dd n_{\SSS}]}=\LexSorted{\SSS}{\Text}$
  (\cref{def:lex-sorted}).
  Let $A_{\SSS}[1 \dd n_{\SSS}]$ denote an array defined such that,
  for every $i \in [1 \dd n_{\SSS}]$, it holds $A_{\SSS}[i] = s^{\rm lex}_i$.
  Let $B_{\SSS}[1 \dd N]$ denote a bitvector defined such that,
  for every $j \in [1 \dd N]$, $B_{\SSS}[j] = \one$ holds if and only if
  $j \in \SSS$. Let $A_{\rm map}[1 \dd n_{\SSS}]$ be an array containing
  a permutation of $[1 \dd n_{\SSS}]$ such that,
  for every $j \in [1 \dd n_{\SSS}]$,
  letting $i = A_{\rm map}[j]$, it holds $s^{\rm text}_{j} = s^{\rm lex}_{i}$.
  Let $A_{\rm str}[1 \dd n_{\SSS}]$ be an array defined such that,
  for any $i \in [1 \dd n_{\SSS}]$, it holds $A_{\rm str}[i] = \revstr{X_i}$,
  where $X_i = \Textinf[s^{\rm lex}_i - \tau \dd s^{\rm lex}_i + 2\tau)$.

  The structures from the claim require sequences of $m$ strings of
  length $\ell$. We obtain such sequences by padding the context strings
  and dividing their sequence into blocks.
  Denote $m' := m \cdot \lceil \tfrac{\max(n_{\SSS}, m)}{m} \rceil$ and let
  $W[1 \dd m']$ denote an array defined such that:
  \begin{itemize}
  \item for $i \in [1 \dd n_{\SSS}]$,
    $W[i]=A_{\rm str}[i]\cdot\zero^{\ell-3\tau}$;
  \item for every $i \in (n_{\SSS} \dd m']$, it holds $W[i] := \zero^{\ell}$.
  \end{itemize}
  Denote $n_{\rm blocks} := m' / m$. For any $i \in [1 \dd n_{\rm blocks}]$,
  let $W_i[1 \dd m]$ be an array defined such that,
  for every $j \in [1 \dd m]$, it holds $W_i[j] = W[(i-1)m + j]$.
  In other words, we split the array $W$ into blocks of size $m$, and
  the strings in the $i$th block are stored in $W_i[1 \dd m]$.

  For every $i\in[1\dd n_{\rm blocks}]$ and $p\in[0\dd\ell]$,
  let $L_{i,p}[0\dd\AlphabetSize^p)$ be an array such that, for every
  $X\in\IntegerAlphabet^p$, it holds
  $L_{i,p}[\Val{\AlphabetSize}{X}]
    =\big|\{j\in[1\dd m]:X\text{ is a prefix of }W_i[j]\}\big|$
  (see \cref{def:val}).

  \DSComponents
  The data structure consists of the following components:
  \begin{enumerate}

  \item The data structure from \cref{th:val-encoding} with parameters
    $N$ and $\AlphabetSize$. It uses $\bigO(\sqrt N)$ words.

  \item The data structure from \cref{th:string-reversal} for reversing
    packed strings of length at most $N$ over alphabet $\IntegerAlphabet$.
    It uses $\bigO(\sqrt N)$ words.

  \item The packed representation
    $\PackedRepresentation{w}{\AlphabetSize}{\Text}$
    (\cref{def:packed-representation}) of $\Text$.
    It needs $\bigO(N \log \AlphabetSize)$ bits.

  \item The structure from \cref{pr:range-beg-and-end} for text $\Text$.
    It uses $\bigO(\sqrt N)$ words of space.

  \item The bitvector $B_{\SSS}[1 \dd N]$ augmented with support for
    $\bigO(1)$-time rank and select queries using \cref{th:bin-rank-select}.
    The bitvector needs $\bigO(N)$ bits, and the augmentation of
    \cref{th:bin-rank-select} does not asymptotically increase the space.

  \item The array $A_{\SSS}[1\dd n_{\SSS}]$ in plain form.
    It uses $\bigO(1+n_{\SSS})$ words.

  \item The array $A_{\rm map}[1\dd n_{\SSS}]$ in plain form.
    It also uses $\bigO(1+n_{\SSS})$ words.

  \item For every $i\in[1\dd n_{\rm blocks}]$, the prefix special
    rank structure from the claim for $W_i[1\dd m]$.
    Their total size is $\bigO(n_{\rm blocks}\cdot S(\AlphabetSize,N))$ bits.

  \item The arrays $L_{i,0},\ldots,L_{i,\ell}$ for all
    $i\in[1\dd n_{\rm blocks}]$, stored in plain form.
    Since $\sum_{p=0}^{\ell}\AlphabetSize^p=\bigO(\AlphabetSize^\ell)$,
    their total size is $\bigO(n_{\rm blocks}\cdot\AlphabetSize^\ell)$ words.
  \end{enumerate}

  We have $m=\Theta(N/\log_{\AlphabetSize}N)$ and
  $n_{\SSS}=\bigO(N/\tau)=\bigO(m)$. Thus,
  $m'=\bigO(m+n_{\SSS})=\bigO(m)$ and $n_{\rm blocks}=\bigO(1)$.
  Also, $\AlphabetSize^\ell\leq\sqrt m\leq\sqrt N$ and
  $\sqrt N=\bigO(N/\log_{\AlphabetSize}N)$.
  Since $w=\bigO(\log N)$, converting the word-space bounds to
  bits gives $\bigO(N\log\AlphabetSize+S(\AlphabetSize,N))$ total space.

  \DSQueries
  Let $j\in[1\dd N]\setminus\RTwo{\tau}{\Text}$.
  If $j>N-3\tau+2$, use
  \cref{pr:packed-representation}\eqref{pr:packed-representation-substring}
  to extract the packed representation
  $\PackedRepresentation{w}{\AlphabetSize}{X}$ of $X=\Text[j\dd N]$.
  Query the structure from \cref{pr:range-beg-and-end} with $|X|$ and
  this representation to obtain $b=\RangeBegTwo{X}{\Text}$, and return
  $b+1$. The uniqueness of $\Text[N]$ implies
  $\OccTwo{X}{\Text}=\{j\}$, and hence $b+1=\ISA{\Text}[j]$.

  Otherwise, $j\leq N-3\tau+2$, and we proceed in four steps.
  \begin{enumerate}
  \item\label{step:nonperiodic-isa-query-succ}
    First, we find $s=\Successor{\SSS}{j}$ and the index $i$
    satisfying $A_{\SSS}[i]=s$. To this end, use the structure from
    \cref{th:bin-rank-select} for $B_{\SSS}$ to compute
    $k=\Rank{B_{\SSS}}{j-1}{\one}$ and
    $s=\Select{B_{\SSS}}{k+1}{\one}$.
    Set $i=A_{\rm map}[k+1]$ and $\deltatext=s-j$.
  \item Next, we compute the packed representation of the distinguishing
    prefix $D=\Text[j\dd s+2\tau)$ and the parameters for the prefix
    special rank query. Compute $\ell_D=2\tau+\deltatext=|D|$ and use
    \cref{pr:packed-representation}\eqref{pr:packed-representation-substring}
    to extract $\PackedRepresentation{w}{\AlphabetSize}{D}$.
    Query the structure from \cref{pr:range-beg-and-end} with $\ell_D$
    and this representation to obtain $b=\RangeBegTwo{D}{\Text}$.
    Apply the structures from \cref{th:string-reversal,th:val-encoding}
    to compute the packed representation
    $\PackedRepresentation{w}{\AlphabetSize}{\revstr{D}}$ of
    $\revstr{D}$ and its integer value
    $d_{\rm rev}=\Val{\AlphabetSize}{\revstr{D}}$.
  \item We now compute
    $r=\PrefixSpecialRank{A_{\rm str}}{i}{\ell_D}$ by combining counts
    from the blocks $W_1,\ldots,W_{n_{\rm blocks}}$, as follows:
    \begin{enumerate}
    \item\label{step:nonperiodic-isa-query-block}
      We first determine the block containing $W[i]$ and count
      the matches in preceding blocks. Compute
      $q=\lceil i/m\rceil$ and $i_q=i-(q-1)m$.
      Sum $L_{t,\ell_D}[d_{\rm rev}]$ over $t\in[1\dd q)$ to obtain
      $r_{\rm base}$.
    \item\label{step:nonperiodic-isa-query-local}
      Use the prefix special rank structure for $W_q$ to compute
      $r_{\rm local}=\PrefixSpecialRank{W_q}{i_q}{\ell_D}$.
      Set $r=r_{\rm base}+r_{\rm local}$.
    \end{enumerate}
  \item\label{step:nonperiodic-isa-query-return}
    Finally, return $b+r$.
  \end{enumerate}

  We now show that the four-step algorithm returns $\ISA{\Text}[j]$.
  By the definition of $A_{\rm map}$,
  Step~\ref{step:nonperiodic-isa-query-succ} gives $A_{\SSS}[i]=s$.
  The density condition in \cref{def:sss}\eqref{def:sss-density}
  guarantees $s\in[j\dd j+\tau)$, and hence
  $0\leq\deltatext<\tau$.
  Since $A_{\SSS}[i]=s$ and $j=s-\deltatext$, the length-$\ell_D$
  prefix of $A_{\rm str}[i]=\revstr{\Textinf[s-\tau\dd s+2\tau)}$
  is $\revstr{D}$. Moreover, $W[t]$ and $A_{\rm str}[t]$ have the
  same prefix of length $3\tau$ for every $t\in[1\dd n_{\SSS}]$.
  Thus, $\PrefixSpecialRank{A_{\rm str}}{i}{\ell_D}$ counts precisely
  the indices $t\in[1\dd i]$ for which
  $\revstr{D}$ is a prefix of $W[t]$. Since $i\leq n_{\SSS}$,
  the appended strings $\zero^\ell$ do not contribute to this count.
  In Step~\ref{step:nonperiodic-isa-query-block}, it holds
  $W_q[i_q]=W[i]$ and $i_q\in[1\dd m]$, and, by the definition of the
  prefix-count arrays, $r_{\rm base}$ counts the strings having prefix
  $\revstr{D}$ in the blocks $W_1,\ldots,W_{q-1}$.
  The query in Step~\ref{step:nonperiodic-isa-query-local} is valid
  because $i_q\in[1\dd m]$ and $\ell_D\leq3\tau-1\leq\ell$.
  The length-$\ell_D$ prefix of $W_q[i_q]=W[i]$ is $\revstr{D}$.
  Therefore, the local query counts exactly the remaining matches
  among the first $i_q$ strings in $W_q$, and
  \[
    \PrefixSpecialRank{A_{\rm str}}{i}{\ell_D}
      =\sum_{t=1}^{q-1}L_{t,\ell_D}[d_{\rm rev}]
         +\PrefixSpecialRank{W_q}{i_q}{\ell_D}
      =r.
  \]
  Finally, since $\ell_D=\deltatext+2\tau$,
  \cref{lm:reduce-inverse-suffix-array-to-prefix-special-rank}
  shows that the value $b+r$ returned in
  Step~\ref{step:nonperiodic-isa-query-return} is $\ISA{\Text}[j]$.

  In both the short-suffix case and the four-step algorithm, the
  strings passed to \cref{pr:range-beg-and-end} have length at most
  $3\tau-1<\tfrac12\log_{\AlphabetSize}N$, as required.
  Substring extraction, reversal, integer encoding, and rank/select
  queries also take constant time. The sum over preceding blocks
  takes $\bigO(n_{\rm blocks})=\bigO(1)$ time.
  The total query time is therefore $\bigO(1+Q(\AlphabetSize,N))$,
  including the single prefix special rank query.

  \DSConstruction
  We construct the components from the packed text in four steps.
  \begin{enumerate}

  \item\label{step:nonperiodic-isa-construct-support} First, prepare the auxiliary structures.
    To obtain $\tau$, $m$, and $\ell$, observe that
    $\tau$ and $\ell$ are the largest integers $t$ satisfying
    $\AlphabetSize^{7t}\leq N$ and $\AlphabetSize^{2t}\leq m$,
    respectively. We thus compute $\tau$ by repeatedly dividing $N$ by
    $\AlphabetSize^7$, set $m=\lfloor N/(64\tau)\rfloor$, and compute
    $\ell$ by repeatedly dividing $m$ by $\AlphabetSize^2$, in
    $\bigO(\log N)$ time in total.
    Construct the integer-value encoder from
    \cref{th:val-encoding} and the reversal structure from
    \cref{th:string-reversal}.
    Store the packed text and construct its suffix-array range
    structure from \cref{pr:range-beg-and-end}.

  \item\label{step:nonperiodic-isa-construct-synchronizing} We next construct the arrays and bitvector for finding the
    next synchronizing position and its lexicographic index.
    \begin{enumerate}
    \item Using \cref{th:sss-packed-construction}, compute an array
      $A[1\dd n_{\SSS}]$ listing a $\tau$-synchronizing set $\SSS$ of
      $\Text$ of size $\bigO(N/\tau)$.
      Using \cref{pr:packed-representation}, initialize
      $B_{\SSS}:=\zero^N$ and set $B_{\SSS}[A[i]]:=\one$ for every
      $i\in[1\dd n_{\SSS}]$. Augment $B_{\SSS}$ with rank and
      selection support using \cref{th:bin-rank-select}.
    \item Sort the entries of $A$ using \cref{th:sss-lex-sort},
      obtaining $A_{\SSS}$.
    \item For every $i\in[1\dd n_{\SSS}]$, compute
      $j=\Rank{B_{\SSS}}{A_{\SSS}[i]}{\one}$ and set
      $A_{\rm map}[j]:=i$.
      Here, $j$ is the index of $A_{\SSS}[i]$ among the positions
      in $\SSS$ in increasing text order. Consequently,
      $s_j^{\rm text}=A_{\SSS}[i]=s_i^{\rm lex}$, as required.
    \end{enumerate}

  \item To build the prefix special rank structures, first compute
    their input sequences and then apply the preprocessing algorithm
    from the claim.
    \begin{enumerate}

    \item\label{step:nonperiodic-isa-construct-sequences} We first construct the packed sequence representations of
      $W_1,\ldots,W_{n_{\rm blocks}}$ as follows. Compute
      $\lceil\log\AlphabetSize\rceil$ and use
      \cref{pr:packed-representation} to initialize the packed
      representations of $\zero^{\ell-3\tau}$ and $\zero^\ell$.
      For every $j\in[1\dd n_{\SSS}]$, denote $s:=A_{\SSS}[j]$.
      Since $s\in\SSS\subseteq[1\dd N-2\tau+1]$, we construct the
      packed representation of $X_j=\Textinf[s-\tau\dd s+2\tau)$
      using the substring and concatenation operations from
      \cref{pr:packed-representation}:
      \begin{itemize}
      \item If $s>\tau$, extract the packed representation of
        $\Text[s-\tau\dd s+2\tau)$ from that of $\Text$.
      \item If $s\leq\tau$, extract the packed representations of
        $\Text[N+s-\tau\dd N]$ and $\Text[1\dd s+2\tau)$ from that of
        $\Text$, and concatenate them in this order. Their lengths are
        $\tau-s+1$ and $s+2\tau-1$, respectively, and their concatenation
        is exactly $X_j$.
      \end{itemize}
      Reverse the packed representation of $X_j$ using the structure from
      \cref{th:string-reversal}
      constructed above, and concatenate the result with the packed
      representation of $\zero^{\ell-3\tau}$ to obtain that of $W[j]$.
      For every $j\in(n_{\SSS}\dd m']$, use the packed representation
      of $\zero^\ell$ for $W[j]$.
      Next, for every $i\in[1\dd n_{\rm blocks}]$, apply the concatenation
      operation to the packed representations of
      $W[(i-1)m+1],\ldots,W[im]$, each of length $\ell$, obtaining
      $\PackedSeqRepresentation{w}{\AlphabetSize}{W_i}$.

    \item\label{step:nonperiodic-isa-construct-prefix-rank} For each $i\in[1\dd n_{\rm blocks}]$, construct the
      prefix special rank structure from the claim using $m$ and
      a copy of $\PackedSeqRepresentation{w}{\AlphabetSize}{W_i}$.
      Retain the original representation.
    \end{enumerate}

  \item\label{step:nonperiodic-isa-construct-prefix-counts} Finally, compute the prefix-count arrays. Apply
    \cref{th:packed-prefix-frequencies} to each
    $\PackedSeqRepresentation{w}{\AlphabetSize}{W_i}$, retained in
    Step~\ref{step:nonperiodic-isa-construct-prefix-rank}. For every $p\in[0\dd\ell]$, the resulting
    array $F_p$ equals $L_{i,p}$ by their definitions.
  \end{enumerate}

  It remains to check the other parameter requirements and the
  complexity bounds. The structure from \cref{pr:range-beg-and-end},
  the algorithms from
  \cref{th:sss-packed-construction,th:sss-lex-sort,th:packed-prefix-frequencies},
  and the prefix special rank structures from the claim use size
  parameter $N$ and alphabet size $\AlphabetSize$, and the rank and
  selection support for $B_{\SSS}$ uses size parameter $N$. The packed
  operations on $B_{\SSS}$ use alphabet size $2$ and length bound $N$.
  Moreover, $1\leq\tau\leq\tfrac17\log_{\AlphabetSize}N$. This gives
  $\tau\leq\lfloor N/2\rfloor$ and $\tau\leq\tfrac16\log_{\AlphabetSize}N$, and
  \cref{th:sss-packed-construction} applies with $\mu=\tfrac17$.
  Since $\ell\geq3\tau\geq1$, we have
  $m\geq\AlphabetSize^2\geq\AlphabetSize$.
  Also, since $\ell\leq\tfrac12\log_{\AlphabetSize}N$ and
  $\log_{\AlphabetSize}N<7(\tau+1)\leq14\tau$, the definition of $m$
  gives $m\ell<(N/(64\tau))\cdot7\tau<N/8$, and hence
  $m\ell\leq\lfloor N/8\rfloor$.
  Thus, every sequence $W_i$ satisfies the assumptions of the
  prefix special rank structure in the claim.
  Every context has length $3\tau\leq\ell$, and every concatenated
  block has length $m\ell<N/8$, permitting length bound $N$
  in the packed operations. Sequence indices fit in one word since
  $m'\leq n_{\SSS}+m\leq17N/16<2N\leq2^w$.
  The prefix-frequency construction also applies, because
  $\AlphabetSize^\ell\leq m$, $m\ell\leq N$, and $w\geq1+\log N$.

  We now analyze the construction costs.
  In Step~\ref{step:nonperiodic-isa-construct-synchronizing}, initializing $B_{\SSS}$ and building its rank/select
  support take $\bigO(N/\log N)$ time. Setting its bits and
  constructing $A_{\rm map}$ take $\bigO(n_{\SSS})$ time.
  The other operations in Steps~\ref{step:nonperiodic-isa-construct-support}--\ref{step:nonperiodic-isa-construct-synchronizing} satisfy the cited bounds.
  Using $\sqrt N=\bigO(N/\log_{\AlphabetSize}N)$ and
  $n_{\SSS}=\bigO(N/\tau)=\bigO(N/\log_{\AlphabetSize}N)$,
  these steps take $\bigO(N/\log_{\AlphabetSize}N)$ time and words.
  In Step~\ref{step:nonperiodic-isa-construct-sequences}, computing $\lceil\log\AlphabetSize\rceil$ takes
  $\bigO(\log N)=\bigO(m)$ time. Each $W[j]$ is constructed in constant time because
  $3\tau\leq\ell\leq\log_{\AlphabetSize}N$.
  Each block concatenation takes
  $\bigO(m+m\ell\log\AlphabetSize/w)=\bigO(m)$ time.
  Since $m'=\bigO(m)$, Step~\ref{step:nonperiodic-isa-construct-sequences} takes $\bigO(m)$ time and words.
  The copies used in Step~\ref{step:nonperiodic-isa-construct-prefix-rank} also take $\bigO(m)$ time and words
  in total. Its $\bigO(1)$ preprocessing calls take $\bigO(P_t(\AlphabetSize,N))$
  time and $\bigO(P_s(\AlphabetSize,N))$ bits of space.
  Finally, Step~\ref{step:nonperiodic-isa-construct-prefix-counts} takes
  $\bigO(m\ell\log\AlphabetSize/\log m)=\bigO(m)$ time and words
  per block, using $\ell\leq\tfrac12\log_{\AlphabetSize}m$.
  As there are $\bigO(1)$ blocks and $w=\bigO(\log N)$,
  the total construction takes
  $\bigO(N/\log_{\AlphabetSize}N+P_t(\AlphabetSize,N))$ time and
  uses $\bigO(N\log\AlphabetSize+P_s(\AlphabetSize,N))$ bits of space.
\end{proof}

%% file: appendix/periodic-preliminaries.tex
\subsubsection{Preliminaries}\label{sec:appendix-periodic-prelim}

\begin{definition}[Parameters of a periodic position]\label{def:pos-root}
  Let $\Textlen\in\Z_{\geq2}$, $\Text\in\Sigma^{\Textlen}$,
  $\tau\in[1\dd\lfloor\Textlen/2\rfloor]$, and
  $j\in\RTwo{\tau}{\Text}$ (\cref{def:periodic-positions}).
  We use the periodic parameters from~\cite{breaking}.
  Denote $p=\per{\Text[j\dd j+3\tau-1)}$. The \emph{root} and
  the \emph{end of the periodic prefix} at $j$ are defined by
  \[
    \RootPos{j}{\tau}{\Text}
      :=\min\{\Text[j+t\dd j+t+p):t\in[0\dd p)\},
    \qquad
    \RunEndPos{j}{\tau}{\Text}
      :=j+p+\LCE{\Text}{j}{j+p}.
  \]
  Let $H=\RootPos{j}{\tau}{\Text}$. The factorization
  $\Text[j\dd\RunEndPos{j}{\tau}{\Text})=H'H^kH''$, where
  $k\in\Z_{\geq0}$, $H'$ is a proper suffix of $H$, and $H''$ is
  a proper prefix of $H$,
  is unique because $H$ is primitive. We define the \emph{head},
  \emph{exponent}, and \emph{tail} at $j$ as
  $\HeadPos{j}{\tau}{\Text}:=|H'|$,
  $\ExpPos{j}{\tau}{\Text}:=k$, and
  $\TailPos{j}{\tau}{\Text}:=|H''|$, respectively.
  The endpoint after the last full copy of $H$ is
  \[
    \RunEndFullPos{j}{\tau}{\Text}
      :=\RunEndPos{j}{\tau}{\Text}-\TailPos{j}{\tau}{\Text}
      =j+\HeadPos{j}{\tau}{\Text}+\ExpPos{j}{\tau}{\Text}\cdot|H|.
  \]
  Finally, the \emph{type} $\TypePos{j}{\tau}{\Text}$ is $+1$ if
  $\RunEndPos{j}{\tau}{\Text}\leq\Textlen$ and
  $\Text[\RunEndPos{j}{\tau}{\Text}]
    \succ\Text[\RunEndPos{j}{\tau}{\Text}-p]$; otherwise, it is $-1$.
  Thus, negative type includes the case in which the periodic prefix
  reaches the end of $\Text$.
\end{definition}

\begin{definition}[Subsets of periodic positions]\label{def:R-subsets}
  Let $\Textlen\in\Z_{\geq2}$, $\Text\in\Sigma^{\Textlen}$, and
  $\tau\in[1\dd\lfloor\Textlen/2\rfloor]$.
  We denote the positions of negative and positive type by
  $\RMinusTwo{\tau}{\Text}
    :=\{j\in\RTwo{\tau}{\Text}:\TypePos{j}{\tau}{\Text}=-1\}$
  and $\RPlusTwo{\tau}{\Text}
    :=\RTwo{\tau}{\Text}\setminus\RMinusTwo{\tau}{\Text}$.
  For $H\in\Sigma^+$ and $s,k\in\Z_{\geq0}$, we further define:
  \begin{itemize}
  \item $\RThree{H}{\tau}{\Text}
      :=\{j\in\RTwo{\tau}{\Text}:\RootPos{j}{\tau}{\Text}=H\}$,
    the positions with root $H$;
  \item $\RFour{s}{H}{\tau}{\Text}
      :=\{j\in\RThree{H}{\tau}{\Text}:\HeadPos{j}{\tau}{\Text}=s\}$,
    the positions with root $H$ and head $s$;
  \item $\RFive{s}{k}{H}{\tau}{\Text}
      :=\{j\in\RFour{s}{H}{\tau}{\Text}:\ExpPos{j}{\tau}{\Text}=k\}$,
    the positions with root $H$, head $s$, and exponent $k$.
  \end{itemize}
  A superscript $-$ or $+$ on any of these sets denotes intersection
  with $\RMinusTwo{\tau}{\Text}$ or $\RPlusTwo{\tau}{\Text}$,
  respectively. For example, $\RSignedFive{-}{s}{k}{H}{\tau}{\Text}$
  consists of the positions of negative type with root $H$, head $s$,
  and exponent $k$.
\end{definition}

\begin{definition}[Starts of maximal blocks of periodic positions]\label{def:R-prim}
  Let $\Textlen\in\Z_{\geq2}$, $\Text\in\Sigma^{\Textlen}$, and
  $\tau\in[1\dd\lfloor\Textlen/2\rfloor]$. We denote the starting
  positions of maximal consecutive blocks in $\RTwo{\tau}{\Text}$ by
  \[
    \RPrimTwo{\tau}{\Text}
      :=\{a\in\RTwo{\tau}{\Text}:a-1\notin\RTwo{\tau}{\Text}\}.
  \]
  Superscripts $-$ and $+$ restrict this set to negative and positive
  type, respectively. A subscript $H$ further restricts it to root $H$.
  In particular, $\RPrimSignedThree{-}{H}{\tau}{\Text}$ denotes the
  starts of maximal consecutive blocks of negative type with root $H$.
\end{definition}

\begin{lemma}[{\cite{breaking}}]\label{lm:periodic-block-properties}
  Let $\Textlen\in\Z_{\geq2}$, $\Text\in\Sigma^{\Textlen}$, and
  $\tau\in[1\dd\lfloor\Textlen/2\rfloor]$. Then:
  \begin{enumerate}
  \item\label{lm:periodic-block-properties-interval}
    For every $a\in\RPrimTwo{\tau}{\Text}$, the maximal consecutive
    block of $\RTwo{\tau}{\Text}$ starting at $a$ is
    $[a\dd\RunEndPos{a}{\tau}{\Text}-3\tau+2)$.
    The root, type, tail, and the two endpoints
    $\RunEndPos{j}{\tau}{\Text}$ and $\RunEndFullPos{j}{\tau}{\Text}$
    are constant for all $j$ in this block.
  \item\label{lm:periodic-block-properties-distinct}
    For $a,a'\in\RPrimTwo{\tau}{\Text}$ with $a<a'$, it holds
    $\RunEndFullPos{a}{\tau}{\Text}
      <\RunEndFullPos{a'}{\tau}{\Text}$.
  \item\label{lm:periodic-block-properties-size}
    It holds $|\RPrimTwo{\tau}{\Text}|\leq\Textlen/(2\tau)$ and
    $\sum_{a\in\RPrimTwo{\tau}{\Text}}
      (\RunEndPos{a}{\tau}{\Text}-a)\leq2\Textlen$.
  \end{enumerate}
\end{lemma}
\begin{proof}
  The first two items, the bound on the sum in the third item, and the
  inequality $a'-a\geq2\tau$ for all $a,a'\in\RPrimTwo{\tau}{\Text}$
  with $a<a'$ are shown in~\cite{breaking}. Since every element of
  $\RPrimTwo{\tau}{\Text}$ belongs to $[1\dd\Textlen-3\tau+2]$
  (\cref{def:periodic-positions}), the latter inequality gives
  $|\RPrimTwo{\tau}{\Text}|\leq\lfloor(\Textlen-3\tau+1)/(2\tau)\rfloor+1
    \leq\Textlen/(2\tau)$.
\end{proof}

\begin{definition}[Sequence of periodic block starts in lex-order]
\label{def:runs-lex-sorted}
  Let $\Textlen\in\Z_{\geq2}$, $\Text\in\Sigma^{\Textlen}$,
  $\tau\in[1\dd\lfloor\Textlen/2\rfloor]$, and $\alpha\in\{-,+\}$.
  Denote $q=|\RPrimSignedTwo{\alpha}{\tau}{\Text}|$.
  By $\RunsLexSorted{\alpha}{\tau}{\Text}$, we denote the sequence
  $(a_t)_{t\in[1\dd q]}$ containing every position in
  $\RPrimSignedTwo{\alpha}{\tau}{\Text}$ exactly once, ordered
  first by its root and, among positions with the same root, by
  the suffix starting at its full endpoint. Thus, the pairs
  \[
    \bigl(\RootPos{a_t}{\tau}{\Text},
      \Text[\RunEndFullPos{a_t}{\tau}{\Text}\dd\Textlen]\bigr)
    \qquad(t\in[1\dd q])
  \]
  are in increasing lexicographic order. Here
  $\Text[\Textlen+1\dd\Textlen]=\emptystring$.
  The full endpoints of distinct block starts are distinct by
  \cref{lm:periodic-block-properties}\eqref{lm:periodic-block-properties-distinct},
  and hence this order is unique.
\end{definition}

\begin{lemma}[{\cite{breaking}}]\label{lm:periodic-suffix-order}
  Let $\Textlen\in\Z_{\geq2}$, $\Text\in\Sigma^{\Textlen}$,
  $\tau\in[1\dd\lfloor\Textlen/2\rfloor]$, $H\in\Sigma^+$,
  $s\in[0\dd|H|)$, and
  $j\in\RFour{s}{H}{\tau}{\Text}$.
  For every $j'\in[1\dd\Textlen]$, it holds
  $\LCE{\Text}{j}{j'}\geq3\tau-1$ if and only if
  $j'\in\RFour{s}{H}{\tau}{\Text}$.
  Moreover, if $j'\in\RFour{s}{H}{\tau}{\Text}$, then, denoting
  $d=\RunEndPos{j}{\tau}{\Text}-j$ and
  $d'=\RunEndPos{j'}{\tau}{\Text}-j'$, we have:
  \begin{enumerate}
  \item If $\TypePos{j}{\tau}{\Text}\neq\TypePos{j'}{\tau}{\Text}$,
    then $\Text[j\dd\Textlen]\prec\Text[j'\dd\Textlen]$ holds if
    and only if $\TypePos{j}{\tau}{\Text}<\TypePos{j'}{\tau}{\Text}$.
  \item If both positions have negative type and $d\neq d'$, then
    $\Text[j\dd\Textlen]\prec\Text[j'\dd\Textlen]$ holds if and
    only if $d<d'$.
  \item If both positions have positive type and $d\neq d'$, then
    $\Text[j\dd\Textlen]\prec\Text[j'\dd\Textlen]$ holds if and
    only if $d>d'$.
  \end{enumerate}
\end{lemma}

\begin{remark}\label{rm:periodic-arbitrary-text}
  Although~\cite{breaking} assumes that the final symbol of $\Text$ is
  unique and smallest, its proofs of the claims in
  \cref{lm:periodic-block-properties,lm:periodic-suffix-order} do not use
  this assumption. These lemmas therefore apply to arbitrary texts.
\end{remark}

\begin{lemma}\label{lm:periodic-exponent-intervals}
  Let $\Textlen\in\Z_{\geq2}$, $\Text\in\Sigma^{\Textlen}$,
  $\tau\in[1\dd\lfloor\Textlen/2\rfloor]$, $H\in\Sigma^+$,
  $s\in[0\dd|H|)$, and $\alpha\in\{-,+\}$. Then:
  \begin{enumerate}
  \item\label{lm:periodic-exponent-intervals-sa}
    The positions in $\RSignedFour{\alpha}{s}{H}{\tau}{\Text}$
    occupy a consecutive interval in $\SA{\Text}$.
    Within this interval, exponents are nondecreasing for $\alpha=-$
    and nonincreasing for $\alpha=+$.
    In particular, for every $k\in\Z_{\geq0}$, the positions in
    $\RSignedFive{\alpha}{s}{k}{H}{\tau}{\Text}$ occupy a consecutive
    interval in $\SA{\Text}$.
  \item\label{lm:periodic-exponent-intervals-no-gaps}
    If $\RSignedFour{\alpha}{s}{H}{\tau}{\Text}\neq\emptyset$,
    then $\{\ExpPos{j}{\tau}{\Text}:
      j\in\RSignedFour{\alpha}{s}{H}{\tau}{\Text}\}$
    is an interval of consecutive integers.
  \end{enumerate}
\end{lemma}
\begin{proof}
  The first claim is immediate if
  $\RFour{s}{H}{\tau}{\Text}=\emptyset$.
  Otherwise, \cref{lm:periodic-suffix-order} shows that
  $\RFour{s}{H}{\tau}{\Text}$ contains precisely the starting
  positions of suffixes with one common length-$(3\tau-1)$ prefix.
  These suffixes occupy a consecutive interval in $\SA{\Text}$,
  with all negative-type suffixes preceding all positive-type
  suffixes. For $j\in\RFour{s}{H}{\tau}{\Text}$, the length of
  its periodic prefix is
  $s+\ExpPos{j}{\tau}{\Text}\cdot|H|+\TailPos{j}{\tau}{\Text}$.
  Since the tail belongs to $[0\dd|H|)$, distinct exponents give
  periodic-prefix lengths ordered in the same way as the exponents.
  The last two parts of \cref{lm:periodic-suffix-order} therefore
  give the claimed order and intervals.

  To prove the second claim, let
  $a\in\RPrimSignedThree{\alpha}{H}{\tau}{\Text}$ and consider the
  block of periodic positions starting at $a$. By
  \cref{lm:periodic-block-properties}\eqref{lm:periodic-block-properties-interval},
  as $j$ ranges over
  this block, the distance $\RunEndFullPos{a}{\tau}{\Text}-j$
  takes precisely the integer values in
  $[3\tau-1-\TailPos{a}{\tau}{\Text}\dd
    \RunEndFullPos{a}{\tau}{\Text}-a]$.
  A position has head $s$ and exponent $k$ if and only if this
  distance equals $s+k|H|$. Thus, the exponents contributed by
  this block form the integer interval
  \[
    \left[
      \left\lceil\tfrac{1}{|H|}\cdot(3\tau-1-\TailPos{a}{\tau}{\Text}-s)\right\rceil
      \dd
      \left\lfloor\tfrac{1}{|H|}\cdot(\RunEndFullPos{a}{\tau}{\Text}-a-s)\right\rfloor
    \right],
  \]
  which may be empty. The tails belong to $[0\dd|H|)$, and hence
  the lower endpoints of these intervals differ by at most one.
  Their nonempty intervals therefore overlap or are adjacent.
  Taking their union over
  $a\in\RPrimSignedThree{\alpha}{H}{\tau}{\Text}$ proves the claim.
\end{proof}

\begin{proposition}[{\cite{breaking}; see also~\cite{hierarchy}}]
\label{pr:periodic-pos}
  Let $N,\AlphabetSize\in\Z_{\geq2}$ satisfy
  $\AlphabetSize\leq N$.
  Consider the word RAM model with word size $w\geq1+\log N$.
  Given $N$, $\AlphabetSize$, and the packed representation
  $\PackedRepresentation{w}{\AlphabetSize}{\Text}$
  (\cref{def:packed-representation}) of a text
  $\Text\in\IntegerAlphabet^{N}$, we can in
  $\bigO(N/\log_{\AlphabetSize}N)$ time
  and using $\bigO(N/\log_{\AlphabetSize}N)$ words
  of space construct a data structure of size
  $\bigO(N/\log_{\AlphabetSize}N)$ words that,
  given an integer $\tau\geq1$ satisfying
  $\tau\leq\tfrac12\log_{\AlphabetSize}N$ and a position
  $j\in\RTwo{\tau}{\Text}$ (\cref{def:periodic-positions}),
  computes all the following in $\bigO(1)$ time (see \cref{def:pos-root}):
  \begin{itemize}
  \item The length and packed representation
    $\PackedRepresentation{w}{\AlphabetSize}{H}$ of
    $H=\RootPos{j}{\tau}{\Text}$.
  \item The values $\HeadPos{j}{\tau}{\Text}$, $\RunEndPos{j}{\tau}{\Text}$,
    $\RunEndFullPos{j}{\tau}{\Text}$, $\ExpPos{j}{\tau}{\Text}$,
    $\TailPos{j}{\tau}{\Text}$, and $\TypePos{j}{\tau}{\Text}$.
  \end{itemize}
\end{proposition}

\begin{theorem}[{\cite{breaking}; see also~\cite{hierarchy}}]
\label{th:periodic-position-intervals}
  Let $N,\AlphabetSize\in\Z_{\geq2}$ satisfy
  $\AlphabetSize\leq N$.
  Consider the word RAM model with word size $w\geq1+\log N$.
  Given $N$, $\AlphabetSize$, an integer $\tau\geq1$ satisfying
  $\tau\leq\tfrac12\log_{\AlphabetSize}N$, and the packed representation
  $\PackedRepresentation{w}{\AlphabetSize}{\Text}$
  (\cref{def:packed-representation}) of any text
  $\Text\in\IntegerAlphabet^N$, we can deterministically compute
  an integer $k\in\Z_{\geq0}$ satisfying $k=\bigO(N/\tau)$ and
  arrays $A_{\rm beg}[1\dd k]$ and $A_{\rm end}[1\dd k]$ such that
  the intervals $[A_{\rm beg}[i]\dd A_{\rm end}[i])$, for
  $i\in[1\dd k]$, are precisely the maximal intervals of
  $\RTwo{\tau}{\Text}$ (\cref{def:periodic-positions}), listed in
  increasing order. The construction takes $\bigO(N/\tau)$ time
  and uses $\bigO(N/\tau)$ words of space.
\end{theorem}

\begin{proposition}[Periodic suffix-array intervals]
\label{pr:periodic-sa-intervals}
  Let $N,\AlphabetSize\in\Z_{\geq2}$ satisfy $\AlphabetSize\leq N$.
  Consider the word RAM model with word size $w\geq1+\log N$.
  Given $N$, $\AlphabetSize$, an integer $\tau\geq1$ satisfying
  $\tau\leq\tfrac16\log_{\AlphabetSize}N$, and the packed representation
  $\PackedRepresentation{w}{\AlphabetSize}{\Text}$
  (\cref{def:packed-representation}) of a text
  $\Text\in\IntegerAlphabet^N$, we can in
  $\bigO(N/\tau)$ time and using $\bigO(N/\tau)$ words of space
  construct a data structure of size $\bigO(N/\log N)$ words that
  supports the following queries in $\bigO(1)$ time:
  \begin{enumerate}
  \item Given $\alpha\in\{-,+\}$, $s\in[0\dd|H|)$,
    $k\in\Z_{\geq0}$, and the length and packed representation of
    $H\in\IntegerAlphabet^+$, where
    $\RSignedFive{\alpha}{s}{k}{H}{\tau}{\Text}\neq\emptyset$
    (\cref{def:R-subsets}), return the integers $b,e$ satisfying
    \[
      (b\dd e]=\{\ISA{\Text}[j]:
        j\in\RSignedFive{\alpha}{s}{k}{H}{\tau}{\Text}\}.
    \]
  \item Given $i\in[1\dd N]$ with
    $\SA{\Text}[i]\in\RTwo{\tau}{\Text}$, return the sign $\alpha$
    of the type, the head $s$, the exponent $k$, and the length
    and packed representation of the root $H$ of position
    $\SA{\Text}[i]$, together with the integers $b,e$ defined
    in the first query.
  \end{enumerate}
\end{proposition}
\begin{proof}
  We adapt the packed construction of exponent boundaries
  from~\cite{breaking}, using
  \cref{pr:periodic-pos,th:periodic-position-intervals}.
  If $\tau\leq2$, then
  $\RTwo{\tau}{\Text}=\emptyset$, and neither query has a valid
  input. In this case, it suffices to store $N$.
  Henceforth, assume that $\tau\geq3$, and let $h=3\tau-1$.

  For every root $H$ and head $s$ with
  $\RFour{s}{H}{\tau}{\Text}\neq\emptyset$, let
  $(b_{s,H}\dd e_{s,H}]$ be its suffix-array interval, and denote
  $z_{s,H}=b_{s,H}+|\RSignedFour{-}{s}{H}{\tau}{\Text}|$.
  By \cref{lm:periodic-suffix-order}, the intervals for negative
  and positive type are, respectively,
  $(b^-_{s,H}\dd e^-_{s,H}]=(b_{s,H}\dd z_{s,H}]$ and
  $(b^+_{s,H}\dd e^+_{s,H}]=(z_{s,H}\dd e_{s,H}]$.
  For each nonempty signed interval, let $f^\alpha_{s,H}$ be
  the exponent of its first suffix: the smallest occurring exponent
  for negative type, and the largest for positive type.
  For an empty signed interval, let $f^\alpha_{s,H}=0$.

  For every $p\in[1\dd\lfloor\tau/3\rfloor]$ and
  $s\in[0\dd p)$, let $L_{p,s}[0\dd\AlphabetSize^p)$ be the
  array defined, for every $H\in\IntegerAlphabet^p$, by
  \[
    L_{p,s}[\Val{\AlphabetSize}{H}]=
    \begin{cases}
      (b_{s,H},e_{s,H},z_{s,H},f^-_{s,H},f^+_{s,H}),
        &\text{if }\RFour{s}{H}{\tau}{\Text}\neq\emptyset,\\
      \bot,&\text{otherwise.}
    \end{cases}
  \]
  Let $(s_j,H_j)_{j\in[1\dd r]}$ list all pairs $(s,H)$ with
  $\RFour{s}{H}{\tau}{\Text}\neq\emptyset$, in increasing order of
  $b_{s,H}$, and define $A[1\dd r]$ by
  \[
    A[j]=\bigl(s_j,|H_j|,
      \PackedRepresentation{w}{\AlphabetSize}{H_j},
      L_{|H_j|,s_j}[\Val{\AlphabetSize}{H_j}]\bigr).
  \]
  Thus, $A[j]$ stores the head, root length, packed root, and five
  interval parameters for the $j$th nonempty root/head interval.
  Let $C[1\dd N]$ be the bitvector with ones exactly at the positions
  $b_{s_j,H_j}+1$, $j\in[1\dd r]$.

  Let $B[1\dd N]$ be the bitvector marking the last rank of every
  nonempty interval for $\RSignedFive{\alpha}{s}{k}{H}{\tau}{\Text}$;
  i.e., $B[i]=\one$ if and only if there exist $\alpha\in\{-,+\}$,
  $H\in\IntegerAlphabet^+$, $s\in[0\dd|H|)$, and $k\in\Z_{\geq0}$
  such that $\RSignedFive{\alpha}{s}{k}{H}{\tau}{\Text}\neq\emptyset$ and
  \[
    i=\max\{\ISA{\Text}[j]:
      j\in\RSignedFive{\alpha}{s}{k}{H}{\tau}{\Text}\}.
  \]
  All other bits are zero.

  \DSComponents
  We store the following components:
  \begin{enumerate}
  \item The integer-value encoder from \cref{th:val-encoding} for
    alphabet size $\AlphabetSize$. It occupies $\bigO(\sqrt N)$ words.
  \item The arrays $L_{p,s}$ and $A$. They occupy
    $\bigO(\tau\AlphabetSize^\tau)$ words, since
    $r=\bigO(\tau\AlphabetSize^\tau)$ and each tuple in $A$ uses
    $\bigO(1)$ words for its integers and packed root.
  \item The packed bitvectors $B$ and $C$, augmented with rank and
    selection support on $B$ and rank support on $C$ using
    \cref{th:bin-rank-select}. They occupy $\bigO(1+N/w)$ words.
  \end{enumerate}
  Since $\AlphabetSize^\tau\leq N^{1/6}$, $\tau=\bigO(\log N)$,
  and $w>\log N$, the total space is $\bigO(N/\log N)$ words.

  \DSQueries
  Within a nonempty signed interval, let
  $q_B=\Rank{B}{b^\alpha_{s,H}}{\one}$. The $t$th exponent group
  in suffix-array order ends at $\Select{B}{q_B+t}{\one}$.
  It begins immediately after $b^\alpha_{s,H}$ if $t=1$, and
  immediately after $\Select{B}{q_B+t-1}{\one}$ otherwise.
  We use these formulas in both queries.
  \begin{enumerate}
  \item Given $\alpha,s,k,|H|$ and the packed representation of $H$,
    compute $y=\Val{\AlphabetSize}{H}$ and read $L_{|H|,s}[y]$.
    Set $t=k-f^-_{s,H}+1$ if $\alpha=-$, and
    $t=f^+_{s,H}-k+1$ if $\alpha=+$. Return the two interval
    boundaries given above. These formulas for $t$ follow from
    the consecutive exponents and their order established in
    \cref{lm:periodic-exponent-intervals}.
  \item Given $i$, compute $j=\Rank{C}{i}{\one}$ and read $A[j]$
    to obtain $s$, $|H|$, the packed representation of $H$, and
    $b_{s,H},e_{s,H},z_{s,H},f^-_{s,H},f^+_{s,H}$.
    The root/head intervals are disjoint, and the query promises
    that $i$ belongs to one of them. Consequently, the last marked
    start at or before $i$ is the start of its interval, even if
    nonperiodic suffixes occur between these intervals.
    Set $\alpha=-$ if $i\leq z_{s,H}$, and $\alpha=+$ otherwise.
    Compute $q_B$ as above and set
    $t=\Rank{B}{i-1}{\one}-q_B+1$.
    This counts the exponent groups preceding rank $i$, plus one.
    Recover $k=f^-_{s,H}+t-1$ if $\alpha=-$, and
    $k=f^+_{s,H}-t+1$ if $\alpha=+$. Return these parameters,
    the stored root, and the interval boundaries given above.
  \end{enumerate}
  Both queries take $\bigO(1)$ time.

  \DSConstruction
  The main task is to construct $B$ from the maximal blocks of
  periodic positions. Each block contains at most one position with
  a given head and exponent. For each root, type, and head, we
  therefore count the blocks containing a position of each exponent.
  We store these counts in unary and use them to mark the ends of
  the exponent intervals in $B$.

  All packed initializations, updates, extractions, and concatenations below
  use \cref{pr:packed-representation}. When several strings must be
  concatenated, we collect their lengths and packed representations
  and perform one concatenation after all of them have been computed.
  \begin{enumerate}
  \item\label{step:periodic-intervals-blocks} We first compute the periodic blocks and the boundaries of
    the subarrays that will be processed together.
    \begin{enumerate}
    \item Compute $\lceil\log\AlphabetSize\rceil$ for the packed
      operations. Apply \cref{th:periodic-position-intervals} with
      $\tau$ and
      $\PackedRepresentation{w}{\AlphabetSize}{\Text}$.
      Let $q$ be the length of the returned array $A_{\rm beg}$,
      which lists $\RPrimTwo{\tau}{\Text}$.
      Construct the structure from \cref{pr:periodic-pos} with
      the same packed text. Construct the two encoders from
      \cref{th:val-encoding} and the reversal structure from
      \cref{th:string-reversal}.
    \item Allocate an array $A_{\rm blocks}[1\dd q]$.
      For each $i\in[1\dd q]$, read $a=A_{\rm beg}[i]$ and query
      the periodic-parameter structure with $(\tau,a)$.
      Let $H$ be the returned root, and compute $p=|H|$,
      $y=\Val{\AlphabetSize}{H}$, the sign $\alpha$ of the returned
      type, $\ell=h-\TailPos{a}{\tau}{\Text}$, and
      $u=\RunEndFullPos{a}{\tau}{\Text}-a$ from the returned values.
      Store
      \[
        A_{\rm blocks}[i]
          =\bigl(p,y,\alpha,
            \PackedRepresentation{w}{\AlphabetSize}{H},a,\ell,u\bigr).
      \]
      Sort $A_{\rm blocks}$ by its first three coordinates
      using radix sort.
    \item We record the boundaries between different sorting keys
      in an array $D$, initially empty. Scan $i=1,\ldots,q$.
      Append $i$ to $D$ if $i=1$ or if the first three coordinates
      of $A_{\rm blocks}[i]$ differ from those of
      $A_{\rm blocks}[i-1]$. After the scan, append $q+1$, and
      denote the resulting array by $D[1\dd d+1]$.
      Thus, $A_{\rm blocks}[D[j]\dd D[j+1])$ contains exactly
      the tuples with one root and type, for each $j\in[1\dd d]$.
      If $q=0$, this gives $D=[1]$ and $d=0$.
    \end{enumerate}

  \item\label{step:periodic-intervals-tables} Next, we construct two small lookup tables used only during
    construction. Let $\lambda=\lfloor\tfrac14\log N\rfloor$.
    \begin{enumerate}
    \item\label{step:periodic-intervals-repetitions} For each $v\in[0\dd\lambda)$, store the packed string
      $(\zero^v\one)^{\lfloor\lambda/(v+1)\rfloor}$.
      To construct $(\zero^v\one)^t$ for $v<\lambda$ and any
      integer $t\geq1$, concatenate copies of the stored string
      and, if needed, one prefix of it.
      For $v\geq\lambda$, initialize $\zero^v\one$ and concatenate
      $t$ copies.
    \item\label{step:periodic-intervals-filter} We next prepare a table for deleting the first bit of a
      binary string and every bit preceded by a one. For every
      $Y\in\BinaryAlphabet^{\leq\lambda}$ and $c\in\BinaryAlphabet$,
      let $t_1<\cdots<t_{q_{Y,c}}$ list the positions $t\in[1\dd|Y|]$
      satisfying $t=1$ and $c=\zero$, or $t>1$ and $Y[t-1]=\zero$.
      Let $Z_{Y,c}=Y[t_1]\cdots Y[t_{q_{Y,c}}]$, with $Z_{Y,c}=\emptystring$ if $q_{Y,c}=0$.
      Let $F$ denote the table defined by
      \[
        F[|Y|,\Val{2}{Y},c]
          =\bigl(q_{Y,c},\PackedRepresentation{w}{2}{Z_{Y,c}}\bigr).
      \]
      Compute the values of $F$ by enumerating the strings $Y$ and
      bits $c$. To apply this table to a binary string $U$, split $U$
      into consecutive blocks of length $\lambda$, except possibly
      the last block.
      For each block $Y$, read $F[|Y|,\Val{2}{Y},c]$, taking
      $c=\one$ for the first block and $c$ equal to the last input
      bit of the preceding block otherwise. Concatenate the returned
      strings. The choice of $c$ ensures that the first bit of $U$
      and every bit preceded by a one are deleted, including at
      the boundaries between consecutive blocks.
    \end{enumerate}

  \item\label{step:periodic-intervals-exponent-bitvectors} We now compute the bitvectors marking the ends of the
    exponent intervals. We store the results in temporary arrays
    $V^\alpha_{p,s}[0\dd\AlphabetSize^p)$, for
    $\alpha\in\{-,+\}$, $p\in[1\dd\lfloor\tau/3\rfloor]$,
    and $s\in[0\dd p)$. Each tuple stores an exponent, a string
    length, and a reference to a packed binary string.
    Initialize the first two coordinates to zero and the third to
    a reference to the packed empty string.
    For each $j\in[1\dd d]$, read $i_1=D[j]$ and
    $i_2=D[j+1]-1$. Read the common values $p,y,\alpha$ and the
    packed root $H$ from $A_{\rm blocks}[i_1]$.
    We process the subarray $A_{\rm blocks}[i_1\dd i_2]$ as follows.
    For $i\in[i_1\dd i_2]$, denote the last three coordinates
    of $A_{\rm blocks}[i]$ by $a_i,\ell_i,u_i$.
    By \cref{lm:periodic-block-properties}, the positions $x$ in
    the block starting at $a_i$ give exactly the distances
    $\RunEndFullPos{a_i}{\tau}{\Text}-x\in[\ell_i\dd u_i]$.
    Scan this subarray to compute
    $k_*=\max_{i\in[i_1\dd i_2]}\lfloor u_i/p\rfloor$.
    Since the positions $a_i$ listed in this subarray are exactly
    $\RPrimSignedThree{\alpha}{H}{\tau}{\Text}$, we have
    \[
      c_{s,k}:=|\RSignedFive{\alpha}{s}{k}{H}{\tau}{\Text}|
        =\bigl|\{i\in[i_1\dd i_2]:\ell_i\leq s+kp\leq u_i\}\bigr|
        \qquad(s\in[0\dd p),\ k\in[0\dd k_*]).
    \]
    For each head $s$, let $E^\alpha_{s,H}$ consist of the blocks
    $\zero^{c_{s,k}-1}\one$ for all $k\in[0\dd k_*]$ with
    $c_{s,k}>0$, in increasing order of $k$ if $\alpha=-$ and
    decreasing order if $\alpha=+$. By
    \cref{lm:periodic-exponent-intervals}, its ones mark the desired
    interval ends. We will store $f^\alpha_{s,H}$, $|E^\alpha_{s,H}|$,
    and a reference to its packed representation in $V^\alpha_{p,s}[y]$.
    We first compute the packed strings
    \[
      U_s=\textstyle\bigodot_{k=0}^{k_*}\zero^{c_{s,k}}\one
      \qquad(s\in[0\dd p)),
    \]
    where concatenation follows increasing order of $k$.
    Every exponent has a block in $U_s$, including those with
    $c_{s,k}=0$. We can therefore locate the block for exponent $k$
    by selecting its ending one and, when $k>0$, the preceding one.
    Allocate an array $A_U[0\dd p)$. Immediately after computing
    each $U_s$, augment it with rank and selection support using
    \cref{th:bin-rank-select}, and store its length and references to
    its packed representation and supporting structure in $A_U[s]$.
    The following substeps operate on the stored tuples of
    $A_{\rm blocks}[i_1\dd i_2]$ and $A_U$.
    \begin{enumerate}
    \item\label{step:periodic-intervals-unary-initial} First, we construct $U_0$. Initialize an empty array
      $A_{\rm pairs}$. Scan $i=i_1,\ldots,i_2$, read $\ell_i,u_i$
      from $A_{\rm blocks}[i]$, and append
      $(\lceil\ell_i/p\rceil,+1)$ and
      $(\lfloor u_i/p\rfloor+1,-1)$ to $A_{\rm pairs}$.
      Append $(0,0)$ and $(k_*+1,0)$, and sort $A_{\rm pairs}$
      by the first coordinate using radix sort.
      Initialize a counter $v=0$ and an empty array $A_{\rm parts}$.
      Scan $A_{\rm pairs}$, comparing adjacent first coordinates to
      find successive maximal intervals with a common first coordinate.
      For each such interval except the last, add to $v$ the sum
      of its second coordinates. Let $x$ be its first coordinate
      and $y'$ the first coordinate in the following interval.
      The counts $c_{0,k}$ for $k\in[x\dd y')$ all equal $v$.
      Construct $(\zero^v\one)^{y'-x}$ using Step~\ref{step:periodic-intervals-repetitions}, and append
      its length and a reference to its packed representation to
      $A_{\rm parts}$. After the scan, concatenate the strings in
      $A_{\rm parts}$ to obtain $U_0$.
      If the block starting at $a_i$ has no position of head zero,
      its two pairs have the same first coordinate and cancel.
    \item\label{step:periodic-intervals-head-changes} Next, we compute the changes between consecutive heads.
      For $s\in[1\dd p)$ and a fixed exponent $k$, the condition
      $\ell_i\leq s+kp\leq u_i$ starts to hold at $s+kp=\ell_i$
      and stops holding at $s+kp=u_i+1$.
      Initialize an empty array $A_{\rm triples}$, and scan
      $A_{\rm blocks}[i_1\dd i_2]$ again. From each stored
      pair $\ell_i,u_i$, compute
      $(\ell_i\bmod p,\lfloor\ell_i/p\rfloor,+1)$ and
      $((u_i+1)\bmod p,\lfloor(u_i+1)/p\rfloor,-1)$.
      Append each triple with nonzero first coordinate to
      $A_{\rm triples}$; the omitted triples concern head zero,
      which was handled in Step~\ref{step:periodic-intervals-unary-initial}.
      Sort $A_{\rm triples}$ by its first two coordinates using
      radix sort, and initialize an empty array $A_{\rm changes}$.
      Scan $A_{\rm triples}$, comparing consecutive pairs of first
      and second coordinates. For each maximal interval with the
      same pair $(s,k)$, add its third coordinates and append
      the resulting triple $(s,k,\delta)$ to $A_{\rm changes}$.
      Denote its final length by $M$. Each stored value $\delta$ equals
      $c_{s,k}-c_{s-1,k}$.
      To locate the triples for each head, allocate $J[1\dd p]$
      and initialize $t=1$. For $s=1,\ldots,p-1$, set $J[s]=t$
      and increase $t$ while $t\leq M$ and the first coordinate
      of $A_{\rm changes}[t]$ equals $s$. Finally, set $J[p]=M+1$.
      Thus, the triples for head $s$ occupy
      $A_{\rm changes}[J[s]\dd J[s+1])$.
    \item\label{step:periodic-intervals-unary-remaining} We now construct $U_1,\ldots,U_{p-1}$ in this order.
      For each $s$, read $U_{s-1}$ and its supporting structure
      from $A_U[s-1]$, and initialize an empty array $A_{\rm parts}$
      of lengths and references to packed strings. Initialize
      $j_0=1$, the next position of $U_{s-1}$ to copy.
      For $t=J[s],\ldots,J[s+1]-1$, read
      $(s,k,\delta)=A_{\rm changes}[t]$.
      Let $j_1=0$ if $k=0$, and
      $j_1=\Select{U_{s-1}}{k}{\one}$ otherwise; let
      $j_2=\Select{U_{s-1}}{k+1}{\one}$.
      The block for exponent $k$ is $U_{s-1}[j_1+1\dd j_2]$,
      with $j_2-j_1-1$ zeros.
      If $j_0\leq j_1$, extract the packed substring
      $U_{s-1}[j_0\dd j_1]$, and append its length and a reference
      to its packed representation to $A_{\rm parts}$.
      Initialize the replacement $\zero^{j_2-j_1-1+\delta}\one$,
      append its length and a reference to its packed representation
      to $A_{\rm parts}$, and set $j_0=j_2+1$.
      After processing these triples, extract the remaining substring
      $U_{s-1}[j_0\dd|U_{s-1}|]$ if it is nonempty, and append
      its length and a reference to its packed representation to
      $A_{\rm parts}$.
      Concatenate the strings in $A_{\rm parts}$ to obtain $U_s$.
      When $J[s]=J[s+1]$, this simply copies $U_{s-1}$.
    \item We next compute the first exponent in each signed
      suffix-array interval. For every $s\in[0\dd p)$, read
      $U_s$ and its supporting structure from $A_U[s]$.
      Count its zeros with a rank query; their number is
      $|\RSignedFour{\alpha}{s}{H}{\tau}{\Text}|$.
      If it is zero, leave the first coordinate of
      $V^\alpha_{p,s}[y]$ equal to zero.
      Otherwise, use selection to find the first and last zero,
      and use rank queries to count the ones preceding them.
      These counts give the smallest and largest occurring exponents.
      Store the smaller count if $\alpha=-$, or the larger count
      if $\alpha=+$, in the first coordinate of $V^\alpha_{p,s}[y]$.
      This is $f^\alpha_{s,H}$.
    \item Finally, for each $s\in[0\dd p)$, read the length and
      packed representation of $U_s$ from $A_U[s]$ and apply
      the table $F$ as described in Step~\ref{step:periodic-intervals-filter}.
      This deletes the first symbol of every block
      $\zero^{c_{s,k}}\one$: a block with $c_{s,k}=0$ disappears,
      and a positive count produces $\zero^{c_{s,k}-1}\one$.
      If $\alpha=-$, the result is already $E^-_{s,H}$.
      If $\alpha=+$ and the result is nonempty, reverse it using
      \cref{th:string-reversal}, remove its first bit, and append a one.
      Reversal puts a one at the start of each block in decreasing
      exponent order; moving the initial one to the end puts the
      ones at the ends of these blocks instead.
      Store the length of the result and a reference to its packed
      representation in the last two coordinates of $V^\alpha_{p,s}[y]$.
    \end{enumerate}

  \item\label{step:periodic-intervals-assemble} Finally, we construct the arrays $L_{p,s}$ and $A$ and
    the bitvectors $B$ and $C$. Initialize all values in $L_{p,s}$
    to $\bot$, initialize $A$ to an empty array, and initialize
    $C:=\zero^N$. Construct the suffix-array range structure from
    \cref{pr:range-beg-and-end} for the packed text.
    \begin{enumerate}
    \item Scan $i=1,\ldots,q$ in $A_{\rm blocks}$.
      Process a tuple only if $i=1$ or its first two coordinates
      differ from those of $A_{\rm blocks}[i-1]$. This visits
      each root once. Read $p,y$ and the packed root $H$ from
      that tuple. For every $s\in[0\dd p)$, form the packed
      representation of the length-$h$ prefix $X$ of
      $H[|H|-s+1\dd|H|]H^h$ by writing its symbols in order.
      Since $H$ is primitive and $h\geq2|H|$, $X$ has shortest
      period $|H|$. As $|X|=h=3\tau-1$ and $|H|\leq\tau/3$, every
      occurrence of $X$ starts at a $\tau$-periodic position
      (\cref{def:periodic-positions}), and its root and head are $H$
      and $s$.
      Conversely, every position in $\RFour{s}{H}{\tau}{\Text}$
      starts an occurrence of $X$ by the definition of the root
      and head. Query the range structure with $X$ to obtain
      $b_{s,H},e_{s,H}$.
      For each $\alpha\in\{-,+\}$, read the first two coordinates
      $f^\alpha_{s,H},|E^\alpha_{s,H}|$ from $V^\alpha_{p,s}[y]$.
      These values are zero if this type does not occur with $H$. Set $z_{s,H}=b_{s,H}+|E^-_{s,H}|$.
      If $b_{s,H}<e_{s,H}$, set
      $L_{p,s}[y]:=(b_{s,H},e_{s,H},z_{s,H},f^-_{s,H},f^+_{s,H})$
      and append $(s,p,\PackedRepresentation{w}{\AlphabetSize}{H},L_{p,s}[y])$
      to $A$.
    \item Sort $A$ by $b_{s,H}$ using radix sort, and denote its
      length by $r$. For every $j\in[1\dd r]$, read
      $b_{s_j,H_j}$ from $A[j]$ and set $C[b_{s_j,H_j}+1]:=\one$.
      Augment $C$ with rank support using \cref{th:bin-rank-select}.
    \item We assemble $B$ using an initially empty array
      $A_{\rm parts}$ of lengths and references to packed strings.
      Scan $A[1\dd r]$. For each tuple, read its head $s$, root length $p$, packed
      root $H$, and interval boundaries. Compute
      $y=\Val{\AlphabetSize}{H}$ using the encoder, and read
      the lengths and packed strings from $V^-_{p,s}[y]$ and
      $V^+_{p,s}[y]$. By \cref{lm:periodic-exponent-intervals},
      these strings satisfy
      $B(b_{s,H}\dd z_{s,H}]=E^-_{s,H}$ and
      $B(z_{s,H}\dd e_{s,H}]=E^+_{s,H}$.
      Append their lengths and references to $A_{\rm parts}$ in
      this order, inserting zero strings for the gaps between
      consecutive intervals and for any initial or final unused ranks.
      Concatenate the strings in $A_{\rm parts}$ to obtain $B$.
      If $r=0$, set $B:=\zero^N$.
      Augment $B$ with rank and selection support using
      \cref{th:bin-rank-select}.
    \end{enumerate}
  \end{enumerate}

  We now analyze the construction. Step~\ref{step:periodic-intervals-blocks} takes $\bigO(N/\tau)$
  time and words of working space, including the auxiliary structures
  and sorting $A_{\rm blocks}$. Constructing $D$ takes $\bigO(q)$ time
  and words.
  The tables in Step~\ref{step:periodic-intervals-tables} take $\bigO(\lambda^2+2^\lambda\lambda)$
  time and words of space to construct. Every stored repetition in
  Step~\ref{step:periodic-intervals-repetitions} has length between $\lambda/2$ and $\lambda$. Thus,
  constructing $(\zero^v\one)^t$ takes
  $\bigO(1+(v+1)t/\log N)$ time: for $v<\lambda$, this bounds
  the number of copies of the stored string; for $v\geq\lambda$,
  we have $t\leq(v+1)t/\lambda$, which bounds the number of
  copies of $\zero^v\one$.
  Applying $F$ to $U$ takes $\bigO(1+|U|/\log N)$ time, since
  each block is processed in constant time.

  To bound the work in Step~\ref{step:periodic-intervals-exponent-bitvectors}, we first bound the total length of
  its unary strings. Let $g=\sum_{j=1}^d p_j$, where $p_j$ is
  the first coordinate of $A_{\rm blocks}[D[j]]$.
  Since each root has length at most $\tau/3$, we have
  $g=\bigO(\tau\AlphabetSize^\tau)
    =\bigO(N^{1/6}\log N)$.
  Consider the subarray processed in one iteration of Step~\ref{step:periodic-intervals-exponent-bitvectors},
  with bounds $i_1,i_2$, root $H$, and sign $\alpha$.
  Let $i_*\in[i_1\dd i_2]$ maximize $u_i$. Then
  \[
    \sum_{s=0}^{|H|-1}|U_s|
      =\sum_{s=0}^{|H|-1}
          |\RSignedFour{\alpha}{s}{H}{\tau}{\Text}|
        +|H|(k_*+1)
      \leq\sum_{s=0}^{|H|-1}
          |\RSignedFour{\alpha}{s}{H}{\tau}{\Text}|
        +u_{i_*}+|H|.
  \]
  The subarrays have disjoint index ranges, and hence select
  different block starts $a_{i_*}$. Moreover,
  $u_{i_*}\leq\RunEndPos{a_{i_*}}{\tau}{\Text}-a_{i_*}$.
  By \cref{lm:periodic-block-properties}, summing over all $d$ subarrays
  bounds the total length of all $U_s$ by $3N+g=\bigO(N)$.

  Each block start contributes two pairs in Step~\ref{step:periodic-intervals-unary-initial} and at most
  two triples in Step~\ref{step:periodic-intervals-head-changes}. There are therefore
  $\bigO(N/\tau+g)$ pairs, triples, and copied or replaced substrings
  in Step~\ref{step:periodic-intervals-exponent-bitvectors}. Merging the triples and constructing all arrays $J$
  takes $\bigO(q+g)$ time and words. The temporary tables
  $V^\alpha_{p,s}$ take $\bigO(\tau\AlphabetSize^\tau)=\bigO(N/\log N)$
  time and words to initialize. Each radix sort uses
  $\bigO(\sqrt N)$ buckets and a constant number of passes.
  Since $d\leq g$, the total sorting overhead is
  $\bigO(g\sqrt N)$. Apart from constructing rank/select support,
  Steps~\ref{step:periodic-intervals-tables}--\ref{step:periodic-intervals-exponent-bitvectors} therefore take
  $\bigO(N/\tau+g\sqrt N+N/\log N+2^\lambda\lambda)$ time.
  To bound the rank/select construction time, apply
  \cref{th:bin-rank-select} separately with each unary string's
  own length bound. For strings of length at least $\sqrt N$,
  the total time is $\bigO(N/\log N)$. There are at most $g$
  shorter strings, and their total construction time is at most
  $\bigO(g\sqrt N)=\bigO(N/\log N)$.

  In Step~\ref{step:periodic-intervals-assemble}, initializing the arrays $L_{p,s}$ takes
  $\bigO(\tau\AlphabetSize^\tau)$ time, and forming the short strings
  for the range queries takes $\bigO(g\tau)$ time. There are at most
  $g$ tuples in $A$; sorting them takes $\bigO(g+\sqrt N)$ time.
  Initializing and marking $C$, assembling $B$, and constructing
  their rank/select support take $\bigO(g+N/\log N)$ time.
  Including the suffix-array range structure, Step~\ref{step:periodic-intervals-assemble} therefore takes
  $\bigO(N/\log_{\AlphabetSize}N+\tau\AlphabetSize^\tau+g\tau)$ time.
  Since $\AlphabetSize^\tau\leq N^{1/6}$ and $\tau=\bigO(\log N)$,
  all terms other than $N/\tau$ and $N/\log_{\AlphabetSize}N$ are
  $\bigO(N/\log N)$. The assumption
  $\tau\leq\tfrac16\log_{\AlphabetSize}N$ therefore bounds the
  total construction time by $\bigO(N/\tau)$.

  It remains to verify the length and word-size requirements of
  the tools used above. Every individual $U_s$ has length less than
  $2N$: it contains at most $N-h+1$ zeros and at most $N+1$ ones,
  and $h\geq8$. Packed operations on these strings can therefore
  use length bound $N+(N-1)$, which satisfies $w>\log(2N-1)$.
  All counts, coordinates, and individual string lengths fit in one
  word. In particular, the boundaries in $D$ and $J$ are at most
  $q+1$ and $2q+1$, respectively, both at most $N$ since
  $q\leq N/(2\tau)$ and $\tau\geq3$. The position $j_0$ in
  Step~\ref{step:periodic-intervals-unary-remaining} is at most $|U_{s-1}|+1<2N$.
  The suffix-array range structure uses size parameter $N$ and
  alphabet size $\AlphabetSize$, and its queries use strings of length
  $h\leq\tfrac12\log_{\AlphabetSize}N$. The applications of
  \cref{pr:periodic-pos,th:periodic-position-intervals} use size
  parameter $N$ and alphabet size $\AlphabetSize$, and they satisfy
  their requirement $\tau\leq\tfrac12\log_{\AlphabetSize}N$.
  The reversal structure is constructed once, with alphabet size
  two and length bound $N$; each $E^\alpha_{s,H}$ has length at
  most $N$. The encoders use bound $N$ and alphabet sizes
  $\AlphabetSize$ and $2$, and all their
  input strings have length at most the corresponding logarithm.
  The rank and selection structures for $B$ and $C$ use size
  parameter $N$.
  No upper bound on $w$ is needed.
  The packed strings and their rank/select structures occupy
  $\bigO(g+N/w)$ words, including the references stored in $A_U$.
  The tables $V^\alpha_{p,s}$ use $\bigO(\tau\AlphabetSize^\tau)$
  additional words. The arrays $A_{\rm blocks}$, $D$, $J$, and the
  other auxiliary structures occupy $\bigO(N/\tau)$ words.
  Thus, the claimed working-space bound holds as well.
\end{proof}

%% file: appendix/periodic-suffix-array.tex
\subsubsection{Suffix Array Queries}
\label{sec:appendix-periodic-sa}

\begin{lemma}\label{lm:periodic-sa-range-select}
  Let $\Textlen\in\Z_{\geq2}$, $\Text\in\Sigma^{\Textlen}$,
  and $\tau\in[1\dd\lfloor\Textlen/2\rfloor]$.
  Let $i\in[1\dd\Textlen]$ satisfy $\SA{\Text}[i]\in\RTwo{\tau}{\Text}$
  (see \cref{def:suffix-array,def:periodic-positions}),
  and denote $j=\SA{\Text}[i]$.
  Let $\alpha\in\{-,+\}$ be the sign of $\TypePos{j}{\tau}{\Text}$,
  and denote $H=\RootPos{j}{\tau}{\Text}$,
  $s=\HeadPos{j}{\tau}{\Text}$, $k=\ExpPos{j}{\tau}{\Text}$,
  and $d=s+k|H|$ (see \cref{def:pos-root}).

  Let $(a_t)_{t\in[1\dd q]}=\RunsLexSorted{\alpha}{\tau}{\Text}$
  (\cref{def:runs-lex-sorted}), and define $A[1\dd q]$ by
  $A[t]=\RunEndFullPos{a_t}{\tau}{\Text}-a_t$.

  Let $u,v,b,e$ be the integers satisfying
  \[
    (u\dd v]=\{t\in[1\dd q]:\RootPos{a_t}{\tau}{\Text}=H\},
    \qquad
    (b\dd e]=\{\ISA{\Text}[x]:
       x\in\RSignedFive{\alpha}{s}{k}{H}{\tau}{\Text}\}.
  \]
  Define
  \[
    r=
    \begin{cases}
      \TwoSidedRangeCount{A}{v}{d}-(e-i),&\text{if }\alpha=-,\\
      \TwoSidedRangeCount{A}{u}{d}+(i-b),&\text{if }\alpha=+.
    \end{cases}
  \]
  Then $t=\RangeSelect{A}{r}{d}$ belongs to $(u\dd v]$, and
  $\SA{\Text}[i]=\RunEndFullPos{a_t}{\tau}{\Text}-d$
  (see \cref{def:range-count,def:range-select}).
\end{lemma}
\begin{proof}
  By the definitions of $H,s,k,\alpha$, we have
  $j\in\RSignedFive{\alpha}{s}{k}{H}{\tau}{\Text}$.
  By \cref{def:runs-lex-sorted}, the indices $t$ with
  $\RootPos{a_t}{\tau}{\Text}=H$ form one consecutive interval,
  which is $(u\dd v]$.
  The interval $(b\dd e]$ exists by
  \cref{lm:periodic-exponent-intervals}\eqref{lm:periodic-exponent-intervals-sa}
  and contains $i=\ISA{\Text}[j]$.

  Let $t\in(u\dd v]$ be the index such that the maximal consecutive
  block of $\RTwo{\tau}{\Text}$ containing $j$ starts at $a_t$.
  The full endpoint is constant throughout this block, and hence
  $j=\RunEndFullPos{a_t}{\tau}{\Text}-d$ and $A[t]\geq d$.
  We first determine which other blocks contain positions in
  $\RSignedFive{\alpha}{s}{k}{H}{\tau}{\Text}$ whose suffixes
  follow $\Text[j\dd\Textlen]$ when $\alpha=-$, or precede it when
  $\alpha=+$.

  Let $h\in(u\dd v]$ satisfy $A[h]\geq d$, with $h>t$ if
  $\alpha=-$ and $h<t$ if $\alpha=+$. Put
  $x=\RunEndFullPos{a_h}{\tau}{\Text}-d$.
  This is the only possible position with head $s$ and exponent
  $k$ in the block starting at $a_h$. The inequality $A[h]\geq d$
  gives $x\geq a_h$. To establish membership in that block, we
  must also show
  $\RunEndPos{a_h}{\tau}{\Text}-x\geq3\tau-1$.
  Write $t_h=\TailPos{a_h}{\tau}{\Text}$ and
  $t_j=\TailPos{j}{\tau}{\Text}$. We show $t_h\geq t_j$
  in each of the following two cases:
  \begin{itemize}
  \item Suppose $\alpha=-$ and $h>t$.
    The order of the blocks gives
    $\Text[\RunEndFullPos{a_h}{\tau}{\Text}\dd\Textlen]
      \succ\Text[\RunEndFullPos{j}{\tau}{\Text}\dd\Textlen]$.
    If $t_h<t_j$, these suffixes share their first $t_h$ symbols,
    namely $H[1\dd t_h]$. At the next symbol, the former suffix
    ends or contains a symbol smaller than $H[t_h+1]$, by its
    negative type. The latter still contains $H[t_h+1]$.
    This contradicts their order.
  \item Suppose $\alpha=+$ and $h<t$.
    Now
    $\Text[\RunEndFullPos{a_h}{\tau}{\Text}\dd\Textlen]
      \prec\Text[\RunEndFullPos{j}{\tau}{\Text}\dd\Textlen]$.
    If $t_h<t_j$, the first $t_h$ symbols again agree, but the
    next symbol of the former suffix is greater than
    $H[t_h+1]$, by its positive type. The next symbol of the
    latter is $H[t_h+1]$, contradicting their order.
  \end{itemize}
  In either case,
  \[
    \RunEndPos{a_h}{\tau}{\Text}-x
       =d+t_h\geq d+t_j
       =\RunEndPos{j}{\tau}{\Text}-j\geq3\tau-1.
  \]
  Thus, \cref{lm:periodic-block-properties}\eqref{lm:periodic-block-properties-interval} places $x$
  in the block starting at $a_h$. It has root $H$ and type
  specified by $\alpha$. Since
  $\RunEndFullPos{x}{\tau}{\Text}-x=d=s+k|H|$ and $0\leq s<|H|$,
  division by $|H|$ gives remainder $s$ and quotient $k$.
  These are precisely the head and exponent at $x$.
  The suffixes starting at $x$ and $j$ have the same length-$d$
  prefix $H'H^k$, where $H'$ is the length-$s$ suffix of $H$.
  Their order is therefore the order of the suffixes starting
  at $\RunEndFullPos{a_h}{\tau}{\Text}$ and
  $\RunEndFullPos{a_t}{\tau}{\Text}$.

  Conversely, every
  $x\in\RSignedFive{\alpha}{s}{k}{H}{\tau}{\Text}$ belongs to a
  unique block with start $a_h$, where $h\in(u\dd v]$.
  The same endpoint identity gives
  $x=\RunEndFullPos{a_h}{\tau}{\Text}-d$ and $A[h]\geq d$.
  Deleting the common prefix $H'H^k$ shows that
  $\Text[x\dd\Textlen]\succ\Text[j\dd\Textlen]$ if and only
  if $h>t$, and
  $\Text[x\dd\Textlen]\prec\Text[j\dd\Textlen]$ if and only
  if $h<t$.
  The distinct endpoints in
  \cref{lm:periodic-block-properties}\eqref{lm:periodic-block-properties-distinct} ensure that
  different indices $h$ give different positions $x$.

  For $\alpha=-$, we have therefore counted exactly
  $e-i$ positions using the indices
  $h\in(t\dd v]$ with $A[h]\geq d$.
  Their number is
  $\TwoSidedRangeCount{A}{v}{d}-\TwoSidedRangeCount{A}{t}{d}$.
  For $\alpha=+$, the indices $h\in(u\dd t)$ with $A[h]\geq d$
  count exactly the $i-(b+1)$ preceding positions.
  Their number is
  $\TwoSidedRangeCount{A}{t-1}{d}-\TwoSidedRangeCount{A}{u}{d}$.
  By the definition of $r$, these two equalities give
  $r=\TwoSidedRangeCount{A}{t}{d}$ when $\alpha=-$, and
  $r=\TwoSidedRangeCount{A}{t-1}{d}+1$ when $\alpha=+$.
  Since $A[t]\geq d$, both values are the rank of index $t$
  among the indices whose values in $A$ are at least $d$.
  Hence, $\RangeSelect{A}{r}{d}=t$, and
  $\SA{\Text}[i]=j=\RunEndFullPos{a_t}{\tau}{\Text}-d$,
  as claimed.
\end{proof}

\begin{proposition}\label{pr:periodic-sa}
  Let $\AlphabetSize,N\in\Z_{\geq2}$ satisfy
  $\AlphabetSize<N^{1/7}$, and let
  $\tau=\lfloor\tfrac17\log_{\AlphabetSize}N\rfloor\geq1$.
  Consider the word RAM model with word size
  $w\geq1+\log N$.
  Assume that there exists a data structure answering range counting
  queries (\cref{def:range-count}) that, given $N$, integers $m\in[2\dd N]$
  and $m'\in[0\dd m]$, and an array $A[1\dd m']$ of positive integers satisfying
  $\sum_{i=1}^{m'}A[i]\leq N$ and
  $\sum_{i=1}^{m'}A[i]\leq m\log m$,
  achieves the following complexities:
  \begin{itemize}
  \item space usage $S(m,N)$ words,
  \item preprocessing time $P_t(m,N)$,
  \item preprocessing space $P_s(m,N)$ words,
  \item query time $Q(m,N)$.
  \end{itemize}
  Then, there exists $m=\Theta(N/\log_{\AlphabetSize}N)$ with $m\leq N$
  and a data structure that, given $N$, $\AlphabetSize$, and the packed representation
  $\PackedRepresentation{w}{\AlphabetSize}{\Text}$
  (\cref{def:packed-representation}) of any text
  $\Text\in\IntegerAlphabet^N$, answers queries that, given
  $i\in[1\dd N]$ satisfying $\SA{\Text}[i]\in\RTwo{\tau}{\Text}$
  (\cref{def:periodic-positions}), return
  $\SA{\Text}[i]$ (\cref{def:suffix-array}), and achieves:
  \begin{itemize}
  \item space usage $\bigO(N/\log_{\AlphabetSize}N+S(m,N))$ words,
  \item preprocessing time $\bigO(N/\log_{\AlphabetSize}N+P_t(m,N))$,
  \item preprocessing space $\bigO(N/\log_{\AlphabetSize}N+P_s(m,N))$ words,
  \item query time $\bigO(1+Q(m,N))$.
  \end{itemize}
\end{proposition}
\begin{proof}
  We apply the identity from \cref{lm:periodic-sa-range-select},
  using the structure from the claim for counting and
  \cref{th:range-count-for-small-sum} for selection.
  Let $m=\lceil N/\tau\rceil$. If $\tau\leq2$, then
  $\RTwo{\tau}{\Text}=\emptyset$, and it suffices to store $N$.
  Henceforth, assume that $\tau\geq3$, and denote $L=\lfloor\tau/3\rfloor$.

  For each $\alpha\in\{-,+\}$, let
  $(a^\alpha_t)_{t\in[1\dd q_\alpha]}
    =\RunsLexSorted{\alpha}{\tau}{\Text}$
  (\cref{def:runs-lex-sorted}).
  Let $F^\alpha[1\dd q_\alpha]$ and $A^\alpha[1\dd q_\alpha]$
  be the arrays defined by
  \[
    F^\alpha[t]=\RunEndFullPos{a^\alpha_t}{\tau}{\Text},
    \qquad
    A^\alpha[t]=F^\alpha[t]-a^\alpha_t
    \qquad\text{for }t\in[1\dd q_\alpha].
  \]
  For every $p\in[1\dd L]$, let
  $U^\alpha_p[0\dd\AlphabetSize^p)$ be the array defined as follows.
  For every $H\in\IntegerAlphabet^p$, let
  $U^\alpha_p[\Val{\AlphabetSize}{H}]=(u,v)$, where
  $(u\dd v]=\{t\in[1\dd q_\alpha]:\RootPos{a^\alpha_t}{\tau}{\Text}=H\}$
  if this set is nonempty; otherwise, let
  $U^\alpha_p[\Val{\AlphabetSize}{H}]=\bot$.
  Thus, these arrays give the boundaries of the group with root $H$.

  By \cref{def:pos-root,lm:periodic-block-properties}, each $A^\alpha[t]$
  belongs to $[1\dd N]$, and
  $\sum_{t=1}^{q_\alpha}A^\alpha[t]\leq2N$.
  To keep each range-query instance within the sum bound $N$, we use
  the partition of each $A^\alpha$ obtained by repeatedly taking the
  longest prefix of the remaining array whose sum is at most $N$.
  Let $0=o^\alpha_0<\cdots<o^\alpha_{r_\alpha}=q_\alpha$ be the
  boundaries of this partition, and let
  $A^\alpha_h[j]=A^\alpha[o^\alpha_{h-1}+j]$ for
  $h\in[1\dd r_\alpha]$ and $j\in[1\dd o^\alpha_h-o^\alpha_{h-1}]$.
  If $q_\alpha=0$, let $r_\alpha=0$ and $o^\alpha_0=0$.
  There are at most three subarrays for each $\alpha$:
  the sums of any two consecutive subarrays together exceed $N$,
  whereas four subarrays would then have total sum greater than $2N$.

  \DSComponents
  We store the following components:
  \begin{enumerate}
  \item The integers $N$, $\tau$, $r_-$, and $r_+$, the interval
    structure from \cref{pr:periodic-sa-intervals}, and the
    integer-value encoder from \cref{th:val-encoding} for
    $N,\AlphabetSize$. They occupy $\bigO(N/\log N+\sqrt N)$ words.
  \item The arrays $F^\alpha$ and $U^\alpha_p$, for
    $\alpha\in\{-,+\}$ and $p\in[1\dd L]$.
    They occupy $\bigO(q_-+q_++\AlphabetSize^L)$ words.
  \item For each $\alpha\in\{-,+\}$ and $h\in[1\dd r_\alpha]$,
    the range-counting structure from the claim and the range-selection
    structure from \cref{th:range-count-for-small-sum} for $A^\alpha_h$,
    both with parameters $m,N$ (and $c=1$ for the latter), together with their input arrays and
    the boundaries $o^\alpha_{h-1},o^\alpha_h$.
    The arrays, boundaries, and selection structures use $\bigO(m)$
    words in total, and the counting structures use $\bigO(S(m,N))$ words.
  \end{enumerate}
  Since $q_-+q_+\leq N/(2\tau)$, $\AlphabetSize^L\leq N^{1/21}$,
  and $\tau=\Theta(\log_{\AlphabetSize}N)$, the auxiliary size is
  $\bigO(m)$ words, and the total size is
  $\bigO(m+S(m,N))$ words.

  \DSQueries
  We first explain how the subarrays support counting and selection
  on $A^\alpha$. A counting query
  $\TwoSidedRangeCount{A^\alpha}{z}{d}$ with $z=0$ returns zero.
  Otherwise, add the counts for all complete subarrays contained in
  $A^\alpha[1\dd z]$ and, if $z$ lies inside the next subarray
  $A^\alpha_h$, its prefix count with arguments
  $(z-o^\alpha_{h-1},d)$. The boundaries identify these subarrays,
  and their counting structures answer at most three queries.

  For a selection query $\RangeSelect{A^\alpha}{r}{d}$, compute
  $n_h=\TwoSidedRangeCount{A^\alpha_h}{o^\alpha_h-o^\alpha_{h-1}}{d}$
  for every $h\in[1\dd r_\alpha]$, using the same counting structures.
  Compute the cumulative counts $C[0\dd r_\alpha]$, where
  $C[0]=0$ and $C[h]=C[h-1]+n_h$.
  If $r>C[r_\alpha]$, return $\infty$. Otherwise, find the unique
  $h$ satisfying $C[h-1]<r\leq C[h]$ and return
  $o^\alpha_{h-1}+\RangeSelect{A^\alpha_h}{r-C[h-1]}{d}$,
  using the selection structure for $A^\alpha_h$.
  The indices in earlier subarrays precede those in later subarrays,
  which proves this formula. Since $r_\alpha\leq3$, both operations
  take $\bigO(1+Q(m,N))$ time.

  Given $i\in[1\dd N]$ with $\SA{\Text}[i]\in\RTwo{\tau}{\Text}$,
  let $j=\SA{\Text}[i]$, and denote $H=\RootPos{j}{\tau}{\Text}$,
  $s=\HeadPos{j}{\tau}{\Text}$, and $k=\ExpPos{j}{\tau}{\Text}$.
  Let $\alpha\in\{-,+\}$ be the sign of $\TypePos{j}{\tau}{\Text}$,
  and let $b,e$ be the boundaries of the suffix-array interval for
  positions with this type, root, head, and exponent; i.e.,
  \[
    (b\dd e]=\{\ISA{\Text}[x]:
      x\in\RSignedFive{\alpha}{s}{k}{H}{\tau}{\Text}\}.
  \]
  We answer the suffix-array query in three steps.
  \begin{enumerate}
  \item\label{step:periodic-sa-query-interval} Use the second query from \cref{pr:periodic-sa-intervals}
    with $i$ to compute $\alpha,s,k,b,e$, $|H|$, and
    $\PackedRepresentation{w}{\AlphabetSize}{H}$.
    Use the encoder from \cref{th:val-encoding} to compute
    $x=\Val{\AlphabetSize}{H}$, read $(u,v)=U^\alpha_{|H|}[x]$,
    and compute $d=s+k|H|$.
  \item\label{step:periodic-sa-query-rank} If $\alpha=-$, compute
    $r=\TwoSidedRangeCount{A^-}{v}{d}-(e-i)$.
    If $\alpha=+$, compute
    $r=\TwoSidedRangeCount{A^+}{u}{d}+(i-b)$.
  \item\label{step:periodic-sa-query-position} Compute $t=\RangeSelect{A^\alpha}{r}{d}$ and return
    $F^\alpha[t]-d$.
  \end{enumerate}
  The interval structure returns the parameters of
  $j=\SA{\Text}[i]$, and the root of its block start is also $H$.
  Hence, the lookup in Step~\ref{step:periodic-sa-query-interval} returns its root-group boundaries.
  By \cref{lm:periodic-sa-range-select}, the rank $r$ is valid,
  $t\in(u\dd v]$, and $\SA{\Text}[i]=F^\alpha[t]-d$.
  This proves correctness. Step~\ref{step:periodic-sa-query-interval} takes constant time, and Steps~\ref{step:periodic-sa-query-rank}
  and~\ref{step:periodic-sa-query-position} use the counting and selection operations described above,
  giving query time $\bigO(1+Q(m,N))$.

  \DSConstruction
  We construct the components in four steps.
  \begin{enumerate}
  \item\label{step:periodic-sa-construct-support} First, we construct the supporting structures and list the
    starts of the maximal periodic blocks. Construct the interval
    structure from \cref{pr:periodic-sa-intervals} using
    $\tau$ and the packed text. Construct the periodic-parameter structure
    from \cref{pr:periodic-pos} using the packed text,
    and the encoder from \cref{th:val-encoding}.
    Apply \cref{th:periodic-position-intervals} with
    $\tau$ and the packed text to obtain the block starts
    $c_1<\cdots<c_q$, where $q=q_-+q_+$.
  \item\label{step:periodic-sa-construct-sort-endpoints} Next, we sort the suffixes starting at the full endpoints.
    Query the periodic-parameter structure at each $c_h$ to obtain
    its type, root, and full endpoint $F[h]=\RunEndFullPos{c_h}{\tau}{\Text}$.
    Apply \cref{th:sparse-suffix-sorting} to $q$,
    the packed text, and the array $F[1\dd q]$.
    Denote the resulting array by $E[1\dd q]$. Form the pairs $(E[r],r)$
    for $r\in[1\dd q]$ and sort them by their first coordinate
    using radix sort. Since $F[1]<\cdots<F[q]$ by
    \cref{lm:periodic-block-properties}, the second coordinate of
    the $h$th resulting pair gives the rank $\rho[h]$ of the suffix
    starting at $F[h]$ among these suffixes.
  \item\label{step:periodic-sa-construct-arrays} We now construct the arrays $A^\alpha,F^\alpha,U^\alpha_p$.
    \begin{enumerate}
    \item Compute the powers $\AlphabetSize^0,\ldots,\AlphabetSize^L$.
      For each $h\in[1\dd q]$, let $H_h$ be the root of $c_h$.
      Compute $p_h=|H_h|$, $y_h=\Val{\AlphabetSize}{H_h}$ using
      the encoder, and $x_h=y_h\cdot\AlphabetSize^{L-p_h}$.
      Ordering the pairs $(x_h,p_h)$ is equivalent to ordering the
      roots lexicographically: padding on the right with zeros
      preserves the first differing symbol, and the length breaks
      a tie when a proper prefix is extended only by zeros.
    \item For each $\alpha\in\{-,+\}$, initialize all values in
      $U^\alpha_1,\ldots,U^\alpha_L$ to $\bot$.
      Collect in an array the indices $h\in[1\dd q]$ with
      $c_h\in\RPrimSignedTwo{\alpha}{\tau}{\Text}$, and sort them
      by $(x_h,p_h,\rho[h])$ using radix sort.
      If $h$ is the $t$th index in the resulting array, set
      $F^\alpha[t]=F[h]$ and $A^\alpha[t]=F[h]-c_h$.
      Scan the sorted array to find each maximal interval
      $[t_1\dd t_2]$ with equal pairs $(x_h,p_h)$.
      Read the index $h$ at position $t_1$ of the sorted array, and set
      $U^\alpha_{p_h}[y_h]=(t_1-1,t_2)$.
    \end{enumerate}
  \item\label{step:periodic-sa-construct-range-structures} Finally, we partition the arrays and construct their counting
    and selection structures. For each $\alpha\in\{-,+\}$,
    initialize the boundary array with zero. If $q_\alpha=0$,
    store $r_\alpha=0$ and make no call. Otherwise, scan $A^\alpha$
    from left to right, maintaining the sum $s_{\rm cur}$ of the current
    subarray, initially zero. Before appending $A^\alpha[t]$,
    test whether $A^\alpha[t]>N-s_{\rm cur}$. If so, end the current subarray
    immediately before $t$, store its boundary, and reset $s_{\rm cur}$ to zero.
    Append $A^\alpha[t]$ and add it to $s_{\rm cur}$. Store the final boundary
    and the number $r_\alpha$ of subarrays after the scan.
    For each subarray, construct the counting structure from the claim
    and the selection structure from \cref{th:range-count-for-small-sum}
    with $c=1$, each on a separate copy of the input array.
  \end{enumerate}

  We now check the tool hypotheses and the construction bounds.
  The choice of $\tau$ gives $\tau=\Theta(\log_{\AlphabetSize}N)$,
  $m=\Theta(N/\log_{\AlphabetSize}N)\leq N$, and
  $\log m=\Theta(\log N)$. Moreover, $N\leq m\tau\leq m\log m$, where the last inequality holds because
  $m\geq N/\tau\geq7N/\log N$ gives $\log m\geq\tfrac12\log N\geq\tau$.
  The same bound gives $m\geq2$.
  Each subarray has at most $q\leq N/(2\tau)\leq m$ positive
  integers and sum at most $N$, and hence satisfies both sum
  requirements of the counting and selection structures.
  The interval and periodic-parameter structures, the computation of
  the maximal periodic blocks, and sparse suffix sorting use size
  parameter $N$ and alphabet size $\AlphabetSize$.
  The periodic tools accept $\tau$, since
  $\tau\leq\tfrac17\log_{\AlphabetSize}N<\tfrac16\log_{\AlphabetSize}N$.
  The full endpoints are distinct and increasing and belong to
  $[1\dd N+1]$. Thus, the array $F[1\dd q]$ used in Step~\ref{step:periodic-sa-construct-sort-endpoints}
  is a valid input set for \cref{th:sparse-suffix-sorting}.
  Every root has length at most $L\leq\tau/3$, within the
  input-length limit of the encoder.

  Step~\ref{step:periodic-sa-construct-support} takes $\bigO(N/\tau+\sqrt N)=\bigO(m)$ time and words,
  and Step~\ref{step:periodic-sa-construct-sort-endpoints} takes $\bigO(N/\log_{\AlphabetSize}N+q)=\bigO(m)$.
  In Step~\ref{step:periodic-sa-construct-arrays}, computing the powers and root values takes
  $\bigO(L+q)$ time. Initializing all the root-group tables takes
  $\bigO(\sum_{p=1}^{L}\AlphabetSize^p)=\bigO(\AlphabetSize^L)$
  time and words. The radix sorts have a constant number of
  coordinates in $[0\dd N+1]$; a constant number of passes with
  $\bigO(\sqrt N)$ buckets takes $\bigO(q+\sqrt N)=\bigO(m)$
  time and words. Filling the remaining arrays takes $\bigO(q)$ time.
  The scans in Step~\ref{step:periodic-sa-construct-range-structures} take $\bigO(m)$ time and words.
  There are at most six subarrays, and constructing all their selection
  structures takes $\bigO(m)$ time and words by
  \cref{th:range-count-for-small-sum}.
  The counting constructions take $\bigO(P_t(m,N))$ time in total
  and use $\bigO(P_s(m,N))$ words, including completed outputs,
  since preprocessing space includes the output (\cref{rm:space}).

  All powers and padded root values are at most
  $\AlphabetSize^L\leq N^{1/21}$, all counts and array indices are
  at most $N$, and full endpoints are at most $N+1<2N\leq2^w$.
  Compute $m$ using division and remainder, and test the subarray
  sum before each addition, so that these operations stay within one word.
  The radix sorts process their coordinates separately.
  In a query, $d=s+k|H|\leq N$ by \cref{def:pos-root};
  evaluate $e-i$ or $i-b$ before computing $r$.
  The validity of $r$ gives $1\leq r\leq q_\alpha$, and the cumulative
  counts used for selection are also at most $q_\alpha$.
  Thus, $w\geq1+\log N$ suffices throughout.
  Including the counting structures, the construction takes
  $\bigO(m+P_t(m,N))$ time and uses
  $\bigO(m+P_s(m,N))$ words of space, as claimed.
\end{proof}

%% file: appendix/periodic-inverse-suffix-array.tex
\subsubsection{Inverse Suffix Array Queries}
\label{sec:appendix-periodic-isa}

\begin{lemma}\label{lm:periodic-isa-range-count}
  Let $\Textlen\in\Z_{\geq2}$, $\Text\in\Sigma^{\Textlen}$,
  and $\tau\in[1\dd\lfloor\Textlen/2\rfloor]$.
  Let $j\in\RTwo{\tau}{\Text}$ (\cref{def:periodic-positions}).
  Let $\alpha\in\{-,+\}$ be the sign of $\TypePos{j}{\tau}{\Text}$,
  and denote $H=\RootPos{j}{\tau}{\Text}$,
  $s=\HeadPos{j}{\tau}{\Text}$, $k=\ExpPos{j}{\tau}{\Text}$,
  and $d=s+k|H|$ (see \cref{def:pos-root}).

  Let $(a_t)_{t\in[1\dd q]}=\RunsLexSorted{\alpha}{\tau}{\Text}$
  (\cref{def:runs-lex-sorted}), and define $A[1\dd q]$ by
  $A[t]=\RunEndFullPos{a_t}{\tau}{\Text}-a_t$.

  Let $u,v,b,e$ be the integers satisfying
  \[
    (u\dd v]=\{t\in[1\dd q]:\RootPos{a_t}{\tau}{\Text}=H\},
    \qquad
    (b\dd e]=\{\ISA{\Text}[x]:
       x\in\RSignedFive{\alpha}{s}{k}{H}{\tau}{\Text}\}.
  \]
  Let $t\in(u\dd v]$ be the index such that the maximal consecutive
  block of $\RTwo{\tau}{\Text}$ containing $j$ starts at $a_t$.
  Then (see \cref{def:range-count})
  \[
    \ISA{\Text}[j]=
    \begin{cases}
      e-\bigl(\TwoSidedRangeCount{A}{v}{d}
                -\TwoSidedRangeCount{A}{t}{d}\bigr),
        &\text{if }\alpha=-,\\
      b+1+\TwoSidedRangeCount{A}{t-1}{d}
                -\TwoSidedRangeCount{A}{u}{d},
        &\text{if }\alpha=+.
    \end{cases}
  \]
\end{lemma}
\begin{proof}
  Let $i=\ISA{\Text}[j]$, so that $i\in(b\dd e]$.
  Define $r=\TwoSidedRangeCount{A}{v}{d}-(e-i)$ if $\alpha=-$,
  and $r=\TwoSidedRangeCount{A}{u}{d}+(i-b)$ if $\alpha=+$.
  Since $\SA{\Text}[i]=j$, \cref{lm:periodic-sa-range-select}
  applies with the same parameters $H,s,k,\alpha$. Thus, the index
  $t'=\RangeSelect{A}{r}{d}$ belongs to $(u\dd v]$ and satisfies
  $j=\SA{\Text}[i]=\RunEndFullPos{a_{t'}}{\tau}{\Text}-d$.
  On the other hand, $j$ has head $s$ and exponent $k$, and
  its full endpoint equals that of its block start $a_t$.
  Thus, $j=\RunEndFullPos{a_t}{\tau}{\Text}-d$.
  The full endpoints of distinct block starts are distinct by
  \cref{lm:periodic-block-properties}\eqref{lm:periodic-block-properties-distinct},
  and hence $t'=t$.

  Since $\RangeSelect{A}{r}{d}=t$, the definition of range
  selection gives $A[t]\geq d$ and
  $r=\TwoSidedRangeCount{A}{t}{d}
    =\TwoSidedRangeCount{A}{t-1}{d}+1$.
  For $\alpha=-$, substituting the definition of $r$ gives
  $e-i=\TwoSidedRangeCount{A}{v}{d}-\TwoSidedRangeCount{A}{t}{d}$.
  For $\alpha=+$, it gives
  $i-b=1+\TwoSidedRangeCount{A}{t-1}{d}-\TwoSidedRangeCount{A}{u}{d}$.
  Rearranging these equalities proves the two formulas.
\end{proof}

\begin{proposition}\label{pr:periodic-isa}
  Let $\AlphabetSize,N\in\Z_{\geq2}$ satisfy
  $\AlphabetSize<N^{1/7}$, and let
  $\tau=\lfloor\tfrac17\log_{\AlphabetSize}N\rfloor\geq1$.
  Consider the word RAM model with word size
  $w\geq1+\log N$.
  Assume that there exists a data structure answering range counting
  queries (\cref{def:range-count}) that, given $N$, integers $m\in[2\dd N]$
  and $m'\in[0\dd m]$, and an array $A[1\dd m']$ of positive integers satisfying
  $\sum_{i=1}^{m'}A[i]\leq N$ and
  $\sum_{i=1}^{m'}A[i]\leq m\log m$,
  achieves the following complexities:
  \begin{itemize}
  \item space usage $S(m,N)$ words,
  \item preprocessing time $P_t(m,N)$,
  \item preprocessing space $P_s(m,N)$ words,
  \item query time $Q(m,N)$.
  \end{itemize}
  Then, there exists $m=\Theta(N/\log_{\AlphabetSize}N)$ with $m\leq N$
  and a data structure that, given $N$, $\AlphabetSize$, and the packed representation
  $\PackedRepresentation{w}{\AlphabetSize}{\Text}$
  (\cref{def:packed-representation}) of any text
  $\Text\in\IntegerAlphabet^N$, answers queries that, given
  $j\in\RTwo{\tau}{\Text}$ (\cref{def:periodic-positions}), return
  $\ISA{\Text}[j]$ (\cref{def:inverse-suffix-array}), and achieves:
  \begin{itemize}
  \item space usage $\bigO(N/\log_{\AlphabetSize}N+S(m,N))$ words,
  \item preprocessing time $\bigO(N/\log_{\AlphabetSize}N+P_t(m,N))$,
  \item preprocessing space $\bigO(N/\log_{\AlphabetSize}N+P_s(m,N))$ words,
  \item query time $\bigO(1+Q(m,N))$.
  \end{itemize}
\end{proposition}
\begin{proof}
  We use the counting identity from \cref{lm:periodic-isa-range-count}
  and construct its arrays using sparse suffix sorting.
  Let $m=\lceil N/\tau\rceil$. If $\tau\leq2$, then
  $\RTwo{\tau}{\Text}=\emptyset$, and it suffices to store $N$.
  Henceforth, assume that $\tau\geq3$, and denote $L=\lfloor\tau/3\rfloor$.

  Let $c_1<\cdots<c_q$ list $\RPrimTwo{\tau}{\Text}$
  (\cref{def:R-prim}), and let $B[1\dd N]$ be the bitvector with
  ones exactly at $c_1,\ldots,c_q$. For each $\alpha\in\{-,+\}$,
  let $(a^\alpha_t)_{t\in[1\dd q_\alpha]}
    =\RunsLexSorted{\alpha}{\tau}{\Text}$
  (\cref{def:runs-lex-sorted}), and define
  \[
    A^\alpha[t]=\RunEndFullPos{a^\alpha_t}{\tau}{\Text}-a^\alpha_t
    \qquad\text{for }t\in[1\dd q_\alpha].
  \]
  Let $I[1\dd q]$ be the following array. For each $h\in[1\dd q]$,
  let $I[h]=(t,u,v)$, where $\alpha$ is the sign of the type of $c_h$,
  $a^\alpha_t=c_h$, and
  $(u\dd v]=\{r\in[1\dd q_\alpha]:
    \RootPos{a^\alpha_r}{\tau}{\Text}=\RootPos{c_h}{\tau}{\Text}\}$.
  Thus, $I[h]$ gives the index of the block starting at $c_h$ in the
  array for its type, together with the boundaries of its root group.

  By \cref{def:pos-root,lm:periodic-block-properties}, each $A^\alpha[t]$ is
  a positive integer at most $N$, and
  $\sum_{t=1}^{q_\alpha}A^\alpha[t]\leq2N$.
  To keep each range-query instance within the sum bound $N$, we use
  the partition of each $A^\alpha$ obtained by repeatedly taking
  the longest prefix of the remaining array whose sum is at most $N$.
  Let $0=o^\alpha_0<\cdots<o^\alpha_{r_\alpha}=q_\alpha$ be their
  boundaries, and let
  $A^\alpha_i[x]=A^\alpha[o^\alpha_{i-1}+x]$ for
  $i\in[1\dd r_\alpha]$ and $x\in[1\dd o^\alpha_i-o^\alpha_{i-1}]$.
  If $q_\alpha=0$, let $r_\alpha=0$ and $o^\alpha_0=0$.
  There are at most three subarrays for each type. Indeed, the sums
  of any two consecutive subarrays together exceed $N$, by the
  maximality of the first. Four subarrays would therefore have
  total sum greater than $2N$.

  \DSComponents
  We store the following components:
  \begin{enumerate}
  \item The integers $N$, $\tau$, $r_-$, and $r_+$, the
    periodic-parameter structure from \cref{pr:periodic-pos},
    and the interval structure from \cref{pr:periodic-sa-intervals}.
    They use $\bigO(N/\log_{\AlphabetSize}N)$ words in total.
  \item The packed representation of $B$, augmented with rank
    support using \cref{th:bin-rank-select}, and the array $I$.
    They use $\bigO(1+N/w)$ and $\bigO(q)$ words, respectively.
  \item For each $\alpha\in\{-,+\}$ and $i\in[1\dd r_\alpha]$,
    the range-counting structure from the claim for $A^\alpha_i$,
    with common parameters $m,N$, its input array, and the boundaries
    $o^\alpha_{i-1},o^\alpha_i$. Together, they use $\bigO(m)$
    words for the arrays and boundaries and $\bigO(S(m,N))$ words
    for the structures.
  \end{enumerate}
  Since $q\leq N/(2\tau)$ by \cref{lm:periodic-block-properties},
  $\tau=\Theta(\log_{\AlphabetSize}N)$, and $w\geq1+\log N$,
  the size is $\bigO(m+S(m,N))$ words.

  \DSQueries
  We first explain how to answer a range-counting query
  $\TwoSidedRangeCount{A^\alpha}{z}{d}$ using the stored subarrays.
  If $z=0$, return zero. Otherwise, add the counts for all complete
  subarrays contained in $A^\alpha[1\dd z]$ and, if $z$ lies inside
  the next subarray $A^\alpha_i$, its prefix count with arguments
  $(z-o^\alpha_{i-1},d)$. The stored boundaries identify these
  subarrays in constant time. This uses at most three range-counting
  queries and takes $\bigO(1+Q(m,N))$ time.

  Given $j\in\RTwo{\tau}{\Text}$, we answer the inverse suffix
  array query in two steps.
  \begin{enumerate}
  \item\label{step:periodic-isa-query-parameters} Use the structure from \cref{pr:periodic-pos} with $(\tau,j)$ to obtain
    the sign $\alpha$ of its type, head $s$, exponent $k$, full endpoint $f$,
    and the length and packed representation of its root $H$.
    Use the structure from \cref{pr:periodic-sa-intervals} with $(\alpha,s,k,H)$
    to obtain $b,e$ for the suffix-array interval of
    $\RSignedFive{\alpha}{s}{k}{H}{\tau}{\Text}$.
    Compute $h=\Rank{B}{j}{\one}$ and read $(t,u,v)=I[h]$.
    Since $j$ is periodic, $c_h$ is the start of the maximal block
    containing $j$.
  \item\label{step:periodic-isa-query-count} Compute $d=f-j$. If $\alpha=-$, return
    $e-(\TwoSidedRangeCount{A^-}{v}{d}
       -\TwoSidedRangeCount{A^-}{t}{d})$.
    If $\alpha=+$, return
    $b+1+(\TwoSidedRangeCount{A^+}{t-1}{d}
       -\TwoSidedRangeCount{A^+}{u}{d})$.
  \end{enumerate}
  By \cref{def:pos-root}, $d=s+k|H|$. The parameters $t,u,v,b,e$
  therefore satisfy the conditions of \cref{lm:periodic-isa-range-count},
  which proves that the returned value is $\ISA{\Text}[j]$.
  Step~\ref{step:periodic-isa-query-parameters} takes constant time, and Step~\ref{step:periodic-isa-query-count} uses two range-counting
  queries on $A^\alpha$, implemented as above. The query time is
  $\bigO(1+Q(m,N))$.

  \DSConstruction
  We construct the components in four steps.
  \begin{enumerate}
  \item\label{step:periodic-isa-construct-support} First, we prepare the periodic navigation structures and
    list the starts of the maximal periodic blocks.
    Construct the structure from \cref{pr:periodic-pos} using
    the packed text, and the structure from
    \cref{pr:periodic-sa-intervals} using $\tau$ and the packed text.
    Apply \cref{th:periodic-position-intervals} to the same inputs
    to obtain $c_1,\ldots,c_q$ in increasing order.
    Initialize the packed representation of $B$ to $\zero^N$,
    and set $B[c_h]=\one$ for every $h\in[1\dd q]$, using
    \cref{pr:packed-representation}. Augment $B$ with rank support
    using \cref{th:bin-rank-select}.
  \item\label{step:periodic-isa-construct-sort-endpoints} Next, we compute the order of the suffixes starting at the
    full endpoints of these blocks. Query the periodic-parameter
    structure at each $c_h$ to obtain its type, root, and full
    endpoint $F[h]=\RunEndFullPos{c_h}{\tau}{\Text}$.
    Apply \cref{th:sparse-suffix-sorting} to $q$,
    the packed text, and the array $F[1\dd q]$.
    Denote the resulting array by $E[1\dd q]$.
    Form the pairs $(E[r],r)$ for $r\in[1\dd q]$ and sort them
    by their first coordinate using radix sort. Since
    $F[1]<\cdots<F[q]$ by \cref{lm:periodic-block-properties},
    the second coordinate of the $h$th resulting pair is the rank
    $\rho[h]$ of the suffix starting at $F[h]$ among these suffixes.
  \item\label{step:periodic-isa-construct-arrays} We now construct the two arrays $A^-,A^+$ and the map $I$.
    \begin{enumerate}
    \item Compute the powers
      $\AlphabetSize^0,\ldots,\AlphabetSize^L$, and construct the
      integer-value encoder from \cref{th:val-encoding}.
      For each $h\in[1\dd q]$, let $H_h$ be the root of $c_h$.
      Compute $p_h=|H_h|$ and
      $x_h=\Val{\AlphabetSize}{H_h}\cdot\AlphabetSize^{L-p_h}$.
      This is the integer value of $H_h$ padded on the right with
      zeros to length $L$. Ordering the pairs $(x_h,p_h)$ is
      equivalent to ordering the roots lexicographically: padding
      preserves the first differing symbol, and the length breaks
      a tie when a proper prefix is extended only by zeros.
    \item For each $\alpha\in\{-,+\}$, collect in an array the indices
      $h\in[1\dd q]$ with $c_h\in\RPrimSignedTwo{\alpha}{\tau}{\Text}$.
      Sort this array by $(x_h,p_h,\rho[h])$ using radix sort.
      If $h$ is the $t$th index in the resulting array, set
      $a^\alpha_t=c_h$ and $A^\alpha[t]=F[h]-c_h$.
      Scan the sorted array to find each maximal interval
      $[t_1\dd t_2]$ with equal pairs $(x_h,p_h)$.
      For every index $h$ at position $t$ in this interval, set
      $I[h]=(t,t_1-1,t_2)$.
    \end{enumerate}
  \item\label{step:periodic-isa-construct-range-structures} Finally, we construct the range-counting structures.
    For each $\alpha\in\{-,+\}$, initialize its boundary array with zero.
    If $q_\alpha=0$, store $r_\alpha=0$ and make no call. Otherwise,
    scan $A^\alpha$ from left to right,
    maintaining the sum $s_{\rm cur}$ of the current subarray, initially zero.
    Before appending the next integer $A^\alpha[t]$, test whether
    $A^\alpha[t]>N-s_{\rm cur}$. If so, end the current subarray immediately
    before $t$, store its boundary, and reset $s_{\rm cur}$ to zero.
    Append $A^\alpha[t]$ and add it to $s_{\rm cur}$. Store the final boundary
    after the scan and store the number $r_\alpha$ of subarrays.
    Store each subarray in its own array and construct its
    range-counting structure from the claim.
  \end{enumerate}

  We now verify the hypotheses of the tools and analyze the construction.
  The choice of $\tau$ gives
  $\tau=\Theta(\log_{\AlphabetSize}N)$ and
  $m=\Theta(N/\log_{\AlphabetSize}N)\leq N$.
  Moreover, $\log m=\Theta(\log N)$ and $N\leq m\tau\leq m\log m$, where the last inequality holds because
  $m\geq N/\tau\geq7N/\log N$ gives $\log m\geq\tfrac12\log N\geq\tau$.
  The same bound gives $m\geq2$. Each subarray $A^\alpha_i$ has at most
  $q\leq N/(2\tau)\leq m$ positive integers and sum at most $N$,
  and hence satisfies both sum bounds in the claim.

  The periodic-parameter and interval structures, the computation of
  the maximal periodic blocks, sparse suffix sorting, and the encoder
  use size parameter $N$ and alphabet size $\AlphabetSize$, and the
  rank support for $B$ uses size parameter $N$.
  The periodic structures accept this $\tau$, since
  $\tau\leq\tfrac17\log_{\AlphabetSize}N
    <\tfrac16\log_{\AlphabetSize}N$.
  By \cref{lm:periodic-block-properties}, the full endpoints are
  strictly increasing and belong to $[1\dd N+1]$.
  Thus, $F[1\dd q]$ lists a valid input set for
  \cref{th:sparse-suffix-sorting}.
  Every root has length at most $L\leq\tau/3$, within the
  input-length limit of \cref{th:val-encoding}.

  Step~\ref{step:periodic-isa-construct-support} takes $\bigO(N/\tau)$ time and words, including the
  $\bigO(q+N/\log N)$ work to initialize and augment $B$.
  Step~\ref{step:periodic-isa-construct-sort-endpoints} takes $\bigO(N/\log_{\AlphabetSize}N+q)=\bigO(m)$
  time and words. In Step~\ref{step:periodic-isa-construct-arrays}, the powers and encoder take
  $\bigO(\log N+\sqrt N)=\bigO(m)$ time and words to construct.
  All radix sorts process $\bigO(q)$ integers with a constant
  number of coordinates in $[0\dd N+1]$. Using a constant number
  of passes with $\bigO(\sqrt N)$ buckets gives $\bigO(m)$ time
  and words. The scans and array construction also take $\bigO(m)$.
  Step~\ref{step:periodic-isa-construct-range-structures} takes $\bigO(m)$ time for the partition and invokes the
  range-counting construction at most six times, always with the
  same parameters $m,N$. These calls take $\bigO(P_t(m,N))$ time
  in total and use $\bigO(P_s(m,N))$ words of space, including
  the structures already built, since preprocessing space includes
  the output (\cref{rm:space}).

  All text positions, array indices, and nonnegative counts used by
  the construction lie in $[0\dd N]$. Full endpoints may also equal
  $N+1$, which fits in one word because $N+1<2N\leq2^w$.
  The powers and padded root values are at most
  $\AlphabetSize^L\leq N^{1/21}$. The radix sorts process the coordinates separately.
  Compute $m$ using division and remainder, and perform the
  partition test before adding the next integer, so that these
  computations also stay within one word. Evaluate the differences
  between range counts before the final addition or subtraction
  in a query. All resulting values are at most $N$.
  Thus, $w\geq1+\log N$ suffices throughout. Including the
  range-counting structures, the construction takes
  $\bigO(m+P_t(m,N))$ time and uses
  $\bigO(m+P_s(m,N))$ words of space, as claimed.
\end{proof}

%% file: appendix/stronger-lower-bound.tex
Below, we provide an alternative proof of \cref{th:suffix-array-asymptotic-space-lower-bound,th:inverse-suffix-array-asymptotic-space-lower-bound} using the insights of \cite{SchurmannS08}.
The main advantage of that approach is that it yields high-quality explicit lower bounds.

\begin{proposition}\label{pr:number-of-distinct-suffix-arrays}
  Let $\AlphabetSize,\Textlen \in \Z_{\geq 2}$ be such that $\AlphabetSize \leq \Textlen$.
  The number of distinct suffix arrays realized by strings in
  $[0 \dd \AlphabetSize)^{\Textlen}$ is at least
  \[
    \AlphabetSize!\cdot \AlphabetSize^{\Textlen-\AlphabetSize}
      \ge \max\!\left(
        \tfrac{\AlphabetSize^{\Textlen}}{e^{\AlphabetSize-1}},\
        \AlphabetSize^{\Textlen/2}
      \right).
  \]
\end{proposition}
\begin{proof}
  Let $\mathcal N(\Textlen,\AlphabetSize)$ denote the number of distinct suffix arrays
  realized by strings in $[0 \dd \AlphabetSize)^{\Textlen}$. As shown in
  \cite{SchurmannS08,KucherovTV13}, $\mathcal N(\Textlen,\AlphabetSize)$ equals the
  number of $\Textlen$-element permutations with strictly fewer than
  $\AlphabetSize$ \emph{descents}, defined for a permutation
  $\pi \colon [1 \dd \Textlen] \to [1 \dd \Textlen]$ as indices
  $i \in [1 \dd \Textlen-1]$ such that $\pi(i) > \pi(i+1)$.
  We prove the claim in four steps:
  \begin{enumerate}

  \item\label{step:number-of-distinct-sa-valid}
    First, we construct a family of permutations with strictly fewer than
    $\AlphabetSize$ descents. Let $\rho$ be an arbitrary permutation of
    $[1 \dd \AlphabetSize]$, and let
    $f \colon [\AlphabetSize+1 \dd \Textlen] \to [1 \dd \AlphabetSize]$ be
    an arbitrary function. For every $j \in [1 \dd \AlphabetSize]$, let
    $B_j$ denote the block obtained by first listing the element $\rho_j$ and
    then all $x \in [\AlphabetSize+1 \dd \Textlen]$ satisfying
    $f(x)=\rho_j$, in increasing order. Let $\pi_{\rho,f}$ denote the sequence
    \[
      \pi_{\rho,f}
        = \langle \pi_{\rho,f}(1), \ldots, \pi_{\rho,f}(\Textlen) \rangle
        = B_1 \cdot B_2 \cdots B_{\AlphabetSize}.
    \]
    Every element of $[1 \dd \Textlen]$ appears exactly once in
    $\pi_{\rho,f}$. Thus, $\pi_{\rho,f}$ is a permutation of
    $[1 \dd \Textlen]$. Moreover, every block $B_j$ is increasing: it
    starts with $\rho_j \in [1 \dd \AlphabetSize]$, while all remaining
    elements of the block lie in $[\AlphabetSize+1 \dd \Textlen]$ and are
    listed in increasing order. Therefore, descents can occur only at the
    $\AlphabetSize-1$ boundaries between consecutive blocks, and hence
    $\pi_{\rho,f}$ has strictly fewer than $\AlphabetSize$ descents.

  \item\label{step:number-of-distinct-sa-count}
    Next, we prove
    $\mathcal N(\Textlen,\AlphabetSize) \geq
      \AlphabetSize!\cdot\AlphabetSize^{\Textlen-\AlphabetSize}$.
    The map $(\rho,f)\mapsto\pi_{\rho,f}$ is injective. Indeed, in
    $\pi_{\rho,f}$, the elements of $[1 \dd \AlphabetSize]$ are exactly the
    first elements of the blocks. Reading them from left to right therefore recovers
    $\rho$. Once $\rho$ is known, every element
    $x \in [\AlphabetSize+1 \dd \Textlen]$ lies in exactly one block, and
    the first element of that block is precisely $f(x)$. Thus,
    $\pi_{\rho,f}$ determines both $\rho$ and $f$. There are
    $\AlphabetSize!$ choices for $\rho$ and
    $\AlphabetSize^{\Textlen-\AlphabetSize}$ choices for $f$. By
    Step~\ref{step:number-of-distinct-sa-valid}, all resulting permutations
    have strictly fewer than $\AlphabetSize$ descents, proving
    \[
      \mathcal N(\Textlen,\AlphabetSize)
        \geq \AlphabetSize!\cdot
          \AlphabetSize^{\Textlen-\AlphabetSize}.
    \]

  \item\label{step:number-of-distinct-sa-first-bound}
    We now prove
    $\AlphabetSize!\cdot\AlphabetSize^{\Textlen-\AlphabetSize}
      \geq \AlphabetSize^{\Textlen}/e^{\AlphabetSize-1}$.
    Using the standard lower bound
    $\AlphabetSize! > e\cdot(\AlphabetSize/e)^{\AlphabetSize}
      = \AlphabetSize^{\AlphabetSize}/e^{\AlphabetSize-1}$, we obtain the
    claimed inequality after multiplying both sides by
    $\AlphabetSize^{\Textlen-\AlphabetSize}$.

  \item Finally, we prove
    $\AlphabetSize!\cdot\AlphabetSize^{\Textlen-\AlphabetSize}
      \geq \AlphabetSize^{\Textlen/2}$.
    For every $i \in [1 \dd \AlphabetSize]$, we have
    $i(\AlphabetSize+1-i)
      = \AlphabetSize+(i-1)(\AlphabetSize-i) \geq \AlphabetSize$.
    Taking the product over $i \in [1 \dd \AlphabetSize]$ gives
    \[
      (\AlphabetSize!)^2
        = \prod_{i=1}^{\AlphabetSize} i(\AlphabetSize+1-i)
        \geq \AlphabetSize^{\AlphabetSize}.
    \]
    Since $\Textlen \geq \AlphabetSize$, it follows that
    $\AlphabetSize!\cdot\AlphabetSize^{\Textlen-\AlphabetSize}
      \geq \AlphabetSize^{\Textlen-\AlphabetSize/2}
      \geq \AlphabetSize^{\Textlen/2}$.\qedhere
  \end{enumerate}
\end{proof}

\begin{theorem}\label{th:suffix-array-explicit-space-lower-bound}
  Let $\AlphabetSize,\Textlen \in \Z_{\geq 2}$ satisfy
  $\AlphabetSize \leq \Textlen$.
  Any data structure supporting suffix array queries
  (\cref{def:suffix-array}) for every text
  $\Text \in \IntegerAlphabet^{\Textlen}$ has worst-case space usage at least
  \[
    \max\!\left(
      \Textlen \log \AlphabetSize - (\AlphabetSize-1)\log e - 1,
      \tfrac{1}{2}\Textlen \log \AlphabetSize - 1
    \right)
  \]
  bits, and at least $\tfrac14\Textlen\log\AlphabetSize$ bits. The same lower
  bounds hold for any data structure supporting inverse
  suffix array queries (\cref{def:inverse-suffix-array}) for every text
  $\Text \in \IntegerAlphabet^{\Textlen}$.
\end{theorem}
\begin{proof}
  Let $\mathcal N(\Textlen,\AlphabetSize)$ denote the number of distinct suffix arrays
  realized by strings in $[0 \dd \AlphabetSize)^{\Textlen}$. A data structure
  supporting suffix array queries for every such text must have at least
  $\mathcal N(\Textlen,\AlphabetSize)$ distinct stored bit strings, because querying all
  suffix array entries determines the suffix array. The same statement holds
  for a data structure supporting inverse suffix array queries, because querying
  all inverse suffix array entries determines the inverse suffix array, whose
  inverse permutation is the suffix array.

  Let $S$ denote the worst-case space usage of the suffix-array or
  inverse-suffix-array data structure under consideration. Distinct realizable
  suffix arrays must correspond to distinct stored bit strings. A data structure
  using at most $S$ bits corresponds to a bit string of length at most $S$, and
  the number of such bit strings is $\sum_{i=0}^{S} 2^i < 2^{S+1}$. Therefore,
  $2^{S+1} > \mathcal N(\Textlen,\AlphabetSize)$.

  By \cref{pr:number-of-distinct-suffix-arrays},
  \[
    2^{S+1} > \mathcal N(\Textlen,\AlphabetSize) \geq \max\!\left(\tfrac{\AlphabetSize^{\Textlen}}{e^{\AlphabetSize-1}},\AlphabetSize^{\Textlen/2}\right).
  \]
  Taking logarithms yields
  \[
    S+1 > \max\!\left(\Textlen\log\AlphabetSize-(\AlphabetSize-1)\log e,\tfrac{\Textlen}{2}\log \AlphabetSize\right).
  \]
  Thus,
  \[
    S \geq \max\!\left(\Textlen\log\AlphabetSize-(\AlphabetSize-1)\log e-1,\tfrac{\Textlen}{2}\log \AlphabetSize-1\right).
  \]
  It remains to show $S\geq\tfrac14\Textlen\log\AlphabetSize$. If
  $\Textlen\log\AlphabetSize\geq4$, this follows from
  $\tfrac12\Textlen\log\AlphabetSize-1\geq\tfrac14\Textlen\log\AlphabetSize$.
  Otherwise, $\Textlen\log\AlphabetSize<4$, and the inequality
  $S+1>\tfrac12\Textlen\log\AlphabetSize\geq1$ shows that the integer $S$ is
  positive. Hence, $S\geq1>\tfrac14\Textlen\log\AlphabetSize$.
\end{proof}